\documentclass[11pt,a4paper]{report}
\usepackage{mythesis}

\usepackage[utf8]{inputenc}

\usepackage{amsmath, amsthm, amsfonts, amssymb, latexsym}
\usepackage{graphicx}
\usepackage{color}
\usepackage{appendix}

\usepackage{caption}
\usepackage[perpage,symbol*]{footmisc}

\usepackage{paralist}

\usepackage{epigraph}

\usepackage{url}

\usepackage[pagebackref,unicode]{hyperref}
\hypersetup{
    unicode=true,          
    pdftitle={New frontiers in Numerical Relativity},
    pdfauthor={Miguel Zilh\~ao},
    pdfsubject={PhD Thesis},
    pdfcreator={LaTeX},          
    pdfproducer={LaTeX},         
    pdfkeywords={General Relativity, Numerical Relativity, Black Holes},
    colorlinks=false,            
    bookmarks=true,              
    linktoc=page,                
  }
\renewcommand*{\backref}[1]{}
\renewcommand*{\backrefalt}[4]{%
  \ifcase #1 { } \or (Cited on page~#2.) \else %
  (Cited on pages~#2.) \fi %
}

\let\origdoublepage\cleardoublepage
\newcommand{\clearemptydoublepage}{%
  \clearpage
  {\pagestyle{empty}\origdoublepage}%
}

\let\cleardoublepage\clearemptydoublepage

\newcommand{\del}[3]{\frac{ \partial^{#1} #2}{\partial #3}}

\newcommand{\diag}{\mathrm{diag}}
\newcommand{\dd}{\mathrm{d}}
\newcommand{\F}[2]{\mathcal{F}^{#1}{}_{#2} }
\newcommand{\GG}[2]{\Gamma^{#1}{}_{#2} }
\newcommand{\half}{\frac{1}{2}}

\title{New frontiers in Numerical Relativity}
\author{Miguel Rodrigues Zilhão Nogueira}
\date{Setembro de 2012}

\submitionplace{Dissertação apresentada ao Programa Doutoral MAP-fis para
  cumprimento dos requisitos necessários à obtenção do grau de Doutor em Física,
  realizada sob a orientação científica de Carlos Herdeiro, Professor do
  Departamento de Física da Universidade de Aveiro, e co-orientação de Vitor
  Cardoso, Professor do Instituto Superior Técnico.}

\begin{document}

\pagestyle{plain}

\coverp
\titlep


\chapter*{Agradecimentos}

\epigraph{For long you live and high you fly\\
  And smiles you'll give and tears you'll cry\\
  And all you touch and all you see\\
  Is all your life will ever be}{Pink Floyd\\ \emph{Breathe}}

Estes últimos quatro anos foram extraordinariamente enriquecedores para mim e tenho, felizmente, várias razões para estar grato por esta jornada que agora finda.
Começo por deixar aqui um agradecimento aos meus amigos e orientadores Carlos e Vitor, com quem muito aprendi e com quem é um prazer trabalhar.
O ritmo de trabalho e entusiasmo de ambos por Física, e Ciência em geral, são para mim fontes de inspiração.
O Carlos merece ainda especial destaque uma vez que, entre Mecânica Quântica e Relatividade Geral, com ele aprendi as bases da Física moderna.
Só tenho pena de esta pequena nota não reflectir todo o meu reconhecimento a ambos.

O trabalho que tenho vindo a desenvolver, e que esta tese resume, resulta na sua maior parte de trabalho feito em colaboração também com o Uli, a Helvi, o Leonardo e o Andrea, a quem queria deixar também o meu reconhecimento.
Não poderia ter escolhido um melhor grupo de trabalho.

Durante os últimos quatro anos tive a felicidade de poder passar períodos de tempo a visitar outros grupos e instituições, o que muito enriqueceu este trabalho.
Gostaria em particular de agradecer ao Vitor e ao pessoal do CENTRA por ter sido sempre bem acolhido nas minhas várias idas ao Instituto Superior Técnico; 
ao Emanuele, ao Marco, e ao Departamento de Física e Astronomia da Universidade do Mississippi pela hospitalidade durante a minha visita; 
ao Uli, Carlos e grupo do CSIC-IEEC pelo período que com eles estive em Barcelona;
ao Luis e ao Perimeter Institute for Theoretical Physics pela óptima experiência que foram os quatro meses que lá passei;
e por último, queria também agradecer ao Crispino e ao grupo em Belém do Pará pelas duas magníficas semanas em que nos receberam na Amazónia, mesmo na recta final deste trabalho.

Sendo o conteúdo desta tese um resumo de trabalho feito nos últimos quatro anos, a realidade é que esta acaba por ser o culminar de todo o meu processo educativo formal---um percurso escolar iniciado há vinte anos na Escola Primária de Lavadores---que tem aqui um ponto final, e para o qual várias pessoas foram importantes.
Assim, deixo aqui também uma palavra de reconhecimento a quem neste percurso me acompanhou, com destaque para os meus colegas e professores na Escola Secundária Inês de Castro.
Gostaria em particular de mencionar os meus colegas de turma, o Patrício, o Norberto, o Ricardo, a Annie, a Mariana, e especialmente a Joana.
Da interacção com todos eles resulta a pessoa que hoje sou, e estou-lhes por isso grato.

A minha formação em Física começa depois no Departamento de Física da Faculdade de Ciências da Universidade do Porto, onde passei a maior parte do meu percurso escolar, e considero ter tido muito boa formação.
Fica então aqui o meu obrigado aos meus professores e, em especial, aos meus colegas de curso, Mariana, Joana, Raul, Lino, Coelho, João Nuno, e com destaque particular para o Carlos---tanto por tudo o que fomos aprendendo a estudar para os exames durante a licenciatura, como pelas conversas que fomos conseguindo manter nestes últimos anos.

A maior parte deste trabalho foi efectuado no Centro de Física do Porto (CFP), onde tive sempre boas condições e ambiente para trabalhar.
Agradeço a todos os elementos do Centro, e destaco em particular o João Nuno, meu colega de gabinete, e o Luís Filipe, tanto pela colaboração num outro projecto, como por nunca haver um momento aborrecido na sua companhia.

O meu obrigado também ao Departamento de Física da Universidade de Aveiro onde também nada me faltou.
À Carmen, ao Marco e ao Flávio agradeço o bom ambiente que sempre lá temos.

A temporada que passei em Waterloo tornou-se muito mais agradável na companhia dos amigos que lá fiz, com as nossas longas discussões ao jantar.
Fica aqui então o meu obrigado ao Federico, ao Francesco, ao João, ao Jon, ao Mehdi, ao Siavash, ao Farbod, ao Anton, e especialmente à Yasaman.

Aos portugueses que pagam os seus impostos e à Fundação para a Ciência e a Tecnologia (FCT) agradeço o apoio financeiro para todo este tra\-ba\-lho, a\-tra\-vés da bol\-sa SFRH/\allowbreak BD/\allowbreak 43558/\allowbreak 2008 com co-financiamento do POPH/FSE.

Nem só de ciência vive um aluno de doutoramento.
Assim, e apesar de não ter frequentado aulas durante estes quatro anos, não posso deixar de agradecer ao Mota, ao Cesário, ao Paulo, e ao resto do pessoal da Escola de Música de Canidelo.
Lá tive a minha formação em música---que foi extremamente importante para mim---formação essa que me acompanha e sempre acompanhará.
Fica também uma palavra para os meus companheiros da Associação Porto Céltico, com quem nunca nos aborrecemos.

Para último deixo a minha gratidão à minha família, que é a minha constante do movimento:
à Nair---que faz parte de mim; aos meus pais, Adriano e Alexandra---que sempre me apoiaram; aos meus irmãos, Nuno e Luís---que nunca mudem; aos meus avós, Manuel, Aurora, Domingos e em especial à Alice---que foi a minha primeira professora; e à Rosa.
Nada que eu possa aqui escrever alguma vez faria justiça à importância que estas pessoas tiveram neste meu percurso, pelo que posso apenas terminar dedicando-lhes este trabalho.



\cleardoublepage

\chapter*{Abstract}

The first attempts at solving a binary black hole spacetime date back to the 1960s, with the pioneering works of Hahn and Lindquist.
In spite of all the computational advances and enormous efforts by several groups, 
the first stable, long-term evolution of the orbit and merger of two black holes was only accomplished over 40 years later, in 2005.
Since then, the field of Numerical Relativity has matured, and been extensively used to explore and uncover a plethora of physical phenomena in various scenarios.

In this thesis, we take this field to new frontiers by exploring its extensions to higher dimensions, non-asymptotically flat spacetimes and Einstein-Maxwell theory.
We start by reviewing the usual formalism and tools, including the ``3+1'' decomposition, initial data construction, the BSSN evolution scheme and standard wave extraction procedures.
We then present a dimensional reduction procedure that allows one to use existing numerical codes (with minor adaptations) to evolve higher-dimensional systems with enough symmetry, and show corresponding results obtained for five-dimensional head-on collisions of black holes.
Finally, we show evolutions of black holes in non-asymptotically flat spacetimes, 
and in Einstein-Maxwell theory.

\cleardoublepage

\chapter*{Resumo}

As primeiras tentativas de evolução da geometria
de um sistema binário de buracos negros datam da década de 60, com o trabalho pioneiro de Hahn e Lindquist.
Apesar de todos os avanços computacionais e enormes esforços por parte de vários grupos, as primeiras evoluções estáveis da órbita e coalescência de dois buracos negros foram conseguidas apenas 40 anos depois, em 2005.
Desde então, o campo da Relatividade Numérica amadureceu, e tem sido usado extensivamente para explorar e descobrir fenómenos físicos em vários cenários.

Nesta tese, levamos este campo a novas fronteiras e exploramos extensões a dimensões extra, espaços não-assimptoticamente planos e teoria de Einstein-Maxwell.
Começamos por rever o formalismo e ferramentas usuais, incluindo a decomposição ``3+1'', construção de dados iniciais, o esquema de evolução BSSN e os procedimentos padrão de extracção de radiação.
Seguidamente apresentamos um procedimento de redução dimensional que permite o uso de códigos numéricos existentes (com adaptações menores) para evoluir sistemas em dimensões mais elevadas com simetria suficiente, apresentando ainda os correspondentes resultados obtidos para colisões frontais de buracos negros em cinco dimensões.
Finalmente, mostramos resultados em espaços não-assimptoticamente planos e teoria de Einstein-Maxwell.



\cleardoublepage
\tableofcontents
\cleardoublepage
\listoftables
\cleardoublepage
\listoffigures
\cleardoublepage

\pagestyle{headings}
\numberwithin{equation}{section}

\chapter{Introduction}
\label{ch:intro}

\renewcommand{\textflush}{flushepinormal}

\epigraph{Why you care about small things? World very simple place. World only
  have two things: things you can eat and things you no can eat.}{Quina\\
  \textsc{Final Fantasy IX}}

\section{Motivation and historical background}

Formulated by Einstein in 1915~\cite{Einstein:1915by,Einstein:1915bz,Einstein:1915ca}, general relativity is one of the most beautiful theories ever discovered.
Its very elegance, however, can also be a disadvantage.
We are able to do purely analytical calculations---sometimes using just pen and paper---in highly symmetrical ideal examples where exact solutions are known, or limits where gravity is weak. 
Alas, Nature is not that simple.

To attack more complicated problems---such as those with strong and dynamical gravitational fields---one will eventually need to perform numerical computations. 
The quintessential example is the two-body problem.
With well known solutions in terms of conics in Newtonian gravity, the equivalent problem in general relativity---the evolution of a black hole binary---posses no (known) closed-form solution.
Perturbative analytical techniques do exist and some are very well suited to study certain stages of this problem. 
In particular, the inspiral phase (before the merger) is well modelled by post-Newtonian methods; the ringdown phase (after merger) can be described by perturbative methods using the \emph{quasi-normal modes} of the final black hole.
Full numerical simulations are required, however, to evolve the system during the merger.

Much of the motivation to understand the nature of such systems and the corresponding energy emitted via gravitational radiation originally came from the gravitational wave astronomy field.
A first generation of highly sensitive gravitational waves detectors---LIGO~\cite{Abramovici:1992ah}, Virgo~\cite{Caron:1997hu}, GEO~\cite{Luck:1997hv} and TAMA~\cite{Ando:2001ej}---have been operational and a second generation of even more sensitive detectors is under construction.
These detectors may allow us to study signals produced from strong-field systems, which carry the specific signature of the system that produced them. 
The analysis of these signals may then provide us with a new window to the universe. 
For that, however, we rely on source modelling: templates of theoretical waveforms from likely sources are needed if one wishes to reconstruct the signal.

Numerical relativity can be regarded as a tool 
to study spacetimes that cannot be studied by analytical means. 
It dates back to the mid 1960s with Hahn and Lindquist's attempts of numerically evolving Einstein's field equations for a binary black hole spacetime~\cite{1964AnPhy..29..304H}. 
Their computer power was very limited, however, and not much physics could be extracted from the simulation. 
More reliable simulations were only performed in the late 1970s by Smarr~\cite{Smarr:1976qy} and Eppley~\cite{Eppley:1977dk}, which again attempted the head-on collision of two black holes.
Though almost a decade after Hahn and Lindquist, the available computer power was still only enough to evolve low resolution simulations.

With the development of faster computers and the extra motivation of returning
to the two-body problem coming from LIGO, the 1990s finally witnessed the
simulation of a head-on collision of two black
holes~\cite{Bernstein:1994wt,Anninos:1994gp} as well as the study of more
complex systems. 
To name just a few: simulations of rapidly rotating neutron
stars~\cite{Stergioulas:1997ja},
the formation of a toroidal event horizon in the collapse of a system containing a toroidal distribution of particles~\cite{Shapiro:1992gr,Hughes:1994ea}
and one of the most influential results, gravitational collapse and its relation with critical phenomena~\cite{Choptuik:1992jv}. 
For a more comprehensive overview 
see, for example,~\cite{Lehner:2001wq}.

In spite of all these successes, the real breakthrough came only in 2005 with
the first simulations of stable, long-term evolutions of the inspiral and merger of
two black holes~\cite{Pretorius:2005gq,Baker:2005vv,Campanelli:2005dd}. For an
overview of the two-body problem in general relativity
see~\cite{Pretorius:2007nq}.

Since then, numerical codes have considerably improved and much progress has
been made. In particular, we have witnessed numerical evolutions of (see
e.g.~\cite{Hinder:2010vn} for a thorough overview of some recent results):
\begin{itemize}

\item binary black hole mergers lasting for 15 orbits before
  merger~\cite{Boyle:2007ft,Scheel:2008rj}. The corresponding waveforms, which
  include the infall, non-linear merger and ringdown phase, have been used in
  comparisons with post-Newtonian results.

\item black hole binaries with mass ratios up to
  $1:100$~\cite{Lousto:2010ut,Sperhake:2011ik}. Waveforms for high mass ratios
  are essential, since they are expected to be the most common,
  astrophysically. Computationally, however, they are much more demanding than
  comparable mass cases.

\item rotating black holes with near extremal spin~\cite{Lovelace:2011nu,Lousto:2012es}.

\item zoom-whirl orbits---characterised by black hole trajectories that
  alternate between a whirling quasi-circular motion and a highly eccentric
  quasi-elliptical zooming out
  ~\cite{Pretorius:2007jn,Healy:2009zm}.

\item so called ``superkick'' configurations---equal mass black holes with
  (initially) opposite spin vectors lying in the orbital plane---where the
  post-merger recoil velocity can reach up to $\sim
  4000$~km/s~\cite{Campanelli:2007cga,Brugmann:2007zj}. Recently~\cite{Lousto:2011kp},
  this effect was combined with the ``hangup''
  configuration~\cite{Campanelli:2006uy}---where the black holes have spin
  aligned with the orbital angular momentum---to predict maximum recoils up to
  $\sim 5000$~km/s.  Given that such velocities are enough to eject a black hole
  from the centre of a galaxy~\cite{Merritt:2004xa}, these results could have
  important consequences for astronomy (such as in structure formation), see
  e.g.~\cite{BoylanKolchin:2004tf,Haiman:2004ve,Madau:2004st,Komossa:2012cy}.

\item high velocity collisions of black holes---head-on collisions up to
  $0.94c$~\cite{Sperhake:2008ga}, where the radiated energy was found to be
  around $14\%$ when extrapolating the relative velocity between the black holes
  to $c$; as well as non-head-on
  collisions~\cite{Shibata:2008rq,Sperhake:2009jz}, where the impact parameter
  for black hole merger was determined in the limit where the relative velocity
  approaches $c$.

\end{itemize}

As we can observe from these previous examples, numerical relativity has now reached a state of maturity and, at least in the four-dimensional asymptotically flat vacuum case, is largely under control allowing us to evolve a large class of different configurations.

Motivation to study gravity in the dynamical/strong field regime is not restricted to the evolution of the two-body problem or variations thereof, and computation of the respective waveform. 
Indeed, incentive to study such systems also comes from fields other than gravity itself. 
In the following we mention some of these topics and briefly describe how numerical relativity can be expected to shed light on some outstanding issues.

\begin{description}

\item[Tests of cosmic censorship hypothesis] \hfill \\
  It has been known for some time from the Penrose-Hawking singularity theorems
  that, quite generically, solutions of Einstein's field equations with
  physically reasonable matter content can develop
  singularities~\cite{HawkingEllis}. If such singularities are visible to the
  rest of spacetime (i.e., no horizon is covering them), predictability may
  break down. Originally formulated by Penrose in 1969~\cite{Penrose:1969pc},
  what is known as the \emph{weak cosmic censorship conjecture} roughly states
  that, generically, singularities of gravitational collapse are covered by an
  event horizon and therefore have no causal contact with distant observers.

  In the absence of a generic proof\footnote{Indeed, the conjecture has not even
    been stated in a rigorous way---as often happens, that is part of the
    task.}, one can try and put the conjecture to the test in specific
  configurations. The ability to perform full blown non-linear numerical
  simulations in arbitrary spacetimes could here prove invaluable.

  With such simulations, the conjecture has been shown to hold under extremely
  violent events---the ultra-relativistic collision of black
  holes~\cite{Sperhake:2008ga}. In higher-dimensional gravity, it was
  shown by Lehner and Pretorius that 
  cosmic censorship does not seem to hold generically, even in
  vacuum~\cite{Lehner:2010pn}. Specifically, it was shown that five dimensional
  black strings (solutions of five dimensional vacuum gravity, known to be
  unstable~\cite{Gregory:1993vy}) display, when perturbed, a self-similar
  behaviour that ultimately gives rise to naked singularities in rather generic
  conditions. Also in higher dimensions, results by Okawa
  et~al.~\cite{Okawa:2011fv} seem to indicate that in a high-velocity scattering
  of five-dimensional black holes, curvature radius shorter than the Planck
  length can be observed (i.e., no horizon is covering this region). This could
  be regarded as an effective singularity in classical gravity.

\item[Stability of (higher-dimensional) black hole solutions] \hfill \\
  Higher dimensional gravity has a much richer diversity of black object
  solutions than its four dimensional counterpart. Spherical topology is not the
  only allowed topology for objects with a horizon and one can also have, e.g.,
  black rings (with a donut-like
  topology)~\cite{Emparan:2001wn,Pomeransky:2006bd} and even regular solutions
  with disconnected horizons, such as the ``black Saturn''~\cite{Elvang:2007rd},
  the ``black di-ring''~\cite{Iguchi:2007is} or the ``bicycling black
  rings''~\cite{Elvang:2007hs}.
  
  The study of these objects is relevant for a number of reasons. Other than the
  obvious intrinsic value that such studies carry and the possibly interesting
  mathematical properties that some of these objects may have, the understanding
  of these solutions can also be helpful for:
  \begin{inparaenum}[(i)]
  \item quantum gravity---the calculations of black hole entropy within string
    theory were first performed in five dimensional spacetimes, and only
    afterwards extended to four dimensions;
  \item gauge/gravity correspondence, which maps properties of $D$-dimensional
    black holes to strongly coupled field theories in $D-1$ dimensions; 
  \item large extra dimensions scenarios, suggesting that (microscopic)
    higher-dimensional black objects could be formed in particle collisions with
    centre of mass energy $\gtrsim \mathrm{TeV}$ (such as at the LHC).
  \end{inparaenum}
  See the review article~\cite{Emparan:2008eg} for more details and further
  motivation.

  The stability of these higher-dimensional black objects is now starting to be
  explored. It had been conjectured that for $D\ge 6$ ultra-spinning Myers-Perry
  black holes will be unstable~\cite{Emparan:2003sy}, and this instability has been
  confirmed by an analysis of linearised axi-symmetric perturbations in
  $D=7,8,9$~\cite{Dias:2009iu}.

  Clearly, the study of the non-linear development of these instabilities and
  determination of the respective endpoint requires numerical methods.  Such
  studies have been recently presented for a non axi-symmetric perturbation in
  $D=5$~\cite{Shibata:2009ad} and $D=6,7,8$~\cite{Shibata:2010wz}, where it was
  found that the single spinning Myers-Perry black hole is unstable, for
  sufficiently large rotation parameter.

\item[AdS/CFT correspondence] \hfill \\
  In 1997--98, a powerful tool known as the AdS/CFT correspondence (or the
  gauge/gra\-vi\-ty duality, or even, more generically, as \emph{holography}) was
  introduced~\cite{Maldacena:1997re}. This holographic correspondence (if true
  in general) is extremely powerful since it maps the dynamics of a
  non-perturbative, strongly coupled regime of certain gauge theories in $D$
  dimensions to $(D+1)$-dimensional classical gravity. This means that, for such
  gauge theories, we can map strongly coupled quantum field theory dynamics to
  systems of partial differential equations, which can in principle be solved
  (numerically, if needed).

  In particular, high energy collisions of black holes are said to have a dual
  description in terms of high energy collisions with balls of de-confined
  plasma surrounded by a confining phase. 
  These are the type of events that may have direct
  observational consequences for the experiments at Brookhaven's Relativistic
  Heavy Ion Collider (RHIC)~\cite{Amsel:2007cw,Gubser:2008pc}.

  Numerical relativity in anti-de Sitter (AdS) is notoriously difficult, and
  therefore pro\-gress has been slow in applying its techniques to the
  aforementioned problems. Nevertheless, some exciting results have been
  recently put forth~\cite{Chesler:2009cy,Chesler:2010bi,Bantilan:2012vu}.

\item[TeV-scale gravity scenarios] \hfill \\
  As first pointed out by {}'t~Hooft~\cite{'tHooft:1987rb}, if two point
  particles collide at energies above the Planck energy, it is expected that
  gravity should dominate the interaction and thus, quite remarkably, the
  process should be well described by general relativity.

  Thorne's \emph{Hoop Conjecture}~\cite{Thorne:1972ji} further tells us that if
  one traps a given amount of Energy~$E$ in a region of space such that a
  circular hoop with radius $R$ encloses this matter \emph{in all directions}, a
  black hole is formed if its Schwarzschild radius $R_S \equiv \left(\frac{16
      \pi G E}{(D-2) \mathcal{A}_{D-2}c^4} \right)^{\frac{1}{D-3}} > R$.  This
  conjecture (or rather, the classical variant thereof) has recently gained more
  support with the work by Choptuik and Pretorius~\cite{Choptuik:2009ww}, where
  it was shown that collisions of boson stars \emph{do} form black holes, for
  sufficiently high boost parameter.

  If this conjecture does hold, it would imply that particle collisions could
  produce black holes~\cite{Dimopoulos:2001hw,Giddings:2001bu}. As argued above,
  the production of black holes at trans-Planckian collision energies (compared
  to the fundamental Planck scale) should be well described by using classical
  general relativity (see also~\cite{Kanti:2008eq} and references
  therein). Writing down the exact solution describing the collision of two
  ultra-relativistic particles in general relativity is not feasible, however,
  and approximations have to be used. One possible approximation (good for its
  simplicity) is to use black holes, and model the scattering of point particles
  by black hole collisions.

  This gains further relevance in the context of the so-called TeV-gravity
  scenarios. Such models were proposed as a possible solution to the
  \emph{hierarchy problem}, i.e., the relative weakness of gravity by about 40
  orders of magnitude compared to the other fundamental interactions.
  It has been proposed that this can be resolved if one adopts the idea that the
  Standard Model is confined to a brane in a higher dimensional space, such that
  the extra dimensions are much larger than the four dimensional Planck scale
  (they may be large up to a sub-millimetre
  scale)~\cite{Antoniadis:1990ew,ArkaniHamed:1998rs,Antoniadis:1998ig}. In a
  different version of the model, the extra dimensions are infinite, but the
  metric has an exponential factor introducing a finite length
  scale~\cite{Randall:1999ee,Randall:1999vf}.

  In such models, the fundamental Planck scale could be as low as 1 TeV. Thus,
  high energy colliders, such as the Large Hadron Collider (LHC), may directly
  probe strongly coupled gravitational
  physics~\cite{Argyres:1998qn,Banks:1999gd,Giddings:2001bu,Dimopoulos:2001hw,
    Ahn:2002mj,Chamblin:2004zg}. In fact, such tests may even be routinely
  available in the collisions of ultra-high energy cosmic rays with the Earth's
  atmosphere~\cite{Feng:2001ib,Ahn:2003qn,Cardoso:2004zi}, or in astrophysical
  black hole environments~\cite{Banados:2009pr,Berti:2009bk,Jacobson:2009zg}
  (for reviews see~\cite{Cavaglia:2002si,Kanti:2004nr,Kanti:2008eq}).

  Numerical simulations of high energy black hole collisions in higher
  dimensional spacetimes, then, could give an accurate estimate of the fractions
  of the collision energy and angular momentum that are lost in the
  higher-dimensional space by emission of gravitational waves; such information
  would be extremely important to improve the modelling of microscopic black hole
  production, and of the ensuing evaporation phase, which might be observed
  during LHC collisions.

  The challenge is then to use the classical framework to determine the cross
  section for production and, for each initial setup, the fractions of the
  collision energy and angular momentum that are lost in the higher dimensional
  space by emission of gravitational waves.  This information will be of
  paramount importance to improve the modelling of microscopic black hole
  production in event generators such as
  \textsc{Truenoir}~\cite{Dimopoulos:2001hw},
  \textsc{Charybdis2}~\cite{Frost:2009cf},
  \textsc{Catfish}~\cite{Cavaglia:2006uk} or
  \textsc{Blackmax}~\cite{Dai:2007ki,Dai:2009by}. The event generators will then
  provide a description of the corresponding evaporation phase, which might be
  observed during LHC collisions.

\end{description}
For a thorough review of these topics, challenges and how tools coming from numerical relativity can help see~\cite{Cardoso:2012qm}.

\section{The new frontiers}

With these motivations in mind, we propose in this thesis to extend current
numerical relativity tools to new frontiers.

\subsection{Higher-dimensional gravity}

The first such frontier, in light of our discussion in the previous section, is
higher-dimensional gravity. We start by emphasising that full blown $4+1$,
$5+1$, etc. numerical simulations of Einstein's field equations without
symmetry are currently (and in the near future) not possible due to the heavy
computational costs. We have thus developed a framework and a numerical code
that can, in principle, be applied to different spacetime dimensions (with
enough symmetry) with little adaptations.  This is achieved by taking the
$D$~dimensional vacuum spacetime to have an isometry group fit to include a
large class of interesting problems. If this isometry group is sufficiently
large, it allows a dimensional reduction of the problem to 3+1 dimensions, where
our originally higher-dimensional problem now appears as (four dimensional)
general relativity coupled to source
terms. 
Thus, the different spacetime dimensions manifest themselves only in the
different ``matter'' content of the four dimensional theory. An obvious
advantage of this approach is that we can use existing numerical codes with
small adaptations: taking a working four-dimensional code, the four dimensional
equations need to be coupled to the appropriate source terms and some issues
related to the chosen coordinates must be addressed. Incidentally, other issues
possibly related with the choice of gauge conditions further complicate the
problem.

We should here point out that alternative approaches have been proposed to
evolve numerically Einstein's equations in higher dimensions, as well as other
codes tailored to study specific problems. In particular we note the previously
mentioned pioneering works concerned with the non-linear development of the
Gregory-Laflamme instability~\cite{Gregory:1993vy} of cosmic
strings~\cite{Choptuik:2003qd,Lehner:2010pn}; studies of gravitational
collapse~\cite{Sorkin:2009bc,Sorkin:2009wh};
static situations~\cite{Headrick:2009pv}; and the alternative approach,
based on the \emph{Cartoon method}~\cite{Alcubierre:1999ab}, that has been developed
and tested by Yoshino and Shibata~\cite{Yoshino:2009xp,Shibata:2010wz}. For a
review of numerical relativity in higher-dimensions see
also~\cite{Yoshino:2011zz,Yoshino:2011zza}.

\subsection{Non-asymptotically flat geometries}

Another frontier has to do with numerical evolutions in non-asymptotically flat
spacetimes. 

\begin{description}
\item[de Sitter] \hfill \\
  Going back to four dimensions, an obvious first choice of a non-asymptotically
  flat spacetime is \emph{de Sitter}, the simplest model for an accelerating
  universe.
  de Sitter is a maximally symmetric solution of Einstein's equations with a
  positive cosmological constant, describing a Friedmann-Robertson-Walker (FRW)
  cosmology with a constant Hubble parameter.  There is now a large body of
  observational evidence for a present cosmological acceleration well modelled
  by a positive cosmological constant $\Lambda$~\cite{Komatsu:2010fb}.

  Cosmological dynamics should leave imprints in gravitational phenomena, such
  as gravitational radiation emitted in a black hole binary
  coalescence. Identifying such signatures can thus be phenomenologically
  relevant in view of ongoing efforts to directly detect gravitational
  radiation.

  Studying the dynamics of black holes in asymptotically de Sitter spacetimes
  can also potentially teach us about more fundamental questions such as cosmic
  censorship, as in the following scenario.  Consider two black holes of
  sufficiently large mass in a de Sitter spacetime. If, upon merger, the final
  black hole is too large to fit in its cosmological horizon the end state of
  such an evolution would be a naked singularity.  This possibility begs for a
  time evolution of such a configuration, which we will show and discuss.

\item[Black holes in a box] \hfill \\
  As argued above, having a framework to solve Einstein's equations in
  asymptotically Anti-de Sitter geometries would be of major help for
  studies of AdS/CFT duality, in particular in dynamical settings. This is no
  easy task, however, and a major reason for that is the ``active role'' played
  by the boundary of AdS spaces. This is easily visualised in the Penrose
  diagram of AdS, which has a timelike boundary. Null geodesics in AdS reach the
  boundary in a finite affine parameter, and one therefore often refers to an
  asymptotically AdS space as a ``box'', having in mind that AdS boundary
  conditions directly affect the bulk
  physics~\cite{Wald:1980jn,Ishibashi:2003jd,Ishibashi:2004wx}.

  As a first step to model the role of the boundary in evolutions, we will
  here give an overview of the work reported in~\cite{Witek:2010qc} where a toy
  model for AdS was considered. Therein, as we will explain, the cosmological
  constant is set to zero and mirror-like boundary conditions are imposed on a
  box containing the dynamical system, which mimics the global AdS geometry.

  Inside this box, black hole binaries are evolved, producing an inspiralling
  merger. Such systems are very well tested in standard asymptotically flat
  spacetimes with purely outgoing boundary conditions, and differences to
  these cases can be clearly seen.

\item[Black holes in cylinders] \hfill \\
  Again in the topic of higher-dimensional gravity, now in scenarios with
  compact extra dimensions, a natural question to ask is how the compactness of
  the extra dimensions changes the dynamics of such scenarios (as opposed to
  the asymptotically flat cases) and understanding the role of the compactness
  of the extra dimensions in the aforementioned TeV gravity scenarios.

  There is considerable literature on Kaluza-Klein black holes and black holes
  on cylinders~\cite{Myers:1986rx,Korotkin:1994dw,Frolov:2003kd,Harmark:2005pp},
  but, to the best of our knowledge, the full non-linear dynamics of black holes
  in such spacetimes remain unexplored.

  In this spirit, using the formalism developed for higher-dimensional
  spacetimes, we have started exploring what happens when one of the directions
  is compactified.

\end{description}

\subsection{Einstein-Maxwell}

Finally, we have started exploring the electrovacuum Einstein-Maxwell system.
We first note that while the dynamics of black holes interacting with
electromagnetic fields and plasmas have been the subject of a number of
numerical studies (e.g.~\cite{Palenzuela:2009yr,Palenzuela:2010nf}), dynamics of
charged (Reissner-Nordstr\"om) black holes have remained rather unexplored.

Studying the dynamics of charged black holes is relevant for a number of
reasons. In the context of astrophysics, charged black holes may actually be of
interest in realistic systems. First, a rotating black hole in an external
magnetic field will accrete charged particles up to a given value,
$Q=2B_0J$~\cite{Wald:1974np}. It is thus conceivable that astrophysical black
holes could have some (albeit rather small) amount of electrical charge. It is
then of interest to understand the role of this charge in the Blandford-Znajek
mechanism~\cite{Blandford:1977ds}, which has been suggested for extracting spin
energy from the hole, or in a related mechanism capable of extracting energy
from a moving black hole~\cite{Palenzuela:2010nf,Palenzuela:2010xn} to power
outflows from accretion disk-fed black holes.
Also of interest is investigating the role of charge in post-merger recoil
velocities of black hole binaries, and see if the recently predicted recoils of
$\sim 5000$~km/s~\cite{Lousto:2011kp} could be exceeded.

Incentive to study such systems also comes from outside of astrophysics. In
particular, as already mentioned above, it was argued by
{}'t~Hooft~\cite{'tHooft:1987rb}, that in trans-Planckian particle collisions,
gravity should dominate the interaction and thus the process should be well
described by general relativity---we can say that, for ultra high energy
collisions, \emph{matter does not matter}~\cite{Choptuik:2009ww}. Calculations
of shock wave collisions, however, seem to suggest that even though other
interactions---say charge---may become irrelevant in the ultra-relativistic
limit, the properties of the final black hole (and of the associated emission of
gravitational radiation) will in fact depend on the amount of charge carried by
the colliding particles~\cite{Yoshino:2006dp,Gingrich:2006qh}. One then wonders
whether the often repeated \emph{matter does not matter} scenario is actually
true.

Light can be shed in this issue by performing highly boosted collisions of charged black holes (analogous to the ones performed in vacuum~\cite{Sperhake:2008ga,Shibata:2008rq}) and comparing the results---in particular the profile of the corresponding waveform---against equivalent electrically neutral systems.

With these incentives, we will report on the first steps taken in the numerical
evolution of Reissner-Nordstr\"om black holes, building on previous numerical
evolutions of the Einstein-Maxwell
system~\cite{Palenzuela:2008sf,Palenzuela:2009yr,Palenzuela:2009hx,Mosta:2009rr}.

\section{Structure}

The structure of this thesis is as follows. We start by reviewing, in
chapter~\ref{chap:d-1+1}, standard differential geometry results, summarise the
formalism of the ``$(D-1)+1$'' decomposition\footnote{Usually found in the
  literature as ``3+1'' decomposition. Here we will keep the $D$ arbitrary, but
  the differences are minimal.} and conformal decomposition to write Einstein's
equations in a dynamical systems form. In chapter~\ref{chap:init_data} we then
review the construction of relevant initial data for the class of problems we
will be interested on and discuss, in chapter~\ref{chap:bssn-gauge}, the
numerical implementation of Einstein's equations: first, we need to re-write the
evolution equations in the so-called BSSN (Baumgarte, Shapiro, Shibata,
Nakamura) form; we next discuss the gauge conditions and finish by giving a very
brief overview of the numerical code we use for the simulations.  In
chapter~\ref{ch:wave-ext_horizon} we review the usual procedures to extract the
physical results from numerical simulations: wave extracting and horizon
finding.  These chapters consist mostly of review material found in the usual
literature
(e.g.~\cite{Gourgoulhon:2007ue,Alcubierre:2008,baumgarte2010numerical}).
Finally, in chapter~\ref{chap:high-dimNR}, we introduce a dimensional reduction procedure that allows us to reduce higher-dimensional systems (with enough symmetry) to effective four-dimensional theories with source terms.
This enables us to perform numerical evolutions of such higher-dimensional systems by adapting existing numerical codes.
We also discuss the construction of initial data and present results.
Chapter~\ref{ch:non-asympt-flat} is dedicated to evolutions in non-asymptotically flat spacetimes: we present the aforementioned collisions of black holes in asymptotically de Sitter spacetimes, black holes in a box and black holes in asymptotically cylindrical spacetimes.
In chapter~\ref{ch:EM} we report on evolutions of charged black holes, in electrovacuum Einstein-Maxwell theory, and we end with some final remarks and future directions in chapter~\ref{ch:final}.

\section{Preliminaries}
\label{sec:prelim}

Let us consider a $D$-dimensional pseudo-Riemannian manifold $(\mathcal{M},g)$, that is, a differentiable manifold $\mathcal{M}$ equipped with a smooth, symmetric metric tensor $g$ with signature $(-+\cdots+)$.
We further assume that the manifold is covered by a set of coordinates $\{x^{\mu}\}$, $\mu=0,\ldots,D-1$.

A \emph{coordinate basis} of the \emph{tangent space} of $\mathcal{M}$ at $p$, $T_p \mathcal{M}$, is given by $\partial_{\mu} \equiv \partial/\partial x^{\mu}$.
A \emph{vector} $V \in T_p \mathcal{M}$ can be written in the form
\begin{equation}
V = V^{\mu} \partial_{\mu} \,,
\end{equation}
where $V^{\mu}$ are the components of $V$ in the basis $\partial_{\mu}$.
When a vector $V$ acts on a function $f$ it produces the directional derivative of $f$ along $V$: 
\begin{equation}
  V(f) = V^{\mu} \partial_{\mu} f \,.
\end{equation}

A \emph{1-form} $\omega \in T_p^{\star} \mathcal{M}$ (the \emph{cotangent space} at $p$) is an object which is dual to a vector, i.e., it produces a number when acting on a vector.
The simplest example of such an object is the differential $\dd f$ of a function $f$.
The action of $\dd f$ on $V$ is defined to be
\begin{equation}
\langle \dd f, V \rangle \equiv V(f) = V^{\mu} \partial_{\mu} f \,.
\end{equation}
Since $\dd f = \partial_{\mu} f \dd x^\mu$, $\{ \dd x^\mu \}$ is a natural choice as a basis of $T_p^{\star} \mathcal{M}$.
We thus naturally expand an arbitrary 1-form $\omega$ as
\begin{equation}
  \omega = \omega_{\mu} \dd x^{\mu} \,.
\end{equation}

The metric tensor $g$ allows us to define a scalar product between two vectors $U$ and $V$
\begin{equation}
  U \cdot V \equiv g(U,V) 
  =  g\left(\partial_{\mu}, \partial_{\nu}\right) U^{\mu} V^\nu 
  \equiv g_{\mu\nu} U^{\mu} V^\nu \,,
\end{equation}
which induces an isomorphism between vectors and 1-forms, corresponding in the index notation to the usual raising and lowering of indices: if $U$ is a vector, one can define a 1-form $U_{\flat}$ through
\begin{equation}
  \langle U_{\flat}, V \rangle \equiv g(U,V) = g_{\mu\nu} U^{\mu} V^\nu 
  \equiv \left( U_{\flat} \right)_{\nu} V^\nu \quad \forall \, V \,.
\end{equation}
Analogously, given a 1-form $\omega$ we can map it to a vector $\omega^{\sharp}$ through
\begin{equation}
  \langle \sigma, \omega^{\sharp} \rangle \equiv g^{-1}(\sigma,\omega) 
  = g^{-1} \left( \sigma_{\mu} \dd x^{\mu}, \omega_{\nu} \dd x^{\nu} \right)
  = g^{\mu\nu} \sigma_{\mu} \omega_{\nu}
  \equiv  \sigma_{\mu} \left( \omega^{\sharp} \right)^{\mu} \quad \forall \, \sigma \,.
\end{equation}
Since we will be working mostly in the index notation, and the placement of the index makes clear whether one is dealing with vectors of 1-forms, we will omit the flat and sharp symbols throughout.

The metric further allows us to determine the distance between two nearby points in the manifold according to
\begin{equation}
  \dd s^2 = g_{\mu\nu} \dd x^{\mu} \dd x^{\nu} \,.
\end{equation}
%
%

Notice that a basis of $T_p \mathcal{M}$ (and of $T_p^{\star} \mathcal{M}$) need not be coordinate.
One can have, for instance, the combination $e_{\alpha} \equiv A_{\alpha}{}^{\mu} \partial_{\mu}$.
$\{e_{\alpha}\}$ is an example of a \emph{non-coordinate basis}.

We now introduce a (generic) non-coordinate basis obeying
\begin{align}
\label{eq:ap_basis}
  \left[e_\alpha, e_\beta \right] = c_{\alpha\beta}{}^\delta e_\delta \,,
\end{align}
where the \emph{Lie bracket} $[X,Y]$ is defined by
\begin{equation}
  \label{eq:lie-bracket}
  [X,Y] f = X \left( Y(f) \right) - Y \left( X(f) \right) \,.
\end{equation}
The connection coefficients $\GG{\alpha}{\beta\gamma}$ then take the form
\begin{align}
  \label{eq:ap_connection}
  \GG{\alpha}{\beta\gamma} = \frac{1}{2} g^{\alpha\delta} \left(
    g_{\delta\beta,\gamma} + g_{\delta\gamma,\beta} - g_{\beta\gamma,\delta} + c_{\delta\beta\gamma} + c_{\delta\gamma\beta} - c_{\beta\gamma\delta}
  \right),
\end{align}
where $\GG{\alpha}{[\beta\gamma]} = -\frac{1}{2} c_{\beta\gamma}{}^\alpha$.
When using coordinate basis ($c_{\beta\gamma}{}^\alpha = 0$), these are usually called the Christoffel symbols.

We now define the Riemann curvature tensor
\begin{align}
  \label{eq:ap_riemann}
  R^\alpha{}_{\beta\gamma\delta} = \GG{\alpha}{\beta\delta,\gamma} - \GG{\alpha}{\beta\gamma,\delta} + \GG{\alpha}{\lambda\gamma} \GG{\lambda}{\beta\delta} - \GG{\alpha}{\lambda\delta} \GG{\lambda}{\beta\gamma} - \GG{\alpha}{\beta\lambda} c_{\gamma\delta}{}^\lambda \,.
\end{align}
Mind the notation
\begin{align}
  \label{eq:ap_deriv}
  T_{\alpha,\beta} \equiv \partial_{e_\beta } T_\alpha \equiv e_\beta \left( T_\alpha \right) 
  .
\end{align}
Thus, $T_{\alpha, \sigma \lambda} = \partial_{e_\lambda} \partial_{e_\sigma} T_\alpha \neq \partial_{e_\sigma} \partial_{e_\lambda} T_\alpha $.

General relativity is a geometric theory of gravity which relates the curvature of spacetime to its matter content via the Einstein field equations, which read
\begin{equation}
  \label{eq:EFE-0}
  G_{\mu\nu} \equiv  R_{\mu\nu} - \frac{1}{2} R \, g_{\mu\nu} 
                       = 8\pi T_{\mu\nu} \,,
\end{equation}
where $R_{\mu\nu} \equiv R^{\lambda}{}_{\mu\lambda\nu}$ is the Ricci curvature tensor, $R$ its trace (the Ricci scalar), $g_{\mu\nu}$ the metric tensor and $T_{\mu\nu}$ the stress-energy tensor.
All these quantities are $D$-dimensional.

Throughout this work we will always use the $(-+\cdots+)$ metric signature and geometrised units $G=1=c$.



\chapter{\texorpdfstring{$(D-1)+1$}{(D-1)+1} decomposition}
\label{chap:d-1+1}

We start by briefly stating some known results from differential geometry that will be of use to us. In this chapter we use the following conventions: Greek indices ($\alpha$, $\beta$, $\gamma$, \dots) run from 0 to $D-1$; Latin indices ($i$, $j$, $k$, \dots) run from 1 to $D-1$.

We work on a $D$-dimensional manifold $\mathcal{M}$ with a metric $g_{\mu\nu}$. 
We denote the torsion-free Levi-Civita connection on $\mathcal{M}$ associated with $g_{\mu\nu}$ by $ ^D \! \nabla$. All quantities defined relative to the manifold $\mathcal{M}$ will have a leading $^D$ superscript, e.g., the Riemann curvature tensor on $\mathcal{M}$ is denoted by $^D \! R^\mu{}_{\alpha\beta\gamma}$.

\section{Hypersurfaces}

\subsection{Definition}

A codimension 1 hypersurface $\Sigma$ is a $(D-1)$-dimensional submanifold of $\mathcal{M}$, defined as the image of a $(D-1)$-dimensional manifold $\hat\Sigma$ by an embedding $\Phi$, $\Sigma = \Phi(\hat \Sigma)$~\cite{Poisson:2004,Gourgoulhon:2007ue}. 
Given a scalar field $t$ on $\mathcal{M}$, we can select a particular hypersurface $\Sigma$ by putting a restriction on the coordinates
\begin{align}
  t \left(x^\alpha \right) = 0,
\end{align}
or by giving parametric equations
\begin{align}
  x^\mu = x^\mu \left( y^i \right),
\end{align}
where $y^i$ are coordinates intrinsic to the hypersurface.

\subsection{Normal vector}
The 1-form $\partial_\mu t$ is normal to the hypersurface. We can introduce a unit normal $n_\mu$ (if the hypersurface is not null) as
\begin{equation}
  n_\mu = \frac{1}{\sqrt{|\partial_\alpha t \partial^\alpha t}|} \partial_\mu t.
\end{equation}
With this definition
\begin{align}
  n^\mu n_\mu \equiv \sigma =
    \begin{cases}
      -1 & \text{if $\Sigma$ is spacelike} \\
      +1 & \text{if $\Sigma$ is timelike}
    \end{cases}.
\end{align}

\subsection{Induced metric}
\label{sec:induced-metric}
We obtain the \emph{induced metric on $\Sigma$} by restricting the line element to displacements confined to the hypersurface. 
Using the parametric equations $x^\mu = x^\mu \left(y^i \right)$, we define the vectors
\begin{align}
  e_i^\mu = \del{}{x^\mu}{y^i}
\end{align}
that are tangent to the curves in $\Sigma$. For displacements confined to $\Sigma$ we have
\begin{align}
  \dd s^2_\Sigma & = g_{\mu\nu} \dd x^\mu \dd x^\nu \notag \\
            & = g_{\mu\nu} \left(
              \del{}{x^\mu}{y^i} \dd y^i
            \right)
            \left(
              \del{}{x^\nu}{y^j} \dd y^j
            \right)    \notag \\
            & = \gamma_{ij} \dd y^i \dd y^j,
\end{align}
where 
\begin{align}
  \gamma_{ij} \equiv  \left(
    \del{}{x^\mu}{y^i} 
  \right)
  \left(
    \del{}{x^\nu}{y^j} 
  \right)
    g_{\mu\nu}
\end{align}
is the induced metric of the hypersurface (also called the \emph{first fundamental form of $\Sigma$}). Notice that if $u, v \in \Sigma$,
\begin{align*}
  u \cdot v = g_{\mu\nu} u^\mu v^\nu = \gamma_{ij} u^{i} v^{j}.
\end{align*}

\subsection{Orthogonal projector}
The \emph{orthogonal projector onto $\Sigma$} is a concept closely related with that of the induced metric. For a hypersurface $\Sigma$ with unit normal $n^\mu$ we define it as
\begin{equation}
  \label{eq:projector}
  P_{\mu\nu} = g_{\mu\nu} - \sigma n_\mu n_\nu.
\end{equation}
To see that this definition makes sense, we note that, for any vector $v^\mu$ in $\mathcal{M}$ (or, more correctly, in $T_p \mathcal{M}$, the \emph{tangent space} of $\mathcal{M}$ at $p$), $P_{\mu\nu} $ will project it tangent to the hypersurface, i.e., orthogonal to $n^\mu$:
\begin{align}
  \left(
    P_{\mu\nu} v^\mu
  \right) n^\nu = 0.
\end{align}
Notice also that the projector is idempotent:
\begin{align}
  P^{\mu}{}_{\lambda} P^\lambda{}_\nu = P^\mu{}_\nu.
\end{align}
Finally, note that for $u, v \in \Sigma$, $P_{\mu\nu}$ acts just like the metric:
\begin{align}
  P_{\mu\nu} u^\mu v^\nu = g_{\mu \nu} u^\mu v^\nu = \gamma_{ij} u^i v^j.
\end{align}
Thus, we see that the orthogonal projector $P_{\mu\nu}$ is the natural extension of the induced metric $\gamma_{ij}$ for all vectors in $T_p\mathcal{M}$. As such, from now on we will no longer make any distinction between these two concepts, and will denote both by $\gamma_{\mu\nu}$ (defined as $ \gamma_{\mu\nu} = g_{\mu\nu} - \sigma n_\mu n_\nu)$. This way we adopt a $D$-dimensional point of view, where we treat all tensor fields defined on $\Sigma$ as if they were defined on $\mathcal{M}$ and we avoid the need to introduce a specific coordinate system on $\Sigma$.

Note that we can project an arbitrary tensor on $\mathcal{M}$ onto $\Sigma$ in the following way. Let $T^{\mu_1 \cdots \mu_p}{}_{\nu_1 \cdots \nu_q}$ be a tensor field on $\mathcal{M}$. Denoting $\left( \gamma \, T \right)^{\alpha_1 \cdots \alpha_p}{}_{\beta_1 \cdots \beta_q}$ another tensor in $\mathcal{M}$ such that
\begin{align}
  \label{eq:proj_operation}
  \left( \gamma \, T \right)^{\alpha_1 \cdots \alpha_p}{}_{\beta_1 \cdots \beta_q}
  = \gamma^{\alpha_1}{}_{\mu_1} \cdots \gamma^{\alpha_p}{}_{\mu_p}
  \gamma^{\nu_1}{}_{\beta_1} \cdots  \gamma^{\nu_q}{}_{\beta_q}
  T^{\mu_1 \cdots \mu_p}{}_{\nu_1 \cdots \nu_q},
\end{align}
we easily see that  $\left( \gamma \, T \right)^{\alpha_1 \cdots \alpha_p}{}_{\beta_1 \cdots \beta_q}$ is in $\Sigma$.

\subsection{Intrinsic curvature}
We now want to define a covariant derivative associated with $\gamma_{\mu \nu}$ on $\Sigma$, $\nabla$, that has the ``usual properties'' of a covariant derivative, in particular that it is torsion-free and satisfies
\begin{equation}
  \label{eq:cov_deriv_gamma}
  \nabla_\alpha \gamma_{\mu\nu} = 0.
\end{equation}
The easiest way to define it is just to project the covariant derivative ${}^D \! \nabla$ onto $\Sigma$ using~\eqref{eq:projector}:
\begin{align}
  \label{eq:cov-deriv}
  \nabla_\rho  T^{\alpha_1 \cdots \alpha_p}{}_{\beta_1 \cdots \beta_q}
  = \gamma^{\alpha_1}{}_{\mu_1} \cdots \gamma^{\alpha_p}{}_{\mu_p}
  \gamma^{\nu_1}{}_{\beta_1} \cdots  \gamma^{\nu_q}{}_{\beta_q} \gamma^\sigma{}_{\rho}
  {}^D \! \nabla_\sigma  T^{\mu_1 \cdots \mu_p}{}_{\nu_1 \cdots \nu_q}.
\end{align}
It can be shown~\cite{Gourgoulhon:2007ue} that this definition of the covariant derivative has all the properties we want (linearity, Leibniz' rule, its torsion vanishes, \dots) and it satisfies~\eqref{eq:cov_deriv_gamma}.

We can now define the Riemann tensor associated with this connection, $R^\alpha{}_{\beta\gamma\delta}$, as the measure of the non-commutativity of this covariant derivative, associated with the $\gamma_{\mu\nu}$ metric on $\Sigma$,
\begin{align}
  \label{eq:riemanndef}
  \nabla_\alpha \nabla_\beta v^\gamma - \nabla_\beta \nabla_\alpha  v^\gamma
  = R^\gamma{}_{\mu\alpha\beta} v^\mu,
\end{align}
where $v^\mu \in \Sigma$. 

$R^\alpha{}_{\beta\gamma\delta}$ represents the \emph{intrinsic curvature of $\Sigma$}.

\subsection{Extrinsic curvature}
The intrinsic curvature of the hypersurface $\Sigma$, as the name implies, is a property of the hypersurface itself. We will now define the \emph{extrinsic curvature}, which depends on how $\Sigma$ is embedded on $\mathcal{M}$. We define it as\footnote{Our definition, with the minus sign, agrees with the standard convention used in the numerical relativity community, but note that some authors use different conventions.}~\cite{Carroll:2004st}
\begin{equation}
  \label{eq:extrinsic_curvature}
  K_{\mu\nu} = - \gamma^\alpha{}_\mu \gamma^\beta{}_\nu \nabla_\alpha n_\beta.
\end{equation}
It can be shown that $K_{\mu\nu} = K_{\nu\mu}$. Defining
\begin{equation}
  a^\mu = n^\nu \nabla_\nu n^\mu,
\end{equation}
we have, after some simple algebra,
\begin{equation}
  \label{eq:K_def}
  K_{\mu\nu} = -\nabla_\mu n_\nu + \sigma n_\mu a_\nu.
\end{equation}
We will always consider spacelike hypersurfaces, so from now on we will work with $\sigma = -1$. Note that, by definition, $K_{\mu\nu}$ lives on $\Sigma$.

\section{Foliations}
We assume that our spacetime can be foliated by a family of spacelike hypersurfaces $\Sigma_t$, that is, there exists a smooth scalar field $\hat t$ on $\mathcal{M}$ such that
\begin{equation}
  \Sigma_t \equiv \left\{
    p \in \mathcal{M}, \, \hat t (p) = t
  \right\}.
\end{equation}
In the following we will not distinguish between $t$ and $\hat t$.

\subsection{The lapse function}
\label{sec:lapse}
We write the unit normal vector as
\begin{equation}
  n_\mu \equiv - \alpha \partial_\mu t, 
\end{equation}
where
\begin{equation}
  \alpha \equiv \frac{1}{ \sqrt{-\partial_\nu t \partial^\nu t} }
\end{equation}
is called the \emph{lapse function}.

\subsection{Normal evolution vector}
We define the \emph{normal evolution vector} as 
\begin{equation}
  m^\mu \equiv \alpha \, n^\mu.
\end{equation}
We can easily see that 
\begin{align*}
  m^\mu {}^D \! \nabla_\mu t = 1,
\end{align*}
which means that $m^\mu$, unlike $n^\mu$, is adapted to the scalar field $t$. It can be shown~\cite{Gourgoulhon:2007ue} that the hypersurfaces $\Sigma_t$ are \emph{Lie dragged} by $m^\mu$. As consequence of this, if $v^\mu$ is in $\Sigma$, $\mathcal{L}_m v$ is also in $\Sigma$.

\subsection{Eulerian observers}
\label{sec:eulerian-observers}

We can identify the unit timelike vector $n^\mu$ as the velocity (or the ``$D$-velocity''\dots) of some observer, that we will call the \emph{Eulerian observer}. The worldlines of these observers are orthogonal to the hypersurfaces $\Sigma_t$, which means that, for a given $t$, the hypersurface $\Sigma_t$ is the set of events that are simultaneous from the point of view of the Eulerian observer.

We define the acceleration of the Eulerian observer the usual way,
\begin{align}
  \label{eq:accel}
  a^\mu = n^\nu \nabla_\nu n^\mu.
\end{align}

Let us now list some formul\ae{} that will be useful for the following sections:
\begin{align}
  a_\mu  & = \nabla_\mu \log \alpha,      \label{eq:formulae1}    \\
  {}^D \! \nabla_\beta n_\alpha & = -K_{\alpha\beta} - n_\beta \nabla_\alpha \log \alpha,    \label{eq:formulae2}   \\
  {}^D \! \nabla_\beta m^\alpha & = -\alpha K^{\alpha}{}_\beta -  n_\beta  \nabla^\alpha \alpha 
  +n^\alpha {}^D \! \nabla_\beta \alpha.  \label{eq:formulae3} 
\end{align}

\subsection{Evolution of \texorpdfstring{$\gamma_{\alpha\beta}$}{gamma}} 
From the definition of Lie derivative and equation~\eqref{eq:formulae3}, one can show
\begin{equation}
\label{eq:evol_gamma}
  \mathcal{L}_m \gamma_{\alpha\beta} = -2 \alpha K_{\alpha\beta},
\end{equation}
and 
\begin{equation}
  \mathcal{L}_m \gamma^{\alpha}{}_{\beta} = 0,
\end{equation}
which means that, for any tensor field $T$ on $\Sigma_t$, its Lie derivative along $m$ is also a tensor field on $\Sigma_t$.

\section{Gauss, Codazzi and Ricci equations}
\label{sec:gauss-codazzi}
We still need a way to relate quantities defined on the hypersurface to those defined on the manifold $\mathcal{M}$; in particular, we would like to have a relation between the $D$-dimensional Riemann curvature tensor, the $(D-1)$-dimensional Riemann tensor on the hypersurface and the extrinsic curvature. Such relations are common in differential geometry---known as the equations of Gauss, Codazzi and Ricci---which we now state without proof (see, e.g.,~\cite{Gourgoulhon:2007ue}).

\subsection{Gauss equation}
\label{sec:gauss}
The starting point is equation~\eqref{eq:riemanndef}. We just need to use equation~\eqref{eq:cov-deriv} to relate $\nabla_\alpha v^\gamma$ with ${}^D \! \nabla_\alpha v^\gamma$. After some algebra we arrive at

\begin{equation}
  \label{eq:gauss}
  \gamma^{\mu}{}_{\alpha} \gamma^{\nu}{}_{\beta}  \gamma^{\gamma}{}_{\rho} \gamma^{\sigma}{}_{\delta} \, {} ^D \! R^\rho{}_{\sigma \mu \nu} 
 = R^\gamma{}_{\delta \alpha \beta} + K^\gamma{}_\alpha K_{\delta\beta} -  K^\gamma{}_\beta K_{\alpha\delta},
\end{equation}
which is called the \emph{Gauss equation}. 
Contracting this equation on $\gamma$ and $\alpha$ we get
\begin{equation}
  \label{eq:contract_gauss}
  \gamma^\mu{}_\alpha \gamma^\nu{}_\beta \, {}^D \! R_{\mu \nu}
  +\gamma_{\alpha \mu} \gamma^{\rho}{}_\beta n^\nu n^\sigma \, {}^D \! R^\mu{}_{\nu\rho\sigma}
= R_{\alpha\beta} + K K_{\alpha\beta} - K_{\alpha\mu} K^{\mu}{}_\beta ,
\end{equation}
where we defined $K\equiv K^\mu{}_\mu = K^i{}_i$ (where the equality comes from the fact that $K_{\mu\nu}$ lives on $\Sigma$).
Taking the trace of this expression we have (noting that $K_{\mu\nu} K^{\mu \nu} = K_{ij} K^{ij} $)
\begin{equation}
  \label{eq:scalar_gauss}
  {}^D \! R + 2 {}^D \! R_{\mu\nu} n^\mu n^\nu = R + K^2 - K_{ij}  K^{ij}.
\end{equation}

\subsection{Codazzi equation}
\label{sec:codazzi}
We now start with the following equation
\begin{align}
  \label{eq:n}
  \nabla_\alpha \nabla_\beta n^\gamma - \nabla_\beta \nabla_\alpha  n^\gamma
  = R^\gamma{}_{\mu\alpha\beta} n^\mu,
\end{align}
and we project it onto $\Sigma$ using~\eqref{eq:proj_operation}. Using equation~\eqref{eq:K_def} and after some algebra we arrive at
\begin{equation}
  \label{eq:codazzi}
  \gamma^\gamma{}_\rho \gamma^\mu{}_\alpha \gamma^\nu{}_\beta n^\sigma \, {}^D \! R^\rho{}_{\sigma\mu\nu} 
= \nabla_\beta K^\gamma{}_\alpha - \nabla_\alpha K^\gamma{}_\beta,
\end{equation}
which is called the \emph{Codazzi equation}. Contracting this equation on $\beta$ and $\gamma$ we have
\begin{equation}
  \label{eq:contract_codazzi}
  \gamma^\mu{}_\alpha n^\nu \, {}^D \! R_{\mu\nu} 
= \nabla_\alpha K - \nabla_\mu K^\mu{}_\alpha.
\end{equation}

\subsection{Ricci equation}
\label{sec:mixed-proj-riem}

We still need one more projection of the Riemann tensor (in fact, the last non-trivial projection). Again, we start with equation~\eqref{eq:n}, but this time we project it only twice onto $\Sigma_t$ and one time along $n^\mu$:
\begin{align}
  \label{eq:twice_sigma_once_n}
  \gamma_{\alpha\mu} n^\sigma \gamma^\nu{}_{\beta}
  \left(
    {}^D \! \nabla_\nu  {}^D \! \nabla_\sigma n^\mu
    - {}^D \! \nabla_\sigma {}^D \! \nabla_\nu  n^\mu
  \right)
  = \gamma_{\alpha\mu} n^\sigma \gamma^\nu{}_\beta n^\rho
  {}^D \! R^{\mu}{}_{\rho\nu\sigma}.
\end{align}
Using equations~\eqref{eq:formulae2},~\eqref{eq:formulae3} and some properties of the Lie derivative we arrive at
\begin{equation}
  \label{eq:ricci_equation}
  \gamma_{\alpha \mu}  \gamma^\nu{}_\beta n^\rho n^\sigma {}^D \! R^\mu{}_{\rho\nu\sigma} 
  = \frac{1}{\alpha} \mathcal{L}_m K_{\alpha\beta} + \frac{1}{\alpha} \nabla_\alpha \partial_\beta \alpha 
  + K_{\alpha\mu} K^\mu{}_\beta,
\end{equation}
which is called the \emph{Ricci equation}.
We can combine equation~\eqref{eq:ricci_equation} with equation~\eqref{eq:contract_gauss} to get
\begin{equation}
\label{eq:ricci-contracted}
  \gamma^\mu{}_{\alpha}  \gamma^\nu{}_\beta  {}^D \! R_{\mu \nu} 
  = - \frac{1}{\alpha} \mathcal{L}_m K_{\alpha\beta} - \frac{1}{\alpha} \nabla_\alpha \partial_\beta \alpha 
  + R_{\alpha\beta} + K  K_{\alpha\beta} - 2 K_{\alpha\mu} K^\mu{}_\beta .  
\end{equation}

\section{Einstein equations}
Our goal now is to write the Einstein equations in an explicit dynamical system
form. Let us start by writing the equations themselves in their ``traditional''
form,
\begin{equation}
\label{eq:EFE_D}
 {}^D \! R_{\mu\nu} - \frac{1}{2}  {}^D \! R \, g_{\mu\nu} = 8 \pi \, T_{\mu\nu}.
\end{equation}
The alternative form
\begin{equation}
\label{eq:EFE_D_2}
  {}^D \! R_{\mu\nu} = 8\pi 
  \left(
    T_{\mu\nu} - \frac{T}{D-2}g_{\mu\nu}
  \right),
\end{equation}
where $T \equiv g^{\mu\nu} T_{\mu\nu} $, will also be useful to us.

\subsection{Decomposition of the stress-energy tensor}
\label{sec:decomp-stress-energy}
We define
\begin{align}
  E        & \equiv T_{\mu\nu} n^\mu n^\nu, \\
  j_\alpha  & \equiv -T_{\mu\nu} n^\mu \gamma^\nu{}_\alpha, \\
  S_{\alpha\beta} & \equiv T_{\mu\nu} \gamma^\mu{}_\alpha \gamma^\nu{}_\beta,
\end{align}
which correspond to the \emph{matter energy density}, the \emph{matter momentum density} and the \emph{matter stress density}, respectively, as measured by the Eulerian observer. We further define $S \equiv g^{\mu\nu}S_{\mu\nu}  = \gamma^{ij} S_{ij} $ and note that $T = S-E$.

\subsection{Projection of the Einstein equations}
\label{sec:proj-einst-equat}

\subsubsection{Projection onto $\Sigma_t$}
\label{sec:proj-onto-sigma}
Using equation~\eqref{eq:ricci-contracted} we project equation~\eqref{eq:EFE_D_2} onto $\Sigma_t$. We get
\begin{align}
  \label{eq:proj_sigma}
  \mathcal{L}_m K_{\alpha\beta} = -\nabla_\alpha \nabla_\beta\alpha
  +\alpha \left[
    R_{\alpha\beta} + K K_{\alpha\beta} - 2K_{\alpha\mu} K^\mu{}_\beta
    +\frac{8\pi}{D-2}
      (S-E) \gamma_{\alpha\beta}
      - 8\pi S_{\alpha\beta}
  \right].
\end{align}
Note that every single term in this equation lives in $\Sigma_t$. Thus, we can restrict the indices to spacial ones,
\begin{align}
  \label{eq:proj_sigma_spacial}
  \mathcal{L}_m K_{ij} = -\nabla_i \nabla_j\alpha
  +\alpha \left[
    R_{ij} + K K_{ij} - 2K_{ik} K^k{}_j
    +\frac{8\pi}{D-2}
      (S-E) \gamma_{ij}
      - 8\pi S_{ij}
  \right]
.
\end{align}

\subsubsection{Projection along $n^\mu$}
\label{sec:projection-along-n}
This step is easy, we just need to contract equation~\eqref{eq:EFE_D} with $n^\mu  n^\nu$ and use~\eqref{eq:scalar_gauss}, yielding

\begin{equation}
  \label{eq:hamiltonian}
  R + K^2 - K_{ij} K^{ij} = 16 \pi E
.
\end{equation}
This equation is called the \emph{Hamiltonian constraint}.

\subsubsection{Mixed projection}
\label{sec:mixed-projection}
Finally, we need to project equation~\eqref{eq:EFE_D} once onto $\Sigma_t$ and once along $n^\mu$. Using equation~\eqref{eq:contract_codazzi} we get
\begin{equation}
  \label{eq:momentum}
  \nabla_j \left( K^{ij} - \gamma^{ij} K  \right) = 8 \pi j^i
.
\end{equation}
This equation is called the \emph{momentum constraint}.

\section{Choice of coordinates}
\label{sec:choice-coordinates}
Equations~\eqref{eq:proj_sigma_spacial} ($(D-1)D/2$ equations),~\eqref{eq:hamiltonian} (1 equation) and~\eqref{eq:momentum} ($D-1$ equations) contain the same information as equation~\eqref{eq:EFE_D} (it can be checked that the number of independent components is the same: $(D-1)D/2 + 1 + (D-1) = D(D+1)/2$). Before we can cast these equations in a dynamical system form, however, we have to introduce a specific coordinate system, something we have not yet done.

We will introduce coordinates adapted to the foliation $\Sigma_t$ in the following way~\cite{Gourgoulhon:2007ue}. On each hypersurface $\Sigma_t$ we have a coordinate system $x^i = x^1, x^2, \ldots, x^{D-1}$ that is varying smoothly between neighbouring hypersurfaces, so that $x^\alpha = t,  x^1, x^2, \ldots, x^{D-1}$ is a well behaved coordinate system on $\mathcal{M}$. In this coordinate system
\begin{align}
  n_\mu = (-\alpha,0, \ldots,0).
\end{align}
We define the \emph{shift vector $\beta$} as
\begin{align}
  \label{eq:shift}
  \beta \equiv \partial_t - m,
\end{align}
or in components, $\beta^\mu \equiv \delta^\mu{}_t - m^\mu$. Note that $ n_\mu \beta^\mu = 0$, so $\beta$ lives on $\Sigma_t$ ($\beta^t = 0)$. Using~\eqref{eq:shift} we can also write
\begin{align}
  n^\mu = \frac{1}{\alpha} \left(
    1,-\beta^i
  \right).
\end{align}
Notice also that
\begin{align*}
  \partial_t \cdot \partial_t = -\alpha^2 + \beta^\mu \beta_\mu = -\alpha^2 + \beta^k \beta_k.
\end{align*}

We are now able to write the metric components $g_{\mu\nu}$ relative to this coordinate system,
\begin{align*}
  g_{0 0}  & = g_{\mu\nu} \left(\partial_t \right)^\mu \left(\partial_t \right)^\nu 
  =   \partial_t \cdot \partial_t = -\alpha^2 + \beta^k \beta_k, \\
  g_{0 i}  & = g_{\mu\nu} \left(\partial_t \right)^\mu \left(\partial_i \right)^\nu 
  =   \left( m + \beta \right) \cdot \partial_i = \beta \cdot \partial_i = \beta_j \delta^j{}_i = \beta_i, \\
  g_{ij} & =  g_{\mu\nu} \left(\partial_i \right)^\mu \left(\partial_j \right)^\nu 
  =  \gamma_{kl} \left(\partial_i \right)^k \left(\partial_j \right)^l = \gamma_{ij} .
\end{align*}
The line element is thus
\begin{align}
  g_{\mu\nu}\dd x^\mu  \dd x^\nu = -\alpha^2 \dd t^2 + \gamma_{ij} 
  \left(\dd x^i + \beta^i \dd t \right)   \left(\dd x^j + \beta^j \dd t \right),
\end{align}
or, in matrix form,
\begin{align}
  g_{\mu\nu} = \left(
    \begin{matrix}
      -\alpha^2 + \beta_k \beta^k & \beta_i \\
      \beta_j                    &  \gamma_{ij}
    \end{matrix}
  \right) \,.
\end{align}
The inverse metric takes the form
\begin{align}
  g^{\mu\nu} = \left(
    \begin{matrix}
      -\frac{1}{\alpha^2}        & \frac{\beta^i}{\alpha^2} \\
      \frac{\beta^j}{\alpha^2}   &  \gamma^{ij} - \frac{\beta^i \beta^j}{\alpha^2}
    \end{matrix}
  \right) \,.
\end{align}
The determinants of $g_{\mu\nu}$ and $\gamma_{ij}$ are related by
\begin{align}
  \sqrt{-g} = \alpha \sqrt{\gamma},
\end{align}
where 
\begin{align*}
  g  \equiv \det g_{\mu\nu}, \qquad \gamma \equiv \det \gamma_{ij}.
\end{align*}

\section{The PDE system}
\label{sec:pde-system}
Using the properties of the Lie derivative and the definition of shift vector, equation~\eqref{eq:shift}, we can write
\begin{align}
  \mathcal{L}_m K_{ij} = \partial_t K_{ij} - \mathcal{L}_\beta K_{ij}.
\end{align}
Equation~\eqref{eq:evol_gamma} can also be put in the form
\begin{align}
  \label{eq:lie}
  \left(
    \partial_t -\mathcal{L}_\beta
  \right) \gamma_{ij}
  = -2 \alpha K_{ij}.
\end{align}

We now have our full system, which we rewrite here
\begin{subequations}
\label{eq:ADM-eq}
  \begin{align}
    \label{eq:lie2}
    & 
      \left(
        \partial_t -\mathcal{L}_\beta \right) \gamma_{ij} = -2 \alpha K_{ij}
    , \\
    \label{eq:proj_sigma_spacial2}
    & 
      \left(
        \partial_t -\mathcal{L}_\beta \right) K_{ij} = -\nabla_i \nabla_j\alpha
      +\alpha \left[ R_{ij} + K K_{ij} - 2K_{ik} K^k{}_j +\frac{8\pi}{D-2} (S-E)
        \gamma_{ij} - 8\pi S_{ij} \right]
    , \\
    \label{eq:hamiltonian2}
    & 
      R + K^2 - K_{ij} K^{ij} = 16 \pi E
    , \\
    \label{eq:momentum2}
    & 
      \nabla_j \left( K^{ij} - \gamma^{ij} K \right) = 8 \pi j^i 
    .
  \end{align}
\end{subequations}
Note that we can write the covariant derivatives $\nabla_k$ and the Lie derivatives $\mathcal{L}_\beta$ in terms of partial derivatives of the coordinates $x^i$, and the Ricci tensor $R_{ij}$ and Ricci scalar $R$ in terms of $\gamma_{ij}$ and its derivatives in the usual way. This way, assuming that the source terms $E$, $j_i$, $S_{ij}$ are given, we have a second-order non-linear system of PDEs with the unknowns $\gamma_{ij}$, $K_{ij}$, 
$\alpha$, $\beta^i$.

The above equations~\eqref{eq:ADM-eq} are known in the numerical relativity community as the \emph{ADM equations}, after the work of Arnowitt, Deser and Misner~\cite{Arnowitt:1962hi} on their Hamiltonian formulation of general relativity.
In this form, however, the equations were in fact first written by York~\cite{York1979} (in four spacetime dimensions), and are thus sometimes referred to as the ADM-York equations.

By now we have cast the Einstein equations on an explicit $(D-1)$-dimensional dynamical system form. Note, however, that whereas equations~\eqref{eq:lie2} and~\eqref{eq:proj_sigma_spacial2} are evolution equations, equations~\eqref{eq:hamiltonian2} and~\eqref{eq:momentum2} are not. These last two equations constitute \emph{constraints} that the system must satisfy at all times. In particular, these constraints must be satisfied at $t=0$. So we would now need to specify the relevant initial conditions, satisfying equations~\eqref{eq:hamiltonian2} and~\eqref{eq:momentum2}, and then evolve them using equations~\eqref{eq:lie2} and~\eqref{eq:proj_sigma_spacial2}, while making sure that equations~\eqref{eq:hamiltonian2} and~\eqref{eq:momentum2} always hold.

The question arises: given a specific physical problem (say, a head-on collision of two black holes), how does one specify the initial conditions that correspond to the problem we have in mind? This is the \emph{initial data} problem
which will be the focus of the next chapter.



\chapter{Initial data}
\label{chap:init_data}

On any dynamical system, to perform an evolution
one needs to supply the initial conditions. In our case, this amounts to providing a snapshot of the gravitational fields on some hypersurface---the \emph{initial data}. Then, one evolves this data to neighbouring hypersurfaces and so on. 

In general relativity initial data cannot be freely specified.
As we have seen in chapter~\ref{chap:d-1+1}, not all of Einstein's equations are evolutions equations. 
We also have a set of constraint equations that must be satisfied at all times, the Hamiltonian~\eqref{eq:hamiltonian2} and momentum~\eqref{eq:momentum2} constraints.
In particular, these equations need to be solved at $t=0$ for the $(\gamma_{ij}, K_{ij})$ that represent the physical system we are interested in evolving. 
We then feed these values to the evolution equations themselves.

In general, this step is far from trivial. There is no unique recipe for the writing of initial data corresponding to an arbitrary gravitational system. 
For some systems, however,---such as vacuum spacetimes with moving black holes---recipes do exist. 
Actually, for the four-dimensional case, several methods for constructing initial data for different systems have been explored over the years (see~\cite{Cook:2000vr} for a review). 
For higher-dimensional systems, however, only recently the ``standard'' way of constructing initial data for moving black holes in the vacuum was generalised~\cite{Yoshino:2005ps,Yoshino:2006kc}.

In this chapter we will give an overview of the procedure of \emph{conformal decomposition} first introduced by York and Lichnerowicz~\cite{Lichnerowicz1944,
  York1971,York1972,York1973} which rearranges the degrees of freedom contained
in the three-metric $\gamma_{ij}$ and extrinsic curvature $K_{ij}$ via a
conformal transformation and a split of the curvature into trace and traceless
part
followed by a transverse-traceless decomposition of the conformally rescaled traceless
extrinsic curvature.

We will focus specifically on initial data for vacuum spacetimes, generalising
the well-known Brill-Lindquist~\cite{Brill:1963yv} and
Bowen-York~\cite{Bowen:1980yu,Brandt:1997tf} initial data along the lines
of~\cite{Yoshino:2005ps,Yoshino:2006kc}.
For alternative procedures to tackle the initial data problem we refer the reader to Cook's review~\cite{Cook:2000vr}, Alcubierre's book~\cite{Alcubierre:2008}, the recent book by Baumgarte \& Shapiro~\cite{baumgarte2010numerical} and references therein.

As in the previous chapter, Greek indices are spacetime indices running from 0 to $D-1$; Latin indices are spatial indices, running from 1 to $D-1$.

\section{Conformal decomposition}
\label{chap:conformal}

\subsection{Conformal transformations}

We start by recalling a known general result: 
given an $N$-dimensional manifold with metric $g_{\mu\nu}$, if one performs the conformal transformation 
\begin{equation}
  g_{\mu\nu}
  = \phi \left( x^\alpha \right) \hat g_{\mu\nu},
\end{equation}
the Ricci scalars relative to the metrics $g_{\mu\nu}$ and $\hat g_{\mu\nu}$ are related by
\begin{equation}
  R = \frac{\hat R}{\phi} + \frac{1-N}{\phi^2} \hat \nabla^\alpha \partial_\alpha \phi
  -\frac{\partial_\alpha \phi \partial^\alpha\phi}{4 \phi^3} (1-N)(6-N),
\end{equation}
where $\hat \nabla$ is the covariant derivative associated with the conformal metric $\hat{g}_{\mu\nu}$.

Let us now consider our case, where we have a $(D-1)$-dimensional spacelike slice with induced metric $\gamma_{ij}$ and ``conformal metric'' $\hat \gamma_{ij}$. We have $N = D-1$, and we make
\begin{align*}
  \phi = \psi^p, \qquad p = \frac{4}{D-3}.
\end{align*}
We have
\begin{equation}
  R = \psi^{-p} \hat R 
  + (2-D) p \, \psi^{-p-1} \hat \nabla^k \partial_k \psi.
\end{equation}

We further decompose the extrinsic curvature in trace and trace-free parts,
\begin{equation}
  \label{eq:Kij-decomposition}
  K_{ij} \equiv A_{ij} + \frac{K}{D-1} \gamma_{ij}, 
\end{equation}
where $ K \equiv \gamma^{ij}K_{ij}$ and, by definition, $\gamma^{ij} A_{ij} = 0$. 
Defining $A^{ij} = \gamma^{ik} \gamma^{jl} A_{kl}$, we can also write
\begin{equation}
  K^{ij} \equiv A^{ij} + \frac{K}{D-1} \gamma^{ij}.
\end{equation}

\subsubsection{Conformal decomposition of the Hamiltonian and  momentum constraint}

Under such a transformation, it is a matter of simple substitution to show that the Hamiltonian constraint equation~\eqref{eq:hamiltonian2} takes the form
\begin{equation}
\label{eq:conform_hamilton}
  \hat \triangle \psi + \frac{\psi}{p(2-D)} \hat R - \frac{\psi^{p+1}}{p(2-D)} A^{ij} A_{ij}
  -\frac{\psi^{p+1}}{p(D-1)} K^2 = 16 \pi E \frac{\psi^{p+1}}{p(2-D)},
\end{equation}
where $\hat \triangle \equiv \hat \nabla^k \hat \nabla_k $.

With a straightforward calculation we can easily show that
\begin{equation}
  \nabla_i A^{ij} =  \psi^{-q} \hat \nabla_k 
  \left(
    \psi^q A^{kj}
  \right),
\end{equation}
with $q \equiv  2 \frac{D+1}{D-3} $. 
Thus, we define
\begin{equation}
  \hat A^{ij} \equiv \psi^q A^{ij} \equiv  \psi^{2 \frac{D+1}{D-3}} A^{ij},
\end{equation}
and we will lower its indices with $\hat \gamma_{ij}$,
\begin{equation}
  \hat A_{ij} \equiv \hat \gamma_{ik} \hat \gamma_{jl} \hat A^{kl} 
  = \psi^2 A_{ij}.
\end{equation}
We thus have
\begin{equation*}
  \nabla_i K^{ij} = \psi^{-q} \hat \nabla_i \hat A^{ij} 
  + \frac{\psi^{-p}}{D-1} \hat \nabla^j K . 
\end{equation*}

Equation~\eqref{eq:momentum} is then written in the form
\begin{equation}
\label{eq:conform_momentum}
  \hat \nabla_i \hat A^{ij} - \frac{D-2}{D-1} \psi^{ 2 \frac{D-1}{D-3}  } \, \hat \nabla^j K
= 8 \pi \psi^{ 2 \frac{D+1}{D-3} } \, j^j.
\end{equation}

All we need now is to write equation~\eqref{eq:conform_hamilton} as function of $\hat A_{ij}$, which is very easy. Our system is now
\begin{align}
\label{eq:conform_hamilton2}
& 
  \hat \triangle \psi - \frac{D-3}{4(D-2)} \psi \hat R 
  + \frac{D-3}{4(D-2)}  \psi^{-\frac{3D-5}{D-3} } \hat A^{ij} \hat A_{ij}
  - \frac{D-3}{4(D-1)} \psi^{ \frac{D+1}{D-3}  }  K^2 
  = - 4 \pi E \frac{D-3}{D-2}  \psi^{ \frac{D+1}{D-3} }
,   \\
\label{eq:conform_momentum2}
&
\hat \nabla_i \hat A^{ij} - \frac{D-2}{D-1} \psi^{ 2 \frac{D-1}{D-3}  } \, \hat \nabla^j K
= 8 \pi \psi^{ 2 \frac{D+1}{D-3} } \, j^j
,
\end{align}
where 
\begin{align*}
  g_{\mu\nu}\dd x^\mu  \dd x^\nu = -\alpha^2 \dd t^2 +  \psi^{ \frac{4}{D-3}  } \hat \gamma_{ij} 
  \left(\dd x^i + \beta^i \dd t \right)   \left(\dd x^j + \beta^j \dd t \right) \,. 
\end{align*}

\section{Initial data for vacuum spacetimes}

Let us now consider the equations~\eqref{eq:conform_hamilton2} and~\eqref{eq:conform_momentum2} for the particular case of vacuum solutions ($j^i = 0 = E$). We further impose that the ``conformal metric'' $\hat \gamma_{ij}$ is flat (and, thus, $\hat R = 0$ ) and the ``maximum slicing condition'', $K = 0 $ (to be discussed in section~\ref{sec:1+log_0}).  
The equations~\eqref{eq:conform_hamilton2} and~\eqref{eq:conform_momentum} greatly simplify, and we are left with
\begin{align}
  &  \partial_i \hat A^{ij} = 0, \label{eq:vacuum_hamilton} \\
  &  \hat \triangle \psi + \frac{D-3}{4(D-2)} \psi^{- \frac{3D -5}{D-3}   } \hat A^{ij} \hat A_{ij} = 0, \label{eq:vacuum_momentum}
\end{align}
where $\hat \triangle \equiv \partial_i \partial^i $ is now the flat space Laplace operator.


Here we make a pause to recall that the Schwarzschild-Tangherlini metric in $D$ dimensions is
\begin{equation}
  \dd s^2 = -\left(1 - \frac{\mu}{r^{D-3}} \right) \dd t^2 + \frac{\dd r^2}{1 - \frac{\mu}{r^{D-3}} } 
    + r^2 \dd \Omega^2_{D-2},
\end{equation}
where $ \mu = \frac{16 \pi M}{(D-2)\mathcal{A}_{D-2}} $, $M$ being the mass of the black hole and $\mathcal{A}_{N-1} = \frac{2\pi^{N/2}}{\Gamma(N/2)} $  the area of the hypersphere. By performing the coordinate transformation
\begin{align*}
  r = R \left(
    1 + \frac{\mu}{4 R^{D-3}}
  \right)^{ \frac{2}{D-3}  }
\end{align*}
we can write it in isotropic coordinates as 
\begin{equation}
  \label{eq:schwarzs}
  \dd s^2 = - \left(1 - \frac{16 R^{3 + D} \mu}{(4 R^D + R^3 \mu)^2}  \right) \dd t^2   
    + \left(1 + \frac{\mu}{4R^{D-3}} \right)^{\frac{4}{D-3}}
    \left( \dd R^2 + R^2 \dd \Omega^2_{D-2}
    \right).
\end{equation}
We will shortly make use of this geometry.

\subsection{Brill-Lindquist initial data}
\label{sec:brill}

We now assume that the extrinsic curvature vanishes identically, $K_{ij} = 0$, a condition that holds for \emph{time-symmetric} initial data. It can be shown~\cite{Gourgoulhon:2007ue} that  if $K_{ij} = 0 $ and we choose coordinates such that $\alpha = 1$, we have 
\begin{align*}
  \mathcal{L}_m g_{\alpha\beta} = 0,
\end{align*}
which means that, locally, $ m^\mu$ is a Killing vector. $m^\mu$ is also orthogonal to the hypersurface $\Sigma_{t=0}$, and as such this configuration is static. This property only holds locally (on $\Sigma_{t = 0}$) and we therefore call this configuration \emph{momentarily static}.

For $K_{ij} = 0$ equation~\eqref{eq:vacuum_hamilton} is automatically satisfied, and~\eqref{eq:vacuum_momentum} reduces to the standard $D-1$-dimensional flat space Laplace equation,
\begin{equation}
  \label{eq:lappsi}
  \hat \triangle \psi = 0.
\end{equation}
We impose the following conditions on $\psi$
\begin{equation}
\label{eq:lappsi_conditions}
  \lim_{ r \to \infty} \psi = 1,
\end{equation}
which 
is the 
asymptotic flatness condition (remember that $  \gamma_{ij} = \psi^{ \frac{4}{D-3}  } \hat \gamma_{ij}$).

Let $r_{(i)} \equiv |r - x_{(i)}| $, where the $x_{(i)}$ are arbitrary points in our spacetime. A solution to equation~\eqref{eq:lappsi} is given by
\begin{equation}
  \label{eq:psi_BL}
  \psi = 1 + \sum_{i=1}^{N} \frac{C_{(i)}}{r_{(i)}^{D-3}},
\end{equation}
where the $C_{(i)}$ are arbitrary constants. Note that equation~\eqref{eq:psi_BL} obeys the condition~\eqref{eq:lappsi_conditions}. The spatial metric takes the form (recall that the conformal metric $\hat \gamma_{ij}$ is flat)
\begin{equation}
  \label{eq:BL-sol}
  \gamma_{ij} \dd x^i \dd x^j 
  = \left(
    1 + \sum_{i=1}^{N} \frac{C_{(i)}}{r_{(i)}^{D-3}} 
  \right)^{ \frac{4}{D-3} }
  \left(
    \dd r^2 + r^2 \dd \Omega_{D-2}^2
  \right).
\end{equation}
This solution is asymptotically flat (by construction), and if we compare this expression with~\eqref{eq:schwarzs}, we can identify $\mu = 4 \sum_{i=1}^{N} C_{(i)} $, which is the mass parameter measured in the ``principal sheet'' (anticipating the interpretation). 

We now have to analyse what happens as $r \to x_{(i)}$, for a given $i$. When  $r \to x_{(i)}$, $ r_{(i)} \to 0$ and $r_{(j)} \to r_{(i)(j)} \equiv |x_{(i)}-x_{(j)}|$. Setting the origin of our coordinate system at $r = r_{(i)}$ ($x_{(i)} = 0$) we have
\begin{equation*}
  \dd s^2 = \left(
    \frac{C_{(i)}}{r_{(i)}^{D-3}}
  \right)^{ \frac{4}{D-3} }
   \left[
     1 + \frac{r_{(i)}^{D-3}}{C_{(i)}}
     \left(
       1 + \sum_{j \neq i}^N  \frac{C_{(j)}}{ r_{(j)}^{D-3} }
     \right)
   \right]^{ \frac{4}{D-3} }
  \left(
    \dd r_{(i)}^2 + r_{(i)}^2 \dd \Omega_{D-2}^2
  \right),
\end{equation*}
and when $r_{(i)} \to 0$
\begin{equation*}
  \dd s^2 \to \left(
    \frac{C_{(i)}}{r_{(i)}^{D-3}}
  \right)^{ \frac{4}{D-3} }
  \left[
    1 + A_{(i)} \, \frac{r_{(i)}^{D-3}}{C_{(i)}}
  \right]^{ \frac{4}{D-3} }
  \left(
    \dd r_{(i)}^2 + r_{(i)}^2 \dd \Omega_{D-2}^2
  \right), 
\end{equation*}
where we defined $A_{(i)} \equiv 1 + \sum_{j \neq i}^N \frac{C_{(j)}}{r_{(i)(j)}^{D-3} } $. 
With the coordinate transformation $r'_{(i)} = \frac{C_{(i)}^{ \frac{2}{D-3} } }{r_{(i)}} $ we have
\begin{equation}
\label{eq:sheet}
  \dd s^2 \mathop{ \xrightarrow[r_{(i)} \to 0]{}   }_{r'_{(i)} \to \infty}
  \left(
    1 + A_{(i)} \, \frac{C_{(i)}}{r'_{(i)}{}^{D-3}}
  \right)
  \left(
    \dd r'_{(i)}{}^2 + r'_{(i)}{}^2 \dd \Omega_{D-2}^2
  \right). 
\end{equation}
This shows that in this limit the space is also asymptotically flat. Thus, our solution~\eqref{eq:BL-sol} describes a space with $N+1$ asymptotically flat regions. Note that all ``lower sheets'' are separate, i.e., there is no way to travel from one sheet to the other except through the ``upper sheet'' (or ``principal sheet''). Equation~\eqref{eq:sheet} shows that each sheet, asymptotically, has a Schwarzschild-Tangherlini form, with the mass measured in the $i$th sheet being given by
\begin{equation}
  \bar \mu_{(i)} = 4 A_{(i)} C_{(i)} 
  = 4 \left(
    C_{(i)} + \sum_{j \neq i}^{N} \frac{C_{(i)} C_{(j)}}{r_{(i)(j)}^{D-3} }
  \right).
\end{equation}
The observer located on the principal sheet (the ($N+1$)th sheet) is the only one that sees a system of $N$ black holes, with total mass $ \mu_{\mathrm{ADM}} =  \mu_{N+1} = 4 \sum_{i=1}^{N} C_{(i)}  $, as we had already mentioned. Thus we identify $ \mu_{(i)} \equiv 4 C_{(i)}$ and rewrite our expressions in terms of $\mu_{(i)}$,
\begin{align*}
  \psi & = 1 + \sum_{i=1}^{N} \frac{\mu_{(i)}}{4 r_{(i)}^{D-3}}, \\
  \bar \mu_{(i)} &  = \mu_{(i)} \left(
    1  + \sum_{j \neq i}^{N} \frac{ \mu_{(j)}}{4 r_{(i)(j)}^{D-3} }
  \right), \\
  \mu_{\mathrm{ADM}} &  = \sum_{i=1}^N \mu_{(i)},
\end{align*}
where $r_{(i)} \equiv |r - x_{(i)}|$ and  $r_{(i)(j)} \equiv |x_{(i)} - x_{(j)}| $. The points $x_{(i)}$ are called \emph{punctures}.

Note that $\mu \neq \sum_i ^N \bar \mu_{(i)} $. This difference can be attributed to the interaction energy between the black holes. 
It is important to note that $\mu_i$, as we have defined it, is just a convenient label for the mass of the $i$th black hole (but is \emph{not} the mass). The mass of the $i$th black hole as measured on the $i$th sheet (its ``bare mass''), is given by $\bar \mu_{(i)}$.\footnote{There seems to be some mismatch in the literature as to the definition of ``bare mass''. Brill and Lindquist~\cite{Brill:1963yv} clearly define it as $\bar \mu_{(i)}$, in our notation, and they even point out that the sum of the bare masses is different from the total mass. However, Brandt and Brügmann~\cite{Brandt:1997tf} seem to define bare mass as $\mu_{(i)}$.  }

\subsection{Bowen-York initial data}
\label{sec:bowen}

Brill-Lindquist initial data is very useful because it provides us with an analytical solution for the constraint equations. However, it also has little physical relevance. Generally, one is interested in solutions with black holes that are spinning and moving and as such Brill-Lindquist data is clearly not enough. 

In order to have a more general configuration, i.e. one that is not momentarily static, we cannot impose $K_{ij} = 0$. Let us recall that our assumptions are: $K = 0$---the maximum slicing condition; $ \hat \gamma_{i j}$ is flat---the conformal flatness condition; and $ \lim_{r \to \infty} \psi = 1$---the asymptotic flatness condition.


We now start by writing $\hat A^{ij} $ in the form
\begin{equation}
  \hat A^{ij} = (  \hat L X  )^{ij}
  + \hat A^{ij}_{\mathrm{TT} },
\end{equation}
where
\begin{equation}
\label{eq:Lij}
  (\hat L X )^{ij} \equiv \hat \nabla^i X^j + \hat \nabla^j X^i 
  - \frac{2}{D-1} \hat \nabla_k X^k \hat \gamma^{ij}.
\end{equation}
By construction, $ (\hat L X )^{ij} \hat \gamma_{ij} = 0 $, and we impose $ \hat \gamma_{ij} \hat A^{ij}_{\mathrm{TT}} = 0 = \hat \nabla_j \hat A^{ij}_{\mathrm{TT}} $. We will also restrict ourselves to the case $ \hat A^{ij}_{\mathrm{TT}} = 0$. The equations~\eqref{eq:vacuum_hamilton} and~\eqref{eq:vacuum_momentum} take the form
\begin{align}
  & \hat \triangle X^j + \frac{D-3}{D-1} \partial^j \partial_i X^i = 0, \label{eq:BY_eq1} \\
  & \hat \triangle \psi + \frac{D-3}{4(D-2)} \psi^{ - \frac{3D-5}{D-3}  } \hat A^{ij} \hat A_{ij} = 0, \label{eq:BY_eq2} \\
  & \hat A^{ij} = (\hat L X )^{ij}. \label{eq:BY_eq3}
\end{align}
Thus, we have to solve~\eqref{eq:BY_eq1}, plug $X^j$ in~\eqref{eq:BY_eq3} and then solve~\eqref{eq:BY_eq2}. We will see that, even though we will be able to solve~\eqref{eq:BY_eq1} analytically, we generally have to use numerical methods to solve~\eqref{eq:BY_eq2}.

To solve~\eqref{eq:BY_eq1} we make the following decomposition~\cite{Yoshino:2006kc}, which introduces functions $\lambda$ and $V_j$,
\begin{align}
  \label{eq:X_j}
  X_j = \frac{3D-5}{D-3} V_j - 
  \left(
    \partial_j \lambda + x^k \partial_j V_k
  \right).
\end{align}
Equation~\eqref{eq:BY_eq1} then gets the form
\begin{align*}
  \frac{3D-5}{D-3} \hat \triangle V_j - x^k \partial_j \hat \triangle V_k
  - 2\frac{D-2}{D-1} \partial_j \hat \triangle \lambda
  - \frac{D-3}{D-1} \partial_j
  \left(
    x^k \hat \triangle V_k
  \right) = 0,
\end{align*}
which is solved if
\begin{align}
    \label{eq:lap_V-lambda}
    \left\{
      \begin{aligned}
        \hat \triangle V_j & = 0 \\
        \hat \triangle \lambda & = 0
      \end{aligned}
      \right. .
\end{align}
We have reduced our problem to solving two flat space Laplace equations, which have known analytical solutions.
In the following we analyse some possible solutions~\cite{Yoshino:2006kc}.

\subsubsection{Moving black holes}
\label{sec:momenta}

We start by choosing a solution for the system~\eqref{eq:lap_V-lambda} of the form
\begin{align}
  \label{eq:V_j-P}
  V_j = - \frac{2\pi}{(D-2) \mathcal{A}_{D-2} }  \frac{P_j}{r^{D-3}}, \qquad
  \lambda = 0.
\end{align}
$\mathcal{A}_N$ stands for the area of the $N$-dimensional hypersphere. $P_j$ are constants that, as we shall see, will be the linear momentum of the black hole in the $j$ direction.

For such an \emph{ansatz} we have, from equation~\eqref{eq:X_j},
\begin{align}
  \label{eq:X_j-P}
  X_j = - \frac{2\pi }{(D-2) A_{D-2} } \frac{1}{r^{D-3}}
  \left( 
    \frac{3D-5}{D-3} P_j + (D-3) n^k P_k n_j 
  \right),
\end{align}
where $n_j \equiv \frac{x_j}{r} $, and from equations~\eqref{eq:BY_eq3} and~\eqref{eq:Lij}
\begin{align}
  \label{eq:A_ij-P}
  \hat A^{ij} = \frac{4 \pi (D-1)}{(D-2)A_{D-2} } \frac{1}{r^{D-2} }
  \left(
    n^i P^j + n^j P^i - n_k P^k \hat \gamma_{ij} + (D-3) n^i n^j P^k n_k
  \right).
\end{align}

The ADM linear momentum is given by~\cite{Bowen:1980yu,Yoshino:2006kc,Gourgoulhon:2007ue}
\begin{align}
  \label{eq:P_ADM}
  P_i^{\mathrm{ADM}} = \frac{1}{8 \pi}
  \int_{r \to \infty} \left(
    K_{ij} n^j - K n_i
  \right) \sqrt{q} \, \dd ^{D-2}y ,
\end{align}
where we perform the integration on a hypersphere at infinity; $ \sqrt{q} \dd ^{D-2}y $ denotes the induced metric on the hypersphere---using spherical coordinates $ \sqrt{q}  \dd ^{D-2}y = r^{D-2} \dd \Omega_{D-2}  $ (we can write the induced metric on a hypersphere $\mathcal{S}$ of radius $r$ as $\dd s^2_{\mathcal{S}} = q_{A B} \dd y^A \dd y^B = r^2 \dd \Omega^2_{D-2}$, and thus $q = \det {q_{AB}} = (r^2)^{D-2} \dd \Omega_{D-2}^2   $); $n^j$ is its unit normal vector.

Reminding ourselves that 
\begin{align*}
  \hat A^{ij} & = \psi^{ 2 \frac{D+1}{D-3} } A^{ij}, \\
  \hat A_{ij} & = \psi^{ 2 } A_{ij}, \\
  K_{ij} & = A_{ij} \qquad \text{(we are considering $K = 0$)}, \\
  \psi & = 1 
  + O \left(
    \frac{1}{r}
  \right),
\end{align*}
we see that we can calculate the ADM linear momentum without knowing $\psi$. Plugging~\eqref{eq:A_ij-P} into~\eqref{eq:P_ADM} we have that $P_i^{\mathrm{ADM}} = P_i$, as expected.

Finally, we note that, as the equation~\eqref{eq:vacuum_hamilton} is linear, we can superimpose $N$ solutions of the type~\eqref{eq:P_ADM} corresponding to $N$ Schwarzschild black holes,
\begin{align}
  \hat A^{ab}_P = \sum_{i = 1}^N \hat A_{P(i)}^{ab},
\end{align}
where
\begin{align}
  \label{eq:Aab_i}
  \hat A^{ab}_{P(i)} = \frac{4 \pi (D-1)}{(D-2)A_{D-2} } \frac{1}{r^{D-2}_{(i)} }
  \left(
    n^a_{(i)} P^b_{(i)} + n^b_{(i)} P^a_{(i)} 
    - (n_{(i)})_k P^k_{(i)} \hat \gamma_{ab} + (D-3) n^a_{(i)} n^b_{(i)} P^k_{(i)} (n_{(i)})_k
  \right),
\end{align}
where $n^a_{(i)} \equiv \frac{x^a - x^a_{(i)} }{r_{(i)}} $ and the parameters $ P^a_{(i)} $ correspond to the ADM momentum of the $i$th black hole when the separation of the holes is very large.

\subsubsection{Spinning black holes}
\label{sec:spinning-black-holes}

Let us now try a solution of~\eqref{eq:lap_V-lambda} of the form
\begin{align}
  \label{eq:V_j-S}
  V_j =  \frac{(D-3)\pi}{(D-2) A_{D-2} }  \frac{J_{jk} \, n^k }{r^{D-2}}, \qquad
  \lambda = 0,
\end{align}
where $J_{jk} = -J_{kj}$ will be the angular momentum tensor of the black hole.  
We have
\begin{align}
  \label{eq:X_j-S}
  X_j = \frac{4 \pi }{A_{D-2}} J_{jk} \frac{x^k}{r^{D-1}},
\end{align}
and 
\begin{align}
  \label{eq:A_ij-S}
  \hat A^{ij} = - \frac{4\pi (D-1)}{A_{D-2}} \frac{1}{r^{D-1}}
  \left(
    J^{jk} n_k n^i + J^{ik} n_k n^j
  \right).
\end{align}

The ADM angular momentum is given by (when $K = 0$)~\cite{Bowen:1980yu,Yoshino:2006kc,Gourgoulhon:2007ue}
\begin{align}
  \label{eq:JADM}
  J_{ik}^{\mathrm{ADM}} = \frac{1}{8 \pi} \int_{r \to \infty}
  \left(
    x_i K_{jk} - x_j K_{ik}
  \right) n^k  \sqrt{q} \, d^{D-2}y .
\end{align}
We can check that $ J_{ik}^{\mathrm{ADM}} = J_{ik} $. 

For $D = 4$ we can define the angular momentum vector in the usual way,
\begin{align}
  J^i = \frac{1}{2} \epsilon^{ijk} J_{kl}.
\end{align}

As in the previous section, we can now superimpose $N$ solutions of the type~\eqref{eq:A_ij-S},
\begin{align}
  \hat A^{ab}_J = \sum_{i = 1}^N \hat A_{J(i)}^{ab},
\end{align}
where
\begin{align}
  \label{eq:Aab-J}
  \hat A^{ab}_{J(i)} = - \frac{4\pi (D-1)}{A_{D-2}} \frac{1}{r^{D-1}_{(i)}}
  \left(
    J^{bk}_{(i)} (n_{(i)})_k n^a_{(i)} + J^{ak}_{(i)} (n_{(i)})_k n^b_{(i)}
  \right).
\end{align}
The parameters $ J^{ab}_{(i)} $ correspond to the ADM angular momentum of the $i$th black hole when the separation of the holes is very large.

\subsubsection{General case}
\label{sec:general-case}

We can now combine the results from the two previous sections to build a solution of $N$ black holes with arbitrary linear momentum and spin, 
\begin{align}
\label{eq:gen_Aab}
  \hat A^{ab} = \sum_{i = 1}^N \left( 
    \hat A_{P(i)}^{ab} +  \hat A_{J(i)}^{ab} 
  \right),
\end{align}
where $\hat A_{P(i)}^{ab}$ and $\hat A_{P(i)}^{ab}$ are given by equations~\eqref{eq:Aab_i} and~\eqref{eq:Aab-J}.

\begin{description}
\item[Note:] This solution reduces to the Brill-Lindquist momentarily static solution ($K_{ij} = 0$) when $P^a_{(i)} = 0 =  J^{ab}_{(i)}$.
For $N = 1$, $ J^{ab} \neq 0$ and $P^a = 0$, however, we do \emph{not} have a slice of a Kerr (or, for the higher-dimensional case, Myers-Perry) spacetime. It has actually been shown~\cite{Garat:2000pn} that there is no foliation of the Kerr spacetime that is axisymmetric, conformally flat, and reduces smoothly to the Schwarzschild solution in the non-rotating limit.\footnote{For the four-dimensional Kerr solution, but there is no reason to believe that the higher-dimensional case is any different.} This means that our Bowen-York solution with $ J^{ab} \neq 0$ does represent a rotating black hole, but not a stationary one. For the four-dimensional case, when we evolve the data, the system emits gravitational radiation and eventually settles down to the Kerr solution~\cite{Brandt:1994db,Gleiser:1997ng} (the higher-dimensional case has not been studied as of yet). This spurious gravitational radiation has no desirable physical properties and is often referred to as ``junk radiation''.
\end{description}

\subsubsection{Conformal factor}
\label{sec:conformal-factor}

We still need to solve equation~\eqref{eq:BY_eq2} to get the full initial data, and now there is no hope of finding an analytical solution. Let us rewrite the equation we need to solve,
\begin{align}
  \label{eq:triangle_psi}
  \hat \triangle \psi + \frac{D-3}{4(D-2)} \psi^{ - \frac{3D-5}{D-3}  } \hat A^{ij} \hat A_{ij} = 0,
\end{align}
with $\hat A_{ij}$ given by~\eqref{eq:gen_Aab}.

Along the lines of~\cite{Brandt:1997tf} and~\cite{Yoshino:2006kc} we write
\begin{align}
  \label{eq:psi-decom}
  \psi = \psi_{\mathrm{BL}} + u,
\end{align}
where
\begin{align}
  \label{eq:psiBL}
  \psi_{\mathrm{BL}} \equiv 1 + \sum_{i = 1}^N \frac{\mu_{(i)} }{4 r_{(i)}^{D-3}} .
\end{align}
Equation~\eqref{eq:triangle_psi} then takes the form
\begin{align}
  \hat \triangle u + \frac{D-3}{4(D-2)} \hat A^{ab} \hat A_{ab}
       \psi^{ -\frac{3D-5}{D-3} } = 0\ . 
  \label{eq:u}
\end{align}
For the four-dimensional case, Brandt and Br\"ugmann~\cite{Brandt:1997tf} were able to show the existence and uniqueness of $C^2$ solutions for the above equations. Furthermore, the solution for $u$ is found on an Euclidean manifold; we do not need to impose inner boundary conditions to avoid singularities. Brandt and Brügmann also show that this solution is the ``natural'' generalisation of the Brill-Lindquist initial data, i.e., each puncture represents the infinity of another asymptotically flat region of the spacetime and there is no way to travel from one sheet to the other except through the ``upper'' sheet. The higher-dimensional case has not been thoroughly studied, but it is believed that the situation is not radically different~\cite{Yoshino:2006kc}.

\section{Final remarks}

In this chapter we introduced tools for the 
construction of initial data for higher-dimensional numerical relativity.
As we mentioned, even though the four-dimensional case has been thoroughly studied, the study of initial data for higher-dimensional systems started only very recently.
As of yet, only Brill-Lindquist and Bowen-York initial data have been generalised, but with these two approaches one is already able to construct quite interesting systems for vacuum spacetimes.
In particular, the Bowen-York approach allows us to write initial data for spacetimes with an arbitrary number of moving and spinning black holes. 

For the four-dimensional case there are also powerful computer codes to solve the elliptic equation~\eqref{eq:u}, such as the spectral method presented by Ansorg et~al.~\cite{Ansorg:2004ds}. 

In upcoming chapters we will present a generalisation of the spectral solver in~\cite{Ansorg:2004ds} that solves~(\ref{eq:u})
for black hole binaries in $D\ge 5$ dimensions with non-vanishing initial boost,
and preserves the spectral convergence properties observed in four
dimensions.



\chapter{Numerical implementation}
\label{chap:bssn-gauge}

In chapter~\ref{chap:d-1+1}, we have written Einstein's field equations
explicitly in a form (usually referred to as ADM equations~\eqref{eq:ADM-eq})
which one could easily give to a computer to evolve.  As can be seen from this
system of equations, though, we are still not quite ready to perform numerical
evolutions: we still need to say what happens with the variables $\alpha$
(lapse) and $\beta^i$ (shift).  The Einstein equations have not imposed any
evolution equation for these variables.  This reflects our coordinate freedom:
fixing the lapse function and shift vector is a gauge choice, which one could in
principle do arbitrarily.  In turns out, though, that a good choice is crucial
to achieve a stable numerical integration.
We will in this chapter briefly discuss why this is the case and write down the equations we will be using throughout this work.

It also turns out, as researchers eventually found out empirically in the 1990s
when full three-dimensional evolutions were attempted using the ADM equations,
that this system of evolution equations is not well suited to obtain long-term
stable numerical simulations.  This is now known to be due to the fact that the
ADM equations are only weakly hyperbolic.\footnote{The ADM equations do allow
  stable evolutions in spherical symmetry, though (see e.g.~\cite{Alcubierre:2008}).}  People started
experimenting with reformulations of the ADM equations and in 1998 Baumgarte and
Shapiro---revisiting an earlier formulation based on conformal transformations
by Nakamura, Oohara and Kojima~\cite{Nakamura:1987zz} and Shibata and
Nakamura~\cite{Shibata:1995we}---showed that this formulation behaved much
better than ADM for all cases considered~\cite{Baumgarte:1998te}.  This
formulation became known as BSSN (Baumgarte, Shapiro, Shibata and Nakamura) and
is today the most popular scheme used to evolve Einstein's equations.

It was later realised that indeed the BSSN scheme can be shown to be strongly hyperbolic, as opposed to only weakly hyperbolic like in the ADM case, and thus well-posed, e.g.~\cite{Sarbach:2002bt,Yoneda:2002kg}.

We should also mention that other successful evolution schemes do exit.
Most notably, the \emph{generalised harmonic coordinates} approach, e.g.~\cite{Friedrich:1996hq}, was successfully used by Pretorius in the first ever evolutions of binary black holes through several orbits~\cite{Pretorius:2005gq}.
Giving a full overview of such topics falls outside of the scope of this work.
We will in this chapter merely introduce the BSSN evolution equations and we refer the interested reader to, e.g.,~\cite{Alcubierre:2008,baumgarte2010numerical} for comprehensive overviews.

We close this chapter by introducing the numerical code itself used for all the simulations to be presented.

In this chapter, we restrict ourselves to the four-dimensional case (for completeness, we present in appendix~\ref{ch:ddim-bssn} the higher-dimensional BSSN scheme), and therefore spatial (Latin) indices are here restricted to $i=1,2,3$.

\section{BSSN formulation}
\label{sec:bssn}

As we have just mentioned, if we were to try and evolve Einstein's equations in the ADM formulation~\eqref{eq:ADM-eq} (supplemented with the gauge conditions we will introduce in the next section)
we would find out that the system is severely unstable.
In this section we recast the evolution equations in the BSSN form, which allows for stable numerical evolutions.

We start by performing a conformal decomposition of the spatial metric $\gamma_{ij}$ in the following way (observe it is going to be a different decomposition from the one performed in the study of the initial data)
\begin{equation}
\label{eq:conformal-metric}
\tilde \gamma_{ij} \equiv \chi \gamma_{ij} \,.
\end{equation} 
The conformal factor $\chi$ can in principle be freely prescribed.
In the BSSN scheme, one imposes that the determinant of the conformal metric be equal to the determinant of the flat metric $\eta_{ij}$,
\begin{equation}
\label{eq:chi-def}
\chi = \left( \frac{\gamma}{\eta} \right)^{-1/3} \,.
\end{equation}
By construction, we have
\begin{equation}
\label{eq:dettildegamma}
\det \tilde \gamma_{ij} = \eta \,.
\end{equation}
Since we will stick to Cartesian coordinate systems throughout this work, we will always have $\eta = 1 = \det \tilde \gamma_{ij}$, which makes $\chi$ a scalar density with weight $-2/3$.

Just like in equation~(\ref{eq:Kij-decomposition}), we decompose the extrinsic curvature $K_{ij}$ into trace and trace-free parts and apply the conformal transformation we used for the metric to the traceless part,
\begin{equation}
\label{eq:Aij-def}
K_{ij} \equiv \chi^{-1} \left( \tilde A_{ij} + \frac{K}{3} \tilde \gamma_{ij} \right) \,.
\end{equation}

Let us now find evolution equations for the variables we have introduced ($\chi, \tilde{\gamma}_{ij}, K, \tilde{A}_{ij}$).
Inserting~(\ref{eq:conformal-metric}) and (\ref{eq:Aij-def}) into~(\ref{eq:lie2}) and (\ref{eq:proj_sigma_spacial2})
we have, taking the trace,
\begin{equation}
\label{eq:chi0}
    \partial_t \chi = \beta^k \partial_k \chi + \frac{2}{3} \chi (\alpha K
    - \partial_k \beta^k), 
\end{equation}
and
\begin{equation}
\label{eq:K0}
    \partial_t K  = \beta^k \partial_k K - \nabla^k \partial_k \alpha + \alpha \left(
      \tilde{A}^{ij} \tilde{A}_{ij} + \frac{1}{3} K^2 \right)
      + 4 \pi \alpha (E + S) ,
\end{equation}
where in this last equation we have already used the constraint~(\ref{eq:hamiltonian2}) to eliminate the Ricci scalar.
Substituting back we can compute the remaining evolution equations, which take the form
\begin{equation}
\label{eq:tilde-gamma0}
    \partial_t \tilde{\gamma}_{ij} = \beta^k \partial_k \tilde{\gamma}_{ij} +
    2\tilde{\gamma}_{k(i} \partial_{j)} \beta^k - \frac{2}{3}
    \tilde{\gamma}_{ij} \partial_k \beta^k -2\alpha \tilde{A}_{ij}, 
\end{equation}
and
\begin{align}
\label{eq:tilde-A0}
   \partial_t \tilde{A}_{ij} & = \beta^k \partial_k \tilde{A}_{ij} +
    2\tilde{A}_{k(i} \partial_{j)} \beta^k - \frac{2}{3} \tilde{A}_{ij}
    \partial_k \beta^k + \chi \left( \alpha R_{ij} - \nabla_i \partial_j \alpha\right)^{\rm
      TF} \notag \\
    & \quad + \alpha \left( K\,\tilde{A}_{ij}
      - 2 \tilde{A}_i{}^k \tilde{A}_{kj} \right)
     - 8 \pi \alpha \left(
          \chi S_{ij} - \frac{S}{3} \tilde \gamma_{ij}
        \right) ,
\end{align}
where ${}^\mathrm{TF}$ denotes the trace-free part, e.g., $R_{ij}^\mathrm{TF} = R_{ij} - \frac{1}{3} \gamma_{ij} R $.

We further need to decompose the Ricci tensor in two parts,
\begin{equation}
\label{eq:ricci-decomp}
R_{ij} = \tilde R_{ij} + R_{ij}^{\chi}
\end{equation}
where $R_{ij}^{\chi}$ only depends on $\chi$ and $\tilde{R}_{ij}$ is the Ricci tensor associated with the metric $\tilde{\gamma}_{ij}$.
This term contains mixed second derivatives of the metric, something that is undesirable.
For stable numerical integration, the following ``conformal connection'' variable was introduced~\cite{Shibata:1995we,Baumgarte:1998te}
\begin{equation}
\label{eq:Gamma-tilde-def}
\tilde{\Gamma}^i \equiv \tilde\gamma^{jk} \tilde{\Gamma}^i_{jk} 
   = -\partial_j \tilde \gamma^{ij} .
\end{equation}
In terms of this conformal connection, the conformal Ricci tensor then takes the form
\begin{equation}
\label{eq:conf-ricci-0}
  \tilde R_{ij} = -\frac{1}{2} \tilde \gamma^{kl} \partial_l \partial_k \tilde \gamma_{ij}
                 + \tilde{\gamma}_{k(i} \partial_{j)} \tilde{\Gamma}^k
                 - \partial_k \tilde{\gamma}_{l(i}  \partial_{j)} \tilde{\gamma}^{kl}
                 + \frac{1}{2} \tilde{\Gamma}^k \partial_{k} \tilde{\gamma}_{ij}
                 - \tilde{\Gamma}^l_{ik} \tilde{\Gamma}^k_{jl} .
\end{equation}
As we can see, the first term in this expression, which involves a Laplacian, is the only explicit second order derivative operator acting on $\tilde{\gamma}_{ij}$.
All the mixed second derivatives have been absorbed in derivatives of $\tilde{\Gamma}^i$.
Since the BSSN scheme considers $\tilde{\Gamma}^i$ to be an independent variable,
we need an evolution equation for it.
Acting on~(\ref{eq:Gamma-tilde-def}) and interchanging the time and space derivatives we get
\begin{equation}
\label{eq:tilde-Gamma00}
    \partial_t \tilde{\Gamma}^{i} = - \partial_{j} \left(
            \beta^k \partial_k \tilde{\gamma}^{ij}
            - 2\tilde{\gamma}^{k(j} \partial_{k} \beta^{i)} 
            + \frac{2}{3} \tilde{\gamma}^{ij} \partial_k \beta^k 
            + 2\alpha \tilde{A}^{ij}
      \right) .
\end{equation}
We further use the momentum constraint~(\ref{eq:momentum2}) to do away with the divergence of the extrinsic curvature and obtain
\begin{align}
  \label{eq:tilde-Aij00}
      \partial_t \tilde{\Gamma}^i & = \beta^k \partial_k \tilde{\Gamma}^i -
    \tilde{\Gamma}^k \partial_k \beta^i + \frac{2}{3}
    \tilde{\Gamma}^i \partial_k \beta^k + 2 \alpha \tilde{\Gamma}^i_{jk}
    \tilde{A}^{jk} + \frac{1}{3} \tilde{\gamma}^{ij}\partial_j \partial_k
    \beta^k
    + \tilde{\gamma}^{jk} \partial_j \partial_k \beta^i \nonumber \\
    & \quad - \frac{4}{3} \alpha \tilde{\gamma}^{ij} \partial_j K -
    \tilde{A}^{ij} \left( 3 \alpha \chi^{-1} \partial_j \chi + 2\partial_j
      \alpha \right) 
    - 16 \pi \alpha \chi^{-1} j^i .
\end{align}

The full system of evolution equations is then
\begin{subequations}
  \label{eq:bssn-gen}
  \begin{align}
    \partial_t \tilde{\gamma}_{ij} & = \beta^k \partial_k \tilde{\gamma}_{ij} +
    2\tilde{\gamma}_{k(i} \partial_{j)} \beta^k - \frac{2}{3}
    \tilde{\gamma}_{ij} \partial_k \beta^k -2\alpha \tilde{A}_{ij},  \\
    \partial_t \chi & = \beta^k \partial_k \chi + \frac{2}{3} \chi (\alpha K
    - \partial_k \beta^k),  \\
    \partial_t \tilde{A}_{ij} & = \beta^k \partial_k \tilde{A}_{ij} +
    2\tilde{A}_{k(i} \partial_{j)} \beta^k - \frac{2}{3} \tilde{A}_{ij}
    \partial_k \beta^k + \chi \left( \alpha R_{ij} - \nabla_i \partial_j \alpha\right)^{\rm
      TF} \notag \\
    & \quad + \alpha \left( K\,\tilde{A}_{ij}
      - 2 \tilde{A}_i{}^k \tilde{A}_{kj} \right)
     - 8 \pi \alpha \left(
          \chi S_{ij} - \frac{S}{3} \tilde \gamma_{ij}
        \right), \\
    \partial_t K & = \beta^k \partial_k K - \nabla^k \partial_k \alpha + \alpha \left(
      \tilde{A}^{ij} \tilde{A}_{ij} + \frac{1}{3} K^2 \right)
      + 4 \pi \alpha (E + S), \\
    \partial_t \tilde{\Gamma}^i & = \beta^k \partial_k \tilde{\Gamma}^i -
    \tilde{\Gamma}^k \partial_k \beta^i + \frac{2}{3}
    \tilde{\Gamma}^i \partial_k \beta^k + 2 \alpha \tilde{\Gamma}^i_{jk}
    \tilde{A}^{jk} + \frac{1}{3} \tilde{\gamma}^{ij}\partial_j \partial_k
    \beta^k
    + \tilde{\gamma}^{jk} \partial_j \partial_k \beta^i \nonumber \\
    & \quad - \frac{4}{3} \alpha \tilde{\gamma}^{ij} \partial_j K -
    \tilde{A}^{ij} \left( 3 \alpha \chi^{-1} \partial_j \chi + 2\partial_j
      \alpha \right) 
    - 16 \pi \alpha \chi^{-1} j^i \label{eq:tilde-Gamma-evol} \,,
  \end{align}
\end{subequations}
%
where $R_{ij} = \tilde R_{ij} + R_{ij}^{\chi}$
\begin{equation}
\begin{aligned}
  \tilde R_{ij} & = -\frac{1}{2} \tilde \gamma^{kl} \partial_l \partial_k \tilde \gamma_{ij}
                 + \tilde{\gamma}_{k(i} \partial_{j)} \tilde{\Gamma}^k
                 - \partial_k \tilde{\gamma}_{l(i}  \partial_{j)} \tilde{\gamma}^{kl}
                 + \frac{1}{2} \tilde{\Gamma}^k \partial_{k} \tilde{\gamma}_{ij}
                 - \tilde{\Gamma}^l_{ik} \tilde{\Gamma}^k_{jl} \,, \\
  R_{ij}^{\chi}  & =   \frac{1}{2} \chi^{-1} \left(
                                               \partial_i \partial_j \chi
                                             - \partial_k \chi \tilde{\Gamma}^k_{ij}
                                         \right)
                - \frac{1}{4} \chi^{-2} \partial_i \chi \partial_j \chi
                - \frac{1}{2} \tilde{\gamma}_{ij} \chi^{-1} \partial_k \chi \tilde{\Gamma}^k
                \\              
               & \quad + \frac{1}{2}\tilde{\gamma}_{ij} \tilde{\gamma}^{kl} \chi^{-1} \left(
                            \partial_k \partial_l \chi
                            - \frac{3}{2} \chi^{-1} \partial_k \chi  \partial_l \chi 
                          \right) \,.
\end{aligned}
\end{equation}
Source terms are determined by
\begin{equation}
\label{eq:source-gen}
\begin{aligned}
  E & \equiv n^\alpha n^\beta T_{\alpha\beta} \, , & \quad
  j_i & \equiv - \gamma_i{}^\alpha n^\beta T_{\alpha \beta} \, , \\
  S_{ij} & \equiv \gamma^\alpha{}_i \gamma^\beta{}_j T_{\alpha \beta}\, , & \quad
  S & \equiv \gamma^{ij} S_{ij} \, .
\end{aligned}
\end{equation}

The above system of evolution equations~\eqref{eq:bssn-gen} is known as the BSSN evolution scheme and has proven to be extremely robust for numerical evolutions of Einstein's field equations.
Numerous other schemes do exist; most, however, offer no substantial advantage over BSSN, which has remained extremely popular.
We will use BSSN for all our numerical evolutions.

\section{Gauge conditions}
\label{sec:gauge-gen}

We now turn our attention to the gauge conditions.
The first question to ask is: what is a ``good'' choice for $\alpha$ and $\beta^i$?
An obvious first choice, also the simplest possible, is to impose the so-called \emph{geodesic slicing} (also known as \emph{Gaussian normal coordinates}),
\begin{equation}
\label{eq:geo-slice-0}
\alpha = 1 \,, \qquad \beta^i = 0 \,.
\end{equation}
This choice was in fact used in the pioneering work by Hahn and Lindquist~\cite{1964AnPhy..29..304H}; it is now known, however, that it is actually a very bad choice for long-term evolutions.
We can intuitively understand why this is the case.
First we note from equation~(\ref{eq:formulae1}) that the Eulerian observers have zero acceleration and thus follow geodesics (hence the name of this slicing).
In the presence of black holes (or other gravitational sources), geodesics tend to focus.
Coordinate observers will then collide with each other, consequently forming coordinate singularities and crashing the numerical evolution.
We thus need better gauge choices.
It falls outside the scope of this work to give an overview on the merits and disadvantages of the different conditions that have been proposed throughout the years.
We will simply state and motivate the conditions we will be using.

\subsection{1+log slicing}
\label{sec:1+log_0}

A famous choice for the lapse function is known as \emph{maximal slicing}, which corresponds to imposing that the trace of the extrinsic curvature vanishes throughout the evolution,
\begin{equation}
  \label{eq:max-slicing}
  K = 0 
  \,.
\end{equation}
A nice property of this condition is its \emph{singularity avoidance}.
We can see this by taking the trace of equation~(\ref{eq:K_def}), which with~(\ref{eq:max-slicing}) implies $\nabla_{\mu} n^{\mu} = 0$, an incompressibility condition on the velocity field of the Eulerian observers.
This prevents the observers from converging and the subsequent appearance of a coordinate singularity, as in the geodesic slicing case.
Such a property is very much desirable, making maximal slicing an attractive choice.
There is an enormous disadvantage, however, which is the need to solve an elliptic equation at every time step during the numerical evolution in order to ensure~(\ref{eq:max-slicing}).
We therefore would like to have conditions that mimic this property of maximal slicing, yet with a hyperbolic character.

Such a choice is the so-called \emph{1+log slicing},
\begin{equation}
\label{eq:1+log_0}
\left(
  \partial_t - \mathcal{L}_{\beta}
\right) \alpha = - 2 \alpha K  \,,
\end{equation}
which, being a hyperbolic equation, is trivial to implement numerically, has been shown to be extremely robust and mimics the singularity avoidance properties of maximal slicing~\cite{Anninos:1995am}.

This condition gets its name from the fact that, when imposing $\beta^i = 0$, equation~(\ref{eq:1+log_0}) can be integrated to give
\begin{equation}
  \label{eq:alpha_1+log}
  \alpha = 1 + \log \gamma \,,
\end{equation}
where we recall that $\gamma \equiv \det \gamma_{ij}$.

\subsection{Gamma driver}
\label{sec:gamma-driver-0}

Having chosen a condition for the lapse function, it remains then to say what happens to the shift.

A possible choice, known as the \emph{Gamma freezing} condition, is the following
\begin{equation}
\label{eq:gamma-freezing}
\partial_t \tilde{\Gamma}^i = 0 \,.
\end{equation}
Using equation~(\ref{eq:tilde-Gamma-evol}), we can write the above condition as an elliptic equation for $\beta^i$.
This condition is related to the ``minimal distortion'' shift condition~\cite{Smarr:1977uf}, which attempts to choose coordinates such that the time derivative of the 3-metric $\partial_t \gamma_{ij}$ is minimised.
The disadvantage is once again the need to solve an elliptic equation at each time step.

Researchers have therefore proposed alternative conditions, using parabolic or hyperbolic equations, that mimic the minimal distortion condition with good approximation.
The following choice (and variations thereof) is now extremely popular
\begin{equation}
\label{eq:gamma-driver-0}
\left(
  \partial_t - \mathcal{L}_{\beta}
\right) \beta^i = \tilde{\Gamma}^i - \eta_{\beta} \beta^i \,,
\end{equation}
where $\eta_{\beta}$ is a function of spacetime.
This is known as the \emph{Gamma driver} condition~\cite{Alcubierre:2002kk}.

Use of these gauge choices proved crucial for the 2005 breakthroughs using the \emph{moving puncture} technique~\cite{Campanelli:2005dd,Baker:2005vv}.

\section{Numerical code}
\label{sec:lean-code}

Having chosen a set of evolution equations~(\ref{eq:bssn-gen}), gauge conditions~(\ref{eq:1+log_0}), (\ref{eq:gamma-driver-0}) and prescriptions for setting initial data (see chapter~\ref{chap:init_data}), it remains then to assemble everything on a numerical code.
Such a task is far from trivial.
One of the main reasons is that the presence of very different scales in the spacetimes that are usually evolved requires the use of \emph{mesh refinement}.
The problem is further complicated by the need to use parallel computing and to store large amounts of data. 

All numerical simulations that will be presented in subsequent chapters have
been performed by adapting the \textsc{Lean} code~\cite{Sperhake:2006cy},
initially designed for $3+1$ vacuum spacetimes by U.~Sperhake. \textsc{Lean} is
based on the \textsc{Cactus} computational toolkit~\cite{cactus}, it employs the
BSSN formulation of the Einstein
equations~\cite{Shibata:1995we,Baumgarte:1998te} (with fourth order
discretisation in the spatial derivatives) with the moving puncture
method~\cite{Campanelli:2005dd,Baker:2005vv},
uses the \textsc{Carpet} package for Berger-Oliger mesh
refinement~\cite{Schnetter:2003rb,carpet}, the spectral solver described
in~\cite{Ansorg:2004ds} for $3+1$ initial data and Thornburg's
\textsc{AHFinderDirect}~\cite{Thornburg:1995cp,Thornburg:2003sf} for horizon
finding (see section~\ref{sec:horizon-finding}).

For a given numerical simulation our numerical grid will consist of two types
of cubic refinement levels: $n$ outer levels centred on the origin (remaining
stationary throughout the simulation), and $m$ inner levels centred around each
black hole (and following these as the simulation progresses).
The following notation (which we will make frequent use of)
\[
  \{ (256,~128,~74,~24,~12,~6)\times (1.5,~0.75),~h=1/48 \}
\]
specifies a grid with six fixed outer components of ``radius'' $256$, $128$, $74$,
$24$, $12$ and $6$ respectively and two refinement levels with two components
each with radius $1.5$ and $0.75$ centred around either black hole. The resolution is
$h=1/48$ on the finest level and successively decreases to $2^7/48 = 8/3$ on the
outermost level.
Further details about \textsc{Lean} may be found in~\cite{Sperhake:2006cy}.

\begin{subappendices}

\section{\texorpdfstring{$D$-dimensional BSSN equations}{D-dimensional BSSN equations}}
\label{ch:ddim-bssn}

For completeness, we here write the full $D$-dimensional BSSN equations, as first written in~\cite{Yoshino:2009xp}.
These equations can be derived in a procedure entirely analogous to the one outlined in section~\ref{sec:bssn}.
\begin{subequations}
  \label{eq:ddim-bssn-gen}
  \begin{align}
    \partial_t \tilde{\gamma}_{ij} & = \beta^k \partial_k \tilde{\gamma}_{ij} +
    2\tilde{\gamma}_{k(i} \partial_{j)} \beta^k - \frac{2}{D-1}
    \tilde{\gamma}_{ij} \partial_k \beta^k -2\alpha \tilde{A}_{ij},
    \\
    \partial_t \chi & = \beta^k \partial_k \chi + \frac{2}{D-1} \chi (\alpha K
    - \partial_k \beta^k), \\
    \partial_t \tilde{A}_{ij} & = \beta^k \partial_k \tilde{A}_{ij} +
    2\tilde{A}_{k(i} \partial_{j)} \beta^k - \frac{2}{D-1} \tilde{A}_{ij}
    \partial_k \beta^k + \chi \left( \alpha R_{ij} - \nabla_i \partial_j \alpha\right)^{\rm
      TF} \notag \\
    & \quad + \alpha \left( K\,\tilde{A}_{ij}
      - 2 \tilde{A}_i{}^k \tilde{A}_{kj} \right)
     - 8 \pi \alpha \left(
          \chi S_{ij} - \frac{S}{D-1} \tilde \gamma_{ij}
        \right), \\
    \partial_t K & = \beta^k \partial_k K - \nabla^k \partial_k \alpha + \alpha \left(
      \tilde{A}^{ij} \tilde{A}_{ij} + \frac{1}{D-1} K^2 \right)
      + \frac{8 \pi}{D-2} \alpha \left[(D-3)E + S\right], \\
    \partial_t \tilde{\Gamma}^i & = \beta^k \partial_k \tilde{\Gamma}^i -
    \tilde{\Gamma}^k \partial_k \beta^i + \frac{2}{D-1}
    \tilde{\Gamma}^i \partial_k \beta^k + 2 \alpha \tilde{\Gamma}^i_{jk}
    \tilde{A}^{jk} + \frac{1}{D-1} \tilde{\gamma}^{ij}\partial_j \partial_k
    \beta^k
    + \tilde{\gamma}^{jk} \partial_j \partial_k \beta^i \nonumber \\
    & \quad - 2\frac{D-2}{D-1} \alpha \tilde{\gamma}^{ij} \partial_j K -
    \tilde{A}^{ij} \left( (D-1) \alpha \frac{\partial_j \chi}{\chi} + 2\partial_j
      \alpha \right) 
    - 16 \pi \alpha \chi^{-1} j^i,
  \end{align}
\end{subequations}
where ${}^\mathrm{TF}$ denotes the trace-free part and
the Ricci tensor $R_{ij}$ is further split into $R_{ij} = \tilde R_{ij} + R_{ij}^{\chi}$, where
\begin{equation}
\begin{aligned}
  \tilde R_{ij} & = -\frac{1}{2} \tilde \gamma^{kl} \partial_l \partial_k \tilde \gamma_{ij}
                 + \tilde{\gamma}_{k(i} \partial_{j)} \tilde{\Gamma}^k
                 - \partial_k \tilde{\gamma}_{l(i}  \partial_{j)} \tilde{\gamma}^{kl}
                 + \frac{1}{2} \tilde{\Gamma}^k \partial_{k} \tilde{\gamma}_{ij}
                 - \tilde{\Gamma}^l_{ik} \tilde{\Gamma}^k_{jl} \\
  R_{ij}^{\chi}  & = \frac{D-3}{2} \chi^{-1} \left(
                                               \partial_i \partial_j \chi
                                             - \partial_k \chi \tilde{\Gamma}^k_{ij}
                                         \right)
                - \frac{D-3}{4} \chi^{-2} \partial_i \chi \partial_j \chi
                - \frac{1}{2} \tilde{\gamma}_{ij} \chi^{-1} \partial_k \chi \tilde{\Gamma}^k
                \\              
               & \quad + \frac{1}{2}\tilde{\gamma}_{ij} \tilde{\gamma}^{kl} \chi^{-1} \left(
                            \partial_k \partial_l \chi
                            - \frac{(D-1)}{2} \chi^{-1} \partial_k \chi  \partial_l \chi 
                          \right)
\end{aligned}
\end{equation}
Equations~(\ref{eq:bssn-gen}) can be recovered with $D=4$.



\end{subappendices}



\chapter{Wave extraction and horizon finding}
\label{ch:wave-ext_horizon}

We have thus far covered, essentially, all the main tools necessary to successfully evolve Einstein's equations on a computer.
Assuming then that we can specify some arbitrary initial configuration and evolve it for as long as we like, we are still faced with the most important task: how to extract the relevant physical information.
Recalling that the coordinate system used throughout the evolution is designed to be well suited to the numerical evolution and not for human-readability, we easily convince ourselves that it is not trivial to read physical information from the numerical output.
For this purpose, tools were developed to enable the extraction of the gravitational wave information from a numerical simulation and, when dealing with black hole spacetimes, information about the black hole's horizon.

In this chapter we will briefly outline the two main methods of extracting gravitational wave information and the corresponding waveforms: the gauge invariant formalism of Kodama and Ishibashi~\cite{Kodama:2000fa,Kodama:2003jz}---itself a generalisation to higher-dimensional spacetimes of the Regge-Wheeler-Zerilli formalism~\cite{Regge:1957td,Zerilli:1970se}, later put in a gauge-invariant form by Moncrief~\cite{Moncrief:1974am}---and the \emph{Newman-Penrose formalism}~\cite{Newman:1961qr}.
We will also mention the very basics regarding finding (apparent) black hole horizons.

\section{Wave extraction}
\label{sec:wave-ext}

Gravitational waves are ripples in the shape of spacetime that propagate information at finite speed, just as water waves are ripples in the shape of an ocean's surface.
They are one of the most important predictions of general relativity.
These waves have never been directly detected; there is, however, strong indirect evidence for its existence since the discovery of the famous binary pulsar PSR 1913+16 (also known as the Hulse-Taylor binary pulsar after its discoverers~\cite{Hulse:1974eb}), whose orbital period change is consistent with the general relativistic prediction for energy loss via gravitational wave emission.
Other systems have since been uncovered, allowing for even more stringent tests, e.g.~\cite{Lyne:2004cj,Kramer:2009zza}.

These waves are generated by dynamical gravitational fields---roughly speaking, accelerated bodies in non-spherically symmetric motion will emit gravitational waves\footnote{For a system to emit gravitational waves the third time derivative of its quadrupole moment has to be non-zero.}.
As they propagate throughout spacetime, they carry with them information about the physical properties of the system that produced them.
By measuring them with gravitational wave detectors---such as the already mentioned LIGO, Virgo, GEO and TAMA---we can have a brand new window opening up to the universe.
The likelihood of such detections is greatly enhanced if one can use theoretical gravitational wave signals coming from possible astrophysical sources as templates.
Our task here is to briefly introduce the techniques used to generate such templates from numerical simulations.

Before we begin, let us make one last comment. 
We have mentioned energy carried away by gravitational radiation,
but as we know there is no notion of local energy of a gravitational field, so some care has to be taken here.
The usual procedure is to write a stress-energy tensor for the metric fluctuations that is second-order in said fluctuations (the same way that the stress-energy tensor associated with a scalar or electromagnetic field is second order in the fields).
Modulo some subtleties, such a quantity can be constructed, and meaningful quantities can be extracted from it.
We will not be giving details on its derivation or the subtleties involved (see, e.g.,~\cite{Isaacson:1967zz,Isaacson:1968zza,Misner:1974qy}), but merely present the relevant formul\ae{} that will be of use to us.

We will start by recalling known four-dimensional results that will be of use.
In the weak-field limit, we can write the metric tensor as the Minkowski metric plus perturbations,
\begin{equation}
\label{eq:weak-field}
g_{\mu\nu} = \eta_{\mu\nu} + h_{\mu\nu} \,, \qquad |h_{\mu\nu}| \ll 1 \,.
\end{equation}
To first order in $h_{\mu\nu}$, the Riemann tensor is
\begin{equation}
  \label{eq:riemann-weak-field}
  {}^{(4)}\! R_{\alpha\beta\mu\nu} = \frac{1}{2} \left(
                                    \partial_{\beta} \partial_{\mu} h_{\alpha\nu}
                                  + \partial_{\alpha} \partial_{\nu} h_{\beta\mu}
                                  - \partial_{\beta} \partial_{\nu} h_{\alpha\mu}
                                  - \partial_{\alpha} \partial_{\mu} h_{\beta\nu}
                                           \right) \,.
\end{equation}
It is useful to introduce the usual \emph{trace reversed perturbation}
\begin{equation}
  \label{eq:trace-reversed-def}
  \bar{h}_{\mu \nu} \equiv h_{\mu\nu} - \frac{h}{2} \eta_{\mu\nu} \,.
\end{equation}
Imposing the \emph{Lorenz gauge}
\begin{equation}
  \label{eq:lorenz-gauge}
  \partial_{\mu} \bar h^{\mu \nu} = 0 \,,
\end{equation}
the linearised field equations reduce to
\begin{equation}
\label{eq:EE-linear}
\square \bar h_{\mu \nu} = -16 \pi T_{\mu\nu} \,,
\end{equation}
where $\square$ is the d'Alembertian operator in flat space.
In vacuum we get the usual wave equation
\begin{equation}
\label{eq:EE-linear-vacuum}
\square \bar h_{\mu \nu} = 0 \,.
\end{equation}
Since the Lorenz gauge does not completely fix our degrees of freedom, we can further impose the \emph{transverse-traceless (TT) gauge}
\begin{equation}
  \label{eq:TTgauge}
  u^{\nu}\bar{h}^{\rm TT}_{\mu \nu} = 0 \,, \qquad \bar{h}^{\rm TT}{}^\mu{}_\mu = 0 \,,
\end{equation}
where, to simplify, we can use a Cartesian coordinate system $\eta_{\mu \nu} = \diag(-1,1,1,1)$ and $u^{\nu}$ is a unit timelike vector.
The second equation reflects the fact that there is no propagating scalar mode in general relativity.
Note that in this gauge $\bar{h}_{\mu\nu} = h_{\mu\nu}$.
With the constraints~(\ref{eq:lorenz-gauge}) and (\ref{eq:TTgauge}) we are left with two degrees of freedom (in four dimensions; generically, in $D$-dimensions, we have $D(D-3)/2$ degrees of freedom).
We can write the plane-wave solution of equation~(\ref{eq:EE-linear-vacuum}) subject to the constraints~(\ref{eq:TTgauge}) in the usual form
\begin{equation}
  \label{eq:hbar-TT-sol}
  h_{\mu\nu}^{\rm TT} = A_{\mu\nu} e^{ik_{\sigma} x^{\sigma}} \,,
\end{equation}
taking $k^{\mu}=(1,0,0,1)$ and where
\begin{equation}
  \label{eq:Amplitudes}
  A_{\mu\nu} =
  \begin{pmatrix}
    0 & 0    & 0        & 0 \\
    0 & h_{+} & h_{\times} & 0 \\ 
    0 & h_{\times} & -h_{+} & 0 \\ 
    0 & 0    & 0        & 0
  \end{pmatrix} \,.
\end{equation}
$h_{+}$ and $h_{\times}$ are the two independent polarisations of the gravitational wave, known as ``plus'' and ``cross'' polarisations.

It can be shown (e.g.~\cite{Misner:1974qy,Carroll:2004st,baumgarte2010numerical}) that the outgoing energy flux carried by the gravitational radiation is given by
\begin{align}
  \label{eq:GW-flux0}
  F_{\rm GW} & = \frac{d E_{\rm GW}}{dt} =
      \lim_{r\to\infty} \frac{r^2}{16 \pi} \int
      \left( \dot{h}_{+}^2 + \dot{h}_{\times}^2
      \right) \dd \Omega  \, .
\end{align}
To derive this formula, we need to expand Einstein's equations up to second order perturbations.
Terms that are quadratic in the first order perturbations of the metric, after suitable averaging, can then be viewed as sources, constituting an effective stress-energy tensor for gravitational waves.
This stress-energy tensor can then be used to compute energy and momentum carried away by the gravitational radiation.

\subsection{Newman-Penrose formalism}
\label{sec:NP-gen}

We now briefly describe the Newman-Penrose formalism.
This formalism (also known as the \emph{spin-coefficient formalism}) introduced by Newman and Penrose in 1962~\cite{Newman:1961qr} is an alternative way to write Einstein's equations which has proven to be extremely useful in many situations in general relativity, such as in searches of exact solutions, black hole perturbation theory and studies of gravitational radiation.
There is a whole literature devoted to this formalism.
For its application in numerical simulations, we mention for instance the books by Alcubierre~\cite{Alcubierre:2008}, Baumgarte \& Shapiro~\cite{baumgarte2010numerical} and references therein.
Here we will only state the basic equations that we will need and refer the reader to relevant publications where appropriate.

In this section we restrict ourselves to four-dimensional spacetimes since this formalism has not been generalised to higher dimensions.\footnote{A related formalism also based on spin-coefficients, the \emph{Geroch-Held-Penrose} (GHP)~\cite{Geroch:1973am} formalism, has been extended to higher dimensions~\cite{Durkee:2010xq}.}

This formalism starts with introducing a null complex tetrad $\left\{l,k,m,\bar m\right\}$ satisfying 
\begin{equation}
\label{eq:NP-tetrad}
-l \cdot k = 1 = m \cdot \bar m \,,
\end{equation}
where $\bar m$ is the complex conjugate of $m$ and all other inner products vanish.

We further note that the four-dimensional Riemann tensor ${}^{(4)}\!R^{\alpha}{}_{\beta\gamma\delta}$ has 20 independent components.
Its trace, the Ricci tensor ${}^{(4)}\!R_{\alpha\beta}$ has 10.
The remaining degrees of freedom are encoded in the Weyl tensor ${}^{(4)}\!C_{\alpha\beta\gamma\delta}$, defined as
\begin{equation}
  \label{eq:weyl-def}
  {}^{(4)} \! C_{\alpha\beta\gamma\delta} \equiv {}^{(4)} \! R_{\alpha\beta\gamma\delta}
                             - {}^{(4)} \! g_{\alpha [\gamma}  {}^{(4)} \! R_{\delta]\beta}
                             + {}^{(4)} \!g_{\beta [\gamma} {}^{(4)} \! R_{\delta]\alpha}
                             + \frac{1}{3} {}^{(4)} \! g_{\alpha [\gamma } {}^{(4)}
                                                   \! g_{\delta] \beta} {}^{(4)} \! R  \,.
\end{equation}
The Newman-Penrose formalism encodes these degrees of freedom in a set of complex scalars, often called \emph{Newman-Penrose scalars}.
The ten independent components of the Weyl tensor are encoded in the five
complex scalars $\Psi_0, \ldots, \Psi_4$ (often also called \emph{Weyl
  scalars}).%
\footnote{The ten independent components of the Ricci tensor are analogously
  written in terms of four scalars and three complex scalars, but we will never
  make use of these quantities in this work.}
All of these scalars are formed by contracting the Weyl tensor (and the Ricci) with the complex null tetrad.
Since there is no unique choice for a null tetrad satisfying~(\ref{eq:NP-tetrad}), the choice of this tetrad will affect the Weyl scalars and their physical interpretation.

For a class of such tetrads, the so-called \emph{quasi-Kinnersley frames}, $\Psi_1$ and $\Psi_3$ both vanish, and we can interpret $\Psi_0$ and $\Psi_4$ as measures of the incoming and outgoing gravitational radiation, whereas $\Psi_2$ can be interpreted as the ``Coulombic'' part.
$\Psi_4$ and $\Psi_0$ are defined as\footnote{Different sign conventions exist in the literature.}
\begin{align}
  \label{eq:Psi0}
  \Psi_0 & \equiv {}^{(4)}\! C_{\alpha\beta\gamma\delta}
      l^{\alpha} m^{\beta} l^\gamma m^{\delta} \,. \\
  \label{eq:Psi4}
  \Psi_4 & \equiv {}^{(4)}\! C_{\alpha\beta\gamma\delta}
      k^{\alpha} \bar m^{\beta} k^\gamma \bar m^{\delta} \,.
\end{align}
Since (for the suitable tetrad we mentioned above) the latter quantity encodes the outgoing gravitational wave signal, this will be of particular use to us ($\Psi_0$ will also be of use in section~\ref{sec:BHbox}).

In practice, we construct $l$, $k$ and $m$ from an orthonormal triad $e_{\hat r}, e_{\hat{\theta}}, e_{\hat{\phi}}$ orthogonal to the unit timelike vector $e_{\hat t}$
\begin{equation}
\begin{aligned}
l & = \frac{1}{\sqrt{2}} \left( e_{\hat{t}} + e_{\hat{r}} \right) \, , \\
k & = \frac{1}{\sqrt{2}} \left( e_{\hat{t}} - e_{\hat{r}} \right) \, , \\
m & = \frac{1}{\sqrt{2}} \left( e_{\hat{\theta}} + i e_{\hat{\phi}} \right) \, .
\end{aligned}
\end{equation}
We refer the reader to~\cite{Lehner:2007ip} for a review of the formalism; here
we merely note that asymptotically the triad vectors $e_{\hat{r}}$,
$e_{\hat{\theta}}$, $e_{\hat{\phi}}$ behave as the unit radial, polar and
azimuthal vectors.

Having chosen our tetrad, we can now compute an explicit expression for $\Psi_4$ using the definition~(\ref{eq:Psi4}).
In the TT gauge, this can be shown to be, for outgoing waves~\cite{Alcubierre:2008,baumgarte2010numerical}
\begin{align}
  \label{eq:psi0-hh}
  \Psi_0 & = 0 \,, \\
  \label{eq:psi4-hh}
  \Psi_4 & = -\ddot{h}_{+} + i \ddot{h}_{\times} \,,
\end{align}
whereas for ingoing waves, we have instead
\begin{align}
  \Psi_0 & = \ddot{h}_{+} - i\ddot{h}_{\times}  \,, \\
  \Psi_4 & = 0 \,,
\end{align}
where $\dot{}$ denotes a time derivative and $h_{+}$ and $h_{\times}$ are the amplitudes of the plus and cross polarisation of the gravitational wave~(\ref{eq:hbar-TT-sol}), (\ref{eq:Amplitudes}).
Herein lies the usefulness of the $\Psi_4$ scalar.

It is useful to perform a multipolar
decomposition by projecting $\Psi_4$ onto spherical
harmonics of spin weight $s=-2$ (cf., e.g., appendix~D of~\cite{Alcubierre:2008}):
\begin{align}
  \Psi_4(t, \theta, \phi) & =
      \sum_{l,m} \psi^{lm}(t) Y_{lm}^{-2}(\theta,\phi) \ .
\end{align}
In terms of these multipoles, the radiated flux is given by the
expressions~\cite{Newman:1961qr,Newman:1981fn} 
\begin{align}
  \label{eq:GW-flux}
  F_{\rm GW} & = \frac{\dd E_{\rm GW}}{\dd t} =
      \lim_{r\to\infty} \frac{r^2}{16 \pi} \sum_{l,m}
      \left| \int_{-\infty}^t \dd t' \psi^{lm} (t') \right|^2 \,.
\end{align}

In Einstein-Maxwell theory, the right-hand-side of Einstein's equations reads
\begin{equation}
  \label{eq:EM-Tmunu}
  T_{\mu \nu} = \frac{1}{4\pi} \left[ F_{\mu}{}^{\lambda} F_{\nu \lambda}
    - \frac{1}{4} g_{\mu \nu} F^{\lambda \sigma} F_{\lambda \sigma}
    \right] \,,
\end{equation}
where $F^{\mu \nu}$ is the Maxwell-Faraday tensor.
In such cases,
we can analogously extract the electromagnetic wave
signal in the form of the scalar functions, $\Phi_1$ and
$\Phi_2$~\cite{Newman:1961qr,Newman:1981fn}, defined as
\begin{align}
  \label{eq:Phi1}
  \Phi_1 & \equiv \frac{1}{2} F_{\mu \nu} \left(
    l^{\mu} k^{\nu} + \bar m^{\mu} m^{\nu}
  \right) \ , \\
  \label{eq:Phi2}
  \Phi_{2} & \equiv F_{\mu \nu} \bar m^{\mu} k^{\nu} \ .
\end{align}
For outgoing waves at infinity, these quantities behave as
\begin{equation}
  \label{eq:Phi_asympt}
  \Phi_1 \sim \frac{1}{2}\left(
    E_{\hat r} + i B_{\hat r}
    \right)\ , \quad \Phi_2 \sim E_{\hat \theta} - i E_{\hat \phi} \ .
\end{equation}

Again, it is useful to perform a multipolar
decomposition by projecting $\Phi_1$ and $\Phi_2$ onto spherical
harmonics of spin weight $0$ and $-1$ respectively:
\begin{align}
  \Phi_1(t, \theta, \phi) & =
      \sum_{l,m} \phi_{1}^{lm}(t) Y_{lm}^{0}(\theta,\phi) \ ,
      \label{eq:multipole_Phi1} \\
  \Phi_2(t, \theta, \phi) & =
      \sum_{l,m} \phi_{2}^{lm}(t) Y_{lm}^{-1}(\theta,\phi) \ .
\end{align}
In terms of these multipoles, the radiated flux is given by the
expressions~\cite{Newman:1961qr,Newman:1981fn} 
\begin{align}
%
  F_{\rm EM} & = \frac{\dd E_{\rm EM}}{\dd t} =
      \lim_{r\to\infty} \frac{r^2}{4 \pi} \sum_{l,m} 
      \left|  \phi^{lm}_{2} (t) \right|^2 \ . \label{eq:EM-flux}
\end{align}
We see from~(\ref{eq:Phi_asympt}) that $\Phi_2$ encodes the radiative modes.

\subsection{Kodama-Ishibashi}
\label{sec:KI-gen}

A different approach to extract gravitational wave perturbations is that of the gauge-invariant Moncrief formalism~\cite{Moncrief:1974am}.
This has been generalised to higher dimensions by Kodama and Ishibashi (KI)~\cite{Kodama:2000fa,Kodama:2003jz}, and we will review this approach in the following.


In the KI formalism, we start by writing the metric element as a background metric plus a perturbation
\begin{equation}
  \label{eq:KI-ansatz}
  \bar{g}_{AB} = \bar{g}^{(0)}_{AB} + \delta\bar{g}_{AB} \,.
\end{equation}
The background spacetime has the form
\begin{equation}
  \label{eq:KI-back}
  \dd\bar s^{2}{}^{(0)} = \bar{g}_{AB}^{(0)} \dd x^A \dd x^B 
                      = g_{ab}^{(0)} \dd x^a \dd x^b + r^2 \dd \Omega_{D-2}
                      = g_{ab}^{(0)} \dd x^a \dd x^b + r^2 \Omega_{\bar{a}\bar{b}} \dd x^{\bar{a}} \dd x^{\bar{b}} \,,
\end{equation}
where the $x^A$ coordinates refer to the whole spacetime ($A=0,\ldots,D-1$), $x^{a} = t,r$ and $\Omega_{\bar{a}\bar{b}}$ is the metric on the unit $(D-2)$-sphere $S^{D-2}$.

The procedure now is to expand the metric perturbations $\bar{g}_{AB}$ into harmonic functions.
These exist in three flavours---scalar, vector and tensor harmonics.
Metric perturbations can then be written in terms of gauge invariant quantities
~\cite{Kodama:2000fa}:
%
%
\begin{description}
\item[Tensor harmonics $\mathbb{T}_{\bar{a}\bar{b}}$] satisfy
  \begin{equation}
    \label{eq:tensor-harm-def}
    \left(
      \hat{\triangle} + k^2
    \right) \mathbb{T}_{\bar{a}\bar{b}} = 0 \,,
  \end{equation}
  with the properties
  \begin{equation}
    \mathbb{T}^{\bar a}{}_{\bar a} = 0 \,, \quad 
    \mathbb{T}^{\bar a}{}_{\bar b : \bar a} = 0 \,.
    \label{eq:tensor-harm-div}
  \end{equation}
  where $\hat{\triangle}$ is the Laplace-Beltrami operator on $S^{D-2}$ and
  ${}_{:\bar a}$ denotes the covariant derivative with respect to the metric
  $\Omega_{\bar{a}\bar{b}}$ on the sphere.
  Note that we will omit the index labelling the harmonic throughout this discussion.

  For tensor-type perturbations, the metric perturbations $\delta \bar{g}_{AB} =
  h_{AB}$ are expanded in the following way
  \begin{equation}
    \label{eq:h-tensor-perturb}
    h_{ab} = 0 \,, \qquad h_{a\bar{a}} = 0 \,, \qquad
    h_{\bar{a}\bar{b}} = 2r^2 H_T \mathbb{T}_{\bar{a}\bar{b}} \,,
  \end{equation}
  where $H_T = H_T(t,r)$ (again, we leave the harmonic labels implicit), and
  note that there is sum over the indices of the harmonics in this expression.

\item[Vector harmonics $\mathbb{V}_{\bar{a}}$] satisfy
  \begin{align}
    \label{eq:vector-harm-def}
    & \left(
        \hat{\triangle} + k^2
      \right) \mathbb{V}_{\bar{a}} = 0 \,, \\
    \label{eq:vector-harm-div}
    & \mathbb{V}^{\bar a}{}_{:\bar a} = 0 \,.
  \end{align}
  from this, vector-type harmonic tensors can further be defined
  \begin{equation}
    \label{eq:vec-tensor-harm}
    \mathbb{V}_{\bar{a}\bar{b}} = -\frac{1}{2k} \left(
      \mathbb{V}_{\bar{a}:\bar{b}} + \mathbb{V}_{\bar{b}:\bar{a}}
    \right) \,.
  \end{equation}

  We expand the vector-type perturbations as
  \begin{equation}
    \label{eq:h-vec-perturb}
    h_{ab} = 0 \,, \qquad h_{a\bar{a}} = r f_{a} \mathbb{V}_{\bar a} \,, \qquad
    h_{\bar{a}\bar{b}} = 2r^2 H_T \mathbb{V}_{\bar{a}\bar{b}} \,,
  \end{equation}
  where $f_a = f_a(t,r)$.

\item[Scalar harmonics $\mathbb{S}$] satisfy
  \begin{align}
    \label{eq:scalar-harm-def}
    & \left(
        \hat{\triangle} + k^2
      \right) \mathbb{S} = 0 \,,
  \end{align}
  from which we can build scalar-type vector harmonics
  \begin{equation}
    \label{eq:scalar-vec-harm}
    \mathbb{S}_{\bar{a}} = -\frac{1}{k} \mathbb{S}_{:\bar{a}} \,,
  \end{equation}
  and scalar-type harmonic tensors
  \begin{equation}
    \label{eq:scalar-tensor-harm}
    \mathbb{S}_{\bar{a}\bar{b}} = \frac{1}{k^2} \mathbb{S}_{:\bar{a}\bar{b}} 
                              + \frac{1}{D-2} \Omega_{\bar{a}\bar{b}} \mathbb{S} \,.
  \end{equation}

  We expand scalar-type perturbations as
  \begin{equation}
    \label{eq:h-scalar-perturb}
    h_{ab} = f_{ab} \mathbb{S} \,, \qquad h_{a\bar{a}} = r f_a \mathbb{S}_{\bar a} \,, \qquad
    h_{\bar{a}\bar{b}} = 2r^2 \left(
                                H_L \Omega_{\bar{a}\bar{b}} \mathbb{S} 
                              + H_T \mathbb{S}_{\bar{a}\bar{b}}
                          \right)\,,
  \end{equation}
  where $f_{ab} = f_{ab}(t,r)$.

  For $l>1$, the metric perturbations can be expressed in terms of the following
  gauge-invariant variables~\cite{Kodama:2000fa}
  \begin{equation}
    \label{gi1}
    \begin{aligned}
      F & = H_L+\frac{1}{D-2}H_T+\frac{1}{r}X_{a} r^{|a}\,, \\
      F_{ab}& = f_{ab}+X_{a|b}+X_{b|a} \,,
    \end{aligned}
  \end{equation}
  where we have defined
  \begin{equation}
    X_{a}=\frac{r}{k}\left(f_{a}+\frac{r}{k}H_{T|a}\right)\,,\label{gi2}
  \end{equation}
  and we denote the covariant derivative with respect to the metric
  $g^{(0)}_{ab}$ with a subscript ${}_{|a}$.

\end{description}

Impressively, a \emph{master function} $\Phi$ can be defined that, from the perturbed Einstein equations, can be shown to obey the simple wave equation~\cite{Kodama:2003jz}
\begin{equation}
  \label{eq:master-func-eq}
  \left(
    \square - V(r)
  \right) \Phi = 0 \,,
\end{equation}
where $\square$ is the d'Alembertian operator with respect to $g_{ab}^{(0)}$, and the form of the potential $V(r)$ depends on whether one is considering scalar, vector or tensor perturbations.

The master function $\Phi$ is specially useful since it encodes the gravitational waveform; the energy emitted via gravitational radiation can also be computed quite effortlessly.
Writing the index $l$ explicitly, the energy flux in each $l$-multipole is~\cite{Berti:2003si}
\begin{equation}
\label{eq:energyflux}
\frac{\dd E_l}{\dd t}=\frac{1}{32\pi}\frac{D-3}{D-2}k^2(k^2-D+2)(\Phi^l_{,t})^2\,.
\end{equation} 
The total energy emitted in the process is then
\begin{equation}
\label{eq:energyrad}
E=\sum_{l=2}^\infty\int_{-\infty}^{+\infty}\dd t\frac{\dd E_l}{\dd t}\,.
\end{equation}

\section{Horizon finding}
\label{sec:horizon-finding}

When evolving black hole spacetimes, besides the wave extraction tools, physical information can also be read from its horizon properties.
A black hole is a region of spacetime from which no future directed null geodesic can reach an outside observer.
Its surface, the \emph{event horizon}, acts therefore as a one-way membrane.
In asymptotically flat spacetimes, the event horizon can be defined as the boundary of the causal past of future null infinity.
It is thus as global concept, requiring information from the whole spacetime to be located.
From the point of view of a numerical evolution, this is not very useful since one would like to know about the location of the black hole as the simulation progresses.

A more useful concept in this regard is that of the \emph{apparent horizon}.
It is defined as the \emph{outermost marginally trapped surface} on a given spatial hypersurface---a closed surface on which the expansion of (outgoing) null geodesics vanishes.
The apparent horizon is a local concept, depending only on information present on the given hypersurface, making it an ideal diagnostic tool for numerical evolutions.

Given a spatial section $\Sigma$ of a spacetime with 3-metric $\gamma_{ij}$ and extrinsic curvature $K_{ij}$, the expansion of null geodesics can be shown to be
\begin{equation}
  \label{eq:expansion-def}
  \Theta_{\pm} = \pm \nabla_i s^i + K_{ij} s^i s^j - K
\end{equation}
where $\nabla$ is the covariant derivative with respect to the 3-metric $\gamma_{ij}$ and $s^i$ is the spatial normal to the apparent horizon surface within $\Sigma$. 
$\Theta_{+}$ is the expansion of the outgoing null geodesics, $\Theta_{-}$ the expansion of the ingoing ones.
The (black hole) apparent horizon is then defined by the following equation
\begin{equation}
  \label{eq:AH-def}
  \nabla_i s^i + K_{ij} s^i s^j - K = 0 \,.
\end{equation}
General purpose tools exist to solve this equation during numerical evolutions; for an overview see e.g.~\cite{Thornburg:2006zb} and references therein.

Finally, we emphasise that apparent horizons are slicing dependent.
It is possible, for instance, to foliate the Schwarzschild spacetime in such a way that there is no apparent horizon~\cite{Wald:1991zz} (the event horizon, being a global quantity, is an intrinsic property of the geometry and is thus always present).
The presence of an apparent horizon, however, does imply the existence of a section of an event horizon exterior to it (assuming cosmic censorship and $R_{\mu \nu} k^{\mu} k^{\nu} \ge 0$ for all null $k^{\mu}$~\cite{wald1984general}).



\chapter{Higher-dimensional numerical relativity}
\label{chap:high-dimNR}

As mentioned in the Introduction, the ability to perform fully non-linear numerical evolutions of Einstein's field equations in higher-dimensional scenarios has tremendous potential to answer fundamental questions in physics, with
possible applications including studies of the AdS/CFT duality, explorations of TeV-gravity scenarios and the study of higher-dimensional black hole solutions.

Numerical relativity in higher dimensions has only recently started being explored, with pioneering works including those in~\cite{Choptuik:2003qd,Yoshino:2009xp,Zilhao:2010sr,Shibata:2010wz,Witek:2010xi}.
In this chapter, we will describe the approach of~\cite{Zilhao:2010sr,Witek:2010xi}.

The formalism we will present allows us to consider two classes of models, which
are generalisations of axial symmetry to higher dimensional spacetimes: a
$D\ge5$ dimensional vacuum spacetime with an $SO(D-2)$ isometry group, and a
$D\ge6$ dimensional vacuum spacetime with an $SO(D-3)$ isometry group. The
former class allows studies of head-on collisions of non-spinning black holes.
The latter class allows to model black hole collisions with impact parameter and
with spinning black holes, as long as all the dynamics take place on a single
plane.
This class includes the most interesting physical configurations relevant to
accelerator---and cosmic ray---physics (in the context of TeV-scale gravity),
and to the theoretical properties of higher-dimensional black objects (such as
stability and phase diagrams).

In section~\ref{sec:dim-red}, we introduce a general dimensional reduction procedure from $D$-dimensional vacuum general relativity to a lower dimension model; in section~\ref{sec:dim-red_split} we specialise the equations obtained to the case where the $D$-dimensional spacetime has an $SO(D-2)$ isometry group, perform the 3+1 splitting of space and time and write down a system of evolution equations; in section~\ref{sec:init_data-hd} we outline the construction of relevant initial data, following the approach of~\cite{Zilhao:2011yc}; in section~\ref{sec:num-evols}, we discuss some code tests, introduce a wave extraction procedure and present results.
We end this chapter with a discussion in~\ref{sec:final-remarks}.

We note that in this chapter, due to the necessity of introducing multiple covariant derivatives, we shall explain the notation as we go along.

\section{Dimensional reduction}
\label{sec:dim-red}

The starting point of our formalism is a dimensional reduction from
$D$-dimensional general relativity in vacuum to a lower 
dimensional model.

The isometry group of a Schwarzschild (or, for $D>4$,
Tangherlini~\cite{Tangherlini:1963bw}) black hole is $SO(D-1)\times
\mathbb{R}$. For a head-on collision of two non-rotating black holes, the
isometry is further reduced to $SO(D-2)$: indeed, neither the time direction nor
the direction of the collision correspond to symmetries, but a rotation of the
remaining $D-2$ spatial directions leaves the spacetime invariant.

One can take advantage of this symmetry to reduce the spacetime
dimensionality. This can be accomplished by writing Einstein's equations in
$D$~dimensions in a coordinate system which makes the symmetry manifest,
allowing for a lower dimensional interpretation of the $D$-dimensional
Einstein's equations (in the spirit of the Kaluza-Klein reduction). We remark,
however, that we do not perform a compactification; rather, we perform a
dimensional reduction by isometry, as first proposed by
Geroch~\cite{Geroch:1970nt}. The extra dimensions manifest themselves in the
lower dimensionality as a source of Einstein's equations, defined on the lower
dimensional manifold.

In the original proposal of Geroch~\cite{Geroch:1970nt} the symmetry space was
$SO(2)$.  This approach has been applied to numerical relativity, see for
instance~\cite{Sjodin:2000zd,Sperhake:2000fe,Choptuik:2003as}; a five
dimensional extension, with the same symmetry space, has been derived in~\cite{Chiang:1985rk}.  A generalisation to coset manifolds (like the sphere
$S^n$) was given by Cho in~\cite{Cho:1986wk,Cho:1987jf}, but in these papers the
complete form of Einstein's equations was not presented. 

Following the approach by Cho~\cite{Cho:1986wk}, we will start by deriving the general equations obtained doing a dimensional reduction by isometry. 
We will afterwards focus on the isometry group of the $S^n$ sphere and present the equations obtained with a dimensional reduction to four dimensions, as well as their numerical implementation.

\subsection{General formalism}
\label{sec:general-metric}

The most general $D$-dimensional metric $\bar g_{AB}$, $A = 0, \dots, d-1, \dots, (D-1)$, can be written in the following form (in the coordinate basis $\partial_A = (\partial_\mu, \partial_{\bar i})$)
\begin{align}
  \label{eq:metric0}
  \dd \bar s^2 = \bar g_{AB} \dd x^A \dd x^B = \left(
    g_{\mu\nu} + e^2 \kappa^2 g_{\bar i \bar j} B^{\bar i}_\mu B^{\bar j}_\nu
  \right) \dd x^\mu \dd x^\nu
  + 2 e \kappa B^{\bar i}_\mu g_{\bar i \bar j} \dd x^\mu \dd x^{\bar j}
  +  g_{\bar i \bar j} \dd x^{\bar i} \dd x^{\bar j}, 
\end{align}
where $\mu = 0, \dots, d-1 $ and $\bar i = d, \dots, D-1 $. $\kappa$ is a scale parameter and $e$ a coupling constant. 
This metric is fully general and \emph{not} an \emph{ansatz}.

Assume that $\bar g_{AB}$ admits an $m$-dimensional isometry $G$, generated by $m$ Killing vector fields which we express as (assuming $D-d$ is large enough)
\begin{align}
  \xi_a = K^{\bar i}_a \partial_{\bar i},
\end{align}
$a = 1, \dots, m \equiv \mathrm{dim}~G$. The Killing vector fields form the Lie algebra of $G$, satisfying 
\begin{align}
   \mathcal{L}_{\xi_a} \bar g_{AB} & = 0, \label{eq:killing} \\
  \left[ 
    \xi_a, \xi_b
  \right] & = \frac{1}{\kappa} f^c{}_{ab} \xi_c. \label{eq:algebra}
\end{align}
%
Defining 
\begin{align*}
  \left[
    \xi_a, \partial_{\bar i}
  \right] \equiv F^{\bar j}_{a \bar i} \partial_{\bar j }
  = - \left(
    \partial_{\bar i} K^{\bar j}_a
  \right) \partial_{\bar j},
\end{align*}
and the ``dual'' form $\phi^a_{\bar i}$ to the Killing fields $\xi_a$ by
\begin{align*}
  \phi^a_{\bar i} K^{\bar j}_a = \delta^{\bar j}_{\bar i},
\end{align*}
we can derive, from~\eqref{eq:killing},
\begin{equation}
  \begin{aligned}
    & \partial_{\bar i} g_{\bar j \bar k} = F^{\bar l}{}_{\bar i \bar j} g_{\bar l \bar k} 
    + F^{\bar l}{}_{\bar i \bar k} g_{\bar j \bar l}, \\
    & \partial_{\bar j} B^{\bar k}_\mu = -F^{\bar k}{}_{\bar j \bar i} B^{\bar i}_\mu, \\
    & \partial_{\bar i} g_{\mu\nu} = 0,
  \end{aligned}
\end{equation}
where $F^{\bar k}{}_{\bar i \bar j} \equiv \phi^a_{\bar i} F^{\bar k}_{a \bar j}$.

Our goal is to compute the Ricci tensor of metric~(\ref{eq:metric0}), which is more easily done in a non-coordinate basis.
Details of the computation can be found in appendix~\ref{sec:ricci-non-coord}; here we mention only the final result.

We first \emph{define} the ``covariant derivatives'' $\nabla_\mu$ and $\nabla_{\bar j}$ as \begin{align}
  \nabla_\sigma T^{\bar i\alpha}{}_{\bar k \mu} &  \equiv D_\sigma T^{\bar i \alpha}{}_{\bar k \mu} + \F{\bar i}{\sigma \bar l} T^{\bar l \alpha}{}_{\bar k \mu}
  -\F{\bar l}{\sigma \bar k} T^{\bar i \alpha}{}_{\bar l \mu} + \GG{\alpha}{\lambda \sigma} T^{\bar i \lambda}{}_{\bar k \mu} 
  - \GG{\lambda}{\mu \sigma} T^{\bar i \alpha}{}_{\bar k \lambda}, \\
  \nabla_{\bar j} T^{\bar i\alpha}{}_{\bar k \mu} &  \equiv \partial_{\bar j} T^{\bar i \alpha}{}_{\bar k \mu} 
  + \GG{\bar i}{\bar l \bar j} T^{\bar l \alpha}{}_{\bar k \mu} 
  - \GG{\bar l}{\bar k \bar j} T^{\bar i \alpha}{}_{\bar l \mu },
\end{align}
where
\begin{equation} \label{eq:Frel}
  \begin{aligned}
  D_\mu & \equiv \partial_\mu - e \kappa B^{\bar i}_{\mu} \partial_{\bar i} \\
    \F{{\bar k}}{\mu {\bar i}} & \equiv e \kappa \partial_{\bar i} B^{\bar k}_\mu = -e \kappa F^{\bar k}{}_{{\bar i} {\bar j}} B^{\bar j}_{\mu}, \\
    \F{{\bar k}}{{\bar i} {\bar j}} & \equiv 0, \\
    \F{{\bar k}}{\mu \nu} & \equiv -e \kappa G^{\bar k}{}_{\mu\nu} 
    \equiv -e \kappa \left( \partial_\mu B^{\bar k}_\nu - \partial_\nu B^{\bar k}_\mu + e \kappa \, t^{\bar k}{}_{{\bar i}{\bar j}} B^{\bar i}_\mu B^{\bar j}_\nu \right), \\
    t^{\bar i}{}_{{\bar j}{\bar k}} & \equiv F^{\bar i}{}_{{\bar j}{\bar k}} - F^{\bar i}{}_{{\bar k}{\bar j}},
  \end{aligned}
\end{equation}
and both connections are metric,
\begin{align*}
   \nabla_\sigma g_{\mu\nu} & = \partial_\sigma g_{\mu\nu} - \GG{\lambda}{\mu\sigma} g_{\lambda\nu} 
  - \GG{\lambda}{\nu\sigma} g_{\mu \lambda} = 0, \\
  \nabla_{\bar k} g_{\bar i \bar j} & \equiv \partial_{\bar k} g_{\bar i \bar j} 
  - \GG{\bar l}{\bar i \bar k} g_{\bar l \bar j} - \GG{\bar l}{\bar j \bar k} g_{\bar i \bar l} = 0.
\end{align*}
Note however that
\begin{align*}
  \nabla_\sigma g_{\bar i \bar j} \equiv D_\sigma g_{\bar i \bar j} - \F{\bar k}{\sigma \bar i}g_{\bar k \bar j} - \F{\bar k}{\sigma \bar j}g_{\bar i \bar k} \neq 0.
\end{align*}

The Ricci tensor of~(\ref{eq:metric0}) is (see appendix~\ref{sec:ricci-non-coord})
\begin{align}
  \bar R_{{\bar i}{\bar j}} & = R_{{\bar i}{\bar j}} - \frac{1}{4} g^{{\bar k}{\bar l}} \nabla_\beta g_{{\bar k}{\bar l}} \nabla^\beta g_{{\bar i}{\bar j}} 
  + \half g^{{\bar k}{\bar l}} \nabla_\beta g_{{\bar i}{\bar k}} \nabla^\beta g_{{\bar j}{\bar l}} 
  + \frac{1}{4} g^{\alpha\lambda} g^{\beta\rho} g_{{\bar j}{\bar k}} g_{{\bar i}{\bar l}} \F{{\bar k}}{\beta\lambda} \F{{\bar l}}{\rho\alpha}
  - \half \nabla^\beta \nabla_\beta g_{{\bar i}{\bar j}}, \label{eq:Rij} \\
  \bar R_{\mu {\bar i}} & = e\kappa \bar R_{{\bar i}{\bar j}} B^{\bar j}_{\mu} 
  + \half g^{\alpha\lambda} \nabla_\alpha \left(
    g_{{\bar i}{\bar k}} \F{{\bar k}}{\lambda\mu}
  \right) + \frac{1}{4} g^{{\bar k}{\bar l}} \nabla_\beta g_{{\bar k}{\bar l}} g^{\beta\lambda} \F{{\bar m}}{\lambda\mu} g_{{\bar i}{\bar m}}
  + \half \nabla_{\bar k} \left(  
    g^{{\bar k}{\bar l}} \nabla_\mu g_{{\bar l}{\bar i}}
  \right) \notag \\
  &{} \quad - \half \nabla_{\bar i} \left(  
    g^{{\bar k}{\bar l}} \nabla_\mu g_{{\bar k}{\bar l}}
  \right) = \bar R_{{\bar i} \mu}, \label{eq:Rmui} \\
  \bar R_{\mu\nu} & = R_{\mu\nu} + 2 e \kappa B^{\bar i}_{(\mu} \bar R_{\nu) {\bar i}} - e^2 \kappa^2 \bar R_{{\bar i}{\bar j}} B^{\bar i}_{\mu} B^{\bar j}_\nu
  -\half g^{\alpha\lambda} g_{{\bar i}{\bar j}} \F{{\bar i}}{\lambda\mu} \F{{\bar j}}{\alpha\nu}
  -\half \nabla_\nu\left(
    g^{{\bar i}{\bar j}} \nabla_\mu g_{{\bar i}{\bar j}}
  \right) \notag \\
  &{} \quad- \frac{1}{4} g^{{\bar i}{\bar j}} g^{{\bar k}{\bar l}} \nabla_\mu g_{{\bar i}{\bar k}} \nabla_\nu g_{{\bar j}{\bar l}}
  -\half \nabla_{\bar k} \F{{\bar k}}{\mu\nu}, \label{eq:Rmunu}
\end{align}
and 
\begin{align}
  \label{eq:Rscalar2}
    \bar R & = R + \tilde R - \frac{1}{4} g_{{\bar k}{\bar l}} g^{\alpha\lambda} g^{\mu\nu} \F{{\bar l}}{\lambda\mu} \F{{\bar k}}{\alpha \nu}
    -\nabla^\mu \left(g^{{\bar k}{\bar l}} \nabla_\mu g_{{\bar k}{\bar l}} \right)
    - \frac{1}{4} g^{{\bar k}{\bar i}} g^{{\bar j}{\bar l}} \nabla^\mu g_{{\bar k}{\bar l}} \nabla_\mu g_{{\bar i}{\bar j}} \notag \\
    &{} \quad - \frac{1}{4} g^{{\bar k}{\bar l}} g^{{\bar i}{\bar j}} \nabla^\mu g_{{\bar k}{\bar l}} \nabla_\mu g_{{\bar i}{\bar j}}
    .
\end{align}
These are the expressions we were looking for.
Equivalent forms can be found in~\cite{Chiang:1985rk,Cho:1986wk}.


\subsection{Examples}

\subsubsection{\texorpdfstring{$S^1$}{S1}}
\label{sec:kaluza-klein}

As a first (trivial) exercise, we can reproduce the standard Kaluza-Klein expressions. Remember that the Kaluza-Klein metric has the form
\begin{equation}
  \label{eq:kaluza}
  \dd \bar s^2 = g_{\mu\nu} \dd x^\mu \dd x^\nu 
  + e^{2\phi} \left(
    \dd x^5 + A_\mu \dd x^\mu
  \right)^2.
\end{equation}
We can easily recover this case from our formalism by making $d = 4$, $D=5$, $g_{\bar i \bar j} \to e^{2\phi}$, $g^{\bar i \bar j} \to e^{-2\phi}$, $e \kappa B^{\bar i}_\mu \to A_\mu$, and $\F{\bar i}{\mu\nu} \to -F_{\mu\nu} \equiv -\left( \partial_\mu A_\nu - \partial_\nu A_\mu \right)$ (cf. equations~\eqref{eq:Frel}). Remember also that for the Kaluza-Klein case nothing depends on the ``fifth'' dimension, and as such $\F{\bar i}{\mu \bar j} = 0$. We get the usual Kaluza-Klein expressions,
\begin{align*}
  \bar R_{\bar i \bar j} & \to e^{2\phi} \left(
    \frac{1}{4} e^{2\phi} F^{\alpha\beta} F_{\alpha\beta} - \partial_\alpha \phi \partial^\alpha \phi - \nabla^\alpha \partial_\alpha \phi
  \right) \equiv \bar R_{55}, \\
  \bar R_{\mu \bar i} & \to A_\mu \bar R_{55} + \frac{3}{2} \partial^{\alpha}e^{2\phi} F_{\mu\alpha} 
  + e^{2\phi} \nabla^\alpha F_{\mu \alpha} \equiv \bar R_{\mu 5}, \\
  \bar R_{\mu\nu} & \to R_{\mu\nu} + 2 A_{(\mu} \bar R_{\nu) 5} - A_{\mu} A_\nu \bar R_{55} 
  - \half e^{2\phi} F^{\alpha}{}_\mu F_{\alpha\nu} - \nabla_\nu\partial_\mu\phi 
  - \partial_\mu \phi \partial_\nu \phi, \\
  \bar R & \to  R -\frac{1}{4} e^{2\phi} F^{\alpha\beta} F_{\alpha\beta} - 2 \nabla^\alpha \partial_\alpha \phi
  - 2 \partial_\alpha \phi \partial^\alpha \phi.
\end{align*}

\subsubsection{\texorpdfstring{$S^n$}{Sn}}
\label{sec:Sn}

A more interesting case is performing the dimensional reduction on the $S^n$ sphere, $n\equiv D-d \ge 2$. 
For such an isometry, the Killing vectors $\xi_a$, $a=1,\dots,(n+1)n/2$ satisfy
\begin{equation}
  \left[\xi_a,\xi_b\right]=\epsilon_{ab}{}^c\xi_c \, , 
  \label{algebra}
\end{equation}
where $\epsilon_{ab}{}^{c}$ are the structure constants of $SO(n+1)$. Because
the fibre has the minimal dimension necessary to accommodate $n(n+1)/2$
independent Killing vector fields, we may assume without loss of generality that
the Killing vector fields have components exclusively along the fibre:
$\xi_a=\xi_a^{\bar{i}}\partial_{\bar{i}}$.  Furthermore, we may normalise the
Killing vectors so that they only depend on the coordinates of the fibre,
i.e. $\partial_{\mu}\xi_a^{\bar i}=0$. 

Equation~\eqref{eq:killing} gives the following conditions
\begin{align}
  \label{Omega}
  \mathcal{L}_{\xi_a}  g_{\bar{i}\bar{j}} & =0 \, , \\
  \label{comm}
  \mathcal{L}_{\xi_a}  B_{\mu}^{\bar{i}} & =0 \, , \\
  \label{g}
  \mathcal{L}_{\xi_a}  g_{\mu \nu} & =0 \, .
\end{align}
These expressions can be interpreted either as Lie derivatives of
rank-$2$ tensors defined on the $D$-dimensional spacetime, or as Lie
derivatives of a rank-$2$ tensor, a vector and a scalar, which are
defined on $S^{n}$.

Together with~(\ref{algebra}), conditions~\eqref{Omega}-\eqref{g} have the following implications:
\begin{itemize}
\item
  \begin{equation}
    g_{\bar{i}\bar{j}}=
    f(x^{{\mu}})h_{\bar{i}\bar{j}}^{S^{n}} \,,
    \end{equation}
because, from~\eqref{Omega},  $g_{\bar{i}\bar{j}}$ admits the maximal number of Killing vector fields and thus must be the metric on a maximally
symmetric space at each $x^{\mu}$.  Due to~\eqref{algebra} this space must be the 
$S^{n}$ sphere.
$h_{\bar{i}\bar{j}}^{S^{n}}$ denotes the metric on an $S^{n}$ with unit
radius;
\item
  \begin{equation}
    g_{\mu\nu}=
  g_{\mu \nu}(x^{\mu}) \,,
  \end{equation}
because the Killing vector fields $\xi_a$ act transitively on the fibre
and therefore the base space metric must be independent of the fibre
coordinates;

\item
  \begin{equation}
  B_{{\mu}}^{\bar{i}}=0\, ,
  \label{novectors}
\end{equation}
because equation~(\ref{comm}) is equivalent to 
\begin{equation}
  [\xi_a,B_{\mu}]=0\label{comm1} \ ,
\end{equation}
and there exist no non-trivial vector fields on $S^{n}$ for $n\ge 2$
that commute with all Killing vector fields on the sphere.

\end{itemize}

We write the metric on the sphere as 
\begin{align}
\label{eq:sphere}
  g_{\bar i \bar j} \dd x^{\bar i} \dd x^{\bar j} = e^{2\phi} h_{\bar i \bar j} \dd x^{\bar i} \dd x^{\bar j},
\end{align}
with 
$\phi = \phi(x^\mu)$. Our $D$-dimensional metric has a block diagonal form. Making $g_{\bar i \bar j} = e^{2\phi} h_{\bar i \bar j} $ and $B^{\bar i}_\mu = 0$ in the expressions~\eqref{eq:Rij}-\eqref{eq:Rscalar2} we get
\begin{equation}
  \label{eq:Ricci_sphere}
  \begin{aligned}
    \bar R_{\bar i \bar j} & = R_{\bar i \bar j} - e^{2\phi} h_{\bar i \bar j} \left(
      n \partial^\alpha\phi \partial_\alpha \phi + \nabla^\alpha \partial_\alpha \phi
    \right), \\
    \bar R_{\mu \bar i} & = 0, \\
    \bar R_{\mu\nu} & = R_{\mu\nu} - n \nabla_\nu \partial_\mu \phi - n \partial_\mu \phi \partial_\nu \phi, \\
    \bar R & = R + \tilde R - 2n \nabla^\mu \partial_\mu \phi - n(n+1) \partial^\mu \phi \partial_\mu\phi,
  \end{aligned}
\end{equation}
where $R_{\bar i \bar j} $ and $\tilde R$ are the Ricci tensor and Ricci scalar for the metric~\eqref{eq:sphere}.
They evaluate to
\begin{align}
  \label{eq:Rijsphere}
  R_{\bar j \bar l} = 
  (n-1)h_{\bar j \bar l}, \qquad
  \tilde R = 
n(n-1)e^{-2\phi} \,.
\end{align}

For $D$-dimensional vacuum spacetimes $\bar R_{AB} = 0 = \bar R_{\mu\nu} = \bar R_{\bar i \bar j} $. Using also~\eqref{eq:Rijsphere} on~\eqref{eq:Ricci_sphere} we get two coupled equations,
\begin{equation}
\label{eq:eq-jordan-frame}
\begin{aligned}
  & e^{2\phi} \left(
    n \partial^\alpha \phi \partial_\alpha \phi + \nabla^\alpha\partial_\alpha \phi
  \right) = n -1 \\
  & R_{\mu\nu}= n \nabla_\nu\partial_\mu \phi + n \partial_\mu \phi \partial_\nu \phi
\end{aligned}.
\end{equation}

These equations can also be obtained from the following action
\begin{align}
  \label{eq:action0}
  \mathcal{S} = \frac{1}{16\pi G_d}\int \dd ^d x \sqrt{-g} e^{n \phi} \left[
    R + n(n-1) e^{-2\phi} + n(n-1) \partial_{\mu} \phi \partial^{\mu} \phi
  \right] \ .
\end{align}

Performing the substitution $e^{2\phi} = \lambda^p$ will be useful for the upcoming numerical implementation. We get
\begin{equation}
\begin{aligned}
  & p \lambda^{p-1} \left[
    \left(
      \frac{np}{2} -1
    \right)
    \lambda^{-1} \partial^\alpha \lambda \partial_\alpha \lambda
    + \nabla^\alpha \partial_\alpha\lambda
  \right] = 2(n-1), \\
  & R_{\mu\nu} = \frac{np}{2} \lambda^{-1} \nabla_\nu \partial_\mu \lambda
  + \frac{np}{4} \lambda^{-2} (p-2) \partial_\mu \lambda \partial_\nu \lambda .
\end{aligned}\label{eq:eq-lambda}
\end{equation}

For completeness, we write in appendix~\ref{sec:einstein-frame} the equations of motion obtained when we write the action~(\ref{eq:action0}) in the Einstein frame.

\section{Dimensional reduction on a \texorpdfstring{$(D-4)$}{(D-4)}-sphere and 3+1 split}
\label{sec:dim-red_split}

In the previous section we were considering a dimensional reduction under the full isometry group of the higher-dimensional spacetime.
In the case of head-on black hole collisions, this would produce a reduction down to 3 spacetime dimensions.
In practice, we are actually interested in performing a $4+(D-4)$ split of the
$D$ dimensional spacetime. This may be done as follows. The metric on a unit
$S^{D-3}$ may always be written in terms of the line element on a unit
$S^{D-4}$, denoted by $\dd \Omega_{D-4}$, as follows,
\begin{equation}
  h_{\bar{i}\bar{j}}^{S^{D-3}} \dd x^{\bar{i}}\dd x^{\bar{j}}=
      \dd \theta^2+\sin^2\theta \dd \Omega_{D-4} \, ,
\end{equation}
where $\theta$ is a polar-like coordinate, $\theta\in [0,\pi]$. Now we
introduce four dimensional coordinates, $x^\mu\equiv(x^{\bar{\mu}},\theta)$,
$\mu=0,1,2,3$, and define a four dimensional metric
\begin{equation}
  g_{\mu\nu}\dd x^\mu \dd x^\nu=g_{\bar{\mu}\bar{\nu}}\dd x^{\bar{\mu}}\dd x^{\bar{\nu}}
      +f(x^{\bar{\mu}}) \dd \theta^2 \, ,
   \label{4d}
\end{equation}
as well as a new conformal factor
\begin{equation}
  \lambda(x^{\mu})=\sin^2\theta g_{\theta\theta} \, .
  \label{conformal}
\end{equation}

As we have seen in the previous sections, the most general $D$-dimensional metric compatible with $SO(D-2)$ isometry is, for $D\ge 5$
\begin{equation}
  \dd \hat{s}^2= g_{\mu\nu}\dd x^\mu \dd x^\nu + \lambda(x^{\mu}) \dd \Omega_{D-4} \, .
  \label{ansatz} 
\end{equation}

The geometry~\eqref{ansatz} has a manifest $SO(D-3)$ symmetry. We will now
perform a dimensional reduction on a $(D-4)$-sphere, which yields, from the
$D$-dimensional vacuum Einstein equations, a set of $3+1$ dimensional Einstein
equations coupled to quasi-matter.
In cases with larger symmetry (if $SO(D-2)$ is the full isometry group, for example), the
quasi-matter terms do not contain independent degrees of freedom and could in principle be fully determined by the $3+1$ dimensional geometry. 
For such cases we could perform the dimensional reduction on a $(D-3)$-sphere instead (which has the full isometry group $SO(D-2)$), which would yield a $2+1$ dimensional
system. The former method allows, however, the use of existing numerical codes,
with small changes, which justifies our choice.

\begin{figure}[ht]
\centering
\includegraphics[width=0.8\textwidth]{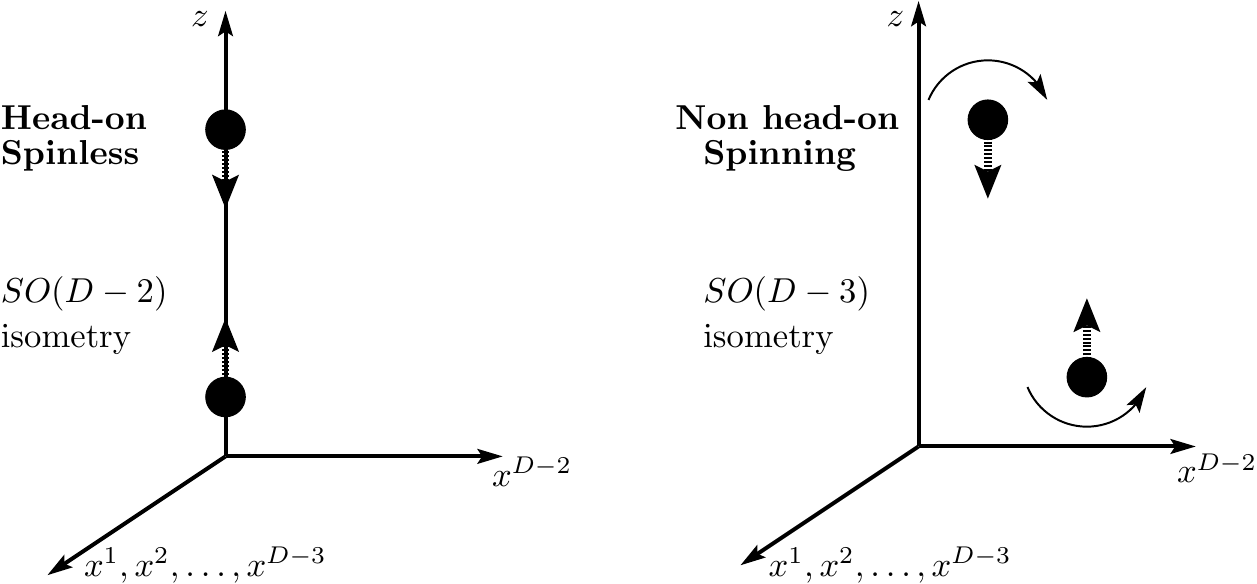}
\caption[$D$-dimensional representation of two types of black hole collisions]{
  $D$-dimensional representation, using coordinates $(t, x^1,x^2,\dots,
  x^{D-3},x^{D-2},z)$, of two types of black hole collisions: (left panel) head-on for
  spinless black holes, for which the isometry group is $SO(D-2)$; (right panel) non
  head-on, with motion on a \textit{single} 2-plane, for black holes spinning in that
  \textit{same} plane only, for which the isometry group is $SO(D-3)$. The
  figures make manifest the isometry group in both cases.}
\label{headon_impact}
\end{figure}

The $SO(D-3)$ isometry group allows the study of a large class of black hole
collisions with impact parameter and with spin: the collisions in which the two
black holes always move on the same 2-plane and the only non trivial components of the
spin 2-form are on that same 2-plane---see figure~\ref{headon_impact}.  
With our framework we are able, therefore, to describe not only head-on collisions of
spinless black holes but also a class of collisions for spinning black holes with impact
parameter.  
As follows from the discussion of~\eqref{novectors}, the ansatz
\eqref{ansatz} describes general spacetimes with $SO(D-3)$ isometry in $D\ge
6$.  We remark that the models with $D\ge6$ are actually the most interesting
for phenomenological studies of large extra dimensions models (see for instance~\cite{Kanti:2004nr}).

Taking~\eqref{ansatz} as an ansatz, we see from~(\ref{eq:action0}) that the $D$-dimensional Einstein-Hilbert action takes the form (for reasons related with the numerical implementation, we now use the variable $\lambda$ instead of the previously used $\phi$)
\begin{equation}
  \mathcal{S} = \frac{1}{16\pi G_4}\int \dd ^4x\sqrt{-g} \lambda^{\frac{D-4}{2}}
  \left[
    R + (D-4)(D-5) \left(
      \lambda^{-1} + \frac{1}{4}\lambda^{-2} \partial_\mu \lambda \partial^\mu \lambda
    \right)
  \right] \, ,
\end{equation}
where the $D$-dimensional Newton's constant $G_D$ is related to the four
dimensional one $G_4$ by the area of the unit ${D-4}$ dimensional sphere:
$G_4=G_D/A^{S^{D-4}}$. Explicitly, the $D$-dimensional Einstein's equations in
vacuum yield the following system of four dimensional equations coupled to a
scalar field:
\begin{align}
  R_{\mu\nu} & =\frac{D-4}{2 \lambda}\left(\nabla_\mu\partial_\nu \lambda
      - \frac{1}{2\lambda} \partial_\mu\lambda \partial_\nu \lambda \right) \, ,
  \label{4deinstein} \\
  \nabla^\mu \partial_\mu \lambda & = 2(D-5)
      - \frac{D-6}{2\lambda} \partial_\mu \lambda \partial^\mu \lambda \, .
  \label{scalar}
\end{align}
In these equations, all operators are covariant with respect to the four
dimensional metric $g_{\mu\nu}$. These could also be obtained from equations~(\ref{eq:eq-lambda}) with $p=1$. The energy momentum tensor is
\begin{equation}
  T_{\mu\nu}=\frac{D-4}{16\pi \lambda}\left[\nabla_\mu\partial_\nu \lambda
      - \frac{1}{2\lambda} \partial_\mu\lambda \partial_\nu\lambda
      - (D-5)g_{\mu\nu} + \frac{D-5}{4\lambda} g_{\mu \nu} \partial_\alpha
      \lambda \partial^\alpha \lambda \right] \, .
  \label{emtensor}
\end{equation}

With this four dimensional perspective, the usual $3+1$ split of spacetime can
be performed, as outlined in section~\ref{sec:choice-coordinates}.  As explained
therein, the projection operator $\gamma_{\mu\nu}$ and the normal to the three
dimensional hypersurface $\Sigma$, $n^\mu$ ($n^\mu n_\mu = -1$), are introduced
\begin{align}
  \gamma_{\mu\nu} = g_{\mu\nu} + n_\mu n_\nu \, ,
\end{align}
as well as the lapse $\alpha$ and shift $\beta^\mu$,
\begin{align}
  \partial_t = \alpha n + \beta \ ,
\end{align}
where $t$ is the time coordinate. The four dimensional metric is then
written in the form
\begin{equation}
  \label{4dinitial}
  \dd s^2=g_{\mu\nu}\dd x^\mu \dd x^\nu=-\alpha^2 \dd t^2+\gamma_{ij}
  (\dd x^i+\beta^i\dd t)(\dd x^j+\beta^j\dd t) \, , \qquad i,j=1,2,3 \, .
\end{equation}

As usual, we introduce the extrinsic curvature $K_{ij} = -\half
\mathcal{L}_n \gamma_{ij}$, which gives the evolution equation for
the $3$-metric~(\ref{eq:lie}).
Defining the variable
\begin{align}
  \label{eq:Kphi0}
  K_\lambda \equiv - \frac{1}{2} \mathcal{L}_n \lambda =
      - \frac{1}{2} n^\mu \partial_\mu \lambda \, ,
\end{align}
we further get an evolution equation for $\lambda$ 
\begin{equation}
  \label{phievo}
  \left(
    \partial_t - \mathcal{L}_\beta \right)
  \lambda  = - 2 \alpha K_\lambda \, .
\end{equation}
Using the relation
\begin{align}
  \label{eq:proj}
  D_\alpha D_\beta \lambda
      = -K_{\alpha\beta} n^\sigma \partial_\sigma \lambda
      +  \gamma^\mu{}_\alpha \gamma^\nu{}_\beta \nabla_\nu \partial_\mu
      \lambda \, ,
\end{align}
where $D_{\alpha}$ denotes now the covariant derivative with respect to the 3-metric $\gamma_{\mu\nu}$ on $\Sigma$,
and equation~\eqref{scalar} we can get an evolution equation for
$K_\lambda$. The contraction of equation~\eqref{eq:proj} with $g^{\alpha\beta}$,
yields
\begin{align}
  \square \lambda = \gamma^{ij}D_i\partial_j \lambda - 2 K K_\lambda
      - n^\mu n^\nu \nabla_\nu \partial_\mu \lambda \, .
\end{align}
Noting that
\begin{align}
  \mathcal{L}_n K_\lambda = n^\mu \partial_\mu K_\lambda 
  = -\half n^\mu \nabla_\mu n^\nu \partial_\nu \lambda
  - \half n^\mu n^\nu \nabla_\mu \partial_\nu \lambda \, ,
\end{align}
and 
\begin{align}
  n^\mu \nabla_\mu n^\nu = \frac{1}{\alpha} D^\nu \alpha \, ,
\end{align}
we obtain
\begin{align}
  - n^\mu n^\nu \nabla_\mu \partial_\nu \lambda =  2 \mathcal{L}_n K_\lambda
  + \frac{1}{\alpha} D^\nu \alpha \partial_\nu \lambda \, .
\end{align}
Noticing also that $D^\nu \alpha \partial_\nu \lambda = \gamma^{ij}\partial_i \alpha \partial_j
\lambda  $, we write
\begin{align}
  \square \lambda = \gamma^{ij}D_i\partial_j \lambda - 2 K K_\lambda
      + 2 \mathcal{L}_n K_\lambda + \frac{1}{\alpha} \gamma^{ij}\partial_i \alpha \partial_j
\lambda   \, .
\end{align}
Moreover, from equation
\begin{align}
  D_\mu \lambda = \gamma^\nu{}_\mu \partial_\nu \lambda 
  = \partial_\mu \lambda - 2 n_\mu K_\lambda \, ,
\end{align}
we get
\begin{align}
  \partial_\alpha \lambda \partial^\alpha \lambda = \gamma^{ij}\partial_i \lambda \partial_j
       \lambda - 4 K_\lambda ^2 \, ,
\end{align}
so that the evolution equation for $K_\lambda$ is
\begin{align}
  \label{eq:evolphi0}
  \frac{1}{\alpha} \left(
    \partial_t - \mathcal{L}_\beta
  \right) K_\lambda
   & = - \frac{1}{2\alpha} \gamma^{ij}\partial_i \lambda \partial_j \alpha
  + (D-5) + K K_\lambda + \frac{D-6}{\lambda} K_\lambda^2
  - \frac{D-6}{4\lambda} \gamma^{ij} \partial_i \lambda \partial_j \lambda
 \notag \\
  & \quad - \half D^k \partial_k \lambda \, .
\end{align}
Equations~\eqref{phievo} and~\eqref{eq:evolphi0} are the evolution
equations for the quasi-matter degrees of freedom.

\subsection{BSSN formulation}

For numerical implementation, we write the evolution equations
in the BSSN formulation, as introduced in section~\ref{chap:bssn-gauge}.
The evolution equations are those of~(\ref{eq:bssn-gen})
with source terms determined by~(\ref{eq:source-gen})
where the energy momentum tensor is given by equation~\eqref{emtensor}.
A straightforward computation shows that
\begin{subequations}
  \label{matterterms}
  \begin{align}
    \begin{split}
    \frac{4 \pi (E + S)}{D-4} & = -(D-5) \lambda^{-1} + \frac{1}{2}
        \lambda^{-1} \chi^{3/2} \tilde \gamma^{ij} \tilde D_i 
        \left(
          \chi^{-1/2} \partial_j \lambda
        \right)  \\
    &{} \quad
    + \frac{D-6}{4} \lambda^{-2} \chi \tilde \gamma^{ij} \partial_i
    \lambda \partial_j \lambda 
    - \lambda^{-1} K K_\lambda - (D-5) \lambda^{-2} K_\lambda^2 \, ,
  \end{split} \\
  \begin{split}
    \frac{8\pi \chi \left( S_{ij} - \frac{S}{3} \gamma_{ij}
      \right)}{D-4} & = \frac{1}{2} \chi \lambda^{-1} \tilde D_i
    \partial_j \lambda + \frac{1}{4} \lambda^{-1} \left(
      \partial_i \lambda \partial_j \chi + \partial_j
      \lambda \partial_i \chi - \tilde \gamma^{kl} \tilde
      \gamma_{ij} \partial_k \lambda
      \partial_l \chi \right) - \frac{1}{4} \chi
    \lambda^{-2} \partial_i
    \lambda \partial_j \lambda  \\
    &{} \quad - \lambda^{-1} K_\lambda \tilde{A}_{ij} -
    \frac{1}{6}\tilde \gamma_{ij} \lambda^{-1} \chi^{3/2} \tilde
    \gamma^{kl} \tilde D_k \left( \chi^{-1/2} \partial_l \lambda
    \right) + \frac{1}{12} \tilde \gamma_{ij} \lambda^{-2} \chi \tilde
    \gamma^{kl} \partial_l \lambda
    \partial_k \lambda \, ,
  \end{split} \\
%
        \frac{16 \pi \chi^{-1} j^i}{D-4} & = 2 \lambda^{-1}
        \tilde \gamma^{ij} \partial_j K_\lambda
        - \lambda^{-2} K_\lambda \tilde \gamma^{ij} \partial_j \lambda 
        - \tilde \gamma^{ik} \tilde \gamma^{lj} \tilde{A}_{kl} \lambda^{-1}
        \partial_j \lambda-\frac{\tilde{\gamma}^{ij}}{3}K\lambda^{-1}
        \partial_j\lambda  \, ,
  \end{align}
\end{subequations}
where $\tilde{D}_i$ is the covariant derivative with respect to
$\tilde{\gamma}_{ij}$.

Finally, the evolution equations for $\lambda$ and $K_\lambda$ are
\begin{subequations}
  \label{kl}
  \begin{align}
    \left( \partial_t - \mathcal{L}_\beta \right) \lambda & = - 2
        \alpha K_\lambda , \\
        \begin{split}
          \left( \partial_t - \mathcal{L}_\beta \right) K_{\lambda} &
          = \alpha \bigg\{ (D-5) + \frac{6-D}{4} \left[ \lambda^{-1}
            \chi \tilde{\gamma}^{ij}\partial_i \lambda
            \partial_j \lambda - 4 \lambda^{-1} K_\lambda^2
          \right] \\
          &{} \quad + K K_\lambda - \half \chi^{3/2} \tilde
          \gamma^{kl} \tilde D_k \left( \chi^{-1/2} \partial_l \lambda
          \right) \bigg\} - \half \chi \tilde{\gamma}^{ij}\partial_j
          \alpha \partial_i \lambda \, .
        \end{split}
  \end{align}
\end{subequations}
As stated before, in the case of head-on collisions of spinless black holes the full
symmetry of the $D$-dimensional system we want to consider makes
equations~\eqref{kl} redundant, by virtue of~\eqref{conformal}.  This allows to
determine the quasi-matter degree of freedom in terms of the three dimensional
spatial geometry, at each time slice.  
The extra symmetry
manifests itself in the fact that $\gamma_{ij}$ possesses, at all times, (at
least) one Killing vector field. If one chooses coordinates adapted to this
Killing vector field, $\partial/\partial \theta$, the metric can then be written
in the form~\eqref{4d}, and then the quasi-matter degree of freedom can be
determined from the spatial geometry by~\eqref{conformal}. In the numerical
implementation, one can either determine, at each time-step, the scalar field
through~(\ref{conformal}), or impose~(\ref{conformal}) only in the initial data,
and then evolve the scalar field using equation~(\ref{kl}).
We have implemented the latter method.

\section{Higher-dimensional initial data}
\label{sec:init_data-hd}

Having written our system of evolution equations, we now need to construct relevant initial data.
Building on the results outlined in chapter~\ref{chap:init_data} (based on~\cite{Yoshino:2005ps,Yoshino:2006kc}), we now present a generalisation of the spectral solver in~\cite{Ansorg:2004ds} that generates initial data for black hole binaries in $D\ge 5$ dimensions with non-vanishing initial boost~\cite{Zilhao:2011yc}.

In this section we need to make some changes to our notation. Early lower case
Latin indices $a,\,b,\,c,\ldots$ will here extend from $1$ to $D-1$, late lower
case Latin indices $i,\,j,\,k,\ldots$ run from $1$ to $3$ and early upper case
Latin indices $A,\,B,\,C,\ldots$ from $4$ to $D-1$.

\subsection{Coordinate transformation}

We start by recalling, from chapter~\ref{chap:init_data} that, for a system of boosted black holes, we can solve the momentum constraint equation~\eqref{eq:momentum2} analytically. 
It remains then to solve equation~(\ref{eq:u}), which we re-write here:
\begin{align}
  \hat \triangle u + \frac{D-3}{4(D-2)} \hat A^{ab} \hat A_{ab}
       \psi^{ -\frac{3D-5}{D-3} } = 0 \ .
  \label{eq:u_chaphd}
\end{align}
The numerical solution of this equation will be our task in this section.

First, it is convenient to transform to a coordinate system adapted to the
generalised axial symmetry $SO(D-2)$ in $D=5$ dimensions and $SO(D-3)$ in
$D\ge6$ dimensions as discussed in section~\ref{sec:dim-red_split}. For this
purpose we consider the (flat) conformal spatial metric in cylindrical coordinates
\begin{equation}
  \hat\gamma_{ab}\dd x^a\dd x^b=\dd z^2 + \dd \rho^2+\rho^2
  \left(\dd \varphi^2 +\sin^2\varphi
    \dd \Omega_{D-4}
  \right) \ ,
  \label{initialspatial}
\end{equation}
where $\dd \Omega_{D-4}$ is the metric on the $(D-4)$-sphere.
Observe that $\varphi$ is a polar rather than an azimuthal coordinate,
i.e.~$\varphi\in [0,\pi]$. Next, 
we introduce ``incomplete'' Cartesian coordinates as
\begin{equation}
  x=\rho \cos\varphi \, ,  \qquad y=\rho \sin\varphi \ ,
  \label{inccartesian}
\end{equation}
where $-\infty<x<+\infty$ and $0\le y<+\infty$.
The $D$~dimensional initial data for the spatial metric is then
\begin{equation}
  \bar \gamma_{ab}\dd x^a\dd x^b=\psi^{\frac{4}{D-3}}\left[\dd x^2+ \dd y^2 + \dd z^2+y^2
    \dd \Omega_{D-4}\right]  \ .
  \label{smhigher}
\end{equation}

We can transform the $D-1$ dimensional Cartesian coordinates
$\mathcal{X}^a = (x^1,
\dots,x^{D-1})$ to the coordinate system $\mathcal{Y}^a = (x,y,z,\xi_1,
\xi_2, \dots, \xi_{D-4})$ with hyperspherical coordinates $\xi_1,\ldots,
\xi_{D-4}$ by
\begin{equation}
  \label{eq:coord-transf}
  \begin{aligned}
    x^1 & = x  & \\
    x^2 & =  y \cos \xi_1 & \\
    x^3 & = z  &\\
    x^4 & =  y \sin \xi_1 \cos \xi_2 \qquad  &(D\ge 6) \\
    x^5 & =  y \sin \xi_1 \sin \xi_2 \cos \xi_3  \qquad  & (D\ge 7) \\
    \vdots \\
    x^{D-3} & = y \sin \xi_1 \cdots \sin \xi_{D-6} \cos \xi_{D-5}  \qquad & (D\ge 7) \\
    x^{D-2} & = y \sin \xi_1 \cdots \sin \xi_{D-5} \cos \xi_{D-4}  \qquad & (D\ge 6) \\
    x^{D-1} & = y \sin \xi_1 \cdots \sin \xi_{D-4}  \qquad & (D\ge 5)
  \end{aligned}\ .
\end{equation}

Without loss of generality, we can always choose coordinates such that the black
holes are initially located on the $z$ axis at $z_1$ and $z_2$ and have momenta
of equal magnitude in opposite directions $P_{(1)}^a = -P_{(2)}^a$.
Inserting the momenta into equation~(\ref{eq:Aab_i}) then provides the conformal
traceless extrinsic cuvature and the differential equation (\ref{eq:u_chaphd})
which is solved numerically for $u$.

The class of symmetries covered by the formalism developed in
this chapter includes head-on and grazing
collisions of non-spinning black holes with initial position and momenta
\begin{align}
  & x_{(1)}^a = (0, 0, z_1,0,\ldots, 0)\,, \quad
     x_{(2)}^a = (0, 0, z_2,0,\ldots, 0) \nonumber \\
  & P_{(1)}^a = (P^x, 0, P^z, 0, \dots, 0) = -P_{(2)}^a \ . \label{eq:pos}
\end{align}
Note that a non-zero $P^y$ is not compatible with the assumed symmetries.
On the other hand, the $x$-axis can always be oriented such that the
collision takes place in the $xz$ plane. Our formalism therefore
covers general grazing collisions of non-spinning black hole binaries
in $D$ dimensions.

\subsection{Four dimensional initial data for a general \texorpdfstring{$D$}{D} head-on collision}

We will now discuss in detail the case of black holes with momenta in the $z$
direction, that is, the case given by setting $P^x=0$ in equation~(\ref{eq:pos}).
The linear momenta are thus given by
\begin{equation}
  P^a_{(1)} = (0,0,P^z,0,\ldots ,0) = -P^a_{(2)}.
  \label{eq:Pheadon}
\end{equation}
%
The rescaled trace-free part of the extrinsic curvature for such a
configuration is
\begin{equation}
  \hat A_{ab} = \hat A_{ab}^{(1)} + \hat A_{ab}^{(2)} \ ,
\end{equation}
where $\hat A_{ab}^{(1)}$ and $\hat A_{ab}^{(2)}$ are given by
equation~\eqref{eq:Aab_i} with \eqref{eq:pos} and \eqref{eq:Pheadon}.
Using equation~\eqref{eq:coord-transf} we can write this in the
coordinate system $\mathcal{Y}^a$ adapted to the spacetime symmetry:
%
%
\begin{equation}
\label{eq:hatAab1}
\hat A_{ab}^{(1)} =
  \displaystyle{
    \frac{4 \pi (D-1) P^z}{(D-2) \mathcal{A}_{D-2}
    (x^2 + y^2 + (z-z_1)^2)^{\frac{D+1}{2}} }
  }
\left(
\begin{array}{c|c}
 \hat a_{i j}^{(1)} & 0  \\
 \hline
 0 & \hat a_{AB}^{(1)}
\end{array}
\right),
\end{equation}
with
\begin{equation}
\label{eq:A}
\hat a_{ij}^{(1)} =
\left(
\begin{smallmatrix}
 - \left[-(D-4)x^2+y^2+(z-z_1)^2\right] (z-z_1) & (D-3)  x y
   (z-z_1) &  x \left[x^2+y^2+ (D-2) (z-z_1)^2\right] \\
 (D-3) x y (z-z_1) & - \left[x^2 - (D-4)y^2+(z-z_1)^2\right]
   (z-z_1) &  y \left[x^2+y^2 + (D-2) (z-z_1)^2\right] \\
  x \left[x^2+y^2+ (D-2) (z-z_1)^2\right] &  y \left[x^2+y^2+ (D-2)
   (z-z_1)^2\right] &  \left[x^2+y^2 + (D-2) (z-z_1)^2\right] (z-z_1)
\end{smallmatrix}
\right) \ ,
\end{equation}
and
\begin{equation}
  \hat a_{AB}^{(1)} = - y^2 (z-z_1) \left[ x^2 + y^2
                      + (z-z_1)^2 \right] h_{AB}\ ,
\end{equation}
where $h_{AB}$ is the metric on the $(D-4)$-sphere.
The expression for $\hat A^{(2)}_{ab}$ is analogous, but with $z_2$ in place of $z_1$ and
$-P^z$ in place of $P^z$ in equation~(\ref{eq:hatAab1}).

We now need to re-express these quantities in terms of our $3+1$ quantities, $(\gamma_{ij}, K_{ij}, \lambda, K_{\lambda})$, as introduced in the previous section.
These are the variables evolved in time
and therefore the variables we ultimately wish to construct from
the initial data calculation. For their extraction we first note that
$\gamma_{ij}$, $K_{ij}$ and $K_{\lambda}$ are related to the $(D-1)$-dimensional
metric $\bar{\gamma}_{ab}$ and extrinsic curvature $\bar{K}_{ab}$ by
\begin{align}
  \bar{\gamma}_{ij} & = \gamma_{ij}\ , \quad
  \bar{\gamma}_{AB} = \lambda h_{AB}\ , \nonumber \\
  \bar{\gamma}_{iA} & = 0\ , \\
  \bar K_{ij} & =  K_{ij}\ , \quad
  \bar K_{AB} = \frac{1}{2} K_\lambda h_{AB}\ , \nonumber \\
  \bar K_{iA} & = 0\ , \quad
  \bar K =  K + \frac{D-4}{2} \frac{K_\lambda}{\lambda}\ .
\end{align}
%
Using these relations and equation~(\ref{ansatz}) we
can express all ``3+1'' variables in terms of those describing the
initial data
\begin{equation}
  \label{eq:init-data-boost}
  \begin{aligned}
    \gamma_{ij} & =  \psi^{\frac{4}{D-3}} 
        \delta_{ij}\ ,\quad 
    %
    \lambda = \psi^{\frac{4}{D-3}} y^2\ , \\
    K_{ij} & = \psi^{-2}(\hat A_{ij}^{(1)} + \hat A_{ij}^{(2)})\ , \quad
    K_\lambda = 2 \psi^{-2} y^2 ( P^{+} + P^{-})\  , \\
    K & = - \frac{(D-4)K_{\lambda}}{2\lambda} \ ,
  \end{aligned}
\end{equation}
where
%
%
%
\begin{equation}
  \begin{split}
P^{+} &\equiv -\frac{4 \pi (D-1) P^z (z-z_1)}{(D-2) \mathcal{A}_{D-2} (x^2 + y^2+ (z-z_1)^2)^{\frac{D-1}{2}}  }\,,\\
P^{-} &\equiv \frac{4 \pi (D-1) P^z (z-z_2)}{(D-2) \mathcal{A}_{D-2} (x^2 + y^2+ (z-z_2)^2)^{\frac{D-1}{2}}  }\,.
\end{split}
\end{equation}
The conformal factor is
\begin{equation}
\begin{aligned}
  \psi = 1 + \frac{\mu_1}{4 \left[x^2+y^2+(z-z_1)^2 \right]^{(D-3)/2}}
          +
  \frac{\mu_2} {4 \left[x^2+y^2+(z-z_2)^2 \right]^{(D-3)/2}}  + u \ ,
\end{aligned}
\end{equation}
and $u$ is the solution of the equation
\begin{equation}
\begin{aligned}
   \left( \partial_{\rho\rho} + \partial_{zz} + \frac{D-3}{\rho}
          \partial_{\rho} \right) u
 = \frac{3-D}{4(D-2)} \hat A^{ab} \hat A_{ab} \psi^{ -\frac{3D-5}{D-3} } \ ,
\end{aligned}\label{eq:elleq0}
\end{equation}
where
\begin{equation}
\begin{aligned}
\hat A^{ab} \hat A_{ab}=(\hat A_{ij}^{(1)} + \hat A_{ij}^{(2)}) ( \hat A^{ij}{}^{(1)}+\hat A^{ij}{}^{(2)}) + (D-4) ( P^{+} + P^{-})^2\ .
\end{aligned}
\end{equation}

Our numerical construction of the function $u$ will be based on the spectral
solver developed in~\cite{Ansorg:2004ds}. This solver employs coordinates
specifically adapted to the asymptotic behaviour of $u$ at spatial infinity. In
order to investigate this behaviour, we next consider a single black hole with
non-zero linear momentum.

\subsection{Single puncture with linear momentum}
\label{sec:single-puncture}

For a single puncture with momentum $P^z$ located at the origin $z=0$,
equation~(\ref{eq:Aab_i}) implies
\begin{equation}
\begin{aligned}
\label{eq:A2single}
 \hat A^{ab} \hat A_{ab} =  \frac{16\pi^2 (D-1)^2 }{(D-2)^2
         \mathcal{A}_{D-2}^2 r^{2(D-2)} } P_z^2
         \left[
         2 + D(D-3) \left(\frac{z}{r}\right)^2
\right] \ ,
\end{aligned}
\end{equation}
so that equation~\eqref{eq:elleq0} takes the form
\begin{equation}
\begin{aligned}
\label{eq:ueqsingle}
\hat \triangle u 
+  \frac{8\pi^2 (D-1)^2 (D-3) }{(D-2)^3 \mathcal{A}_{D-2}^2 r^{2(D-2)} }
     P_z^2 \left[
  1 + \frac{D(D-3)}{2} \left(\frac{z}{r}\right)^2 
\right]
\psi^{ -\frac{3D-5}{D-3} } = 0 \ .
\end{aligned}
\end{equation}
It turns out to be convenient for solving this differential equation to
introduce a hyperspherical coordinate system on the $D-1$ dimensional spatial
slices, such that the flat conformal metric is
\begin{align*}
\dd \hat s^2 = \hat{\gamma}_{ab} \dd x^a \dd x^b 
          = \dd r^{2} + r^2 \left[ \dd \vartheta^2 + \sin \vartheta^2
              \left(\dd \varphi^2
            +  \sin^2 \varphi \dd \Omega_{D-4}\right)\right] \ ,
\end{align*}
with $ \cos \vartheta = \frac{z}{r}$. We further introduce the radial coordinate
\begin{equation}
  \label{eq:Adef}
  X \equiv \left(
  1 + \frac{\mu}{4 r^{D-3}}
  \right)^{-1} \, ,
\end{equation}
which reduces to the coordinate $A$ of equation~(31) in~\cite{Ansorg:2004ds}
for the case of $D=4$ spacetime dimensions.
Expressed in the new coordinate system, equation~\eqref{eq:ueqsingle} becomes
%
  \begin{multline}
    \label{eq:solveu}
      \bigg\{
        \partial_{XX} + \frac{2}{X} \partial_X + \frac{1}{(D-3)^2 X^2 (1-X)^2} \bigg[
          \partial_{\vartheta \vartheta} + (D-3) \cot \vartheta \partial_{\vartheta} \\
      + \frac{1}{\sin^2 \vartheta} \left(
            \partial_{\varphi \varphi} + (D-4) \cot \varphi \partial_{\varphi}
          \right)
        \bigg]
      \bigg\} u  \\
      = -\alpha \left( \frac{P_z}{\mu} \right)^2 X^{-\frac{D-7}{D-3}} 
     \left(
       1 + u X
     \right)^{-\frac{3D-5}{D-3}}
     \left(1 + \frac{D(D-3)}{2} \cos^2 \vartheta \right) \ ,
   \end{multline}
%
with 
\[\alpha \equiv \frac{128 \pi^2 (D-1)^2}{(D-3)(D-2)^3 \mathcal{A}^2_{D-2}} \ .\]
For $D=4$ we recover equation~(40) of~\cite{Ansorg:2004ds}.
In order to study the behaviour of the solution at spatial infinity, we now
perform a Taylor expansion in $v \equiv \frac{P_z}{\mu}$,
\begin{equation}
  \label{eq:utaylor}
  u = \sum_{j=1}^{\infty} v^{2j} u_j \,.
\end{equation}
Odd powers of $v$ have to vanish in order to satisfy equation~(\ref{eq:solveu}). We have the following equation for $u_1$
\begin{multline}
  \label{eq:solveu1}
     \left\{
       \partial_{XX} + \frac{2}{X} \partial_X + \frac{1}{(D-3)^2 X^2 (1-X)^2}
       \left[
        \partial_{\vartheta \vartheta} + (D-3) \cot \vartheta
             \partial_{\vartheta}
      \right]
    \right\} u_1 \\ =
     -\alpha  X^{-\frac{D-7}{D-3}}
    \left(1 + {\textstyle \frac{D(D-3)}{2} } \cos^2 \vartheta \right) \, .
\end{multline}
In order to solve equation~\eqref{eq:solveu1}, we make the \emph{ansatz}
\begin{equation}
\label{eq:u1ansatz}
 u_1 = f(X) + g(X) Q_D(\cos \vartheta) \ ,
\end{equation}
where $Q_D (\cos \vartheta ) = (D-1) \cos^2 \vartheta - 1$. By solving equation (\ref{eq:solveu1}), we find that
the functions $f(X)$ and $g(X)$ take the form
\begin{align}
\label{eq:fg}
f(X)  = \frac{32 \pi^2 (D-3) }{(D-2)^2 \mathcal{A}^2_{D-2}}
\left(  1 - X^{\frac{D+1}{D-3}}\right) \ ,
\end{align}
\begin{align}
g(X) & =  k_1 \left( \frac{X}{1-X} \right)^{\frac{2}{D-3}} +
  k_{2} \left( \frac{1-X}{X} \right)^{\frac{D-1}{D-3}} \notag \\
&  \quad - \alpha {\textstyle \frac{D(D-3)^3}{2(D+1)(D-1)} }
\Bigg[ {\textstyle \frac{1}{D-1}}
  \frac{X^{\frac{D+1}{D-3}}}{(1-X)^{\frac{2}{D-3}}} {}_2 F_1 \left(
    {\textstyle -\frac{D-1}{D-3}, \frac{D-1}{D-3}; 2 \frac{D-2}{D-3};} X
  \right) \nonumber  \\
&  \quad  - {\textstyle \frac{1}{2D} } X^{\frac{D+1}{D-3}} (1-X)^{\frac{D-1}{D-3}} 
  {}_2 F_1 \left( 
    {\textstyle  \frac{2}{D-3}, \frac{2D}{D-3}; 3 \frac{D-1}{D-3};} X \right) \Bigg] \ ,
\end{align}
where ${}_2 F_1(a,b;c;X)$ is the hypergeometric function and $k_{1,2}$ are
constants to be fixed by imposing that $g(X=1) = 0$ and $g(X=0)$ is smooth.
Requiring analyticity at $X=0$ and using the property $F(a,b,c,0)=1$, we
immediately find $k_2=0$.

We are now interested in the large $X\to 1$ limit. Therefore, we use the $z
\rightarrow 1-z$ transformation law for the hypergeometric
functions~\cite{abramowitz},
\begin{multline}
F(a\!-\!c\!+\!1,b\!-\!c\!+\!1,2\!-\!c,z) = \\
(1\!-\!z)^{c-a-b}
\frac{\Gamma(2-c)\Gamma(a+b-c)}{\Gamma(a-c+1)\Gamma(b-c+1)}
 \,F(1\!-\!a,1\!-\!b,c\!-\!a\!-\!b\!+\!1,1\!-\!z)  \\
+ \frac{\Gamma(2-c)\Gamma(c-a-b)}{\Gamma(1-a)\Gamma(1-b)}
 \,F(a\!-\!c\!+\!1,b\!-\!c\!+\!1,-c\!+\!a\!+\!b\!+\!1,1\!-\!z)\ .
\end{multline}
%
Requiring a regular solution we find that $k_1$ has to satisfy
\begin{equation}
k_1=\frac{64 \pi^2 D(D-3)^2}{(D-2)^3(D+1)  \mathcal{A}^2_{D-2}}
\frac{\Gamma\left(\frac{2(D-2)}{D-3}\right)^2}{\Gamma\left(\frac{3D-5}{D-3}\right)}\label{eq:13} \ .
\end{equation}

Let us write these functions explicitly for $D=4,5,7$ (for $D=6$ the
hypergeometric function does not simplify):
\begin{itemize}
\item $D=4:$
  \begin{align}
    \label{eq:fgD4}
    f(X) & = \frac{1}{2} \left( 1 - X^5 \right) \ , \\
    g(X) & = \frac{(1-X)^2}{10 X^3} \Big[
     84 (1-X) \log(1-X) + 84X - 42 X^2 - 14 X^3 - 7 X^4 -4 X^5 - 2 X^6 
  \Big] ;
  \end{align}
These are equations~(42--44) in~\cite{Ansorg:2004ds}, with appropriate redefinitions.
\item $D=5:$
  \begin{align}
    \label{eq:fgD5}
    f(X) & = \frac {16}{9 \pi ^2} \left(1-X^3\right) , \\
    g(X) & = -\frac{80 (1-X)^2 }{81 \pi^2 X^2} 
    \Big[
      4 \log(1-X) + 4X + 2 X^2+X^3
       \Big] ;
  \end{align}
\item $D=7:$
  \begin{align}
    \label{eq:fgD7}
    f(X) & = \frac{128 \left(1-X^2\right)}{25 \pi ^4} \ , \\
    g(X) & = \frac{28}{125 \pi^4 \sqrt{(1-X) X^3}}
    \Big[-30 \sqrt{(1-X) X} + 40 \sqrt{(1-X) X^3}-16 \sqrt{(1-X)  X^7} \notag \\
      & \quad +3 \pi X^2 +6 \left(5-10 X+4 X^2\right) \arcsin \sqrt{X} \Big] \ .
  \end{align}
\end{itemize}
Analysing these expressions, we can anticipate the convergence properties of the numerical solutions obtained in terms of pseudo-spectral methods. 
For instance, analyticity of $f$ and $g$ suggests exponential convergence.
As will become clear in the next section, we are interested in the convergence properties in a coordinate $A$ behaving as $A \sim 1-\frac{1}{r}$, for large $r$. 
We thus introduce a coordinate $A$ that satisfies
\begin{equation}
\label{eq:XtoA}
X=\left(
  1 + (A^{-1}-1)^{D-3}
\right)^{-1} \,.
\end{equation}
In terms of the $A$ coordinate, we find that the functions $f$ are analytical. 
For the function $g$ in the vicinity of $A = 1$, the leading terms behave as follows:
\begin{itemize}
\item $D=5$
  \begin{align}
    \label{eq:singD5}
   g(A) \sim -\frac{80}{81\pi^2} (1-A)^4 \left[ 8\log(1-A) + 7 \right] \,, 
  \end{align}
\item $D=6$
  \begin{align}
    \label{eq:singD6}
   g(A) \sim \frac{19683}{6272\pi^2} (1-A)^5 \,,
  \end{align}
\item $D=7$
  \begin{align}
    \label{eq:singD7}
   g(A) \sim \frac{84}{25\pi^3} (1-A)^6 \,.
  \end{align}
\end{itemize}
From the behaviour of the functions $f$ and $g$ and equation~(\ref{eq:u1ansatz})
we conclude that the first term in the expansion~(\ref{eq:utaylor}) has a
leading-order behaviour $u_1 \sim 1/r^{D-3}$ as $r\rightarrow \infty$.
Iteratively solving equation~\eqref{eq:solveu} for higher powers of $v$ is
complicated by the presence of the source terms on the right hand side, but
under simplifying assumptions indicates that higher-order terms $u_j \ge 2$
acquire additional factors of $1/r$ and therefore the leading-order fall off
behaviour is given correctly by that of $u_1$. This result is confirmed by our
numerical investigation using finite boost parameters as we shall discuss in the
next section.

With regard to the analyticity of the solutions and the resulting expectations
for the convergence properties of a spectral algorithm, we summarise the results
of our analytical study of a single puncture as follows.  In $D=6,7$, the
leading terms are analytic functions in the vicinity of $A=1$.  Actually, for
$D=7$, $g(A)$ is analytic in the vicinity of any point. Therefore, we expect
exponential convergence of the pseudo-spectral code.  For $D=5$, one observes
the presence of a logarithmic term.  This type of term is known to arise in
$D=4$, when punctures have non-vanishing
momenta~\cite{Dain:2001ry,Gleiser:1999hw} and in that case their presence makes
the convergence algebraic in the single puncture case. In the next section we
shall investigate the impact of the logarithmic terms on the convergence
properties of our spectral solver.

\subsection{Two punctures with linear momentum}

\subsubsection{Code changes}

We first explicitly list the modifications applied to the spectral
solver of reference~\cite{Ansorg:2004ds} and demonstrate how these modifications
enable us to generate initial data for boosted black hole binaries with
convergence properties and levels of constraint violation similar to the
$D=4$ case.
For this purpose we start by recalling that the spectral
solver of \cite{Ansorg:2004ds} employs coordinates
\begin{equation}
\label{eq:ABphi}
A \in [0,1]\ , \quad B \in [-1,1]\ , \quad \phi \in [0,2\pi] \ ,
\end{equation}
which are defined by equation~(62) of~\cite{Ansorg:2004ds},
\begin{equation}
\begin{split}
x&=b\frac{2A}{1-A^2}\frac{1-B^2}{1+B^2}\sin\phi\,, \\
y&=b\frac{2A}{1-A^2}\frac{1-B^2}{1+B^2}\cos\phi\,,\\
z&=b\frac{A^2+1}{A^2-1}\frac{2B}{1+B^2}\,,
\end{split}
\end{equation}
where $b$ is half of the coordinate distance between the punctures.
In particular, the coordinate $A$ satisfies
\begin{equation}
  r\to\infty \Leftrightarrow A\to 1\label{eq:rinfty} \ .
\end{equation}

The first modification consist in adapting the source term and Laplace operator
according to~\eqref{eq:elleq0}.

Next, we note that the type of high-energy collisions which form the main
motivation for this work often start from relatively large initial separations
of the holes, $|z_1-z_2|\gg r_S$.
In order to obtain high-precision solutions for such binary configurations,
we found it crucial to introduce a coordinate $A'$ defined as
\begin{equation}
\label{eq:barA}
A = \frac{\sinh\left[\kappa (A'+ 1 )/2 \right]}{\sinh{\kappa}} \ ,
\end{equation}
where $\kappa$ is an adjustable free parameter.
Note that for $\kappa = 0$ we obtain $A = \frac{1}{2}(A' + 1)$
For $\kappa > 0$, however, the new coordinate $A'$ provides the spectral method
with enhanced resolution near $A \sim 0$.

A further modification is related to the asymptotic fall off of the
function $u$ as obtained in the previous section,
\begin{equation}
  \label{eq:ufalloff}
  u \sim \frac{1}{r^{D-3}} \  .
\end{equation}
To naturally accommodate this behaviour with the spectral
coordinates used in the code, we have changed the variable $U$ of equation~(5)
in~\cite{Ansorg:2004ds} to
\begin{equation}
  \label{eq:U}
  u = (A' - 1)^{D-3} U \ .
\end{equation}
Note that this $U$ variable is the variable that the code actually solves for.


Finally, we adjust the calculation of the ADM mass from the numerical
solution. For this purpose, we note that, asymptotically
\begin{equation}
  \label{eq:psiasympt}
  \psi = 1 + \frac{\mu_{+}}{4r_{+}^{D-3}} + \frac{\mu_{-}}{4r_{-}^{D-3}} + u 
  \sim 1 + \frac{\mu}{4r^{D-3}} \ ,
\end{equation}
with $\mu \equiv r_{S_{\rm global}}^{D-3} \equiv \frac{16\pi
M_{\mathrm{ADM}}}{\mathcal{A}_{D-2}(D-2)}$ and $\mu_{\pm} \equiv r_{S_{(\pm)}}^{D-3}$. The ADM mass is then obtained
from
\begin{align}
  \label{eq:rsglobal}
  r_{S_{\rm global}}^{{D-3}} & = r_{S_{(+)}}^{D-3} + r_{S_{(-)}}^{D-3}
      + 4 \lim_{r\to\infty} r^{D-3} u \notag \\
            & = r_{S_{(+)}}^{D-3} + r_{S_{(-)}}^{D-3} + 4
            \left(-2b \frac{\tanh \kappa}{\kappa} \right)^{D-3}
            U(A' = 1) \ ,
\end{align}
where we have used equation~(62) of~\cite{Ansorg:2004ds}, and equation~\eqref{eq:barA}
and~\eqref{eq:U}. We show in table~\ref{tab:admmass} the values obtained for the ADM mass of some cases we considered.
\begin{table}
\caption[ADM mass obtained with equation~\eqref{eq:rsglobal}]{ADM mass obtained with
  equation~\eqref{eq:rsglobal} in units of the ``bare'' Schwarzschild radius 
  $r_S^{D-3} = r_{S_{(+)}}^{D-3} + r_{S_{(-)}}^{D-3}$. The variation of the ADM mass with
  resolution is of the order of $10^{-10}$ for all $D$ and $n \ge 100$ grid points 
  indicating that the accuracy in the ADM mass is limited by round-off 
  errors.\label{tab:admmass} }
\begin{tabular*}{\textwidth}{@{\extracolsep{\fill}}ccccc}
  \hline
  \hline
  $D$ & $b/r_S$ & $P/r_{S}^{D-3}$ & $r_{S_{\rm global}}^{{D-3}}/r_{S}^{D-3}$ & $M_{\rm ADM}/r_{S}^{D-3}$ \\
  \hline 
  4   & 30.185   & 0.8          &  3.555          & 1.78       \\
  5   & 30.185   & 0.8          &  1.931          & 2.27       \\
  6  &  30.185   & 0.8          &  1.415          & 2.96       \\
  7  & 30.185    & 0.8          &  1.236          & 3.81       \\
  \hline
  \hline
\end{tabular*}
\end{table}

\subsubsection{Results}

We now study the numerical results as obtained for $D=4,5,6,7$ with these
adaptations of the spectral solver of~\cite{Ansorg:2004ds}. Throughout the
remainder of this section we will graphically present results in units of the
``bare'' Schwarzschild radius defined as $r_S^{D-3} = r_{S_{(+)}}^{D-3} +
r_{S_{(-)}}^{D-3}$.

\begin{figure}[tbhp]
\centerline{\includegraphics[clip=true,width=0.7\textwidth]{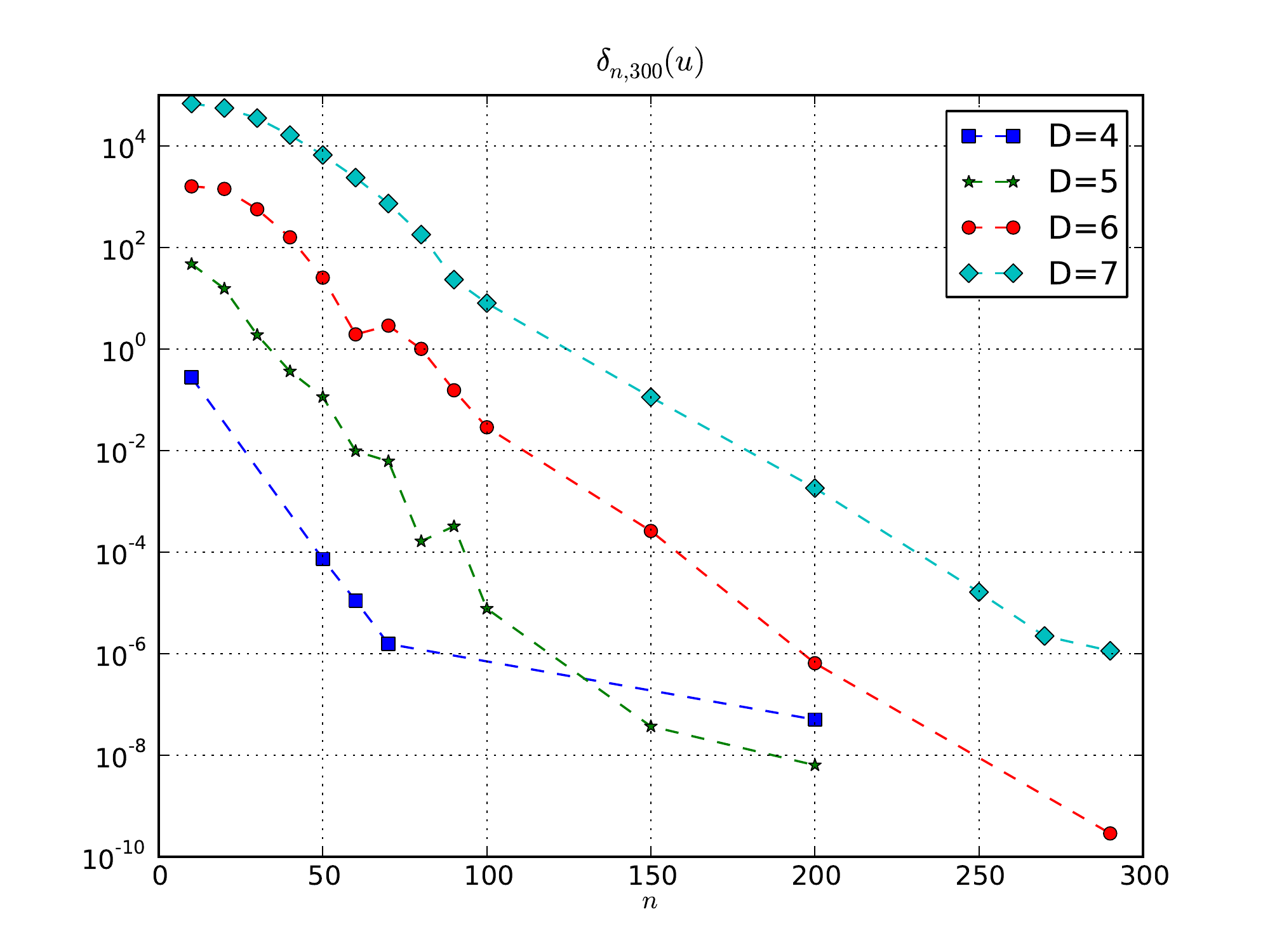}}
\caption{Convergence plot for the $b/r_S=30.185$, $P/r_{S}^{D-3}=\pm0.80$ cases.}
\label{fig:d60_conv} 
\end{figure}

We first address the convergence properties of the numerical algorithm by
evaluating the quantity
\begin{equation}
  \label{eq:conv}
  \delta_{n,m}(u) = \max |1-u_n/u_m| \,,
\end{equation}
where the maximum is obtained along the collision axis, i.e.~$z$-axis in
our case. Here, the index $m$ refers to a reference solution obtained using a
large number $m$ of grid points while $n$ denotes test solutions using a coarser
resolution, $n<m$. The result obtained for black hole binaries with initial
separation $b/r_S=30.185$ and boost $P^z/r_S^{D-3}=0.8$ in $D=4$, $5$, $6$ and
$7$ dimensions is displayed in figure~\ref{fig:d60_conv}. We note from this
figure, that achieving a given target accuracy $\delta_{n,m}$ requires a larger
number of points $n$ as $D$ increases. We emphasise in this context, however,
that this increase in computational cost in higher dimensions is unlikely to
significantly affect the total computational cost of the simulations which
typically are dominated by the time evolution rather than the initial data
calculation.  Most importantly, we observe exponential convergence up to a level
of $\delta_{n,m}(u) \approx 10^{-6}$ for all values of the spacetime
dimensionality $D$. Below that level, the two leftmost curves in
figure~\ref{fig:d60_conv}, corresponding to $D=4$ and $D=5$, respectively, show
that the rate of convergence decreases indicating that the logarithmic terms
become significant and reduce the convergence to algebraic level similar to the
observation in figure~4 of reference~\cite{Ansorg:2004ds}. For $D=6$, the
convergence remains exponential, in agreement with the absence of logarithmic
terms in the analysis of section~\ref{sec:single-puncture}. Irrespective of a
change to algebraic convergence, however, our algorithm is capable of reducing
the quantity $\delta_{m,n}(u)$ for all values of $D$ to a level comparable to
the case $D=4$ and, thus, producing initial data of similar quality as in 3+1
dimensions, provided we use a sufficiently high resolution $n$.

\begin{figure}[tbhp] \centerline{\includegraphics[clip=true,width=0.7\textwidth]{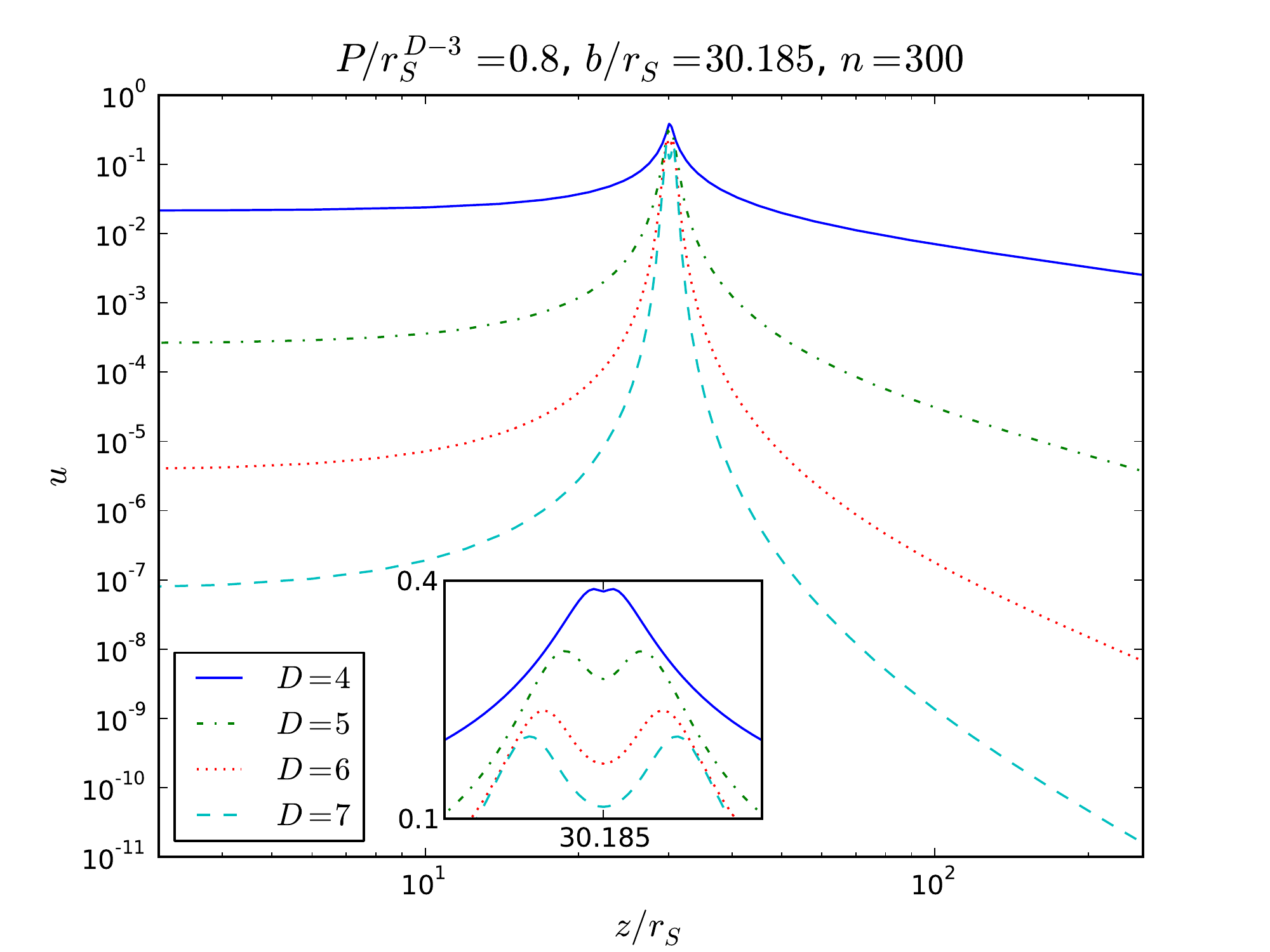}}
  \caption[$u$ function for $D=4,\dots,7$ plotted along the $z$-axis]{$u$
    function for $D=4,\dots,7$ plotted along the $z$-axis, in units of $ r_S
    $. We used $n_A = n_B = n = 300$, $n_{\phi}=4$. We also show a zoom around
    the puncture.}
\label{fig:u_loglog} 
\end{figure}

For illustration, we plot in figure~\ref{fig:u_loglog} the function $u$ obtained
for the case of $b/r_S=30.185$, $P^z/r_{S}^{D-3}=0.8$. The behaviour is
qualitatively similar for all values of $D$, but the figure demonstrates the
faster fall off for larger $D$ as predicted by \eqref{eq:ufalloff}. For this
plot we have used $n_A=300$, $n_B=300$ and $n_\phi=4$ grid points.
The inset in the figure shows the function $u$ in the immediate
vicinity of the puncture. While the profile develops multiple
extrema for $D>4$, the profile remains smooth for all values of $D$.

\begin{figure}[tbhp]
\centerline{\includegraphics[clip=true,width=0.7\textwidth]{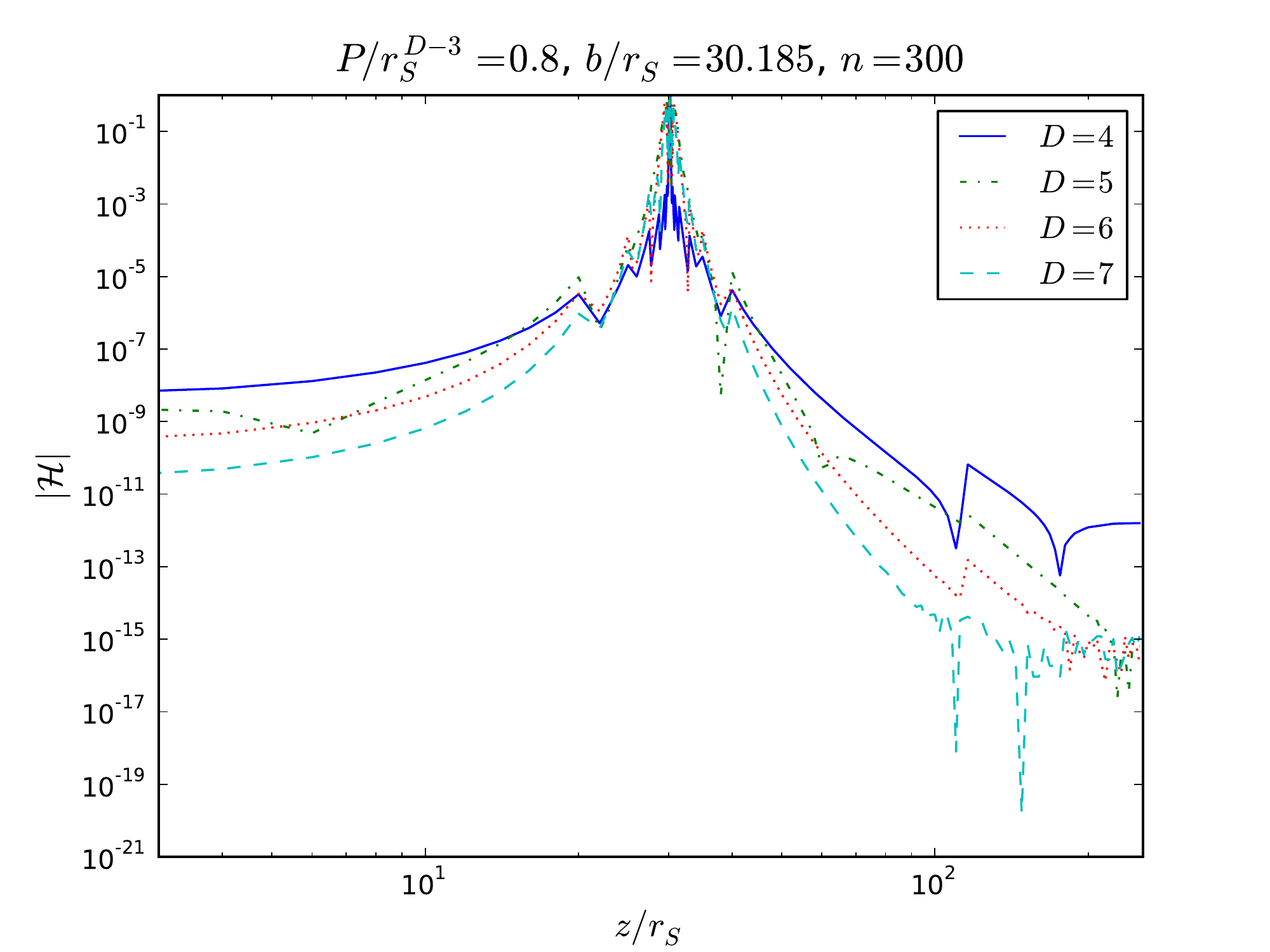}}
\caption[Violation of the Hamiltonian constraint along the
$z$-axis]{Violation of the Hamiltonian constraint along the
  $z$-axis, evaluated with a fourth order finite difference scheme. The growth
  of the constraint violation near the puncture is an artifact of
  finite-differencing across the puncture; see text for details.\label{fig:hc}}
\end{figure}

Finally, we show in figure~\ref{fig:hc}, the Hamiltonian constraint
corresponding to the solutions presented in figure~\ref{fig:u_loglog} as
measured by a fourth-order finite differencing scheme of the \textsc{Lean}
evolution code (section~\ref{sec:lean-code} and
reference~\cite{Sperhake:2006cy}).  We emphasise that the violation of
equation~(\ref{eq:u_chaphd}) inside the spectral initial data solver is
$<10^{-12}$ by construction. The independent evaluation of the constraint
violation in the evolution code serves two purposes. First, it checks that the
differential equation~(\ref{eq:u_chaphd}) solved by the spectral method
corresponds to the Hamiltonian constraint formulated in ADM variables; an error
in coding up the differential equation~(\ref{eq:u_chaphd}) could still result in
a solution for $u$ of the spectral solver, but would manifest itself in
significantly larger violations in figure~\ref{fig:hc}. Second, it demonstrates
that the remaining numerical error is dominated by the time evolution instead of
the initial solver.  Note in this context that the relatively large violations
of order unity near the puncture location in figure~\ref{fig:hc} are an artifact
of the fourth-order discretisation in the diagnostics of the evolution code and
are typical for evolutions of the moving-puncture type; see e.g.~the right panel
in figure~8 in Brown et~al.~\cite{Brown:2008sb}.

The solid (blue) curve obtained for the ``standard'' $D=4$ case serves as
reference.  For all values of $D$ the constraint violations are maximal at the
puncture location $z_1/r_S\approx 15$ and rapidly decrease away from the
puncture. As expected from the higher fall off rate of the grid functions for
larger $D$, the constraints also drop faster for higher dimensionality of the
spacetime.

\section{Numerical evolutions}
\label{sec:num-evols}

Having established that our initial data code is working, we will now show some
numerical results,
obtained by adapting the \textsc{Lean} code introduced in
section~\ref{sec:lean-code}. In this section we will begin by briefly commenting
on numerical issues generated by the quasi-matter terms arising from the
dimensional reduction. We then present some code tests and results.

From the initial data construction of section~\ref{sec:init_data-hd}, we see
that the quasi-matter field $\lambda$ has a $y^2$ fall off as $y\rightarrow 0$,
that is, on the $xz$ plane (cf.~\eqref{eq:init-data-boost}).
From~\eqref{matterterms}, we see that this leads to divisions by zero on the
right-hand side of the BSSN evolution equations; thus, we need to isolate such
irregular terms and re-write the equations in terms of variables which are
explicitly regular at $y=0$.
%
%
In this spirit, we introduce the following evolution variable
\begin{equation}
\label{kappavariable}
  \zeta \equiv  \frac{\chi}{y^2} \lambda \, ,
\end{equation}
%
%
and corresponding auxiliary variable
\begin{eqnarray}
  K_{\zeta} \equiv  -\frac{1}{2\alpha y^2}(\partial_t-\mathcal{L}_\beta)
      (\zeta y^2)=-\frac{1}{2\alpha} \left( \partial_t \zeta
      - \beta^m \partial_m \zeta + \frac{2}{3}\zeta
      \partial_m \beta^m - 2\zeta \frac{\beta^y}{y} \right) \, .
  \label{eq:Kkappa_v2}
\end{eqnarray}

The full quasi-matter terms and evolution equations in terms of these regular variables can be found in Appendices A and B of~\cite{Zilhao:2010sr}.
%
Here we note only that, with the above definition, we have the relation
\begin{eqnarray}
  K_{\lambda} = \frac{y^2}{\chi} K_{\zeta} + \frac{1}{3}
      \frac{y^2\zeta}{\chi} K\, .
      \label{useful}
\end{eqnarray}
%

Finally, for long term evolutions, we employed the following gauge conditions, which are generalisations of conditions~(\ref{eq:1+log_0}), (\ref{eq:gamma-driver-0})
\begin{align}
  \left(
    \partial_t - \beta^k \partial_k
  \right) \alpha &= -2\alpha (\eta_K K + \eta_{K_\zeta} K_{\zeta})\, ,
      \label{eq:dtalpha} \\
  \left(
    \partial_t - \beta^k \partial_k
  \right) \beta^i &= \frac{3}{4} \tilde{\Gamma}^i - \eta \beta^i\, .
      \label{eq:dtbeta}
\end{align}
Note the extra term involving $K_{\zeta}$ in the slicing condition compared with
standard moving puncture gauge in $3+1$ dimensions and the additional freedom we
have introduced in the form of the parameters $\eta_K$ and $\eta_{K_\zeta}$.

\subsection{Code tests}

\subsubsection{Geodesic slicing}

As a first test of our numerical implementation, we have numerically evolved a single $D=5$ Tangherlini black hole in the so-called \emph{geodesic slicing}, which corresponds to fixing the gauge parameters to~(\ref{eq:geo-slice-0})
%
%
%
throughout the evolution.
Such a gauge choice is not adequate to perform long term numerical evolutions.
The advantage of this choice, though, is that one can easily write the Tangherlini metric element in this coordinate system, which we can then match against the numerically obtained solutions~\cite{Yoshino:2009xp}.
This coordinate system may be
achieved by setting a congruence of in-falling radial time-like geodesics, each
geodesic starting from rest at radial coordinate $r_0$, with $r_0$ spanning the
interval $[\mu,+\infty[$, and using their proper time $\tau$ and $r_0$ as
coordinates (instead of the standard $t$, $r$ Schwarzschild-like coordinates).
The line element becomes
\begin{equation}
  \dd s^2=-\dd\tau^2+\frac{\left(r_0(R)^2+\left(\frac{\mu}{r_0(R)}\right)^2
      \tau^2\right)^2}{r_0(R)^2-\left(\frac{\mu}{r_0(R)}\right)^2\tau^2}
      \frac{\dd R^2}{R^2}+\left(r_0(R)^2-\left(\frac{\mu}{r_0(R)}\right)^2
      \tau^2\right)\dd \Omega_3 \, ,
  \label{geodesicmetric}
\end{equation}
where $r_0(R)$ is given by
\begin{equation}
  r_0(R)= R\left(1+\frac{\mu^2}{4R^2}\right) \, .
  \label{rzerotor}
\end{equation}

\begin{figure}[tbhp]
  \centering\includegraphics[width=0.49\textwidth]{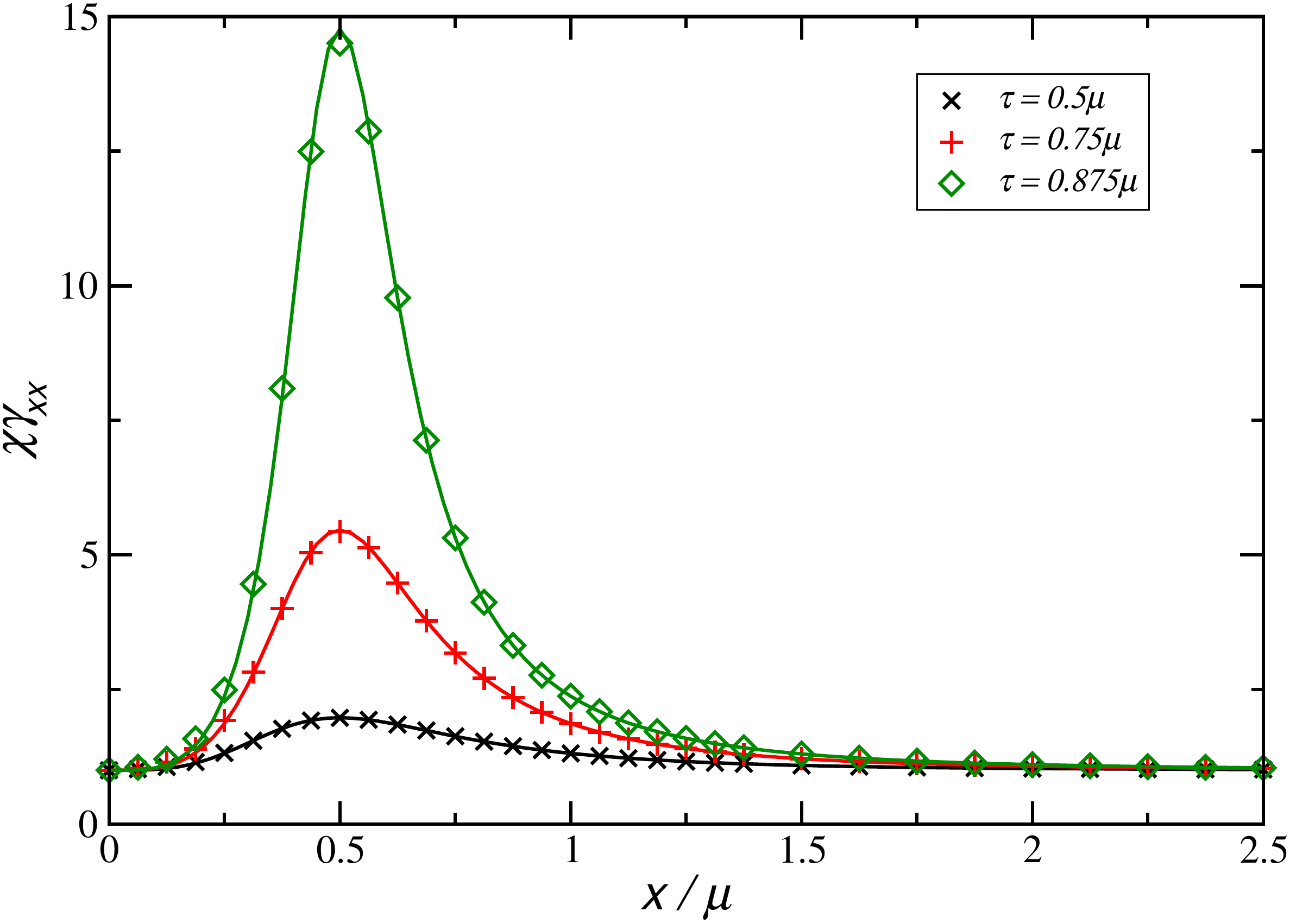}
  \centering\includegraphics[width=0.49\textwidth]{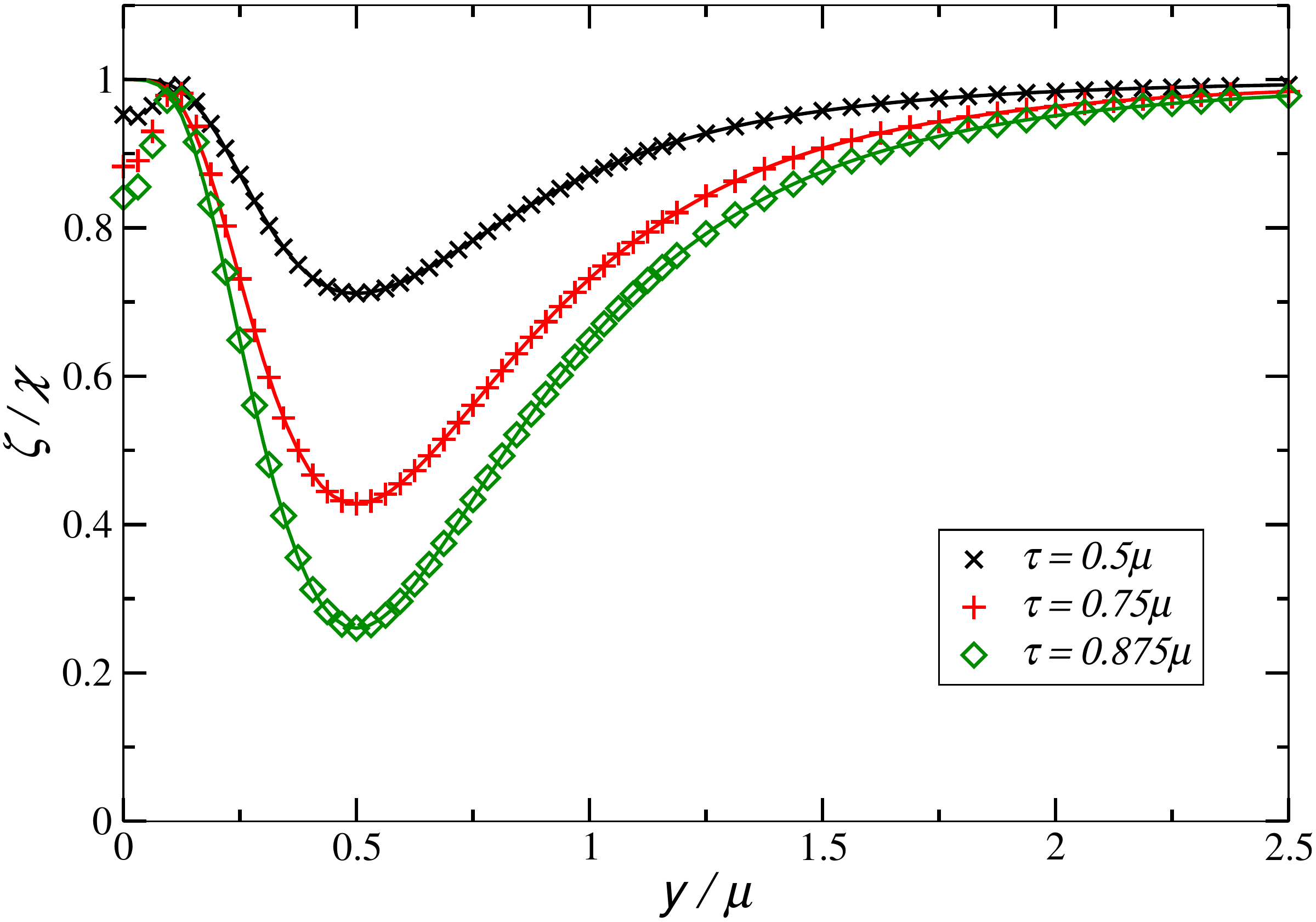}
  \caption[$D=5$ Tangherlini black hole, geodesic slicing]{Numerical values versus
    analytical plot of $\tilde{\gamma}_{xx}$ along the $x$-axis (left panel) and
    of $\zeta/\chi=\lambda/y^2$ along the $y$ axis (right panel), for various
    values of $\tau$, for the single Tangherlini black hole in five dimensions. The
    horizontal axis are in units of $\mu$.  }
  \label{gammaxx}
\end{figure}

Before the breaking down of the numerical evolution, we can compare our numerical results with the above metric element. 
This is shown in
figure~\ref{gammaxx}, where we have plotted one metric component
$\tilde{\gamma}_{xx}$ along the $x$ axis (left) and $\zeta/\chi$ (right), for
various values of $\tau$ using both the analytical solution and numerical
data. The agreement is excellent for $\tilde{\gamma}_{xx}$ and good for
$\zeta/\chi$. The latter shows some deviations very close to the puncture, but
we believe that it is not a problem for two reasons: (i)~the agreement
improves for higher resolution; (ii)~the mismatch does not propagate
outside of the horizon.


\subsubsection{Single black hole evolution}

\begin{figure}[tbhp]
  \centering\includegraphics[width=0.49\textwidth]{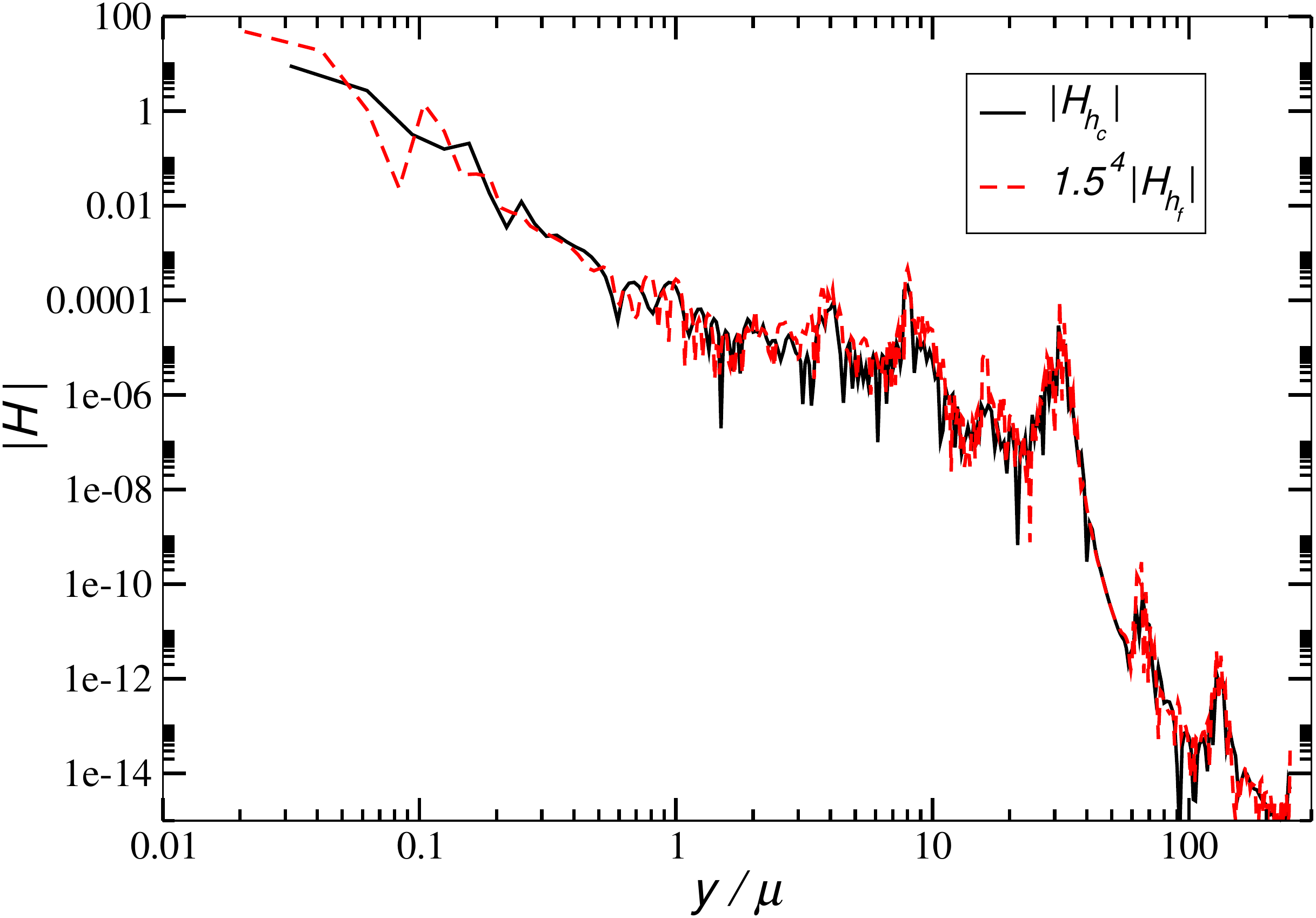}
  \centering\includegraphics[width=0.49\textwidth]{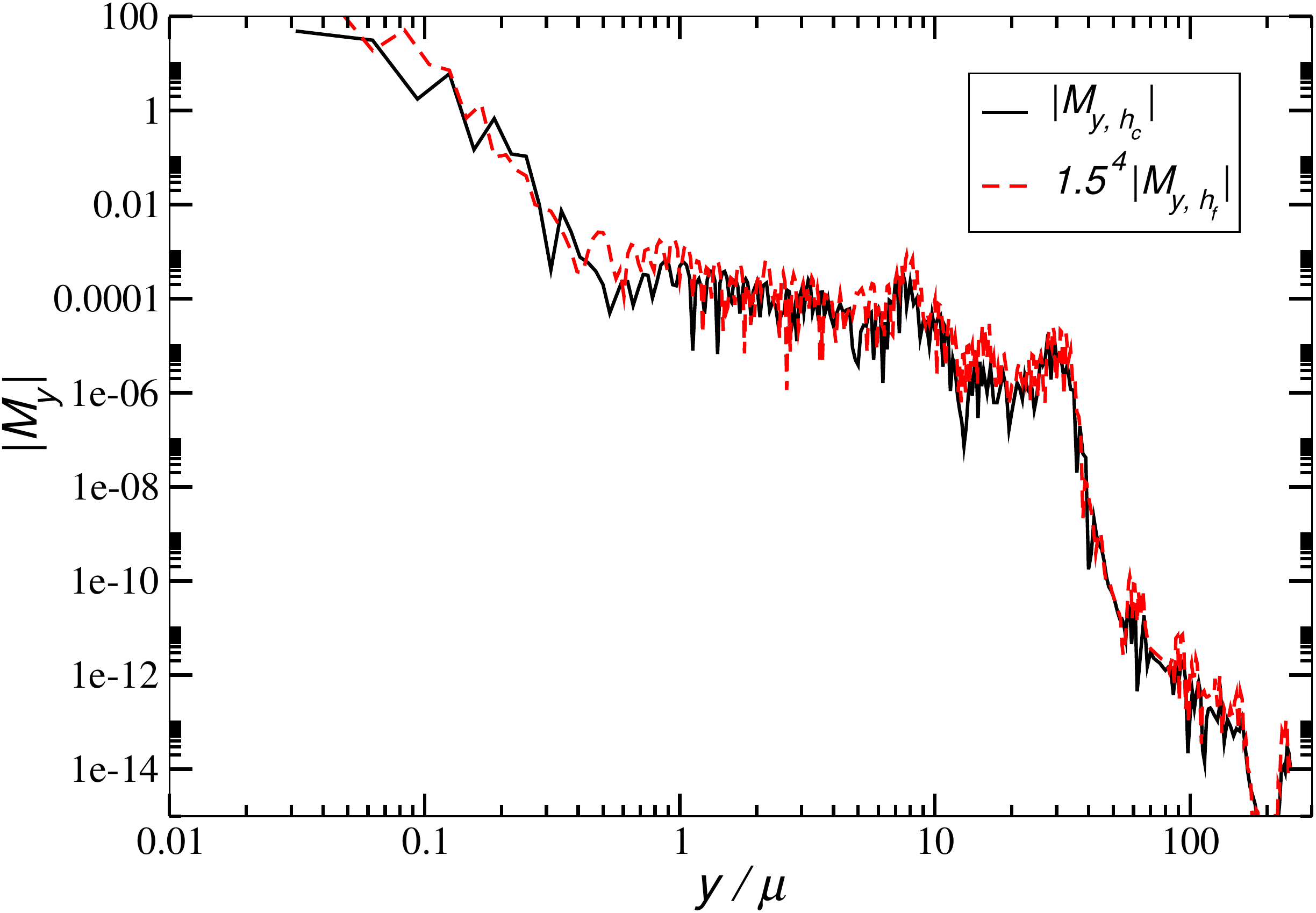}
  \caption[$D=5$ single black hole, constraint violation]{Hamiltonian constraint (left
    panel) and $y$-component of the momentum constraint (right panel) at time
    $t=28\mu$, for the evolution of a single Tangherlini black hole in five dimensions.}
  \label{constraintsplot}
\end{figure}

To further test our numerical framework, we have performed long term simulations
of a single black hole in $D=5$ using the gauge conditions in~(\ref{eq:dtalpha}) and (\ref{eq:dtbeta}), the initial data from equations~(\ref{eq:init-data-boost}) (with $P^{+} = 0 = P^{-}$) and
grid setup (cf.~section~\ref{sec:lean-code})
\begin{equation}
  \left\{(512,~256,~128,~64,~32,~16,~8,~4,~2) \times (),~h\right\} \, ,\nonumber
\end{equation}
in units of $\mu$ with resolutions $h_{\rm c}=1/32$ and $h_{\rm f}=1/48$.  
In figure~\ref{constraintsplot} we show the Hamiltonian constraint and the $y$
component of the momentum constraint at evolution time $t=28\mu$. 
For the Hamiltonian constraint
the convergence is essentially 4th order; for the momentum constraint it
decreases towards 2nd or 3rd order in patches. 
%

%

\subsubsection{Head-on collision}

Finally, we tested the code capability to evolve a head-on collision from rest.
Using again the initial conditions~(\ref{eq:init-data-boost}) with $P^{+} = 0 = P^{-}$, we let two black holes with parameters
\begin{eqnarray}
  && \mu^2_{\rm A} = \mu^2_{\rm B} \equiv \frac{\mu^2}{2} \, , \\
  && z_A = -z_B = 3.185~\mu \, ,
\end{eqnarray}
collide from rest, using the grid setup
\begin{equation}
  \left\{(512,~256,~128,~64,~32,~16,~8) \times (2,~1),~h=1/32\right\} \, , \nonumber
\end{equation}
in units of $\mu$. 
The gauge variables $\alpha$ and $\beta^i$ were evolved
according to equations~\eqref{eq:dtalpha} and
\eqref{eq:dtbeta} with parameters $\eta_K = \eta_{K_\zeta}=1.5$ and $\eta=0.75$.

\begin{figure}[tbhp]
  \centering
    \includegraphics[clip=true,width=0.46\textwidth]{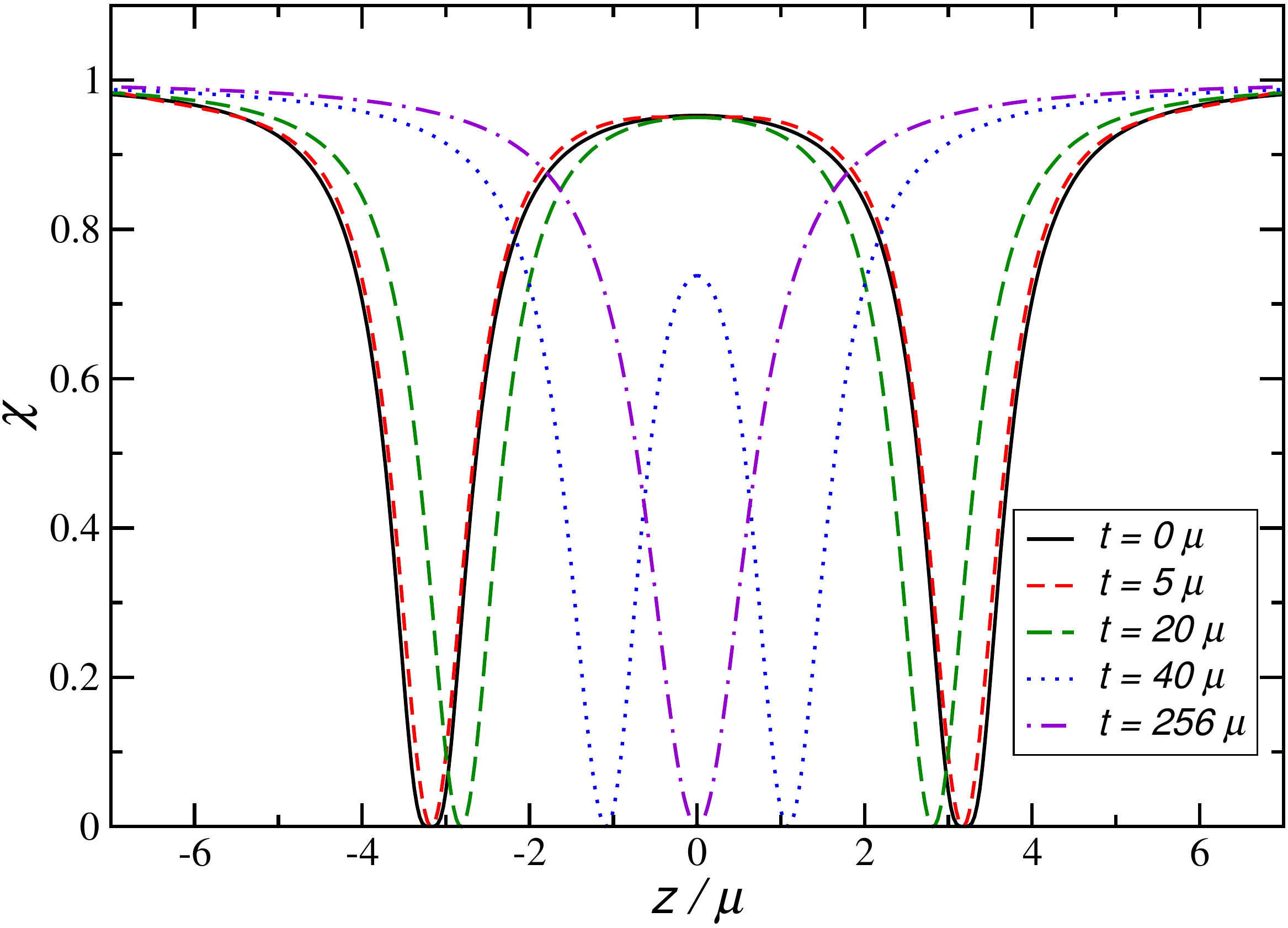}
    \includegraphics[clip=true,width=0.46\textwidth]{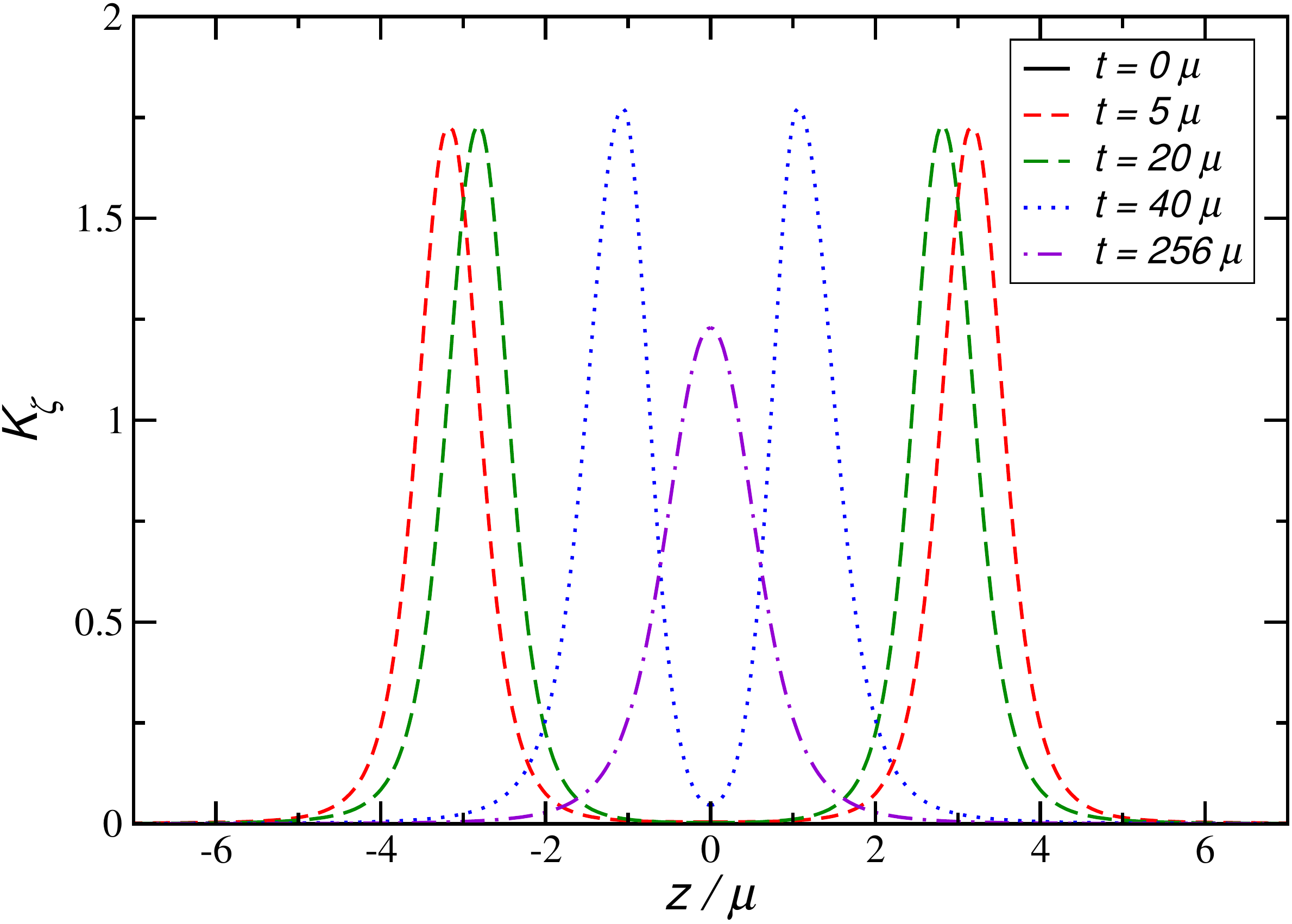}

    \caption[$D=5$ black hole collision from rest]{The BSSN variable $\chi$ (left panel)
      and the quasi-matter momentum $K_{\zeta}$ (right panel) shown along
      the axis of collision for a head-on collision at times $t=0$, $5$, $20$,
      $40$ and $256~\mu$. Note that $K_{\zeta}=0$ at $t=0$. }
  \label{headon_stable}
\end{figure}
%
In figure~\ref{headon_stable} we show the conformal factor $\chi$ and the
momentum $K_{\zeta}$ along the axis of collision at various times for such an
evolution.  At early times, the evolution is dominated by the adjustment of the
gauge (cf.~the solid and short-dashed curves). The two holes next start
approaching each other (long-dashed and dotted curves) and eventually merge and
settle down into a single stationary hole (dash-dotted curves).  No signs of
instabilities were observed.

\subsection{Head-on collisions}

Having established, in the previous section, that our numerical implementation
does work (for the five dimensional case) we will now present some results.

Since the final result of a head-on collision of two $D$~dimensional,
non-spinning black holes approaches, at late times, a $D$~dimensional
Schwarzschild (i.e. Tangherlini) black hole, we can make use of the
Kodama-Ishibashi formalism, presented in section~\ref{sec:KI-gen} to extract the
gravitational wave information. Our remaining task is then to obtain the
relevant gauge-invariant quantities from our numerical data. We will give here
the main steps for this procedure; the full details can be found in
reference~\cite{Witek:2010xi}.

\subsubsection{Coordinate frames}

In the approach developed in section~\ref{sec:dim-red_split}, we perform a
dimensional reduction by isometry on the $(D-4)$-sphere $S^{D-4}$, in such a way
that the $D$~dimensional vacuum Einstein equations are rewritten as an effective
$3+1$ dimensional time evolution problem with source terms that involve a scalar
field.
We focus here on $D\ge5$ dimensional spacetimes with $SO(D-2)$ isometry group,
which allows us to model head-on collisions of non-spinning black holes; we dub
hereafter these spacetimes as \textit{axially symmetric}.  Although the
corresponding symmetry manifold is the $(D-3)$-sphere $S^{D-3}$, the quotient
manifold in our dimensional reduction is its submanifold $S^{D-4}$.  The
coordinate frame in which the numerical simulations are performed is
\begin{equation}
(x^\mu,\phi^1,\dots,\phi^{D-4})=
(t,x,y,z,\phi^1,\dots,\phi^{D-4}) \, ,
\label{frame1}
\end{equation}
where the angles $\phi^1,\dots,\phi^{D-4}$ describe the quotient manifold
$S^{D-4}$ and do not appear explicitly in the simulations.  Here, $z$ is the
symmetry axis, i.e.\ the collision line.

Recall that in the frame~\eqref{frame1}, the spacetime metric has the form
(cf. equation~(\ref{ansatz})) 
\begin{equation}
\dd s^2 
=-\alpha^2\dd t^2+\gamma_{ij}(\dd x^i+\beta^i\dd t)(\dd x^j+\beta^j\dd t)+
\lambda(x^\mu)\dd \Omega_{D-4} \, ,\label{totalmetric}
\end{equation}
%
Also recall, from equations~(\ref{4d}) and (\ref{conformal}), that with an
appropriate transformation of the four dimensional coordinates $x^\mu$, the
residual symmetry left after the dimensional reduction on $S^{D-4}$ can be made
manifest: $x^\mu\rightarrow (x^{\bar\mu},\theta)$ ($\bar\mu=0,1,2$),
\begin{equation}
g_{\mu\nu}(x^\alpha)\dd x^\mu \dd x^\nu=g_{\bar\mu\bar\nu}(x^{\bar\alpha})\dd x^{\bar\mu}\dd x^{\bar\nu}+g_{\theta\theta}(x^{\bar\alpha})\dd \theta^2
\label{thetametric}
\end{equation}
and
\begin{equation}
\lambda(x^\mu)=\sin^2\theta g_{\theta\theta}(x^{\bar\alpha})\,,\label{lambdatheta}
\end{equation}
so that equation~\eqref{totalmetric} takes the form $\dd s^2=g_{\bar\mu\bar\nu}\dd x^{\bar\mu}\dd x^{\bar\nu}+g_{\theta\theta}\dd \Omega_{D-3}$.

To extract the gravitational waves with the KI formalism, 
spacetime, away from the black holes, is required to be approximately
spherically symmetric. In $D$~dimensions this means symmetry with respect to rotations on
$S^{D-2}$,
which is manifest in the coordinate frame:
\begin{equation}
(x^{a},\bar\theta,\theta,\phi^1,\dots,\phi^{D-4})=(t,r,\bar\theta,\theta,\phi^1,\dots,\phi^{D-4}) \,.
\label{frame2}
\end{equation}
Note that $x^{a}=t,r$ and that we have introduced polar-like coordinates
$\bar\theta,\theta\in[0,\pi]$ to ``build up'' the manifold $S^{D-2}$ in the
background, together with a radial spherical coordinate $r$, which is the areal
coordinate in the background.

The coordinate frame \eqref{frame2} is defined in such a way that the metric 
can be expressed as a stationary background $(ds^{(0)})^2$ (i.e., the Tangherlini metric) 
plus a perturbation $(ds^{(1)})^2$ which decays faster than $1/r^{D-3}$ for large $r$, and the formalism from section~\ref{sec:KI-gen} can thus be applied~\cite{Witek:2010xi}.

\subsubsection{Implementation of axisymmetry}\label{implax}

In an axially symmetric spacetime, the metric perturbations are symmetric with
respect to $S^{D-3}$. Therefore, the harmonics in the expansion of $h_{MN}$
depend only on the angle $\bar\theta$ (which does not belong to
$S^{D-3}$). Furthermore, since there are no off-diagonal terms in the metric%
, the only non-vanishing $g_{a\bar i}$ components are
$g_{a\bar\theta}$; the only components $g_{\bar i\bar j}$ are either
proportional to $\gamma_{\bar i\bar j}$, or all vanishing but
$g_{\bar\theta\bar\theta}$. This implies that only scalar spherical harmonics
can appear in the expansion of the metric perturbations. Indeed, if
\begin{equation}
\mathbb{V}^{\bar i}=(\mathbb{V}^{\bar\theta},0,\dots,0)\,,\quad\mathbb{V}^{\bar i}=\mathbb{V}^{\bar i}(\bar\theta) \, ,
\end{equation}
then equation~(\ref{eq:vector-harm-div}) gives
\begin{equation}
\mathbb{V}^{\bar i}_{:\bar i}=\mathbb{V}^{\bar\theta}_{,\bar\theta}=0 \Rightarrow 
\mathbb{V}^{\bar\theta}=0 \Rightarrow \mathbb{V}^{\bar i}=0\,.
\end{equation}
Similarly, from equation~(\ref{eq:tensor-harm-div}) we obtain $\mathbb{T}_{\bar i\bar j}=0$.

The scalar harmonics, solutions of equation~(\ref{eq:scalar-harm-def}) and which depend only on the coordinate $\bar\theta$, are given by the
Gegenbauer polynomials $C_l^{(D-3)/2}$, as discussed in references~\cite{Berti:2003si,Cardoso:2002ay,Yoshino:2005ps}; writing explicitly the index $l$, they take the form
\begin{equation}
\mathbb{S}_l(\bar\theta)=(K^{lD})^{-1/2}C_l^{(D-3)/2}(\cos\bar\theta) \, , \label{gp}
\end{equation} 
where the normalization $K^{lD}$ is chosen such that
\begin{equation}
\int d\Omega^{D-2}\mathbb{S}_l\mathbb{S}_{l'}=\delta_{ll'} \,,
\label{normalK}
\end{equation}
and $k^2=l(l+D-3)$. 

Metric perturbations, and corresponding gauge-invariant functions, can then be computed in terms of these functions~\cite{Witek:2010xi}.
%


\subsubsection{Extracting gravitational waves}
\label{master}
In the KI framework, the emitted gravitational waves are described by the master function $\Phi$, cf.~section~\ref{sec:KI-gen}. 
We can compute directly $\Phi_{,t}$ with~\cite{Kodama:2003jz,Witek:2010xi}\footnote{Note that there is a factor
  $r$ missing in equation~(3.15) of reference~\cite{Kodama:2003jz}.}
\begin{equation} 
\label{eq:KIwavefunction}
\Phi_{,t}=(D-2)r^{(D-4)/2}\frac{-F^r_{~t}+2 r F_{,t}}{k^2-D+2+\frac{(D-2)(D-1)}{2}\frac{r_S^{D-3}}{r^{D-3}}} \, ,
\end{equation}
where $k^2=l(l+D-3)$. 
The energy flux can then be computed from expressions~(\ref{eq:energyflux}), (\ref{eq:energyrad}).




\subsection{Head-on collision from rest in \texorpdfstring{$D=5$}{D=5}}
\label{sec:5D-results}

\begin{table*}[t]
\caption[Grid structure and initial parameters of the head-on collisions
  starting from rest in $D=5$]{Grid structure and initial parameters of the head-on collisions
  starting from rest in $D=5$.
  The grid setup is given in terms of the ``radii'' of the individual refinement levels, in units
  of $r_S$, as well as the resolution near
  the punctures $h$. 
  $d$ is the initial coordinate separation of the two punctures
  and $L$ denotes the proper initial separation.\label{tab:setup}}
\centering
\begin{tabular*}{\textwidth}{@{\extracolsep{\fill}}lccc}
\hline
\hline
Run &  Grid Setup & $d/r_S$ & $L/r_S$  
\\ \hline
%
HD5a &  $\{(256,128,64,32,16,8,4)\times(0.5,0.25),~h=r_S/84\}$ & $1.57$ & $1.42$ \\ 
HD5b &  $\{(256,128,64,32,16,8,4)\times(0.5,0.25),~h=r_S/84\}$ & $1.99$ & $1.87$ \\ 
HD5c &  $\{(256,128,64,32,16,8,4)\times(1,0.5),~h=r_S/84\}$ & $2.51$ & $2.41$ \\ 
HD5d &  $\{(256,128,64,32,16,8,4)\times(1,0.5),~h=r_S/84\}$ & $3.17$ & $3.09$ \\ 
%
HD5e & $\{(256,128,64,32,16,8)\times(2,1,0.5),~h=r_S/84\}$ & $6.37$ & $6.33$ \\ 
HD5f & $\{(256,128,64,32,16,8)\times(2,1,0.5),~h=r_S/84\}$ & $10.37$ & $10.35$ \\ 
\hline
\hline
\end{tabular*}
\end{table*}
Having introduced and tested our formalism and numerical code, we now present
results obtained for head-on collisions of five-dimensional black holes.  The
black holes collide from rest, with initial coordinate separation $d$.  Note
that in five spacetime dimensions the Schwarzschild radius is related to the ADM
mass $M$ via
\begin{equation}
r_S^2=\frac{8M}{3\pi}\,.\label{defadm5d}
\end{equation}
We therefore define the ``total'' Schwarzschild radius $r_S$ such that
$r_S^2=r_{S,1}^2 + r_{S,2}^2$. By using this definition, $r_S$ has physical
dimension of length and provides a suitable unit for measuring both results and
grid setup.

As summarised in table~\ref{tab:setup}, we consider a sequence of binaries with
initial coordinate separation ranging from $d = 3.17 r_{S}$ to $d =
10.37 r_{S}$.  The table further lists the proper separation $L$ along the line
of sight between the holes and the grid configurations used for the individual
simulations.

\subsubsection{Newtonian collision time}

\begin{figure}[tbhp]
  \centering
  \includegraphics[width=0.7\textwidth]{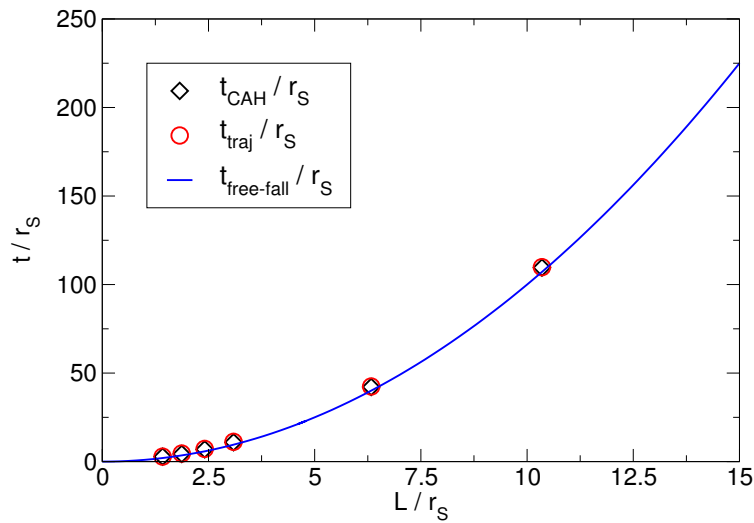}
  \caption[Estimates for the time it takes for two equal-mass black holes to
  collide in $D=5$]{Estimates for the time it takes for two equal-mass black
    holes to collide in $D=5$. The first estimate is given by the time
    $t_{\text{CAH}}$ elapsed until a single common apparent horizon engulfs both
    black holes (diamonds), the second estimate is obtained by using the
    trajectory of the black holes, i.e., the time $t_{\text{traj}}$ at which
    their separation has decreased below the Schwarzschild radius
    (circles). Finally, these numerical results are compared against a simple
    Newtonian estimate, given by equation~(\ref{eq:5dtime}) (blue solid
    line). \label{fig:D5_NewtCollTime} }
\end{figure}

An estimate of the time at which the black holes ``collide'' can be obtained
by considering a Newtonian approximation of two point
particles in $D=5$. 
The Newtonian time it takes for two point-masses (with Schwarzschild
parameters $r_{S,1}$ and $r_{S,2}$) to collide from rest with initial
distance $L$ in $D$ dimensions is given by
\begin{equation}
\label{eq:newt-time}
\frac{t_{\text{free-fall}}}{r_S} = \frac{\mathcal{I}}{D-3}\left(\frac{L}{r_S}\right)^{\frac{D-1}{2}}\,,
\end{equation}
where $r_S^{D-3} =r_{S,1}^{D-3} + r_{S,2}^{D-3}$ and
\begin{equation}
\label{eq:int}
\mathcal{I} = \int_0^1 \sqrt{\frac{z^{\frac{5-D}{D-3}}}{1-z}} \dd z
= \sqrt{\pi} \frac{\Gamma(\frac{1}{2} 
  + \frac{1}{D-3})}{\Gamma(1 + \frac{1}{D-3})}\,.
\end{equation}
For $D=4$, one recovers the standard result
$t_{\text{free-fall}} = \frac{\pi}{2} \sqrt{L^3/r_S^3}r_S\,,$ whereas for $D=5$ we get
\begin{equation}
t_{\text{free-fall}}=\left(L/r_S\right)^2r_S\,.\label{eq:5dtime}
\end{equation}
In general relativity, black hole trajectories and merger times are
intrinsically observer dependent quantities.  For our comparison with Newtonian
estimates we have chosen relativistic trajectories as viewed by observers
adapted to the numerical coordinate system. While the lack of fundamentally
gauge invariant analogues in general relativity prevents us from deriving
rigorous conclusions, we believe such a comparison to serve the intuitive
interpretation of results obtained within the ``moving puncture'' gauge.
Bearing in mind these caveats, we plot in figure~\ref{fig:D5_NewtCollTime} the
analytical estimate of the Newtonian time of collision, together with the
numerically computed time of formation of a common apparent horizon.  Also shown
in the figure is the time at which the separation between the individual hole's
puncture trajectory decreases below the Schwarzschild parameter $r_S$.
The remarkable agreement provides yet another example of how well numerically
successful gauge conditions appear to be adapted to the black hole kinematics.

\subsubsection{Waveforms}
\label{sec:D5waveforms}
%
\begin{figure}[tbhp]
\centering
\includegraphics[width=0.45\textwidth]{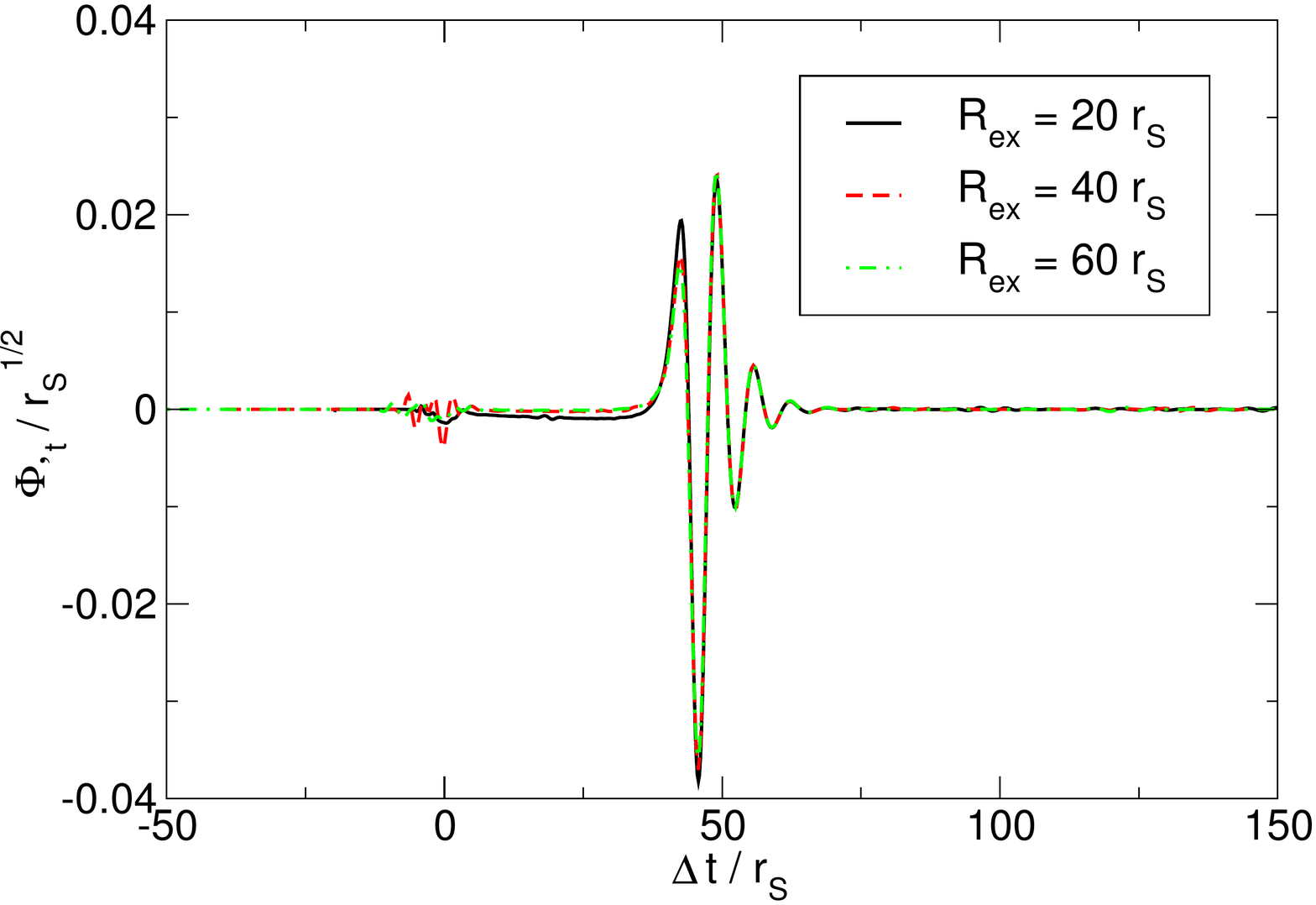} 
\includegraphics[width=0.45\textwidth]{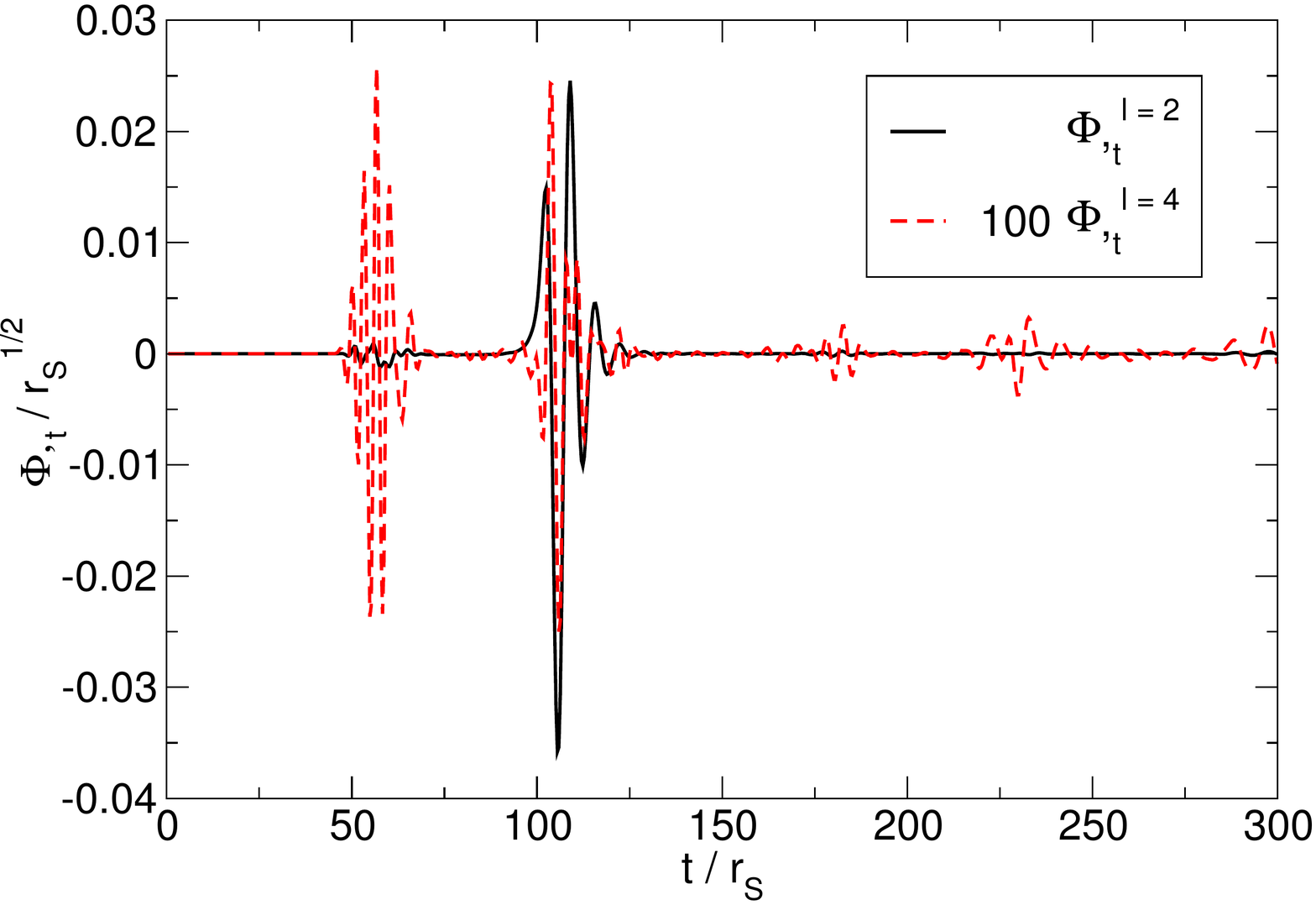} 
\caption[The $l=2$ and $l=4$ mode of the KI function]{Left panel: The $l=2$
  component of the KI waveform for model HD5e extracted at radii $R_{\rm
    ex}/r_{S} = 20, 40$ and $60$ and shifted in time by $R_{\rm ex}/r_S$. Right
  panel: The $l=2$ and $l=4$ mode of the KI function for the same simulation,
  extracted at $R_{\rm ex}/r_S = 60$. For clarity, the $l=4$ component has been
  re-scaled by a factor of $100$. \label{fig:D5_Phir_hf84} }
\end{figure}
%
%
%
Let us now discuss the gravitational wave signal, extracted with the KI
formalism, generated by the head-on collision of two black holes in five
dimensions.

In figure~\ref{fig:D5_Phir_hf84},
the $l=2$ multipole of the KI function $\Phi_{,t}$ for
model HD5e obtained at different extraction radii is plotted.
A small spurious wavepulse due to the initial data construction is visible at
$\Delta t \approx 0$, the so-called ``junk radiation''.
The physical part of the waveform is dominated by the merger signal around
$\Delta t=50r_S$, followed by the (exponentially damped) ringdown, whereas the
infall of the holes before $\Delta t=40r_S$ does not produce a significant
amount of gravitational waves.  Comparison of the waveforms extracted at
different radii demonstrates excellent agreement, in particular for those
extracted at $R_{\rm ex}=40r_S$ and $60r_S$. Extrapolation of the radiated
energy to infinite extraction radius yield a relative error of 5~\% at $R_{\rm
  ex}=60r_S$, indicating that such radii are adequate for the analysis
presented in this work.

Due to symmetry, no gravitational waves are emitted in the $l=3$ multipole, so
that $l=4$ represents the second strongest contribution to the wave signal. As
demonstrated in the right panel of figure~\ref{fig:D5_Phir_hf84},
however, its amplitude is two orders of magnitude below that of the quadrupole.


\begin{figure}[tbhp]
\centering
\includegraphics[width=0.7\textwidth]{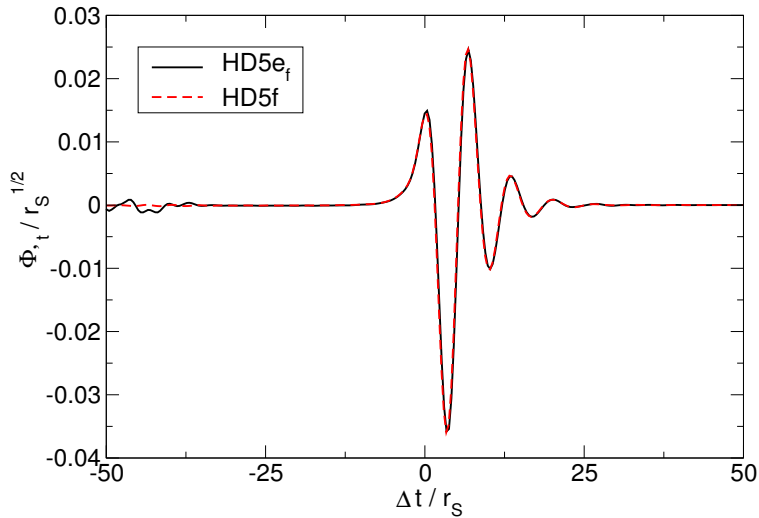} 
\caption[$l=2$ components of the KI function as generated by a head-on collision
of black holes]{$l=2$ components of the KI function as generated by a head-on
  collision of black holes with initial (coordinate) distance $d=6.37r_S$
  (black solid line) and $d=10.37r_S$ (red dashed line).  The wave functions
  have been shifted in time such that the formation of a common apparent horizon
  corresponds to $\Delta t=0$ (and taking into account the time it takes for the
  waves to propagate up to the extraction radius $R_{\rm ex} =
  60r_{S}$). \label{fig:D5_Phit_models}}
\end{figure}
%
In order to assess how accurately we are thus able to approximate an infall from
infinity, we have varied the initial separation for models HD5a to HD5f as
summarised in table~\ref{tab:setup}. 
As demonstrated in figure~\ref{fig:D5_Phit_models}, for models HD5e and HD5f
we can safely neglect the spurious radiation as well as the impact of a
finite initial separation, provided we use a sufficiently large initial distance
$d \gtrsim 6r_S$ of the binary. Here, we compare the radiation emitted
during the head-on collision of black holes starting from rest with initial separations
$6.37r_S$ and $10.37r_S$.  The waveforms have been shifted in time by the
extraction radius $R_{\rm ex} = 60r_{S}$ and such that the formation of a
common apparent horizon occurs at $\Delta t=0$. The merger signal starting
around $\Delta t=0$ shows excellent agreement for the two configurations and is
not affected by the spurious signal visible for HD5e at $\Delta t\approx
-50r_S$.

We conclude this discussion with an analysis of the ringdown. 
After formation of a common horizon, the waveform is dominated by an exponentially
damped sinusoid, as the merged hole {\em rings down} into a stationary state.
By fitting our results with an exponentially-damped sinusoid, we obtain a
characteristic frequency
\begin{equation}
r_{S} \, \omega = 0.955\pm 0.005- i (0.255 \pm 0.005) \,.
\label{qnm}
\end{equation}
This value is in excellent agreement with perturbative calculations, which
predict a lowest quasinormal frequency $r_{S}\,
\omega=0.9477- i 0.2561$ for $l=2$~\cite{Cardoso:2003qd,Yoshino:2005ps,Berti:2009kk}.


\subsubsection{Radiated energy}
\label{sec:D5raden}
%
\begin{figure}[tbhp]
\centering
\includegraphics[width=0.7\textwidth]{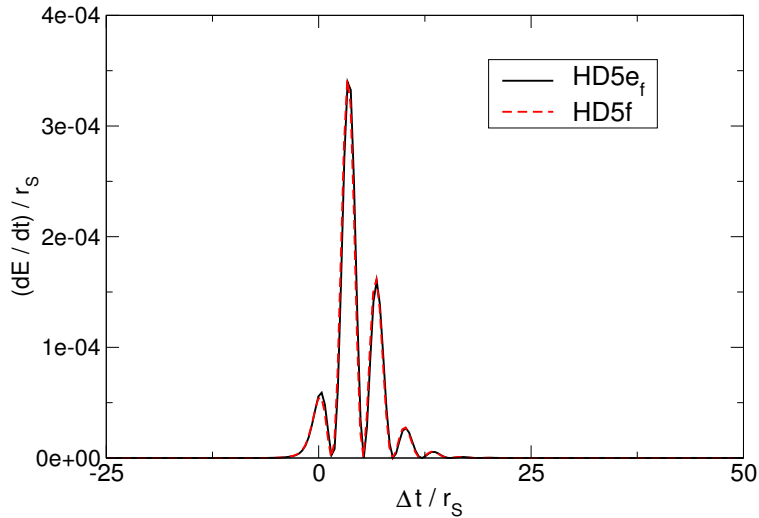} 
\caption[Energy flux in the $l=2$ component of the KI wave function
$\Phi_{,t}$]{Energy flux in the $l=2$ component of the KI wave function
  $\Phi_{,t}$, extracted at $R_{\rm ex} = 60r_{S}$, for models HD5e (black
  solid line) and HD5f (red dashed line) in table~\ref{tab:setup}. The fluxes
  have been shifted in time by the extraction radius $R_{\rm ex} = 60r_{S}$
  and the time $t_{\text{CAH}}$ at which the common apparent horizon
  forms. \label{fig:D5energy}}
\end{figure}
%
%
We now compute the 
energy flux from the KI master function via equation~(\ref{eq:energyflux}). The
fluxes thus obtained for the $l=2$ multipole of models HD5e and HD5f in
table~\ref{tab:setup}, extracted at $R_{\rm ex} = 60r_{S}$, are shown in
figure~\ref{fig:D5energy}.  As in the case of the KI master function in
figure~\ref{fig:D5_Phit_models}, we see no significant variation of the flux for
the two different initial separations.  The flux reaches a maximum value of $\dd
E/\dd t\sim 3.4\times 10^{-4}r_S$, and is then dominated by the ringdown
flux. The energy flux from the $l=4$ mode is typically four orders of magnitude
smaller; this is consistent with the factor of 100 difference of the
corresponding wave multipoles observed in figure~\ref{fig:D5_Phir_hf84}, and the
quadratic dependence of the flux on the wave amplitude.
%
%
%
%
Integrating, we find that a fraction of $E_\mathrm{rad}/M = (8.9\pm0.6)\times 10^{-4}$
of the centre of mass energy is emitted in the form of gravitational radiation.
We have verified for these models that the amount of energy contained in the
spurious radiation is about three orders of magnitude smaller than in the
physical merger signal.





\section{Discussion}
\label{sec:final-remarks}

In this chapter we have presented a framework that allows the generalisation of
the present generation of 3+1 numerical codes to evolve, with relatively minor
modifications, spacetimes with $SO(D-2)$ symmetry in $5$ dimensions and
$SO(D-3)$ symmetry in $D\ge 6$ dimensions. The key idea is a dimensional
reduction of the problem, along the lines of
references~\cite{Geroch:1970nt,Cho:1986wk}, that recasts the $D$-dimensional
Einstein vacuum equations in the form of the standard four dimensional equations
plus some source terms.  The resulting equations can be transformed
straightforwardly into the BSSN formulation that has proved remarkably
successful in numerical evolutions of black hole configurations in 3+1
spacetimes.

The class of problems that may be studied with our framework includes head-on
collisions in $D\ge 5$ and a subset of black hole collisions with impact
parameter and spin in $D\ge 6$.

A procedure to construct initial data and a formalism to extract gravitational radiation observables from the numerical simulations were also introduced.
With these tools, the numerical implementation was done by adapting the {\sc Lean} code and, after a number of tests including the 
convergence of the Hamiltonian and momentum constraints as well as
comparing numerical results with (semi-)analytic expressions for a single
Tangherlini black hole in geodesic slicing, we reported results obtained for evolutions of black hole collisions in five-dimensional spacetimes.

As might be expected, stable evolutions of such spacetimes require some
modifications of the underlying methods of the so-called {\em moving puncture}
technique, especially with regard to the gauge conditions used therein. We have
successfully modified the slicing condition 
in order to obtain long-term stable simulations in $D=5$
dimensions. Unfortunately, these modifications do not appear sufficient to
provide long-term stability for arbitrary values of the dimensionality $D$.
This issue remains under investigation.

Besides obtaining the corresponding waveforms for head-on collision of
five-dimensional black holes, we have further shown that the total energy
released in the form of gravitational waves is approximately $(0.089\pm
0.006)\%$ of the initial centre of mass energy of the system, for a head-on
collision of two black holes starting from rest at very large distances. As a
comparison, the analogous process in $D=4$ releases a slightly smaller quantity:
$(0.055\pm 0.006)\%$.

As yet another test of our implementation, the ringdown part of the waveform was
also shown to yield a quasinormal mode frequency in excellent agreement with
predictions from black hole perturbation theory.

The numbers reported here for the total energy loss in gravitational waves
should increase significantly in high energy collisions, which are the most
relevant scenarios for the applications described in the Introduction. Indeed,
in the four dimensional case, it is known that ultra-relativistic head-on
collisions of equal mass non-rotating black holes release up to 14\% of the
initial centre of mass energy into gravitational
radiation~\cite{Sperhake:2008ga}. The analogous number in higher dimensions is
as yet unknown, and it remains under investigations using the tools here
presented.

Even more energy may be released in high energy collisions with non-vanishing
impact parameter. In~\cite{Shibata:2008rq,Sperhake:2009jz} it was shown that
this number can be as large as 35\% in $D=4$. The formalism here developed
allows, in principle, the study of analogous processes in $D\ge 6$.

\begin{subappendices}


\section{Ricci tensor}
\label{sec:ricci-non-coord}

In this appendix we give the full details about the computation of the Ricci tensor of section~\ref{sec:general-metric}.
We start by writing the metric~\eqref{eq:metric0} in block-diagonal form
\begin{align}
  \label{eq:metric}
  \dd \bar s^2 =  \bar g_{A B} e^A \otimes e^B = g_{\mu\nu} \dd x^\mu \otimes \dd x^\nu 
  + g_{\bar i \bar j} \Theta^{\bar i} \otimes \Theta^{\bar j},
\end{align}
where 
\begin{align}
  \label{eq:non-coord-basis}
  e_A = (D_\mu, \partial_{\bar i}), \quad D_\mu = \partial_\mu - e \kappa B^{\bar i}_{\mu} \partial_{\bar i},
\end{align}
is now the (non-coordinate) basis. Its dual is
\begin{align}
  e^A = (\dd x^\mu, \Theta^{\bar i}), \quad \Theta^{\bar i} = \dd x^{\bar i} 
  + e\kappa B^{\bar i}_\mu \dd x^\mu.
\end{align}
This basis satisfies
\begin{align}
  \label{eq:algebra1}
  [e_A, e_B] = \F{\bar k}{A B} \partial_{\bar k},
\end{align}
where $\F{\bar k}{A B}$ are given by~\eqref{eq:Frel}.
In the following we will not explicitly assume the expressions~\eqref{eq:Frel}, i.e., we will only assume (unless explicitly mentioned otherwise) that $ [e_A, e_B] =  \F{\bar k}{A B} e_{\bar k} = \F{\bar k}{A B} \partial_{\bar k} $, $\partial_{\bar i} g_{\mu \nu} = 0$ and $\F{\bar k}{\bar i \bar j} = 0$. We will not assume the expression for the $e_{\mu}$ in terms of the coordinate basis (even though we will assume that $e_{\bar k} = \partial_{\bar k}$).

\begin{description}
\item[Important remark:] From now on we will work with the non-coordinate basis $e_A$ to simplify the calculations. Note that, for an arbitrary tensor $T_A$, 
  \begin{align*}
    T_{A|B} \equiv \partial_{e_B } T_A \equiv e_B \left( T_A \right) 
    \neq \partial_B T_A \equiv \del{}{T_A}{x^B}.
  \end{align*}
  In particular, $e_\mu \equiv D_\mu \equiv \partial_\mu - e \kappa B^i_\mu \partial_i $, and as such,
  \[
  T_{A|\mu} = \partial_{e_\mu} T_A = D_{\mu} \left( T_A \right) = \partial_\mu T_A - e \kappa B^i_\mu \partial_i T_A.
  \]
  We must keep in mind that $T_{A|\mu \nu} \neq T_{A| \nu\mu} $.
  On the other hand, $e_{\bar k} = \partial_{\bar k}$ and thus
  \[
  T_{A|\bar k} = \partial_{\bar k} T_A.
  \]
\end{description}

We recall our definition of the ``covariant derivatives'' $\nabla_\mu$ and $\nabla_{\bar j}$ as \footnote{Note that now, as we are working on a non-coordinate basis, the order of the indices does matter, i.e., $\GG{\alpha}{\lambda \sigma} \neq \GG{\alpha}{\sigma \lambda} $ and $\GG{\bar i}{\bar j \bar k} \neq \GG{\bar i}{\bar k \bar j}$. } 
\begin{align}
  \nabla_\sigma T^{\bar i\alpha}{}_{\bar k \mu} &  \equiv D_\sigma T^{\bar i \alpha}{}_{\bar k \mu} + \F{\bar i}{\sigma \bar l} T^{\bar l \alpha}{}_{\bar k \mu}
  -\F{\bar l}{\sigma \bar k} T^{\bar i \alpha}{}_{\bar l \mu} + \GG{\alpha}{\lambda \sigma} T^{\bar i \lambda}{}_{\bar k \mu} 
  - \GG{\lambda}{\mu \sigma} T^{\bar i \alpha}{}_{\bar k \lambda}, \\
  \nabla_{\bar j} T^{\bar i\alpha}{}_{\bar k \mu} &  \equiv \partial_{\bar j} T^{\bar i \alpha}{}_{\bar k \mu} 
  + \GG{\bar i}{\bar l \bar j} T^{\bar l \alpha}{}_{\bar k \mu} 
  - \GG{\bar l}{\bar k \bar j} T^{\bar i \alpha}{}_{\bar l \mu },
\end{align}
recall also that both connections are metric,
\begin{align*}
   \nabla_\sigma g_{\mu\nu} & = \partial_\sigma g_{\mu\nu} - \GG{\lambda}{\mu\sigma} g_{\lambda\nu} 
  - \GG{\lambda}{\nu\sigma} g_{\mu \lambda} = 0, \\
  \nabla_{\bar k} g_{\bar i \bar j} & \equiv \partial_{\bar k} g_{\bar i \bar j} 
  - \GG{\bar l}{\bar i \bar k} g_{\bar l \bar j} - \GG{\bar l}{\bar j \bar k} g_{\bar i \bar l} = 0,
\end{align*}
and that
\begin{align*}
  \nabla_\sigma g_{\bar i \bar j} \equiv D_\sigma g_{\bar i \bar j} - \F{\bar k}{\sigma \bar i}g_{\bar k \bar j} - \F{\bar k}{\sigma \bar j}g_{\bar i \bar k} \neq 0.
\end{align*}

We now recall some expressions from section~\ref{sec:prelim}, which we re-write here for convenience: on a non-coordinate basis obeying 
\begin{align}
  \label{eq:basis}
  \left[e_A, e_B \right] = c_{AB}{}^D e_D
\end{align}
the torsion-free connection is given by
\begin{align}
  \label{eq:connection}
  \GG{A}{BC} = \frac{1}{2} g^{AD} \left(
    g_{DB|C} + g_{DC|B} - g_{BC|D} + c_{DBC} + c_{DCB} - c_{BCD}
  \right),
\end{align}
where $\GG{A}{[BC]} = -\frac{1}{2} c_{BC}{}^A$, and the Riemann tensor by
\begin{align}
  \label{eq:riemann}
  R^A{}_{BCD} = \GG{A}{BD|C} - \GG{A}{BC|D} + \GG{A}{EC} \GG{E}{BD} - \GG{A}{ED} \GG{E}{BC} - \GG{A}{BE} c_{CD}{}^E.
\end{align}
We use the convention
\begin{align*}
  \nabla_{e_\alpha} e_{\beta} = \Gamma^{\lambda}{}_{\beta \alpha} e_\lambda.
\end{align*}

Applying these formulas to our case,
\begin{align*}
  & c_{AB}{}^{\bar i} = \F{\bar i}{AB}, \quad c_{AB}{}^\mu = 0, \quad e_\mu = D_\mu, 
  \quad e_{\bar k} = \partial_{\bar k}\,,
\end{align*}
we get from equation~\eqref{eq:connection}
\begin{equation}
  \label{eq:gamma}
  \begin{aligned}
    \bar \Gamma^{\alpha}{}_{\mu\nu} &  = \frac{1}{2} g^{\alpha\lambda} \left(
      g_{\lambda\mu| \nu} + g_{\lambda\nu| \mu} - g_{\mu\nu| \lambda}
    \right) =  \GG{\alpha}{\mu\nu}, \\
    \bar \Gamma^{\bar i}{}_{\mu\nu} & = -\frac{1}{2} \F{\bar i}{\mu\nu}, \\
    \bar \Gamma^\mu{}_{\bar i \nu} & = \half g^{\mu\lambda} g_{\bar i \bar k} \F{\bar k}{\lambda\nu} = \bar \Gamma^\mu{}_{\nu \bar i}, \\
    \bar \Gamma^\mu{}_{\bar i \bar j} &  = -\half g^{\mu\lambda} \nabla_\lambda g_{\bar i \bar j}, \\
    \bar \Gamma^{\bar i}{}_{\mu \bar j} & = \half g^{\bar i \bar k} \nabla_\mu g_{\bar k \bar j}, \\
    \bar \Gamma^{\bar i}{}_{\bar j \mu} & = \half g^{\bar i \bar k} \nabla_\mu g_{\bar k \bar j} + \F{\bar i}{\mu \bar j}, \\
    \bar \Gamma^{\bar i}{}_{\bar j \bar k} & = \half g^{\bar i \bar l} \left(
      g_{\bar l \bar j| \bar k} + g_{\bar l \bar k | \bar j} - g_{\bar j \bar k| \bar l}
    \right) = \GG{\bar i}{\bar j \bar k}\,.
  \end{aligned}
\end{equation}
From~\eqref{eq:riemann} we compute the Ricci tensor,
\begin{align}
  \label{eq:Rmunu1}
    \bar R_{\mu\nu} & = R_{\mu\nu} -\half g^{\alpha\lambda} g_{\bar i \bar j} \F{\bar i}{\lambda\mu} \F{\bar j}{\alpha\nu}
    -\half \nabla_\nu\left(
      g^{\bar i \bar j} \nabla_\mu g_{\bar i \bar j}
    \right) - \frac{1}{4} g^{\bar i \bar j} g^{\bar k \bar l} \nabla_\mu g_{\bar i \bar k} \nabla_\nu g_{\bar j \bar l}
    -\half \nabla_{\bar k} \F{\bar k}{\mu\nu}
  , \\
 \label{eq:Rmui1}
    \bar R_{\mu \bar i} & = \half g^{\alpha\lambda} \nabla_\alpha \left(
      g_{\bar i \bar k} \F{\bar k}{\lambda\mu}
    \right) + \frac{1}{4} g^{\bar k \bar l} \nabla_\beta g_{\bar k \bar l} g^{\beta\lambda} \F{\bar m}{\lambda\mu} g_{\bar i \bar m}
    + \half \nabla_{\bar k} \left(  
      g^{\bar k \bar l} \nabla_\mu g_{\bar l \bar i}
    \right)
    - \half \nabla_{\bar i} \left(  
      g^{\bar k \bar l} \nabla_\mu g_{\bar k \bar l}
    \right)
  , \\
  \label{eq:Rimu1}
  \bar R_{\bar i \mu} & = \bar R_{\mu \bar i} + \F{\bar k}{\mu \bar i| \bar k} 
  - \F{\bar k}{\mu \bar k| \bar i}, \\
 \label{eq:Rij1}
    \bar R_{\bar i \bar j} & = R_{\bar i \bar j} - \frac{1}{4} g^{\bar k \bar l} \nabla_\beta g_{\bar k \bar l} \nabla^\beta g_{\bar i \bar j} 
    + \half g^{\bar k \bar l} \nabla_\beta g_{\bar i \bar k} \nabla^\beta g_{\bar j \bar l} 
    + \frac{1}{4} g^{\alpha\lambda} g^{\beta\rho} g_{\bar j \bar k} g_{\bar i \bar l} \F{\bar k}{\beta\lambda} \F{\bar l}{\rho\alpha}
    - \half \nabla^\beta \nabla_\beta g_{\bar i \bar j}
  \,,
\end{align}
where
\begin{align*}
  \nabla_{\bar k} \F{{\bar k}}{\mu\nu} & \equiv \F{{\bar k}}{\mu\nu|{\bar k}} + \GG{{\bar k}}{{\bar j}{\bar k}}  \F{{\bar j}}{\mu\nu},  \\
  \nabla_{\bar i} \left(  
    g^{{\bar k}{\bar l}} \nabla_\mu g_{{\bar k}{\bar l}}
  \right) & \equiv  \partial_{\bar i} \left(  
    g^{{\bar k}{\bar l}} \nabla_\mu g_{{\bar k}{\bar l}}
  \right), \\
  \nabla_{\bar k} \left(  
    g^{{\bar k}{\bar l}} \nabla_\mu g_{{\bar l}{\bar i}}
  \right) & \equiv \partial_{\bar k} \left(
    g^{{\bar k}{\bar l}} \nabla_\mu g_{{\bar l}{\bar i}}
  \right)
  + \GG{{\bar k}}{{\bar j}{\bar k}} g^{{\bar j}{\bar l}} \nabla_{\mu} g_{{\bar l}{\bar i}} - \GG{{\bar j}}{{\bar i}{\bar k}}g^{{\bar k}{\bar l}} \nabla_\mu g_{{\bar l}{\bar j}},
\end{align*}
and we also used $\GG{\bar k}{[\bar i \bar j]} = -\half c_{\bar i \bar j}{}^{\bar k} = -\half \F{\bar k}{\bar i \bar j} = 0$ and $\GG{\alpha}{[\mu\beta]} = -\half c_{\mu\beta}{}^\alpha = -\half \F{\alpha}{\mu\beta} = 0$. 

The Ricci scalar is given by
\begin{align*}
  \bar R = \bar g^{AB} \bar R_{AB} = g^{\mu\nu} \bar R_{\mu\nu} + g^{\bar i \bar j} \bar R_{\bar i \bar j}.
\end{align*}
We have
\begin{align}
  \label{eq:Rscalar}
    \bar R & = R + \tilde R - \frac{1}{4} g_{\bar k \bar l} g^{\alpha\lambda} g^{\mu\nu} \F{\bar l}{\lambda\mu} \F{\bar k}{\alpha \nu}
    -\nabla^\mu \left(g^{\bar k \bar l} \nabla_\mu g_{\bar k \bar l} \right)
    - \frac{1}{4} g^{\bar k \bar i} g^{\bar j \bar l} \nabla^\mu g_{\bar k \bar l} \nabla_\mu g_{\bar i \bar j} \notag \\
    &{} \quad - \frac{1}{4} g^{\bar k \bar l} g^{\bar i \bar j} \nabla^\mu g_{\bar k \bar l} \nabla_\mu g_{\bar i \bar j}
    ,
\end{align}
where $\tilde R = g^{\bar i \bar j} R_{\bar i \bar j} $.

We still need to write the components of the Ricci tensor in the coordinate basis $(\partial_\mu, \partial_{\bar i})$, so from here on we need to use the specific form of $e_\mu$ and, thus, also the algebra in~\eqref{eq:Frel}. We perform the basis transformation the usual way,
\begin{align*}
  \mathbf{R} & = \bar R_{AB} e^A \otimes e^B \\
  & = \left(
    \bar R_{\mu\nu} + e\kappa \bar R_{{\bar i}\nu} B^{\bar i}_\mu + e\kappa \bar R_{\mu {\bar i}} B^{\bar i}_\nu + e^2 \kappa^2 \bar R_{{\bar i}{\bar j}} B^{\bar i}_\mu B^{\bar j}_\nu
  \right) \dd x^\mu \otimes \dd x^\nu \\
  &{} \quad + \left( 
    \bar R_{{\bar i} \mu} + e \kappa \bar R_{{\bar i}{\bar j}} B^{\bar j}_\mu
  \right) \dd x^{\bar i} \otimes \dd x^\mu
  + \left( 
    \bar R_{\mu {\bar i}} + e \kappa \bar R_{{\bar j}{\bar i}} B^{\bar j}_\mu
  \right) \dd x^\mu \otimes \dd x^{\bar i} 
  + \bar R_{{\bar i}{\bar j}} \dd x^{\bar i} \otimes \dd x^{\bar j},
\end{align*}
and thus, in the basis $(\partial_\mu, \partial_{\bar i})$ where the metric takes the form~\eqref{eq:metric0}, we have
\begin{align}
  \bar R_{{\bar i}{\bar j}} & = R_{{\bar i}{\bar j}} - \frac{1}{4} g^{{\bar k}{\bar l}} \nabla_\beta g_{{\bar k}{\bar l}} \nabla^\beta g_{{\bar i}{\bar j}} 
  + \half g^{{\bar k}{\bar l}} \nabla_\beta g_{{\bar i}{\bar k}} \nabla^\beta g_{{\bar j}{\bar l}} 
  + \frac{1}{4} g^{\alpha\lambda} g^{\beta\rho} g_{{\bar j}{\bar k}} g_{{\bar i}{\bar l}} \F{{\bar k}}{\beta\lambda} \F{{\bar l}}{\rho\alpha}
  - \half \nabla^\beta \nabla_\beta g_{{\bar i}{\bar j}}, \label{eq:Rij-app} \\
  \bar R_{\mu {\bar i}} & = e\kappa \bar R_{{\bar i}{\bar j}} B^{\bar j}_{\mu} 
  + \half g^{\alpha\lambda} \nabla_\alpha \left(
    g_{{\bar i}{\bar k}} \F{{\bar k}}{\lambda\mu}
  \right) + \frac{1}{4} g^{{\bar k}{\bar l}} \nabla_\beta g_{{\bar k}{\bar l}} g^{\beta\lambda} \F{{\bar m}}{\lambda\mu} g_{{\bar i}{\bar m}}
  + \half \nabla_{\bar k} \left(  
    g^{{\bar k}{\bar l}} \nabla_\mu g_{{\bar l}{\bar i}}
  \right) \notag \\
  &{} \quad - \half \nabla_{\bar i} \left(  
    g^{{\bar k}{\bar l}} \nabla_\mu g_{{\bar k}{\bar l}}
  \right) = \bar R_{{\bar i} \mu}, \label{eq:Rmui-app} \\
  \bar R_{\mu\nu} & = R_{\mu\nu} + 2 e \kappa B^{\bar i}_{(\mu} \bar R_{\nu) {\bar i}} - e^2 \kappa^2 \bar R_{{\bar i}{\bar j}} B^{\bar i}_{\mu} B^{\bar j}_\nu
  -\half g^{\alpha\lambda} g_{{\bar i}{\bar j}} \F{{\bar i}}{\lambda\mu} \F{{\bar j}}{\alpha\nu}
  -\half \nabla_\nu\left(
    g^{{\bar i}{\bar j}} \nabla_\mu g_{{\bar i}{\bar j}}
  \right) \notag \\
  &{} \quad- \frac{1}{4} g^{{\bar i}{\bar j}} g^{{\bar k}{\bar l}} \nabla_\mu g_{{\bar i}{\bar k}} \nabla_\nu g_{{\bar j}{\bar l}}
  -\half \nabla_{\bar k} \F{{\bar k}}{\mu\nu}, \label{eq:Rmunu-app}
\end{align}
and 
\begin{align}
  \label{eq:Rscalar2-app}
    \bar R & = R + \tilde R - \frac{1}{4} g_{{\bar k}{\bar l}} g^{\alpha\lambda} g^{\mu\nu} \F{{\bar l}}{\lambda\mu} \F{{\bar k}}{\alpha \nu}
    -\nabla^\mu \left(g^{{\bar k}{\bar l}} \nabla_\mu g_{{\bar k}{\bar l}} \right)
    - \frac{1}{4} g^{{\bar k}{\bar i}} g^{{\bar j}{\bar l}} \nabla^\mu g_{{\bar k}{\bar l}} \nabla_\mu g_{{\bar i}{\bar j}} \notag \\
    &{} \quad - \frac{1}{4} g^{{\bar k}{\bar l}} g^{{\bar i}{\bar j}} \nabla^\mu g_{{\bar k}{\bar l}} \nabla_\mu g_{{\bar i}{\bar j}}
    .
\end{align}

\begin{description}
\item[Note:] We have used the same indices to label components in both the
  coordinate and non-coordinate basis. No confusion shall arise, however, since
  the non-coordinate basis was used merely to simplify the previous
  computations. In the main text we deal exclusively with the coordinate basis,
  to which all components refer.
\end{description}


\section{Equations of motion: Einstein frame}
\label{sec:einstein-frame}
We here write the equations of motion obtained when we write the action~(\ref{eq:action0}) in the Einstein frame.
To do so, we perform the usual conformal transformation 
\begin{equation}
  \label{eq:conf-metric}
  g_{\mu\nu} = \psi^{-\frac{2}{d-2}} \tilde{g}_{\mu\nu}
\end{equation}
where in our case $\psi \equiv e^{n\phi}$.
The action then takes the form
\begin{equation}
  \label{eq:action-einstein}
  \mathcal{S} = \frac{1}{16\pi G_d}\int \dd ^d x \sqrt{-\tilde{g}} \left[
    \tilde{R} + k \partial_{\mu} \tilde{\phi} \partial^{\mu} \tilde{\phi}
    + n(n-1) e^{-2\tilde{\phi}} \right]
\end{equation}
where $k\equiv n \frac{d(d-4)(n-1) - 2(n+2)}{(d-2+n)^2}$ and $\tilde{\phi} = \frac{d-2+n}{d-2} \phi$.
From action~(\ref{eq:action-einstein}) we obtain the equations of motion
\begin{equation}
  \label{eq:eq-einstein-frame}
  \begin{aligned}
  & k \tilde{\nabla}^{\alpha} \partial_{\alpha} \tilde{\phi} + n(n-1) e^{-2\tilde{\phi}} = 0 \\
  & \tilde{R}_{\mu\nu} = -k \partial_{\mu} \tilde{\phi} \partial_{\nu} \tilde{\phi}
  - \frac{n(n-1)}{d-2} e^{-2\tilde{\phi}} \tilde{g}_{\mu\nu}
  \end{aligned}.
\end{equation}


\end{subappendices}



\chapter{Non-asymptotically flat spacetimes}
\label{ch:non-asympt-flat}

\section{de Sitter}
\label{sec:dS}

Nonlinear dynamics in cosmological backgrounds has the potential to teach us
immensely about our universe, and also to serve as prototype for nonlinear
processes in generic curved spacetimes. 
\emph{de Sitter} spacetime, as already mentioned in the Introduction, is the simplest accelerating universe---a maximally symmetric solution of Einstein's equations with a positive cosmological constant---which seems to model quite well the present cosmological acceleration~\cite{Komatsu:2010fb}.

Key questions concerning the evolution towards a de Sitter, spatially homogeneous universe are how inhomogeneities develop in time and, in particular, if they are washed away by the cosmological expansion~\cite{Shibata:1993fx}. Answering them requires controlling the imprint of the gravitational interaction between localised objects on the large-scale expansion. Conversely, the cosmological dynamics should leave imprints in strong gravitational phenomena like primordial black hole formation~\cite{Shibata:1999zs} or the gravitational radiation emitted in a black hole binary coalescence, which carry signatures of the cosmological acceleration as it travels across the universe. Identifying these signatures is not only of conceptual interest but also phenomenologically relevant, in view of the ongoing efforts to directly detect gravitational radiation. 

Finally, dynamics in asymptotically de Sitter spacetimes could also teach us
about more fundamental questions such as cosmic censorship: two black holes of
sufficiently large mass in de Sitter spacetime would, upon merger, give rise to
too large a black hole to fit in its cosmological horizon.  In this case the end
state would be a naked singularity. This possibility begs for a time evolution
of such a configuration. Does the time evolution of non-singular data containing
two black holes result in a naked singularity, or are potentially offending
black holes simply driven away from each other by the cosmological expansion?

In this section, following~\cite{Zilhao:2012bb}, we report on numerical evolutions of black hole binaries in an
asymptotically de Sitter geometry. Even though we consider a range of values for
the cosmological constant far larger than those which are phenomenologically
viable, these results provide useful insight on the general features of
dynamical black hole processes in spacetimes with a cosmological constant, which
can improve our understanding of our universe.

\subsection{Evolution equations}
\label{sec:evol-eq-lambda}

The Einstein equations with cosmological constant $\Lambda$ are
\begin{align}
  \label{eq:ee-lam}
  R_{\mu\nu} - \frac{1}{2}g_{\mu\nu} R = - \Lambda g_{\mu\nu} \,,
\end{align}
and we will always consider $\Lambda > 0$.
We perform the 3+1 decomposition by introducing the projection operator
$\gamma_{\mu\nu}$ and the normal to the three dimensional hyper-surface
$\Sigma$, $n^\mu$ ($n^\mu n_\mu = -1$), as outlined in section~\ref{sec:choice-coordinates} and write the evolution equations in the BSSN form~(\ref{eq:bssn-gen}).

From~\eqref{eq:ee-lam}, we straightforwardly compute the source terms
\begin{equation}
  \label{eq:matterterms}
  \begin{aligned}
    8 \pi E     & = \Lambda  \,, &
    8 \pi j_i   & = 0  \,, \\
    8 \pi S_{ij} & = - \Lambda \chi^{-1} \tilde \gamma_{ij}  \,, &
    8 \pi S     & = -3\Lambda \,.
  \end{aligned}
\end{equation}
A new evolution variable $\bar \chi = \exp(2 \sqrt{\Lambda/3} t) \chi$ has been introduced
instead of the usual BSSN variable $\chi$~\cite{Shibata:1993fx}. The reason is
that for black hole evolutions it is crucial to impose a floor value on $\chi$,
typically $10^{-4}$ or $10^{-6}$, which is inconsistent with the natural
behaviour of this variable in a de Sitter spacetime (as we will see below): $\chi^{-1} \sim
\exp(2 \sqrt{\Lambda/3} t)$. In contrast $\bar \chi \to 1$ when $r\to\infty$ for all times.
The evolution equations are thus
%
%
\begin{subequations}
  \begin{align}
    \partial_t \tilde \gamma_{ij} & = [\cdots] \,, \\
    \partial_t \bar \chi & = 2 \bar \chi (\alpha K - \partial_i \beta^i )/3
      + \beta^i \partial_i \bar \chi + 2\sqrt{\frac{\Lambda}{3}} \bar \chi \,, \\
%
%
    \partial_t   K & =
        [\cdots] - \alpha \Lambda \, , \\
    \partial_t \tilde A_{ij} & =
        [\cdots] 
        \,, \\
   \partial_t \tilde \Gamma^i & =
        [\cdots] 
        \, .
  \end{align}
\end{subequations}
where $ [\cdots] $ denotes the
right-hand side of the BSSN equations~\eqref{eq:bssn-gen}
in the absence of source terms.

\subsection{Schwarzschild-de Sitter}
\label{sec:mcvittie}

The Schwarzschild-de Sitter spacetime, solution of~(\ref{eq:ee-lam}), written in static coordinates reads
\begin{equation}
\dd s^2=-f(R)\dd T^2+f(R)^{-1}\dd R^2+R^2\dd \Omega_2 \,. 
\end{equation}
The solution is characterised by two parameters: the black hole mass $m$ and the Hubble parameter $H$,
\begin{equation}
f(R)=1-2m/R-H^2R^2 \ , \qquad H \equiv \sqrt{\Lambda/3} \ . 
\end{equation}
$f(R)$ has two zeros, at $R=R_\pm$, $R_-<R_+$, if 
\begin{equation} 
0<mH<mH_{\rm crit} \ , \qquad m H_{\rm crit}\equiv \sqrt{1/27} \ . 
\end{equation}
These zeros are the location of the black hole event horizon ($R_-$) and of a cosmological horizon ($R_+$). If $H=0$, then $R_-=2m$; if $m=0$, then $R_+=1/H$. If $H,m\neq 0$, then  $R_->2m$ and $R_+<1/H$. Since $R$ is the areal radius, the area of the spatial sections of the cosmological horizon decreases in the presence of a black hole; and the area of the spatial sections of the black hole horizon increases in the presence of a cosmological constant, as one would intuitively anticipate.

The basic dynamics in this spacetime may be inferred by looking at radial timelike geodesics. They obey the equation $\left(\dd R/\dd \tau\right)^2=E^2-f(R)$,  where $\tau$ is the proper time and $E$ is the conserved quantity associated to the Killing vector field $\partial/\partial T$. In the static patch ($R_-<R<R_+$), $E$ can be regarded as energy. From this equation we see that $f(R)$ is an effective potential. This potential has a maximum at 
\begin{equation}
R_{\rm max}=(m/H^2)^{1/3} \label{eq:rcrit_geo}\,.
\end{equation}
Geodesics starting from rest (i.e. $\dd R/\dd \tau(\tau=\tau_0)=0$) will fall into the black hole if $R_-<R<R_{\rm max}$ or move away from the black hole if $R_{\rm max}<R<R_+$.

As we will discuss in the next section, the initial data for an evolution in the de Sitter  universe can be computed 
in a similar manner as has been done 
in asymptotically flat space as long as one chooses a foliation with extrinsic curvature $K_{ij}$ having only a trace part. Such a coordinate system is known for Schwarzschild-de Sitter: \textit{McVittie coordinates}~\cite{McVittie:1933zz}. These are obtained from static coordinates by the transformation $(T,R)\rightarrow (t,r)$ given by
\begin{equation}
\label{eq:rtoR}
R = (1+\xi)^2 a(t) r \ , \qquad  T=t+H \int \frac{R \, \dd R}{f(R)\sqrt{1-2m/R}} \ ,
\end{equation}
where $a(t)= \exp(H t)$ and $ \xi \equiv \frac{m}{2 a(t) r}$.
One obtains McVittie's form for Schwarzschild-de Sitter:
\begin{equation}
\label{eq:mcvittie}
\dd s^2 = - \left( \frac{1-\xi}{1+\xi} \right)^2 \dd t^2 
+ a(t)^2 (1+\xi)^4 (\dd r^2 + r^2\dd \Omega_2 ) \ .
\end{equation}
For $t=\mathrm{constant}$, one can show that 
indeed $K^i_j=-H\delta^i_j$.

By setting $m=0$ in McVittie coordinates one recovers an FRW cosmological model with $k=0$ (flat spatial curvature) and an exponentially growing scale factor. 
The cosmological horizon $\mathcal{H}_C$ discussed above, located at $R=1/H$, stands at $r_{\mathcal{H}_C}=1/(He^{Ht})$.
The spatial sections of $\mathcal{H}_C$ seem to be shrinking down in this coordinate system. What happens, in fact, is that the exponentially fast expansion is taking any observer to the outside of $\mathcal{H}_C$. This is a well known phenomenon in studies of inflation and, as we shall see, has important consequences for the numerical evolution.

\subsection{Numerical Setup}
The cosmological constant introduces a new term (when compared with the vacuum
case) in the Hamiltonian constraint obtained after the canonical 3+1
decomposition~(\ref{eq:hamiltonian2}), 
\begin{equation}
  R-K_{ij} K^{ij} + K^2 = 2\Lambda
\end{equation}
In references~\cite{Nakao:1990vq,Nakao:1992zc} it was observed that imposing
a spacetime slicing obeying $K^i{}_j=-H\delta^i_j$, and a spatial metric
of the form $\dd l^2 = \psi^4 \tilde \gamma_{ij} \dd x^i \dd x^j$, the equations
to be solved in order to obtain initial data are equivalent to those
in vacuum.  In particular, for a system of $N$ black holes momentarily at rest
(with respect to the given spatial coordinate patch), the conformal
factor $\psi$ takes the form
\begin{equation}
\label{eq:psigen}
\psi = 1 + \sum_{i=1}^N \frac{m_i}{2|r-r_{(i)}|} \,.
\end{equation}
There are $N+1$ asymptotically de Sitter regions, as $|r-r_{(i)}|\rightarrow 0,+\infty$;  the total mass for observers in the common asymptotic region ($|r-r_{(i)}|\rightarrow +\infty$) is $\sum_i m_i$~\cite{Nakao:1992zc}.

Boundary conditions for all quantities are imposed by looking at the behaviour of massless perturbing fields in a pure de Sitter background.
Accordingly, we impose the following asymptotic behaviour for all BSSN variables
\begin{equation}
\label{eq:bcs2}
\partial_t f-\partial_t f_0+\frac{1}{a(t)}\partial_{r}f+\frac{f-f_0}{a(t) r}-H\left (f-f_0\right )=0\,.
\end{equation}
We should note that we also performed evolutions using different sets of boundary conditions, to test the independence of the results on boundary conditions imposed in a region with no causal contact with the interaction region. As far as the behaviour and location of the horizons and all quantities discussed in this paper are concerned, no noticeable difference could be found.

Our numerical simulations use the {\sc Lean} code~\cite{Sperhake:2006cy}, see section~\ref{sec:lean-code}.
The calculation of Black hole Apparent Horizons (BAHs) and Cosmological Apparent Horizons (CAHs) is performed with {\sc AHFinderDirect}~\cite{Thornburg:1995cp, Thornburg:2003sf}. We remark that BAHs, found as marginally trapped surfaces, indicate in de Sitter space (with the same legitimacy as in asymptotically flat space) the existence of an event horizon  \cite{Shiromizu:1993mt}. CAHs are surfaces of zero expansion for  \textit{ingoing} null geodesics. In a single black hole case, in McVittie coordinates, the black hole event horizon and cosmological horizon are indeed foliated by apparent horizons.

The ``expanding'' behaviour of the coordinate system led us to add a new innermost refinement level at periodic time intervals so as to keep the number of points inside the cosmological horizon approximately unchanged.
The necessity for adding extra refinement levels effectively limits our ability to follow the evolution on very long timescales, as the number of time steps to cover a fixed portion of physical time grows exponentially. This feature resembles in many ways the recently reported work by Pretorius and Lehner on the follow-up of the black string instability~\cite{Lehner:2010pn}.

\subsection{Numerical Results}

\begin{figure}[htbp]
\centering
\includegraphics[width=0.8\textwidth,clip=true]{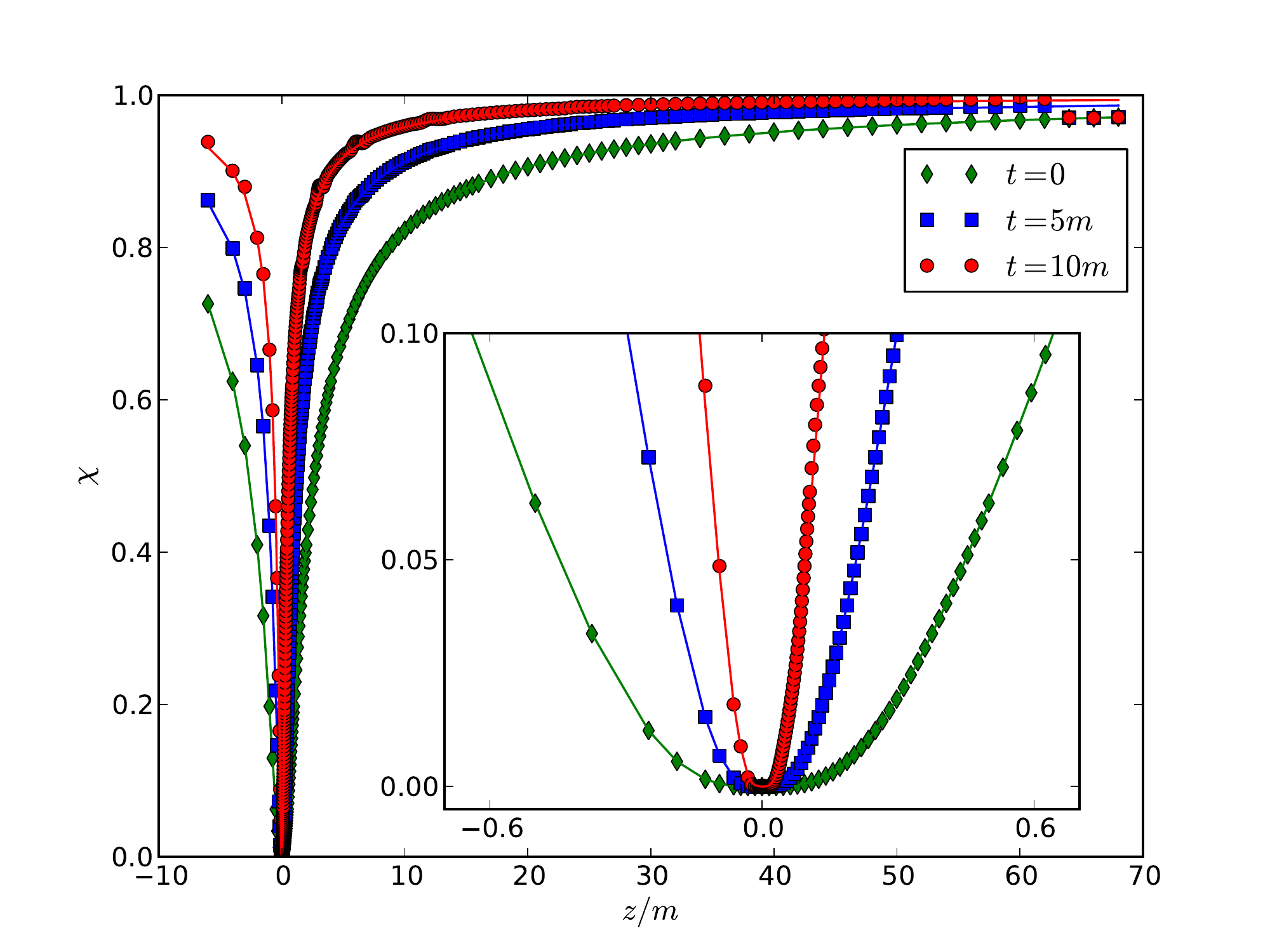}
\caption[Conformal factor for a single black hole evolution with $H=0.8H_{\rm crit}$]{
Conformal factor $\chi$ for a single black hole evolution with $H=0.8H_{\rm crit}$
using the McVittie slicing condition, equation~\eqref{eq:mcvittie_slicing}. The
obtained numerical results are plotted, along the $z$ coordinate
(symmetry $\chi(-z)=\chi(z)$ imposed at $z=0$),
against the expected analytical solutions (solid lines). }
\label{fig:mcvittie}
\end{figure}

As a first test of the numerical implementation, we performed evolutions of a single black hole imposing the McVittie slicing condition; that is, we use~\eqref{eq:mcvittie} as initial data and impose
\begin{equation}
\label{eq:mcvittie_slicing}
\begin{aligned}
\partial_t \alpha &  = {4 m r H e^{Ht}}/{(m + 2 r e^{Ht})^{2}} \,, \qquad 
\partial_t\beta^i & = 0 \, , 
\end{aligned}
\end{equation}
throughout the evolution. The analytical solution \eqref{eq:mcvittie} can be compared with the numerical results. 
For a single black hole evolution with $m=1$ and $H=0.8H_{\rm crit}$, the results are displayed in figure~\ref{fig:mcvittie}. 
Using this slicing, the runs eventually crash (at $t \sim 12 m$). By contrast, the standard ``1+log'' slicing condition~(\ref{eq:1+log_0})
\begin{equation}
\label{eq:1+log}
\partial_t \alpha = \beta^i \partial_i \alpha - 2\alpha \left(
  K - K_0
\right) \,,
\end{equation}
where $K_0 = -3H = -\sqrt{3\Lambda} $, enables us to have long term stable evolutions. As consistency checks, the areal radii at the apparent horizons (both black hole horizon and cosmological horizon) are constants in time and have the value expected from the analytical solution in a single black hole spacetime. Moreover, the areal radius at fixed coordinate radius evolves with time in the way expected from the exact solution.

\begin{figure}[tbhp]
  \centerline{\includegraphics[width=0.8\textwidth,clip=true]{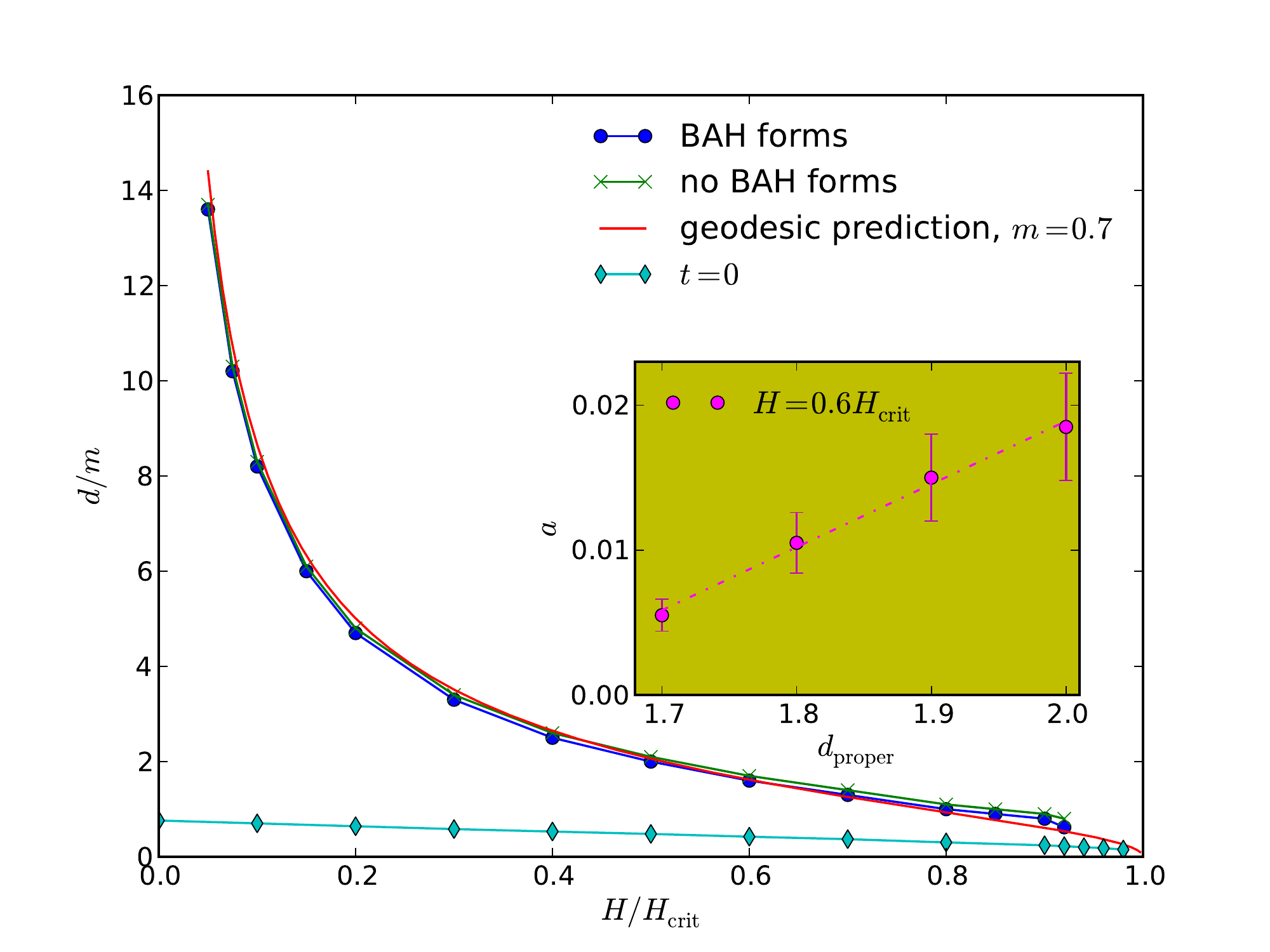}}
  \caption[Critical coordinate distance for small mass
  binaries]{Critical coordinate distance for small mass
    binaries, from both initial data and dynamical evolutions, as well as a
    point particle estimate, as a function of $H/H_{\rm crit}$. We obtain this
    estimate from the coordinate distance to the horizon,
    equation~\eqref{eq:rcrit_geo}, for a particular value of $m$. The $t=0$ line
    refers to the critical separation between having or not having a common BAH
    in the initial data. The inset shows details of the approach to the critical
    line for $H=0.6 H_{\rm crit}$, where $a$ is an acceleration parameter. 
    \label{fig:crit_sep_P0}}
\end{figure}

For binary black hole initial data, we start by reproducing the results of Nakao et.~al~\cite{Nakao:1992zc}, where the critical distance between two black holes for the existence of a common BAH already at $t=0$ was studied. We thus prepare initial data \eqref{eq:psigen} with $m_1 = m_2$ and take all quantities in units of the total mass $m = m_1 + m_2$. The two punctures are set initially at symmetric positions along the $z$ axis. The critical value for the cosmological constant, for which the black hole and cosmological horizon coincide is now $mH_{\rm crit} = {1}/{\sqrt{27}}$. We call \textit{small (large) mass binaries} those, for which $H<H_{\rm crit}$ ($H>H_{\rm crit}$). Our results for the critical separation in small mass binaries, at $t=0$, as function of the Hubble parameter are shown in figure~\ref{fig:crit_sep_P0}. The line (diamond symbols) agrees, after a necessary normalisation, with figure~14 of~\cite{Nakao:1992zc}. 

\begin{figure}[tbhp]
  \centering
  \includegraphics[width=0.8\textwidth,clip=true]{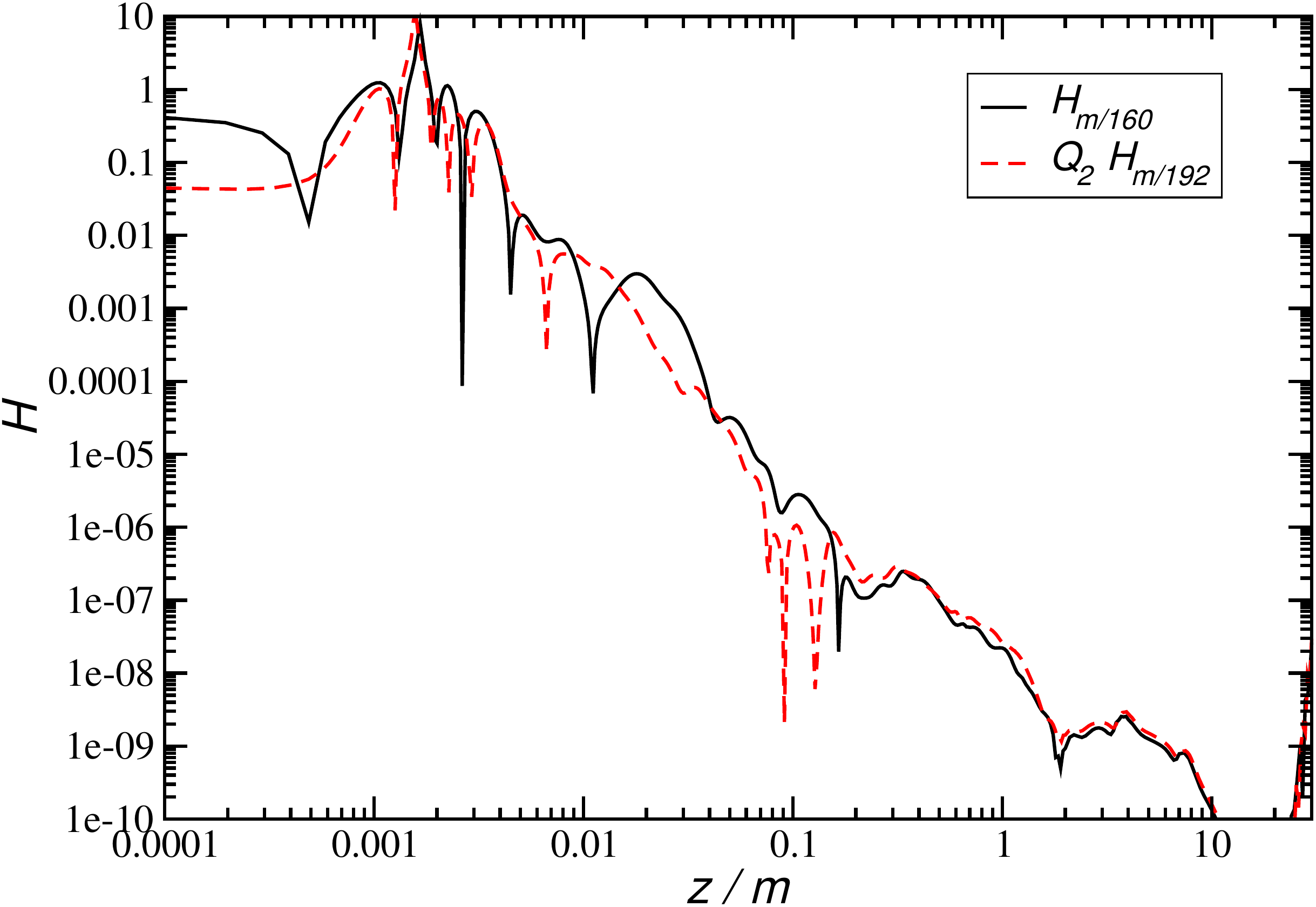}
\caption[Hamiltonian constraint violation]{
  Hamiltonian constraint violation along the $z$-axis at time $t=48m$ for a
  simulation with $H=0.9H_{\rm crit}$ and initial distance $d=0.8m$. 
  \label{fig:ham-constraint}}
\end{figure}

We now consider head-on collisions of two black holes with no initial momentum, i.e.~the time evolution of these data. We have monitored the
Hamiltonian constraint violation level for cases with and without
cosmological constant.  We observe that the constraint violations are
comparable in the two cases and plot in figure~\ref{fig:ham-constraint}
a snapshot of the Hamiltonian constraint violation at $t=48m$
for parameters $H=0.9H_{\rm crit}$
and $d=0.8m$, a typical case with non-zero
cosmological constant. We have used two resolutions, $m/160$ and
$m/192$ (on the innermost refinement level) and have rescaled the
dashed curve by $Q_2=(192/160)^2$ as expected for second-order
convergence.
%

For subcritical Hubble constant $H<H_{\rm crit} = {1}/{(\sqrt{27}m)}$, we monitor the evolution of the areal radius of the BAHs and that of the CAH of an observer at $z=0$. For instance, for $H=0.9H_{\rm crit}$ and proper (initial) separation $3.69m$ 
we find that the areal radii of the BAH and CAH are approximately constant and equal to $R_{\rm BAH}\simeq 2.36m$ and $R_{\rm CAH}\simeq 4.16m$, respectively. As expected the two initial BAHs, as well as the final horizon, are inside the CAH. As a comparison, a Schwarzschild-de Sitter spacetime with the same $H$ has $R_{\rm BAH}\simeq 2.43m$ and $R_{\rm CAH}\simeq 4.16m$. This suggests that the interaction effects (binding energy and emission of gravitational radiation) are of the order of a few per cent for this configuration.

\begin{figure}[tbhp]
  \centering
  \includegraphics[width=0.8\textwidth]{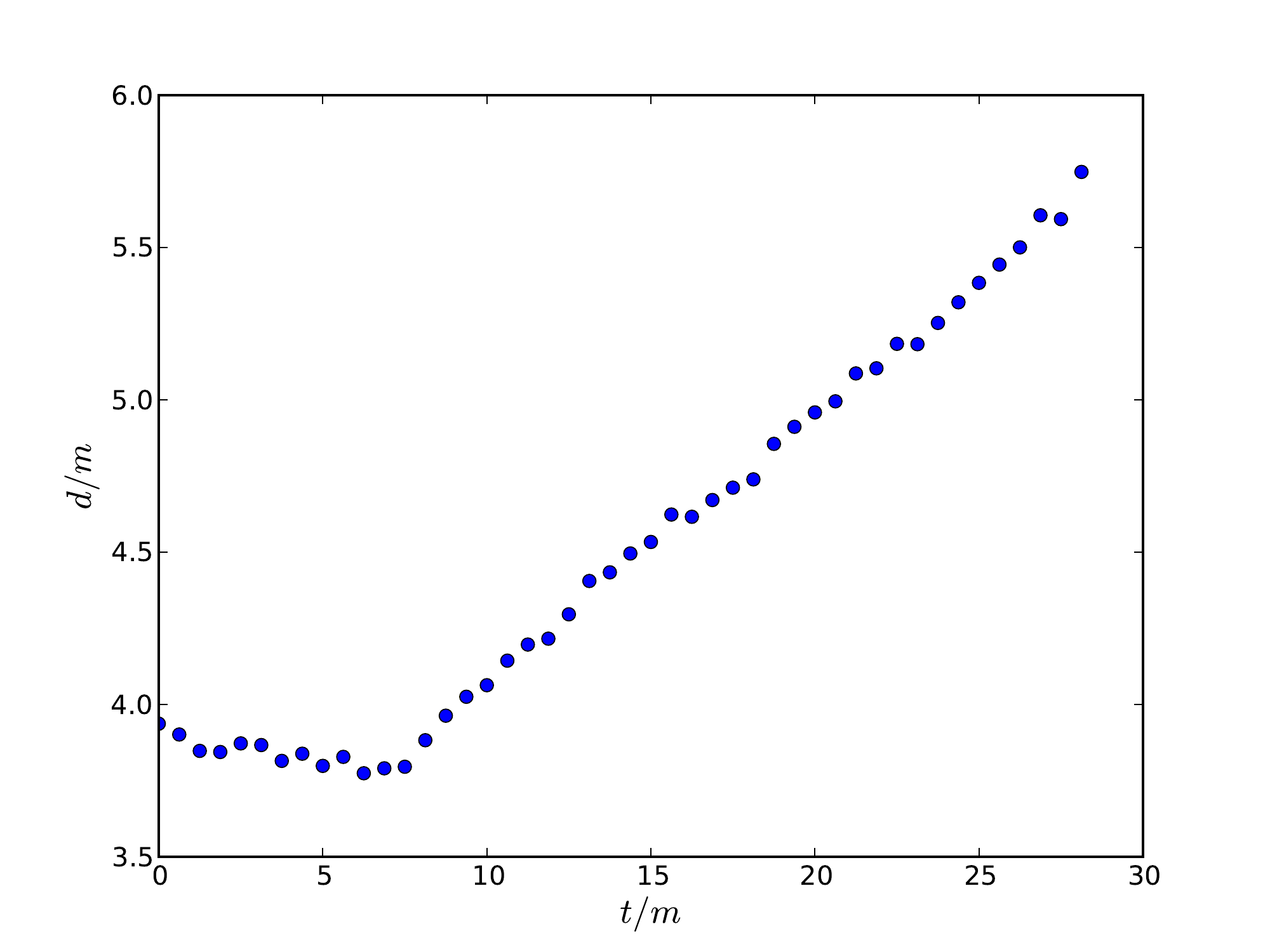}
\caption[Proper distance between the black hole horizons as a function of
time]{Proper distance between the black hole horizons as a function of
time for the $H=0.9H_{\rm crit}$, and initial (coordinate) distance
$d\simeq 0.9m$. The two holes stay at approximately constant distance
up to $t\approx 8m$ after which cosmological expansion starts
dominating.
\label{fig:proper-dist} }
\end{figure}

As the initial separation grows, so does the total time for merger. For
separations larger than a critical value, the two black holes do not merge, but
scatter to infinity. For such scattering configurations, the simulations
eventually exhibit a regime of exponentially increasing proper distance between
the BAH.  Just as in scatters of high energy black holes~\cite{Sperhake:2009jz},
here we find that the immediate merger/scatter regimes are separated by a
blurred region, where the holes sit at an almost fixed proper distance for some
time; cf.~figure~\ref{fig:proper-dist}.  By performing a large set of simulations
for various cosmological parameters $H$ and initial distance $d$, we have
bracketed the critical distance for the merger/scatter region as a function of
the Hubble parameter $H$ for the ``dynamical'' case, i.e., the initial
\emph{coordinate} distance between the black holes such that no common BAH
forms.  The results are displayed in figure~\ref{fig:crit_sep_P0} (circles and
$\times$ symbols).

As expected the critical distance becomes larger as compared to the initial data value (``$t=0$'' line): there are configurations for which a common BAH is absent in the initial data but appears during the evolution (just as in asymptotically flat spacetime). The numerical results can be qualitatively well approximated by a point particle prediction---from equation~\eqref{eq:rcrit_geo}. To do such comparison a transformation to McVittie coordinates needs to be done; we have performed such transformation at McVittie time $t=0$. Intriguingly, for a particular value of $m\simeq 0.7$, the point particle approximation matches quantitatively very well the numerical result; the curve obtained from the
geodesic prediction in figure~\ref{fig:crit_sep_P0} is barely distinguishable
from the numerical results.

A further interesting feature concerns the approach to the critical line. For an initially static binary close to the critical initial separation, the coordinate distance $d$ scales as $d=d_0+a t^2$. In general the acceleration parameter scales as $\log a=C+\Gamma \log(d-d_0)$,
where $\Gamma=1$ in the geodesic approximation. A fit to our numerical results for $H=0.6 H_{\rm crit}$ (dashed curve in the inset of figure~\ref{fig:crit_sep_P0})
for example yields $C=-3.1,\,\Gamma=0.9$ in rough agreement with this expectation. Details of this regime are given in the inset of figure~\ref{fig:crit_sep_P0}.

\begin{figure}[tbhp]
  \centering
  \includegraphics[width=1.0\textwidth,clip=true]{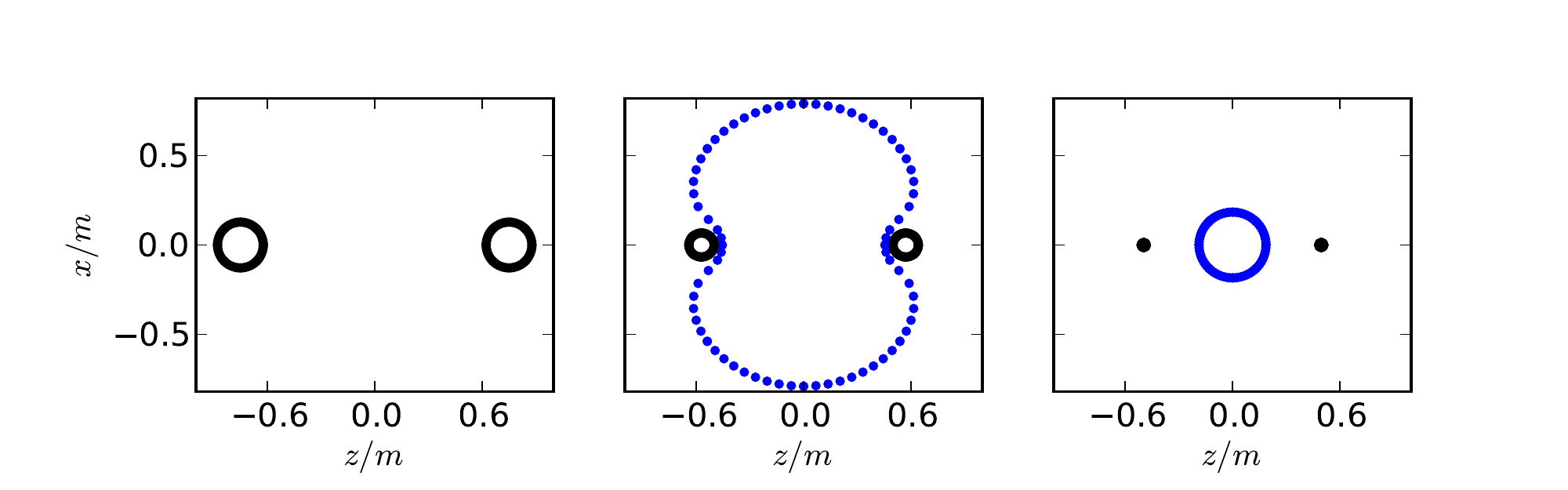}
  \caption[Snapshots at different times of a simulation with $H= 1.05 H_{\rm
    crit}$]{Snapshots at different times (from left to right $t/m=0.0, 8.0156,
    20.016$) of a simulation with $H= 1.05 H_{\rm crit}$, and an initial
    coordinate distance $d/m=1.5002$.  The dotted blue line denotes the CAH (for
    an observer at $z=0$) which is first seen in this simulation at
    $t/m=8.0156$, highly distorted.  At late times, the CAH has an areal radius
    of $R = 4.94876$, while the ``theoretical value'' for pure dS is
    $R=1/H=4.94872$, a remarkable agreement showing that the spacetime is accurately
    empty dS for the observer at $z=0$ and the black holes are not in causal
    contact.  }
  \label{horizon_shape}
\end{figure}

Finally, we have performed evolutions with $H>H_{\rm crit}$. On the assumption of weak gravitational wave release, such evolutions can test the cosmic censorship conjecture since the observation of a merger in such case would reveal a violation of the conjecture~\cite{Hayward:1993tt}. From general arguments and from the simulations with $H<H_{\rm crit}$, we know the cosmological repulsion will dominate for sufficiently large initial distance and in that case we can even expect that a CAH for the observer at $z=0$ will not encompass the BAHs. This indicates the black holes are no longer in causal contact and therefore can never merge. Our numerical results confirm this overall picture. To test the potentially dangerous configurations, we focus on the regime in which the black holes are initially very close. A typical example is depicted in figure~\ref{horizon_shape}, for a supercritical cosmological constant $H= 1.05 H_{\rm crit}$, and an initial coordinate distance $d/m=1.5002$. Even though the initial separation is very small, we find that the holes move {\it away} from each other, with a proper separation increasing as the simulation progresses. In fact, further into the evolution, a distorted CAH appears, and remains for as long as the simulation lasts. At late times, this CAH is spherically symmetric, and has an areal radius which agrees, to within $10^{-5}$, with that of an empty de Sitter spacetime with the same cosmological constant.
The evolution therefore indicates that the spacetime becomes, to an excellent approximation, empty de Sitter space for the observer at $z=0$ and that the black holes are not in causal contact. 
Observe that qualitatively similar evolutions can be found in small mass binaries
when the initial distance is larger than the critical value
%

\subsection{Final Remarks}

We have presented evidence that the numerical evolution of black hole spacetimes
in de Sitter universes is under control.  Our results open the door to new
studies of strong field gravity in cosmologically interesting scenarios. In
closing, we would like to mention that our results are compatible with cosmic
censorship in cosmological backgrounds. However, an analytic solution with
multiple (charged and extremal) black holes in asymptotically de Sitter
spacetime is known, and has been used to study cosmic censorship violations
\cite{Brill:1993tm}. In \textit{collapsing} universes a potential violation of
the conjecture has been reported, although the conclusion relied on singular
initial data. To clarify this issue, it would be of great interest to perform
numerical evolution of large mass black hole binaries, analogous to those
performed herein, but in collapsing universes. This will require adaptations of
our setup, since the ``expanding'' behaviour discussed of the coordinate system
will turn into a ``collapsing'' one, which raises new numerical challenges.



\section{Black holes in a box}
\label{sec:BHbox}
  
\emph{Anti-de Sitter (AdS)} is a non-globally hyperbolic spacetime, which essentially means that it is not enough to prescribe a set of evolution equations and some initial configuration in order to predict what will happen in the future.
On such spacetimes, the boundary plays an \emph{active role}, and in order to have a well-defined Cauchy problem the initial data (and evolution equations) must be supplemented by appropriate boundary conditions at the time-like conformal boundary.

In this section, we will give a brief overview of the work presented in~\cite{Witek:2010qc} where a ``toy model'' for Anti-de Sitter was considered by imprisoning a black hole binary in a box with mirror-like boundary conditions and thus exploring the active role that boundary conditions play in the evolution of a bulk black hole system.

\subsection{Numerical setup}

The vacuum Einstein equation are written in the BSSN scheme~(\ref{eq:bssn-gen}), introduced in section~\ref{sec:bssn}, with the gauge conditions~(\ref{eq:1+log_0}), (\ref{eq:gamma-driver-0}).
We evolve these equations with the {\sc Lean} code~\cite{Sperhake:2006cy}, see section~\ref{sec:lean-code}.

\begin{figure}[tbh]
  \centering
  \includegraphics[width=0.7\textwidth]{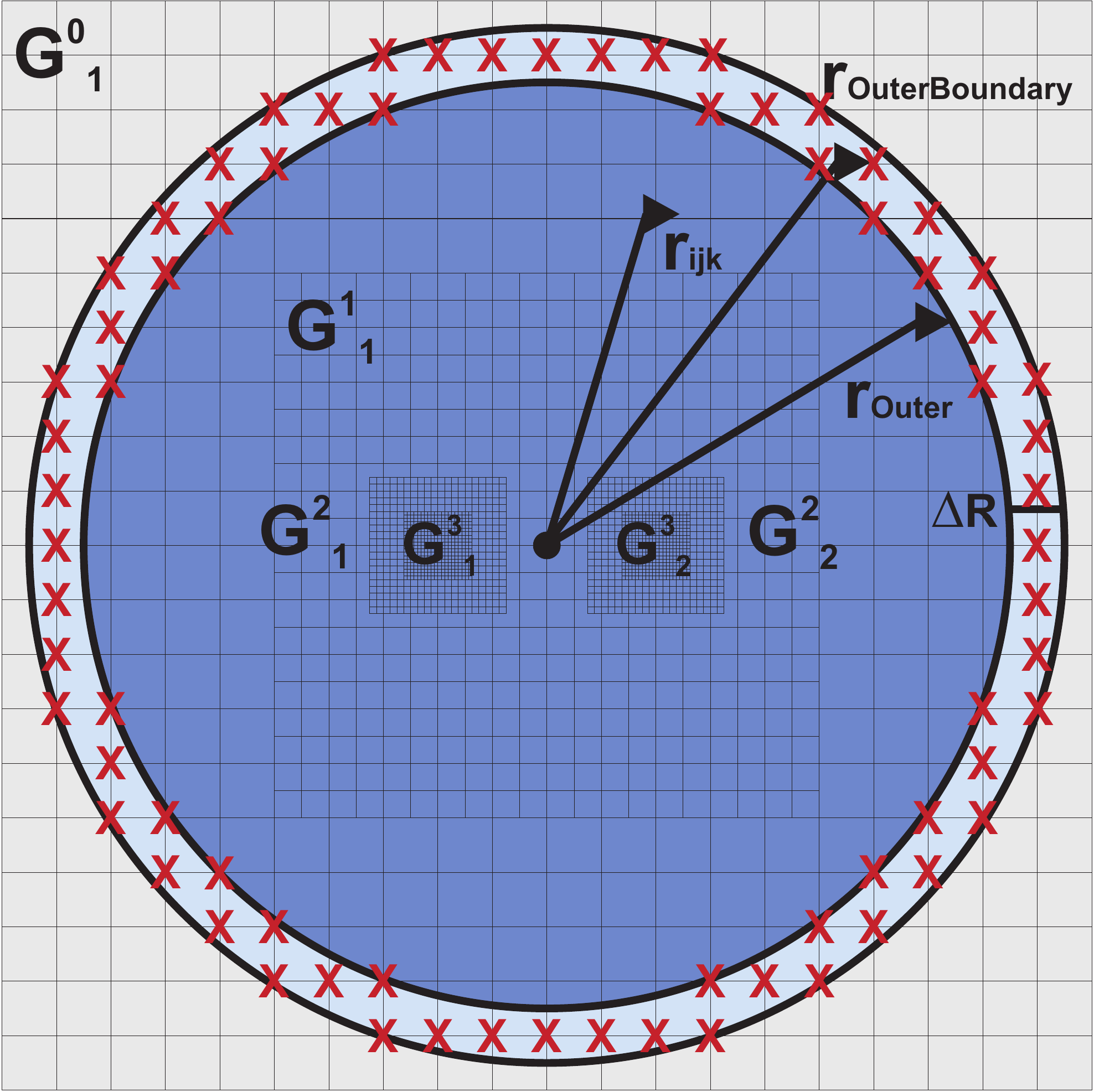}
  \caption{Illustration of a (Lego-)spherical outer boundary. \label{fig:SphBD}}
\end{figure}

The work here presented differs from previous implementations of the \textsc{Lean} code (and most other codes) in the treatment of the outer boundary conditions, which is herein considered to be a reflecting sphere or, rather, an approximation of it by using so-called {\em Lego} spheres; cf.~section~3 in~\cite{Shoemaker:2003td}.
The numerical implementation of such boundary conditions 
is schematically illustrated in figure~\ref{fig:SphBD}. 

Points outside the outer circle of radius $R_B+\Delta R$ are not required for
updating regular points and are simply ignored in the numerical evolution. In
practice, we ensure that the boundary shell is always of sufficient thickness to
accommodate discretization stencils required for the update of regular
gridpoints. The specific boundary condition is then determined by the manner in
which we update grid functions on the boundary points marked as $\times$ in the
figure.

To mimic the global structure of an AdS spacetime we thus enclose the black hole binary inside the spherical mirror and set
\begin{equation}
\label{eq:refbc}
\frac{\partial}{\partial t} f = 0,
\end{equation}
at each boundary point with $f$ denoting any of the BSSN variables listed in equations~(\ref{eq:bssn-gen}).
The final ingredient needed for our numerical implementation is related with the spurious radiation present when evolving black hole binaries---often called \emph{junk radiation}---which can be traced back to the methods used to compute the initial data.
To avoid contamination of our simulations by such spurious radiation
being trapped inside our reflective boundary we employ standard outgoing
radiation boundary conditions at early times and only switch on our reflective
condition at
\begin{equation}
  t_{\rm ref} = R_B +\Delta t_{\rm pulse},
\end{equation}
where we estimate the duration of the spurious wave pulse $\Delta t_{\rm pulse}$ from
previous simulations of similar setups in asymptotically flat spacetimes as for
example presented in~\cite{Baker:2006yw,Sperhake:2006cy,Berti:2007fi}.
The spurious radiation is thus given sufficient time to leave the computational
domain.

Wave extraction is employed in the fashion outlined in section~\ref{sec:NP-gen}, where we will herein also measure the $\Psi_0$ Weyl scalar, which encodes the incoming gravitational wave signal.

The results we will report in the following sections all refer to an inspiral simulation with total mass $M=M_1+M_2$, where the black hole punctures were set with an initial coordinate distance of $d = 6.517 M$ and with Bowen-York momentum parameter $P_i=\pm 0.133 M$.
The grid structure used was
\begin{equation}
  \{(48,24,12,6)\times (1.5, 0.75),~h=1/56\} \,,
\end{equation}
and the Weyl scalars have been extracted at $r_{\rm ex} = 35M$.

\subsection{Gravitational wave signal}
\label{sec:waveforms-box}

The nature of our specific configuration is ideal to study both the outgoing ($\Psi_4$) as well as the ingoing ($\Psi_0$) gravitational wave pulses.

\begin{figure}[tbhp]
  \centering
  \includegraphics[clip=true,width=0.8\textwidth]{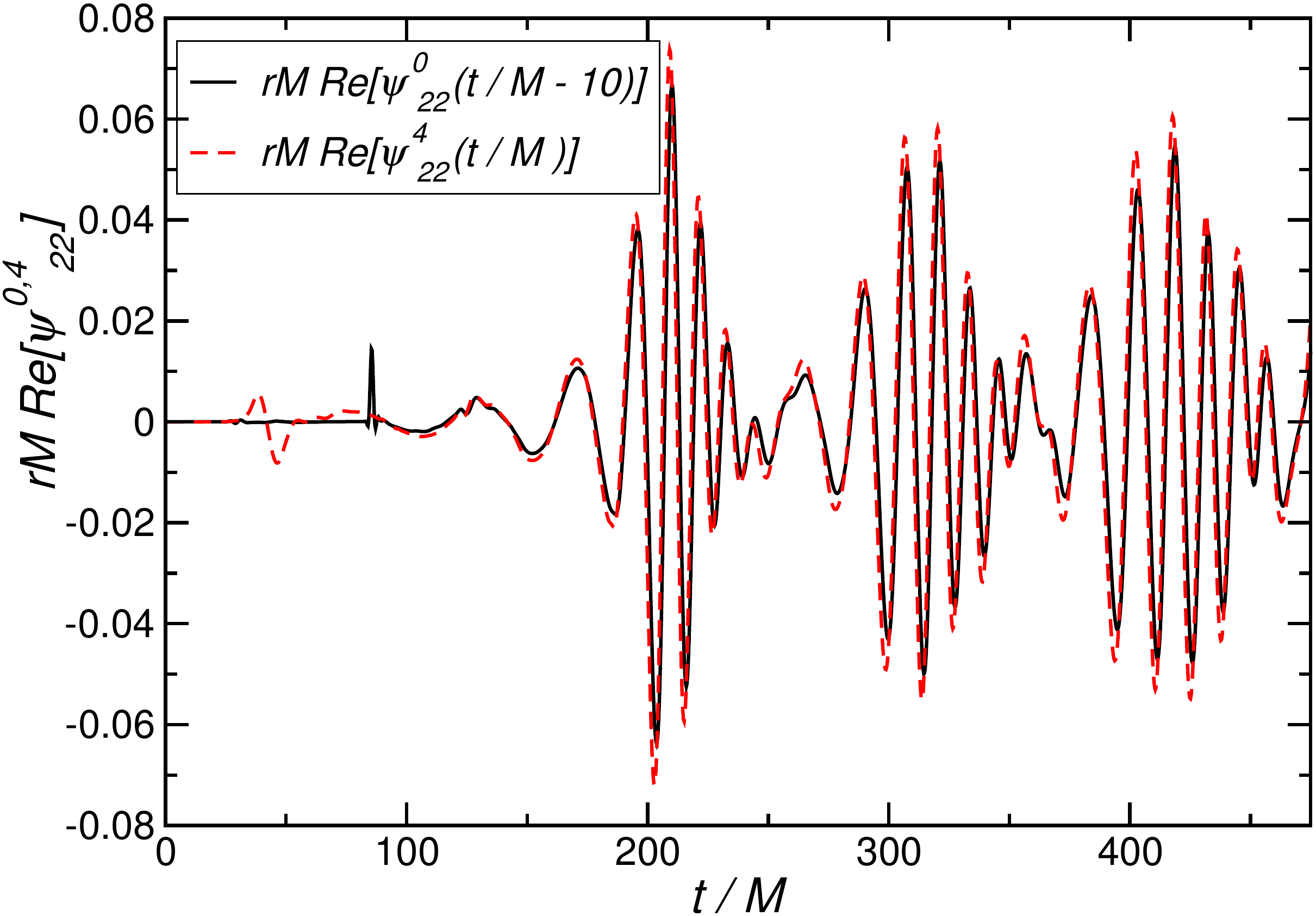}
  \caption[Real part of the $l=m=2$ mode of $rM\Psi_0$ and $rM\Psi_4$]{Real part
    of the $l=m=2$ mode of $rM\Psi_0$ and $rM\Psi_4$. The ingoing signal
    $rM\Psi_0$ has been shifted in time by $\Delta t = 10M$ and in phase by
    $\pi$ (thus equivalent to an extra minus sign) to account for the additional
    propagation time and the reflection.
    \label{fig:comparePsi0Psi4_l2m2} }
\end{figure}

The gravitational wave signal is dominated by the quadrupole contributions which
is shown in figure~\ref{fig:comparePsi0Psi4_l2m2}. The ingoing
signal $\psi^0_{22}$ has been shifted in time by $\Delta t=10~M$ to
compensate for the additional propagation time from the extraction radius
$r_{\rm ex}=35~M$ to the boundary $R_B=40~M$ and back after reflection. The
reflection introduces an additional phase shift of $\Delta \phi = \pi$ which has
also been taken into account in the figure. Within numerical errors, we find the
resulting outgoing and subsequent ingoing pulses to overlap.

\subsection{Interaction of the wave pulse with the remnant black hole}

\begin{figure}[tbhp]
  \centering
  \includegraphics[width=0.8\textwidth]{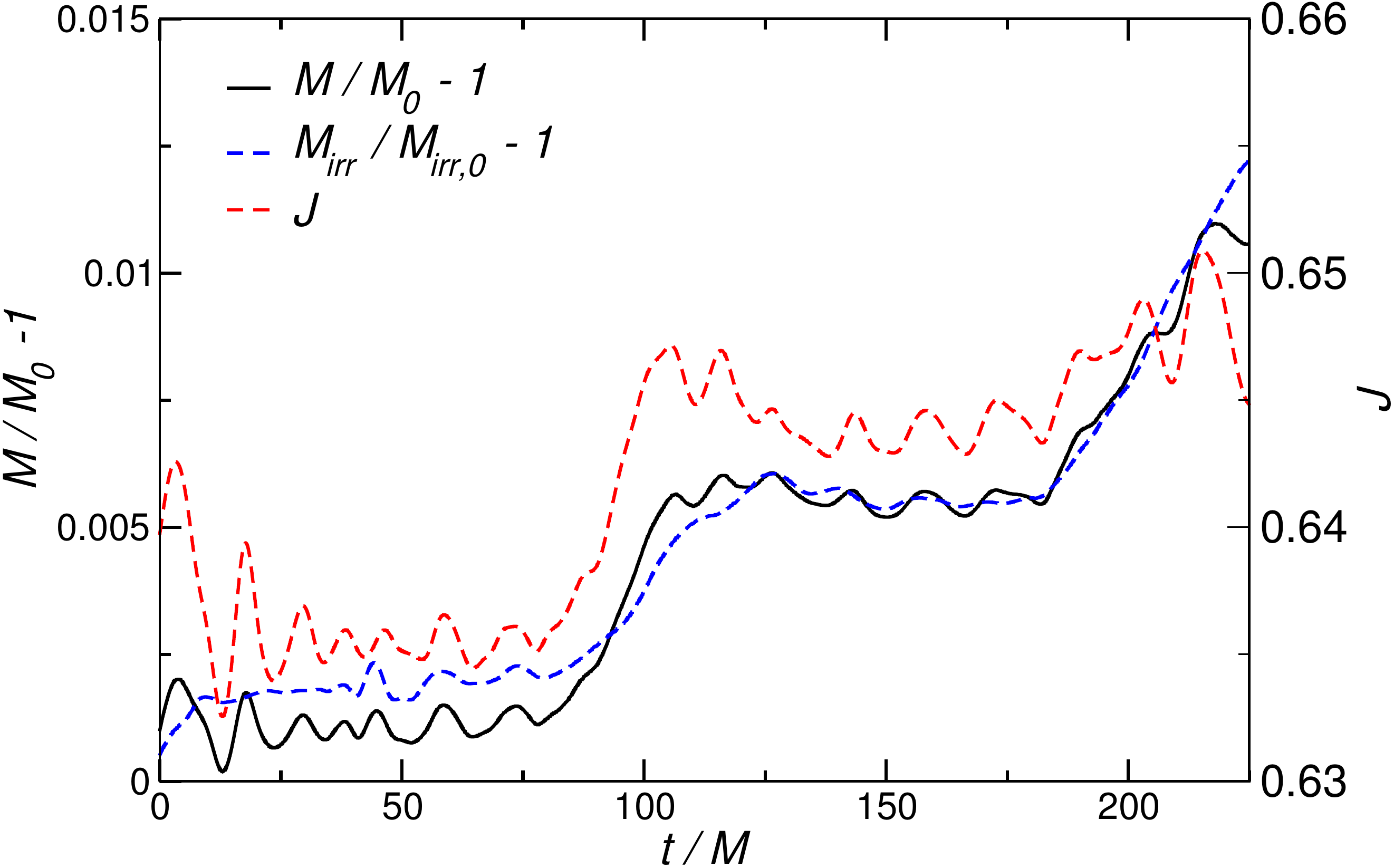}
  \caption[Time evolution of the (relative) mass of the black hole]{Time
    evolution of the (relative) mass of the black hole (solid) computed by $M =
    C_e/4\pi$, the irreducible mass (long dashed) and the total spin $J = j M^2$
    (dashed curve). \label{fig:AHmass}}
\end{figure}

We define black hole mass in terms of the equatorial radius of the horizon
$C_e$ by~\cite{Kiuchi:2009jt}
\begin{equation}
  M=\frac{C_e}{4\pi} \, .
\end{equation}
In figure~\ref{fig:AHmass} it is shown the fractional deviation $(M-M_0)/M_0$ of
the mass of the final black hole from its value immediately after merger
together with the irreducible mass and the black hole spin $J$. 
The mass remains approximately constant until the pulse returns
after its first reflection, then increases, remains constant during the second
passage of the pulse and so on. In contrast, the spin only shows a significant
increase during the first scattering of the pulse off the black hole.


Comparing the increase in the horizon mass with the amount of gravitational wave
energy radiated during the last stages of the inspiral, plunge and merger of a
corresponding binary system in an asymptotically flat spacetime---which is about
$3.5\%$ of the total energy of the
system~\cite{Berti:2007fi,Sperhake:2006cy}---we estimate that about $15\%$ of
the energy emitted during the merger is absorbed by the central spinning black
hole per interaction.

\subsection{Final remarks}

In this section, we have given just a brief overview of the work presented
in~\cite{Witek:2010qc} where the global structure of an AdS background was
mimicked by introducing a reflecting wall at a finite radius. Inside this cavity
a black hole inspiral was evolved.

The results presented are consistent with the intuitive expectations for a
wavepacket of radiation (generated during inspiral plus merger) travelling back
and forth between the mirror-like wall and the black hole: part of this
radiation is absorbed when interacting with the black hole (especially
high-frequencies). We estimate that about $15\%$ of the wavepacket's energy is
absorbed by the black hole per interaction, at least during the first cycles.

It would be extremely interesting to extend this work to implement the evolution
of black holes in real AdS backgrounds, following the recent works
in~\cite{Chesler:2009cy,Chesler:2010bi,Bantilan:2012vu}.


\section{Black holes in cylinders}
\label{sec:cylinders}

From the gauge/gravity duality to braneworld scenarios, black holes in
compactified spacetimes play an important role in fundamental physics. Our
current understanding of black hole solutions and their dynamics in such
spacetimes is rather poor because analytical tools are capable of handling a
limited class of idealised scenarios, only.

In this section, following~\cite{Zilhao:2011zz}, we wish to study how the compactness of extra dimensions changes
the dynamics of such higher-dimensional gravity scenarios. There is considerable
literature on Kaluza-Klein black holes and black holes on
cylinders~\cite{Korotkin:1994dw,Frolov:2003kd,Myers:1986rx,Harmark:2005pp}; the
full non-linear dynamics of black holes on such spacetimes, however, seems to
remain unexplored.

\subsection{Setup}
\begin{figure}[tbhp]
  \centering
  \includegraphics[width=0.8\textwidth]{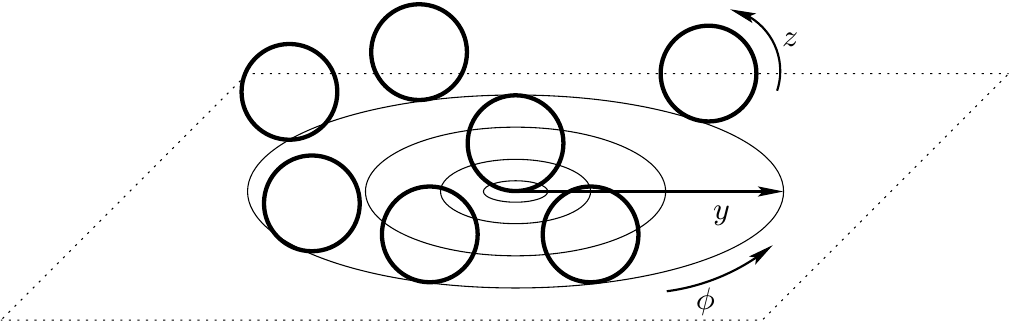}
  \caption[Illustration of the coordinate system for the Minkowski space-time
  ${\mathbb M}^{1,3}\times S^1$]{Illustration of the coordinate system for the
    Minkowski spacetime ${\mathbb M}^{1,3}\times S^1$. A slice with $t={\rm
      constant}$ and $x={\rm constant}$ is shown. $y,\phi$ parameterise a plane,
    wherein $y$ is a radial direction and $\phi$ an azimuthal coordinate. At
    each point in this plane there is a non-contractible circle parameterised by
    $z$. This is illustrated by exhibiting this circle on various points along
    an orbit of $\partial/\partial \phi$ and also at $y=0$. Space-time is a
    (trivial) $S^1$ bundle over ${\mathbb M}^{1,3}$. \label{fig:coordinates}}
\end{figure}

We are interested in describing the evolution of black holes in a five
dimensional spacetime with one periodic direction.  For a five dimensional
cylindrical Minkowski spacetime, ${\mathbb M}^{1,3}\times S^1$, the metric can
be written as
\begin{equation}
  \dd s^2 = \underbrace{-\dd t^2 + \dd x^2  + \dd y^2 + y^2 \dd \phi^2}_{\mathbb{M}^{1,3}} 
  + \underbrace{\dd z^2}_{{S}^1} \, .
\end{equation}
The $S^1$ direction is parameterised by $z$, which takes values in the interval $[-L, L]$, with the two endpoints identified and $L\in \mathbb{R}^+$. 
The coordinate $\phi$ also parameterises a circle, this circle is, however, homotopic to a point, since it shrinks down to zero size at $y=0$, where $y$ is a radial coordinate in the $y-\phi$ plane
which is part of  ${\mathbb M}^{1,3}$---figure~\ref{fig:coordinates}. 

Following the approach outlined in section~\ref{sec:dim-red_split}, we take our five dimensional metric ansatz to be
\begin{equation}
\dd s^2=g_{\mu\nu}\dd x^{\mu}\dd x^\nu+\lambda(x^\mu)\dd \phi^2 \ ,
\end{equation}
where $x^{\mu}=(t,x,y,z)$. 
We perform a dimensional reduction by isometry on $\partial_{\phi}$ 
and end up with a four dimensional model of gravity coupled to a scalar field. 
Performing the standard $3+1$ decomposition and writing the equations in the BSSN scheme, the evolution equations of the resulting system are those of~(\ref{eq:bssn-gen}) with matter terms given by~(\ref{matterterms}).
We here use periodic boundary conditions along the $z$ direction and Sommerfeld radiative boundary conditions along $x$ and $y$.

\subsection{Initial data}
Following the approach of section~\ref{sec:init_data-hd}, 
the four-dimensional Brill-Lindquist initial data appropriate to describe non-spinning, non-rotating black holes momentarily at rest, take the form
\[
\gamma_{ij}\dd x^i \dd x^j=\psi^2[\dd x^2+\dd y^2+\dd z^2] \ , \qquad \lambda=y^2\psi^2 \ , \qquad K_{ij}=0=K_\lambda \ . 
\]
In a spacetime with standard topology (wherein $z$ parameterises a line), the initial data for two black holes with horizon radius $r_S^{1,2}$ and punctures placed at $(x,y,z)=(0,0,\pm a)$, takes 
the form
\begin{equation}
\psi = 1 + \frac{(r_S^1)^2}{4[x^2+y^2+(z-a)^2]} + \frac{(r_S^2)^2}{4[x^2+y^2+(z+a)^2]}
\ . \label{psi_initial}
\end{equation}

\begin{figure}[tbhp]
\centering
\includegraphics[width=0.7\textwidth]{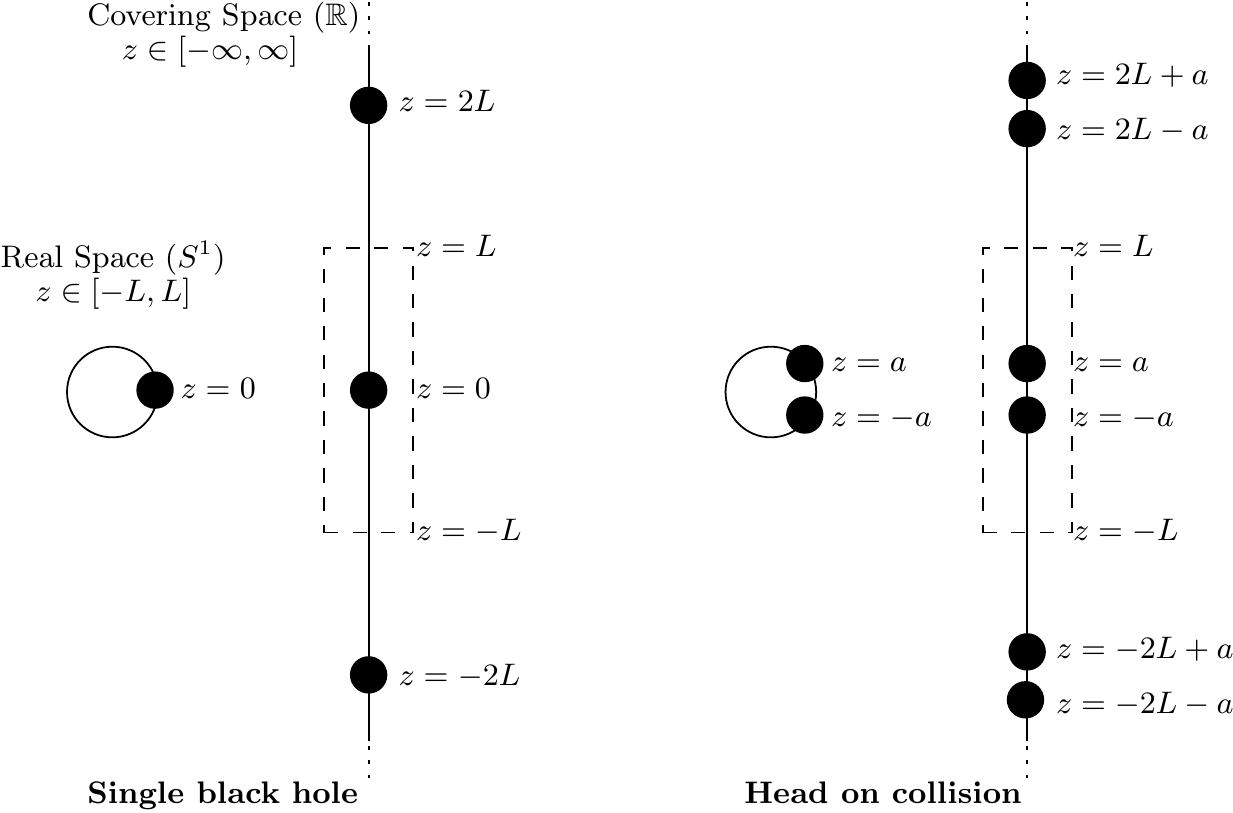}
\caption[Illustration of the correspondence between real space and covering
space for a single black hole and a situation of head on collision]{Illustration
  of the correspondence between real space and covering space for a single black
  hole (left panel) and a situation of head on collision (right panel). The
  dashed boxes drawn in the covering space contain a single copy of the real
  space setup and correspond also to what is contained in the numerical
  grid. \label{fig:array1}}
\end{figure}
The appropriate initial data to describe a black hole in $S^1$ can be viewed as having an infinite array of black holes, all with the same mass, separated by coordinate distance $\Delta z=2L$---figure~\ref{fig:array1}. 
Since the superposition of various black holes in a line is described by adding up the corresponding initial data, for the infinite array of two black holes in the circle located at $z=\pm a$ ($0<a<L$) with horizon radii $r_S^i$, $i=1,2$ (or, equivalently, for two black holes in $S^1$) the initial data is given by 
\begin{align}
\psi & = 1 + \sum_{n=-\infty}^{+\infty}\frac{(r^1_S)^2}{4[x^2+y^2+(z-a-2Ln)^2]}
           + \sum_{n=-\infty}^{+\infty}\frac{(r^2_S)^2}{4[x^2+y^2+(z+a-2Ln)^2]} \notag \\
     & = 1 + \frac{\pi (r^1_S)^2}{8L\rho}
             \frac{\sinh{\frac{\pi \rho}{L}}}
                  {\cosh{\frac{\pi\rho}{L}} - \cos{\frac{\pi (z-a)}{L}}}
           + \frac{\pi (r^2_S)^2}{8L\rho}
             \frac{\sinh{\frac{\pi \rho}{L}}}
                  {\cosh{\frac{\pi\rho}{L}}-\cos{\frac{\pi (z+a)}{L}}}  \ .
\label{psi_initial_s1_2}
\end{align}
where $\rho^2\equiv x^2+y^2$ and in the last equality we have used the result in~\cite{Myers:1986rx}.

\subsection{Results}

Again, we use the \textsc{Lean} code, introduced in section~\ref{sec:lean-code}, for the numerical evolutions.
The main difference to standard implementations of \textsc{Lean} is the use of periodic boundary conditions, which is also non-trivial to implement in a parallel code.

We will now show some results obtained for a head-on collision (from rest) of black holes with an initial separation of $10.37~r_S$, i.e., $a=5.185~r_S$. 
The $z$ ($z\in [-L, L]$) coordinate has been compactified with $L / r_S=64,~32,~16$. 
For comparison purposes, we have also performed a simulation with ``standard'' outgoing boundary conditions ($L \to \infty$), which 
will be here referred to as ``outgoing''. 

All results will be presented in units of the Schwarzschild radius $r_S = r_{S,1} + r_{S,2}$.

\subsubsection{Hamiltonian constraint}

\begin{figure}[tbhp]
\centering
\includegraphics[width=0.45\textwidth]{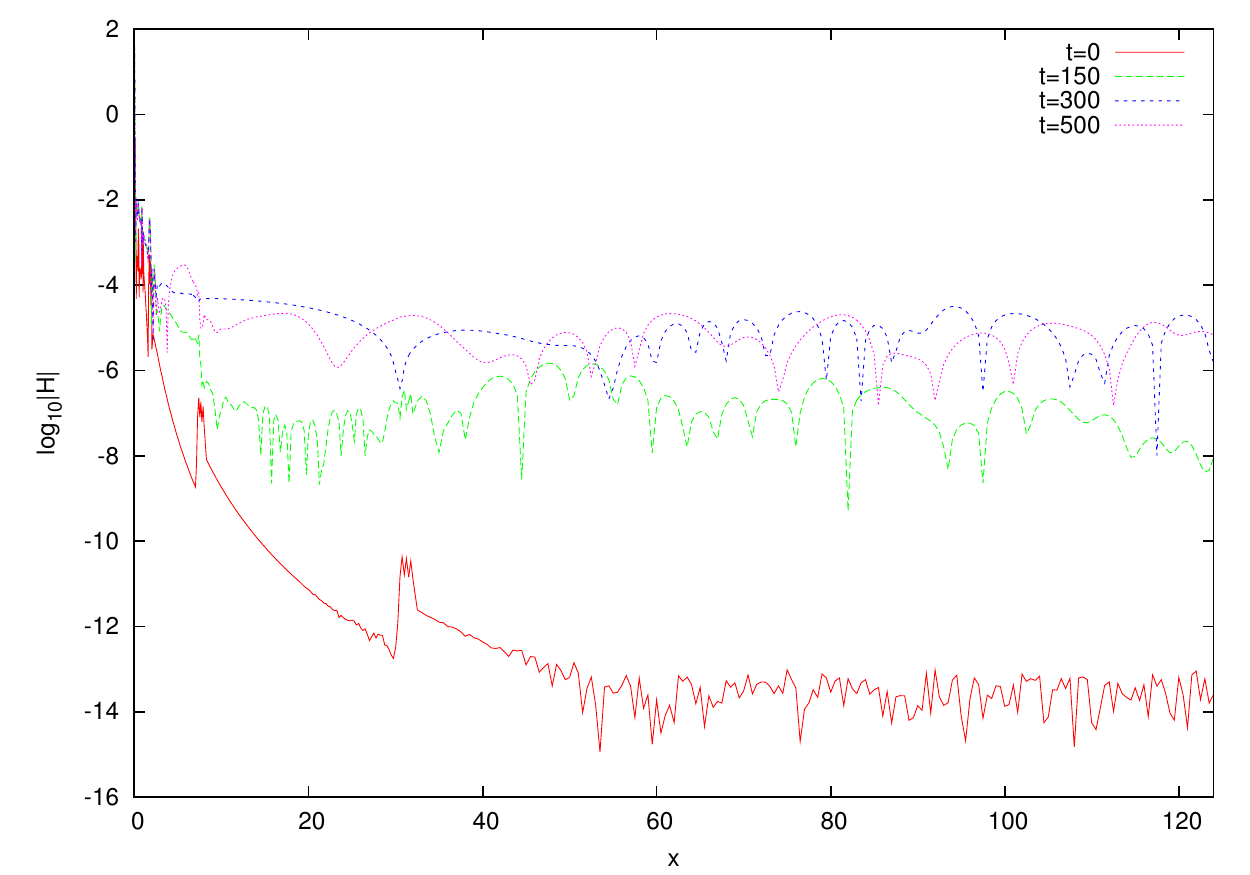}
\includegraphics[width=0.45\textwidth]{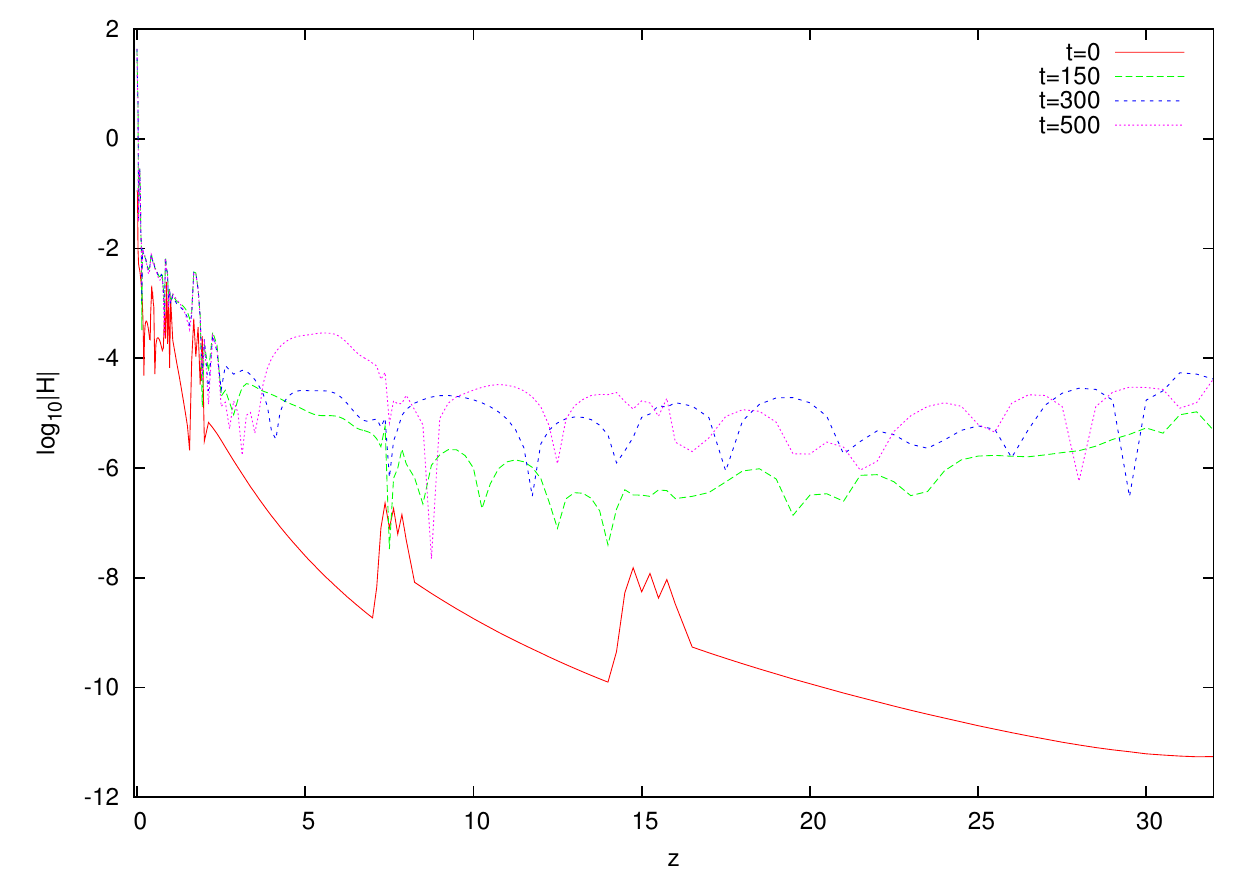}
\caption{Hamiltonian constraint along the $x$ and $z$ axis, for the $L/ r_S=32$ case. \label{fig:hcL32}}
\end{figure}

Figure~\ref{fig:hcL32} shows the Hamiltonian constraint along the $x$ and $z$ axis, respectively, for several time steps for the $L/ r_S=32$ case.
As we can see, the constraint is indeed being satisfied with high accuracy.

\subsubsection{Collision time}
\begin{figure}[tbhp]
  \centering
  \includegraphics[width=0.8\textwidth]{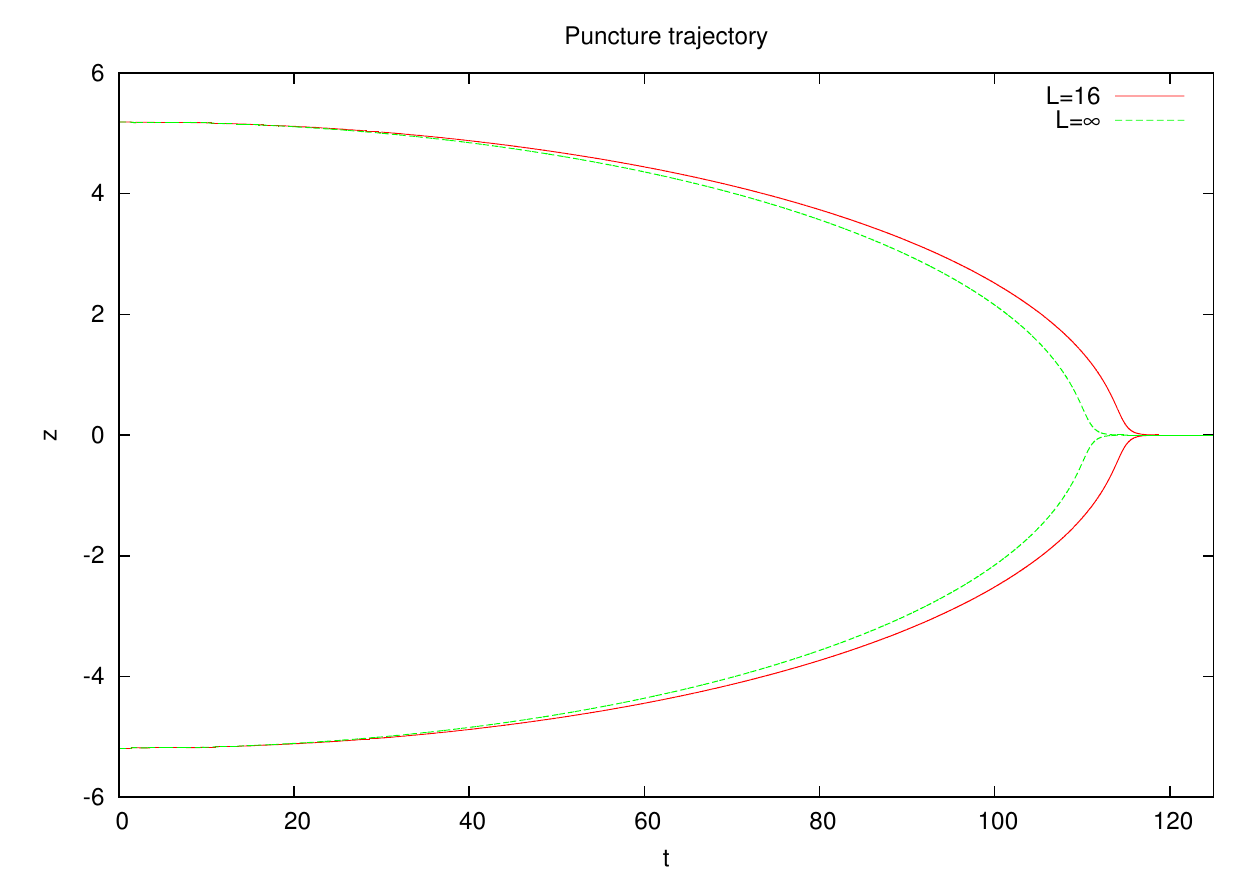}
  \caption{Puncture $z$ coordinate as function of time for the outgoing and $L/ r_S=16$ cases.
  \label{fig:traj}}
\end{figure}

Next, we study the changes in collision time for different compactification radii.
Whereas we do not observe any (noticable) difference for $L/ r_S=64$ and $L/ r_S=32$ as compared to the outgoing case, the case $L/ r_S=16$ shows already a noticable difference.
The puncture trajectories for these cases are plotted in figure~\ref{fig:traj}.

Recall that one can think of a black hole in a cylindrical space as an array of infinite black holes.
Therefore, for a head-on collision on such a cylindrical space, each black hole will also feel the gravitational pull of all the other black holes.
Na\"ively, one thus expects that for this cylindrical case it will take longer 
for the black holes to collide, which is what we observe in figure~\ref{fig:traj}.

\subsection{Final remarks}

Using the formalism introduced in section~\ref{sec:dim-red}, we were able to reduce the head-on collision of (non-spinning) black holes on cylindrical spacetimes (in any dimension) to an
effective $3 + 1$ system with a scalar field, and used this procedure to successfully evolve a head-on collision of two black holes on a five-dimensional cylindrical spacetime.

Further issues that we wish to investigate include monitoring the deformation of the black holes' apparent horizon and computing the energy radiated, along the lines of section~\ref{master}. 
We also further plan to perform simulations with smaller
compactification radii and study the equivalent six-dimensional system.



\chapter{Einstein-Maxwell}
\label{ch:EM}

In this last chapter we go back to four dimensions once again, this time in
Einstein-Maxwell theory, to perform fully non-linear numerical simulations of
charged black hole collisions~\cite{Zilhao:2012gp}.

As mentioned in the Introduction, the dynamics of binary systems of charged,
i.e.~Reissner-Nordstr\"om (RN), black holes have remained rather unexplored
territory. Perhaps this is due to the expectation that astrophysical black
holes carry zero or very small charge; in particular, black holes with mass $M$,
charge $Q$ and angular momentum $aM^2$ are expected to discharge very quickly if
$Q/M \gtrsim
10^{-13}(a/M)^{-1/2}(M/M_{\odot})^{1/2}$~\cite{Gibbons:1975kk,Blandford:1977ds}.
There is, nevertheless, a good deal of motivation for detailed investigations of
the dynamics of charged black holes.

In the context of astrophysics, charged black holes may actually be of interest
in realistic systems. First, a rotating black hole in an external magnetic field
will accrete charged particles up to a given value,
$Q=2B_0J$~\cite{Wald:1974np}. Thus it is conceivable that astrophysical black
holes could have some (albeit rather small) amount of electrical charge. Then it
is of interest to understand the role of this charge in the Blandford-Znajek
mechanism~\cite{Blandford:1977ds}, which has been suggested for extracting spin
energy from the hole, or in a related mechanism capable of extracting energy
from a moving black hole~\cite{Palenzuela:2010nf,Palenzuela:2010xn} to power
outflows from accretion disk-fed black holes. Numerical simulations of charged
black holes interacting with matter and surrounding plasma will enable us to
study such effects.

Motivation for the numerical modelling of charged black holes also arises in the
context of high energy collisions. It is expected that trans-Planckian particle
collisions form black holes; moreover, well above the fundamental Planck scale
such processes should be well described by general relativity and other
interactions should become negligible~\cite{'tHooft:1987rb}, an idea poetically
stated as \textit{matter does not matter} for ultra high energy
collisions~\cite{Choptuik:2009ww}. But is this expectation really correct?
Calculations of shock wave collisions suggest that even though other
interactions---say charge---may become irrelevant in the ultra-relativistic
limit, the properties of the final black hole (and of the associated emission of
gravitational radiation) do depend on the amount of charge carried by the
colliding particles~\cite{Yoshino:2006dp,Gingrich:2006qh}. This issue can be
clarified by the simulation of high-energy collisions of charged black holes and
the subsequent comparison of the results to those obtained for electrically
neutral systems.

Finally we note a variety of conceptual aspects that merit a more detailed
investigation of charged black hole systems.  In head-on collisions with small
velocity, the intuition borrowed from Larmor's formula in Minkowski space
suggests a steady growth of the emitted power with the acceleration. However, it
is by now well established that for uncharged black holes the gravitational
radiation strongly peaks near around time of formation of a common apparent
horizon.  Does the electromagnetic radiation emission follow a similar pattern?
And what is the relative fraction of electromagnetic to gravitational wave
emissions? Moreover, a non-head on collision of charged non-spinning black holes
will allow us to study, as the end state, a (perturbed) Kerr-Newman geometry,
which would be extremely interesting: linearised perturbations around
Kerr-Newman black holes do not decouple~\cite{MTB,Berti:2009kk} and so far close
to nothing is known about their properties. Among others, the stability of the
Kerr-Newman metric is an outstanding open issue.  Furthermore, it has been
observed that the inspiral phase of an orbiting black-hole-binary system can be
well understood via post-Newtonian methods~\cite{Blanchet2006} (see also
e.g.~\cite{Boyle:2007ft, Sperhake:2011zz}).  The additional radiative channel
opened by the presence of electric charge provides additional scope to probe
this observation.

With the above motivations in mind we here initiate the numerical study of
non-linear dynamics of binary systems of charged black holes, building on
previous numerical evolutions of the Einstein-Maxwell
system~\cite{Palenzuela:2008sf,Palenzuela:2009yr,Palenzuela:2009hx,Mosta:2009rr}. For
reasons of simplicity, we focus in this study on binary systems for which
initial data can be constructed by purely analytic
means~\cite{Brill:1963yv,Alcubierre:2009ij}: head-on collisions, starting from
rest, of non-spinning black holes with equal charge-to-mass ratio. This implies
in particular that the black holes carry a charge of the same sign, so the
electromagnetic force will always be repulsive. We extract both gravitational
and electromagnetic radiation and monitor their behaviour as the
charge-to-mass-ratio parameter of the system is varied.

\section{Evolution equations}
\label{sec:evol-eq}
We will adopt the approach outlined
in~\cite{Komissarov:2007wk,Palenzuela:2009hx} to evolve the electro-vacuum
Einstein-Maxwell equations which incorporates suitably added additional fields
to ensure the evolution will preserve the constraints.  This amounts to
considering an enlarged system of the form
\begin{equation}
  \label{eq:EFE}
  \begin{aligned}
    R_{\mu \nu} - \frac{R}{2} g_{\mu \nu} & = 8\pi T_{\mu \nu} \  ,\\
    \nabla_{\mu}\left( F^{\mu \nu} + g^{\mu\nu} \Psi
    \right) & = -\kappa n^{\nu} \Psi \ , \\
    \nabla_{\mu}\left(
    \star \!{}F^{\mu \nu} + g^{\mu\nu} \Phi
    \right) & = -\kappa n^{\nu} \Phi \ ,
  \end{aligned}
\end{equation}
where $\star \!{}F^{\mu \nu}$ denotes the Hodge dual of the
Maxwell-Faraday tensor $F^{\mu \nu}$, $\kappa$ is a constant and
$n^\mu$ the four-velocity of the Eulerian observer.  We recover the
standard Einstein-Maxwell system of equations when $\Psi = 0 = \Phi$.
With the scalar field $\Psi$ and pseudo-scalar $\Phi$ introduced in
this way, the natural evolution of this system drives $\Psi$ and $\Phi$
to zero (for positive $\kappa$), thus ensuring the magnetic and electric
constraints are controlled~\cite{Komissarov:2007wk,Palenzuela:2008sf}.
The electromagnetic stress-energy tensor takes the usual form
\begin{equation}
  \label{eq:Tmunu}
  T_{\mu \nu} = \frac{1}{4\pi} \left[ F_{\mu}{}^{\lambda} F_{\nu \lambda}
    - \frac{1}{4} g_{\mu \nu} F^{\lambda \sigma} F_{\lambda \sigma}
    \right] \ .
\end{equation}
%

\subsection{3+1 decomposition}
We employ the $3+1$ decomposition, as explained in section~\ref{sec:choice-coordinates}, where we introduced the 3-metric
\begin{equation}
  \label{eq:3metric}
  \gamma_{\mu\nu} = g_{\mu \nu} + n_{\mu} n_{\nu} \ ,
\end{equation}
and further decompose the Maxwell-Faraday tensor into the more familiar electric
and magnetic fields measured by the Eulerian observer moving with
four velocity $n^{\mu}$
\begin{equation}
  \begin{aligned}
    F_{\mu \nu} & = n_{\mu} E_{\nu} - n_{\nu} E_{\mu}
      + \epsilon_{\mu\nu\alpha\beta} B^{\alpha} n^{\beta}  \ ,\\
      \star \! F_{\mu \nu} & = n_{\mu} B_{\nu} - n_{\nu} B_{\mu}
      - \epsilon_{\mu\nu\alpha\beta} E^{\alpha} n^{\beta}  \ ,
  \end{aligned}
  \label{eq:faraday}
\end{equation}
where we use the convention $\epsilon_{1230} = \sqrt{-g}$,
$\epsilon_{\alpha \beta \gamma}
= \epsilon_{\alpha \beta \gamma \delta} n^{\delta}$,
$\epsilon_{123} = \sqrt{\gamma}$.

We write the evolution equations in the BSSN form~(\ref{eq:bssn-gen}) 
where, for the case of the electromagnetic energy-momentum tensor of
equation~(\ref{eq:Tmunu}), the source terms are given by
\begin{equation}
  \label{eq:source}
   \begin{aligned}
  E &\equiv T^{\mu \nu}n_{\mu}n_{\nu}
       = \frac{1}{8\pi} \left( E^i E_i + B^i B_i \right) \ , \\
  j_i  &\equiv -\gamma_{i\mu} T^{\mu \nu}n_{\nu}
       = \frac{1}{4\pi} \epsilon_{ijk} E^j B^k \ ,\\
  S_{ij} &\equiv \gamma^{\mu}{}_i \gamma^{\nu}{}_j T_{\mu \nu} \\
       &= \frac{1}{4\pi} \left[
         -E_i E_j - B_i B_j + \frac{1}{2} \gamma_{ij} \left( E^k E_k + B^k B_k \right)
         \right] \ ,
   \end{aligned}
\end{equation}
and $S\equiv \gamma^{ij}S_{ij}$.
The evolution of the electromagnetic fields is determined by
equation~(\ref{eq:EFE}) whose 3+1 decomposition
becomes~\cite{Mosta:2009rr}
\begin{equation}
  \label{eq:maxwell-bssn}
  \begin{aligned}
    \left(
    \partial_t - \mathcal{L}_{\beta}
    \right) E^i & = \alpha K E^i + \epsilon^{ijk} \chi^{-1}
    \left[
    \tilde \gamma_{kl} B^l \partial_j \alpha 
    + \alpha \left(B^l \partial_j \tilde \gamma_{kl}
    + \tilde \gamma_{kl} \partial_j B^l
    - \chi^{-1} \tilde\gamma_{kl}B^l \partial_j \chi \right)
    \right] \\
    & \quad
    - \alpha \chi \tilde \gamma^{ij} \partial_j \Psi \ , \\
    \left(
    \partial_t - \mathcal{L}_{\beta}
    \right) B^i & = \alpha K B^i - \epsilon^{ijk} \chi^{-1}
    \left[
    \tilde \gamma_{kl} E^l \partial_j \alpha
    + \alpha \left(E^l \partial_j \tilde \gamma_{kl}
    + \tilde \gamma_{kl} \partial_j E^l
    - \chi^{-1} \tilde\gamma_{kl} E^l \partial_j \chi \right)
    \right] \\
    & \quad
    - \alpha \chi \tilde \gamma^{ij} \partial_j \Phi \ , \\
    \left(
    \partial_t - \mathcal{L}_{\beta}
    \right) \Psi & = -\alpha \nabla_i E^i - \alpha \kappa \Psi \ , \\
    \left(
    \partial_t - \mathcal{L}_{\beta}
    \right) \Phi & = -\alpha \nabla_i B^i - \alpha \kappa \Phi \ .
  \end{aligned}
\end{equation}
Here, $\mathcal{L}_{\beta}$ denotes the Lie derivative along
the shift vector $\beta^i$.
The Hamiltonian and momentum constraint are
\begin{equation}
  \label{eq:constraints}
  \begin{aligned}
    \mathcal{H} & \equiv R + K^2 - K^{ij}K_{ij} - 16 \pi E=0 \ , \\
    \mathcal{M}_i & \equiv D_j A_i{}^j - \frac{3}{2} A_i{}^j
        \chi^{-1} \partial_j \chi 
        - \frac{2}{3} \partial_i K -8\pi j_i = 0 \ ,
  \end{aligned}
\end{equation}
where $D_i$ is the covariant derivative associated with the three-metric
$\gamma_{ij}$.

\subsection{Initial data}
We focus here on black hole binaries with equal charge and mass colliding from
rest. For these configurations, it is possible to construct initial using the
Brill-Lindquist construction (outlined in section~\ref{sec:brill} for the vacuum
spacetimes; see~\cite{Brill:1963yv,Alcubierre:2009ij} for the charged case). The
main ingredients of this procedure are as follows.

For a vanishing shift $\beta^i$, time symmetry implies
$K_{ij}=0$. Combined with the condition of an initially vanishing
magnetic field,
the magnetic constraint $D_i B^i=0$ and momentum constraint
are automatically satisfied. By further assuming the spatial metric
to be conformally flat
\begin{equation}
  \gamma_{ij} \dd x^i \dd x^j = \psi^4 \left( \dd x^2 + \dd y^2 + \dd z^2 \right) \ ,
  \label{eq:inigamma}
\end{equation}
the Hamiltonian constraint reduces to
\begin{equation}
  \triangle \psi = - \frac{1}{4} E^2 \psi^5 \ ,
\end{equation}
where $\triangle $ is the flat space Laplace operator. The electric
constraint, Gauss's law, has the usual form
\begin{equation}
  D_i E^{i} = 0 \ .
\end{equation}
%
Quite remarkably, for systems of black holes with equal charge-to-mass ratio,
these equations have known analytical solutions~\cite{Alcubierre:2009ij}.
For the special case of two black holes momentarily at rest with
``bare masses'' $m_1$, $m_2$ and ``bare charges'' $q_1$, $q_2=q_1 m_2/m_1$
this analytic solution is given by
\begin{equation}
  \begin{aligned}
    \psi^2 & = \left( 1 + \frac{m_1}{2|\vec x - \vec x_1|}
      + \frac{m_2}{2|\vec x - \vec x_2|} \right)^2 
    - \frac{1}{4} \left( \frac{q_1}{|\vec x - \vec x_1|}
      + \frac{q_2}{|\vec x - \vec x_2|} \right)^2  \ , \\
    E^i & = \psi^{-6} \left( q_1
        \frac{ (\vec x - \vec x_1)^i }{|\vec x - \vec x_1|^{3}}
      + q_2 \frac{ (\vec x - \vec x_2)^i }{|\vec x - \vec x_2|^{3}}
    \right) \ ,
  \end{aligned}
  \label{eq:ini_psi_E}
\end{equation}
where $\vec x_i$ is the coordinate location of the $i$th ``puncture''.\footnote{We note that this foliation, in isotropic coordinates, only covers the outside of the external horizon.}

The initial data is thus completely specified in terms of the independent
mass and charge parameters $m_1$, $m_2$, $q_1$ and the initial
coordinate separation $d$ of the holes. These uniquely determine the
remaining charge parameter $q_2$ via the condition
of equal charge-to-mass ratio. In this study we always choose $m_1=m_2$
and, without loss of generality, position the two holes symetrically
around the origin such that $z_1=d/2=-z_2$. The resulting initial
three metric $\gamma_{ij}$
follows from equations.~(\ref{eq:inigamma}), (\ref{eq:ini_psi_E})
while the extrinsic curvature $K_{ij}$ and magnetic field $B^i$
vanish on the initial slice.

We use the same gauge conditions and outer boundary conditions for the BSSN variables as used in vacuum simulations, cf. equations~(\ref{eq:1+log_0}) and (\ref{eq:gamma-driver-0}).
As outer boundary condition for the electric and magnetic fields we have imposed a falloff as $1/r^2$---from \eqref{eq:ini_psi_E}. For the additional scalar fields a satisfactory behaviour is observed by imposing a falloff as $1/r^3$
(which is the expected falloff rate from dimensional grounds).

\section{Wave Extraction}
\label{wave_extraction}

For a given set of initial parameters $m_1=m_2$, $q_1=q_2$, $d$, the time
evolution provides us with the spatial metric $\gamma_{ij}$, the extrinsic
curvature $K_{ij}$ as well as the electric and magnetic fields $E^i$, $B^i$ as
functions of time. These fields enable us to extract the gravitational and
electromagnetic radiation as explained in section~\ref{sec:NP-gen}. Details
concerning the numerical implementation can be found in~\cite{Sperhake:2006cy}.

We recall that the radiated flux and energy are given by the expressions~(\ref{eq:GW-flux}) and (\ref{eq:EM-flux}):
\begin{align}
  \label{eq:GW-flux2}
  F_{\rm GW} & = \frac{\dd E_{\rm GW}}{\dd t} =
      \lim_{r\to\infty} \frac{r^2}{16 \pi} \sum_{l,m}
      \left| \int_{-\infty}^t \dd t' \psi^{lm} (t') \right|^2 \ , \\
  F_{\rm EM} & = \frac{\dd E_{\rm EM}}{\dd t} =
      \lim_{r\to\infty} \frac{r^2}{4 \pi} \sum_{l,m} 
      \left|  \phi^{lm}_{2} (t) \right|^2 \ . \label{eq:EM-flux2}
\end{align}
As is well known from simulations of uncharged black-hole binaries,
initial data obtained from the Brill-Lindquist construction contains
``spurious'' radiation, which is an artifact of the conformal-flatness
assumption. In calculating properties of the radiation, we account for
this effect by starting the integration of the radiated flux
in equations~(\ref{eq:GW-flux2}), (\ref{eq:EM-flux2}) at some finite time $\Delta t$
after the start of the simulation, thus allowing the spurious pulse
to first radiate off the computational domain. In practice, we obtain
satisfactory results by choosing $\Delta t = R_{\rm ex}+50~M$.
Because the physical radiation is very weak for both the gravitational
and electromagnetic channel in this early infall stage, the error
incurred by this truncation is negligible compared with the uncertainties
due to discretization; cf.~section~\ref{sec:fluxes}.

\section{Analytic predictions}
\label{classical_expectations}
Before discussing in detail the results of our numerical simulations,
it is instructive to discuss the behaviour of the binary system as
expected from an analytic approximation. Such an analysis not only serves
an intuitive understanding of the binary's dynamics, but also provides
predictions to compare with the numerical results presented below.

For this purpose we consider the electrodynamics of a system of
two equal point charges in a Minkowski background spacetime. As in
the black hole case, we denote by $q_1=q_2\equiv Q/2$ and $m_1=m_2\equiv M/2$
the electric charge and mass of the particles which are initially
at rest at position $z=\pm d/2$.

It turns out useful to first consider point charges in Minkowski spacetime
in the static limit. The expected behaviour
of the radial component of the resulting electric field is given
by~\cite{Jackson1998Classical}
\begin{equation}
  \label{eq:Er_asympt}
  E_{\hat r} = 4\pi\sum_{l,m} \frac{l+1}{2l+1} q_{lm}
      \frac{Y_{lm}(\theta,\varphi)}{r^{l+2}} \ ,
\end{equation}
which for a system of two charges of equal magnitude at $z=\pm d/2$ becomes
\begin{equation}
  \label{eq:Er_2q}
  E_{\hat r} \simeq \sqrt{4\pi} Q \frac{Y_{00}}{r^2} 
     + \sqrt{\frac{9\pi}{20}} Q d^2 \frac{Y_{20}}{r^4} \ .
\end{equation}
The dipole vanishes in this case due to the reflection symmetry across
$z=0$. This symmetry is naturally preserved during the time evolution
of the two-charge system. Furthermore, the total electric charge $Q$
is conserved so that the leading-order behaviour of the
electromagnetic radiation is given by variation of the electric
quadrupole, just as for the gravitational radiation.
Notice that in principle other radiative contributions can arise
from the accelerated motion of the charged black holes.
From experience with gravitational radiation generated in the collision
of electrically neutral black-hole binaries, however, we expect
this ``Bremsstrahlung'' to be small in comparison with the merger
signal and hence ignore its contributions in this simple approximation.
The good agreement with the numerical results presented in the
next section bears out the validity of this {\em quadrupole approximation}.
In consequence, it appears legitimate to regard the ``strength'' of the
collision and the excitation of the black-hole ringdown to be
purely kinematic effects.

An estimate for the monopole and quadrupole amplitudes in the limit
of two static point charges is then obtained from inserting the
radial component of the electric field (\ref{eq:Er_2q}) into
the expression (\ref{eq:Phi_asympt}) for $\Phi_1$ and its
multipolar decomposition (\ref{eq:multipole_Phi1})
\begin{align}
  r^2\phi_1^{00}&=\sqrt{\pi} Q\approx 1.77 Q\,,\label{eq:mono}\\
  r^4\phi_1^{20}&=\sqrt{\frac{9\pi}{80}} Q d^2\approx 0.59 Qd^2\,.
  \label{eq:dipole}
\end{align}
The expectation is that these expressions provide a good approximation for
the wave signal during the early infall stage when the black holes are
moving with small velocities. Equation~(\ref{eq:mono}) should also provide a
good approximation for $\phi_1^{00}$ after the merger and ringdown
whereas the quadrupole $\phi_1^{20}$ should eventually approach zero
as a single merged hole corresponds to the case $d=0$ in
equation~(\ref{eq:dipole}).

In order to obtain analytic estimates for the collision time and the
emitted radiation, we need to describe the dynamic behaviour of the
two point charges. Our starting point for this discussion is the
combined gravitational and electromagnetic potential energy for two charges
$i=1,\,2$ in Minkowski spacetime
with mass and charge $m_i$, $q_i$ at distance $r$ from each other
\begin{equation}
  V=-\frac{Gm_1m_2}{r}+\frac{1}{4\pi\epsilon_0}\frac{q_1q_2}{r}\ .
\end{equation}
For the case of two charges
with equal mass and charge $m_i=M/2$,
$q_i=Q/2$ and
starting from rest at $z_0=\pm d/2$,
conservation of energy implies
\begin{equation}
  M\dot{z}^2-\frac{M^2{\cal B}}{4z}=-\frac{M^2{\cal B}}{2d} \ , \label{eqm2}
\end{equation}
where we have used units with $G=4\pi \epsilon_0 = 1$ and
\begin{equation}
  {\cal B}\equiv 1-Q^2/M^2\ .
\end{equation}
The resulting equation of motion for $z(t)$ is obtained
by differentiating equation~(\ref{eqm2}) which results in
\begin{equation}
  M\ddot{z}=-\frac{M^2}{8z^2}+\frac{Q^2}{8z^2}=
      -M^2\frac{{\cal B}}{8z^2}\ . \label{eqm1}
\end{equation}
An estimate for the time for collision follows from
integrating equation~(\ref{eqm2}) over $z\in [d/2,0]$
\begin{equation}
  \left(\frac{t_{\rm collision}}{M}\right)^2
      =\frac{\pi^2 d^3}{2^3M^3 {\cal B}}\ . \label{eq:timecollision}
\end{equation}

From the dynamic evolution of the system we can derive an approximate
prediction for the electromagnetic radiation by evaluating the (traceless)
electric quadrupole tensor $Q_{ij}=\int d^3\vec{x} \rho(\vec{x})(3x_ix_j-r^2\delta_{ij})$~\cite{Jackson1998Classical}.
In terms of
this quadrupole tensor, the total power radiated is given by~\cite{Jackson1998Classical}
\begin{equation}
F_{\rm EM}=\sum_{ij}\frac{1}{4\pi\epsilon_0}
      \frac{1}{360c^5}\dddot{Q}_{ij}^2\ .\label{eq:power}
\end{equation}
For clarity we have reinstated the factors $4\pi \epsilon_0$ and $c^5$ here.
Using
\begin{equation}
\frac{d^3}{dt^3}(z^2)=6\dot{z}\ddot{z}+2z\dddot{z}\ ,
\end{equation}
and the equations of motion (\ref{eqm2}), (\ref{eqm1}) we find
\begin{equation}
F_{\rm EM}=\frac{{\cal B}^3M^3Q^2(1/z-2/d)}{1920z^4}\ .
\end{equation}
Using $\int dt (\ldots ) = \int dz/\dot{z} (\ldots)$, we can evaluate
the time integral up to some cutoff separation, say
$z_{\rm min} = \alpha_b b$, where $b$ is the horizon radius of the
initial black hole,
$b=M(1+\sqrt{{\cal B}})/2$ and $\alpha_b={\cal O}(1)$ is a constant. This
gives,
\begin{equation}
  \label{EMquadprediction}
  \frac{E^{\rm EM}_{\rm rad}}{M} =
      {\cal B}^{5/2}M^{3/2}Q^2
   \frac{(d-2\alpha_b b)^{3/2}
      (15d^2+24d\alpha_b b+32\alpha_b^2b^2)}{50400(d\alpha_b b)^{7/2}}\ .
\end{equation}

Emission of gravitational radiation follows from the quadrupole formula,
which is a numerical factor $4$ times larger, and where the charge is
be replaced by the mass,
\begin{equation}
  \label{GWquadprediction}
  \frac{E^{\rm GW}_{\rm rad}}{M} = {\cal B}^{5/2}M^{7/2}
   \frac{(d-2\alpha_b b)^{3/2}
      (15d^2+24d\alpha_b b+32\alpha_b^2b^2)}{12600(d\alpha_b b)^{7/2}}\ .
\end{equation}
For $Q=0, \alpha_b=1, d=\infty$ we thus obtain
\begin{equation}
  \frac{E^{\rm GW}_{\rm rad}}{M}=\frac{1}{840}\sim 0.0012 \ ,
\end{equation}
in agreement to within a factor of $2$ with numerical simulations (see
\cite{Witek:2010xi} and table~\ref{tab:runs} below; the agreement could
be improved by assuming $\alpha_b\sim 1.3$). As a general result
of this analysis we find in this approximation,
\begin{equation}
  \frac{E^{\rm EM}_{\rm rad}}{E^{\rm GW}_{\rm rad}}=\frac{Q^2}{4M^2}\ .
\label{prediction_ratio}
\end{equation}
For non-extremal holes $Q<M$, our analytic considerations therefore predict that
the energy emitted in electromagnetic radiation is at most $25\%$ of the energy
lost in gravitational radiation. As we shall see below, this turns out to be a
remarkably good prediction for the results obtained from fully numerical
simulations.

\section{Numerical Results}
\label{numerical_results}

The numerical integration of the Einstein-Maxwell equations (\ref{eq:bssn-gen}),
(\ref{eq:maxwell-bssn}) has been performed using fourth-order spatial
discretisation with the \textsc{Lean} code, originally presented
in~\cite{Sperhake:2006cy} for vacuum spacetimes, see
section~\ref{sec:lean-code}.

The initial parameters as well as the grid setup and the radiated gravitational
and electromagnetic wave energy for our set of binary configurations is listed
in table~\ref{tab:runs}. All binaries start from rest with a coordinate distance
$d/M\simeq 8$ or $d/M\simeq 16$ while the charge-to-mass ratio has been varied
from $Q/M=0$ to $Q/M=0.98$.  Note that identical coordinate separations of the
punctures for different values of the charge $Q/M$ correspond to different
horizon-to-horizon proper distances.  This difference is expected and in fact
analysis of the RN solution predicts a divergence of the proper distance in the
limit $Q/M\rightarrow 1$.

\begin{table*}[ht]
  \centering
  \caption[Grid structure, coordinate distance, proper horizon-to-horizon
  distance, charge, gravitational and electromagnetic radiated energy for our set of
  simulations]{Grid structure in the notation of section~II~E of~\cite{Sperhake:2006cy},
    coordinate distance $d/M$, proper horizon-to-horizon
    distance $L/M$, charge $Q/M$, gravitational ($E_{\rm
      rad}^{\mathrm{GW}}$) and electromagnetic ($E_{\rm rad}^{\mathrm{EM}}$)
    radiated energy for our set of simulations. The radiated energy has
    been computed using
    only the $l=2$, $m=0$ mode; the energy contained in higher-order
    multipoles such as $l=4$, $m=0$ is negligible for all configurations.
    \label{tab:runs}}

  \small\addtolength{\tabcolsep}{-5pt}
  \begin{tabular*}{\textwidth}{@{\extracolsep{\fill}}lcccccc}
    \hline
    \hline
    Run     &      Grid                                       &  $d/M$  & $L/M$ & $Q/M$ & $E_{\rm rad}^{\mathrm{GW}}$    & $E_{\rm rad}^{\mathrm{EM}}$ \\
    \hline
    d08q00   & $\{(256,128,64,32,16,8)\times(2,1,0.5), 1/80\}$ & 8.002   & 11.56 & 0     & $5.1\times10^{-4}$ &  --            \\
    d08q03   & $\{(256,128,64,32,16,8)\times(2,1,0.5), 1/80\}$ & 8.002   & 11.60 & 0.3   & $4.5\times10^{-4}$ & $1.3\times10^{-5}$ \\
    d08q04   & $\{(256,128,64,32,16,8)\times(2,1,0.5), 1/80\}$ & 8.002   & 11.65 & 0.4   & $4.0\times10^{-4}$ & $2.1\times10^{-5}$ \\
    d08q05c  & $\{(256,128,64,32,16,8)\times(2,1,0.5), 1/64\}$ & 8.002   & 11.67 & 0.5   & $3.3\times10^{-4}$ & $2.7\times10^{-5}$ \\
    d08q05m  & $\{(256,128,64,32,16,8)\times(2,1,0.5), 1/80\}$ & 8.002   & 11.70 & 0.5   & $3.4\times10^{-4}$ & $2.7\times10^{-5}$ \\
    d08q05f  & $\{(256,128,64,32,16,8)\times(2,1,0.5), 1/96\}$ & 8.002   & 11.67 & 0.5   & $3.4\times10^{-4}$ & $2.7\times10^{-5}$ \\
    d08q055  & $\{(256,128,64,32,16,8)\times(2,1,0.5), 1/80\}$ & 8.002   & 11.70 & 0.55  & $3.0\times10^{-4}$ & $2.89\times10^{-5}$ \\
    d08q06   & $\{(256,128,64,32,16,8)\times(2,1,0.5), 1/80\}$ & 8.002   & 11.75 & 0.6   & $2.6\times10^{-4}$ & $2.97\times10^{-5}$ \\
    d08q07   & $\{(256,128,64,32,16,8)\times(2,1,0.5), 1/80\}$ & 8.002   & 11.87 & 0.7   & $1.8\times10^{-4}$ & $2.7\times10^{-5}$ \\
    d08q08   & $\{(256,128,64,32,16,8)\times(2,1,0.5), 1/80\}$ & 8.002   & 12.0  & 0.8   & $9.8\times10^{-5}$ & $1.8\times10^{-5}$ \\
    d08q09   & $\{(256,128,64,32,16,8)\times(2,1,0.5), 1/80\}$ & 8.002   & 12.3  & 0.9   & $2.6\times10^{-5}$ & $5.5\times10^{-6}$ \\
    d08q098cc & $\{(256,128,64,32,16,8)\times(2,1,0.5), 1/64\}$ & 8.002   & 12.3  & 0.98  & $7.0\times10^{-7}$ & $2.1\times10^{-7}$ \\
    d08q098c & $\{(256,128,64,32,16,8)\times(2,1,0.5), 1/80\}$ & 8.002   & 13.1  & 0.98  & $4.3\times10^{-7}$ & $1.4\times10^{-7}$ \\
    d08q098m & $\{(256,128,64,32,16,8)\times(2,1,0.5), 1/96\}$ & 8.002   & 13.1  & 0.98  & $3.4\times10^{-7}$ & $1.0\times10^{-7}$ \\
    d08q098f & $\{(256,128,64,32,16,8)\times(2,1,0.5), 1/112\}$ & 8.002 & 13.0  & 0.98  & $4.0\times10^{-7}$ & $9.5\times10^{-8}$ \\
    d08q098ff & $\{(256,128,64,32,16,8)\times(2,1,0.5), 1/128\}$ & 8.002 & 13.0 & 0.98  & $4.05\times10^{-7}$ & $8.75\times10^{-8}$ \\
    d08q098fff & $\{(256,128,64,32,16,8)\times(2,1,0.5), 1/136\}$ & 8.002 & 13.1 & 0.98  & $3.73\times10^{-7}$ & $8.41\times10^{-8}$ \\
    d16q00   & $\{(256,128,64,32,16)\times(4,2,1,0.5), 1/64\}$ & 16.002  & 20.2  & 0     & $5.5\times10^{-4}$ & --                \\
    d16q05   & $\{(256,128,64,32,16)\times(4,2,1,0.5), 1/64\}$ & 16.002  & 20.3  & 0.5   & $3.6\times10^{-4}$ & $2.9\times10^{-5}$ \\
    d16q08   & $\{(256,128,64,32,16)\times(4,2,1,0.5), 1/80\}$ & 16.002  & 20.7  & 0.8   & $1.05\times10^{-4}$ & $1.9\times10^{-5}$ \\
    d16q09   & $\{(256,128,64,32,16)\times(4,2,1,0.5), 1/80\}$ & 16.002  & 21.0  & 0.9   & $2.7\times10^{-5}$  & $5.9\times10^{-6}$ \\
 
    \hline
    \hline
  \end{tabular*}
\end{table*}

\subsection{Code tests}
\label{sec:tests}

Before discussing the obtained results in more detail, we present two tests to
validate the performance of our numerical implementation of the evolution
equations: (i) single black-hole evolutions in {\em geodesic slicing} which is
known to result in numerical instabilities after relatively short times but
facilitates direct comparison with a semi-analytic solution and (ii) convergence
analysis of the radiated quadrupole waveforms for simulation d08q05 of
table~\ref{tab:runs}.

\begin{figure}[tbh]
  \centering
  \includegraphics[width=0.45\textwidth]{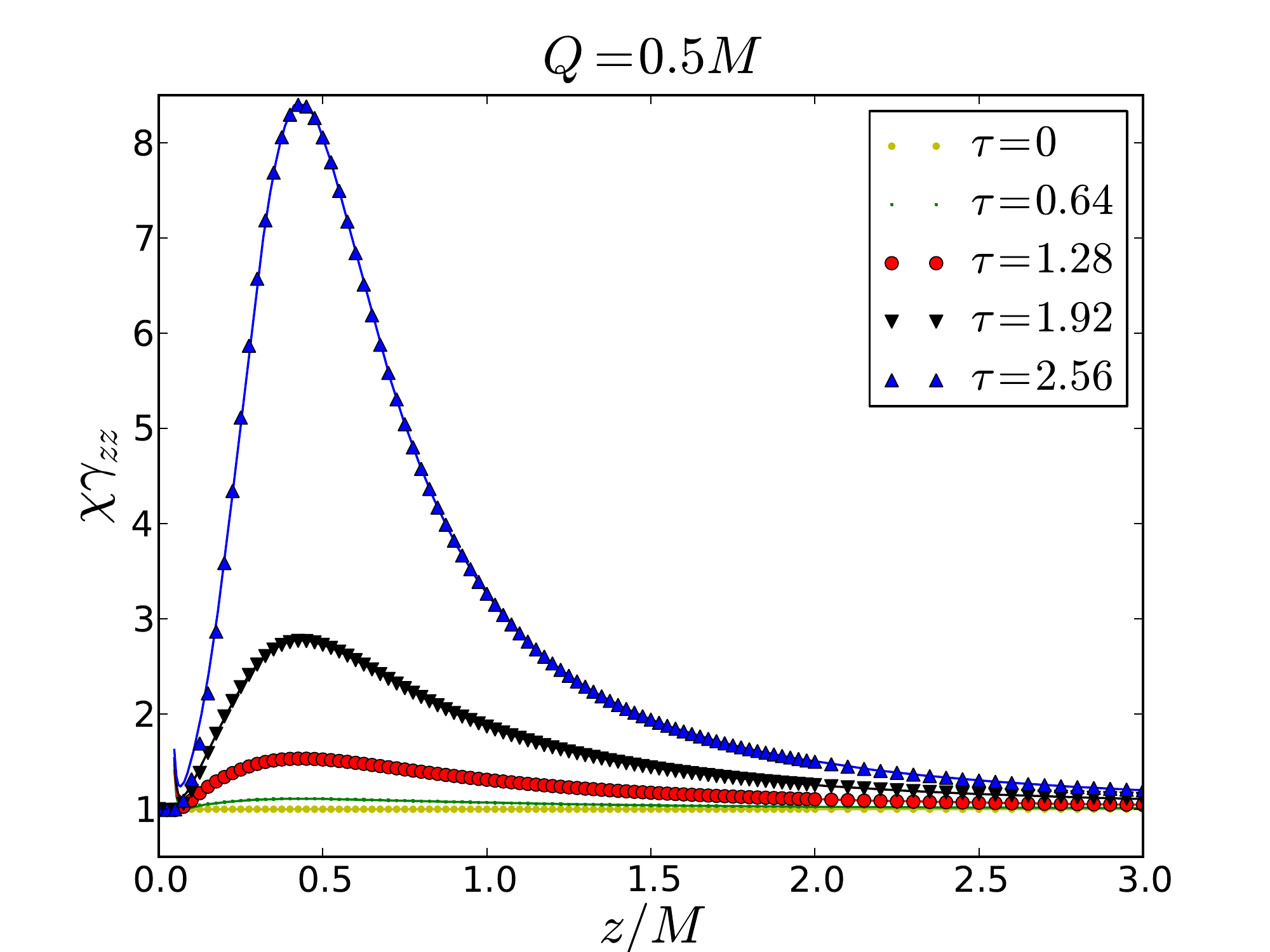}
  \includegraphics[width=0.45\textwidth]{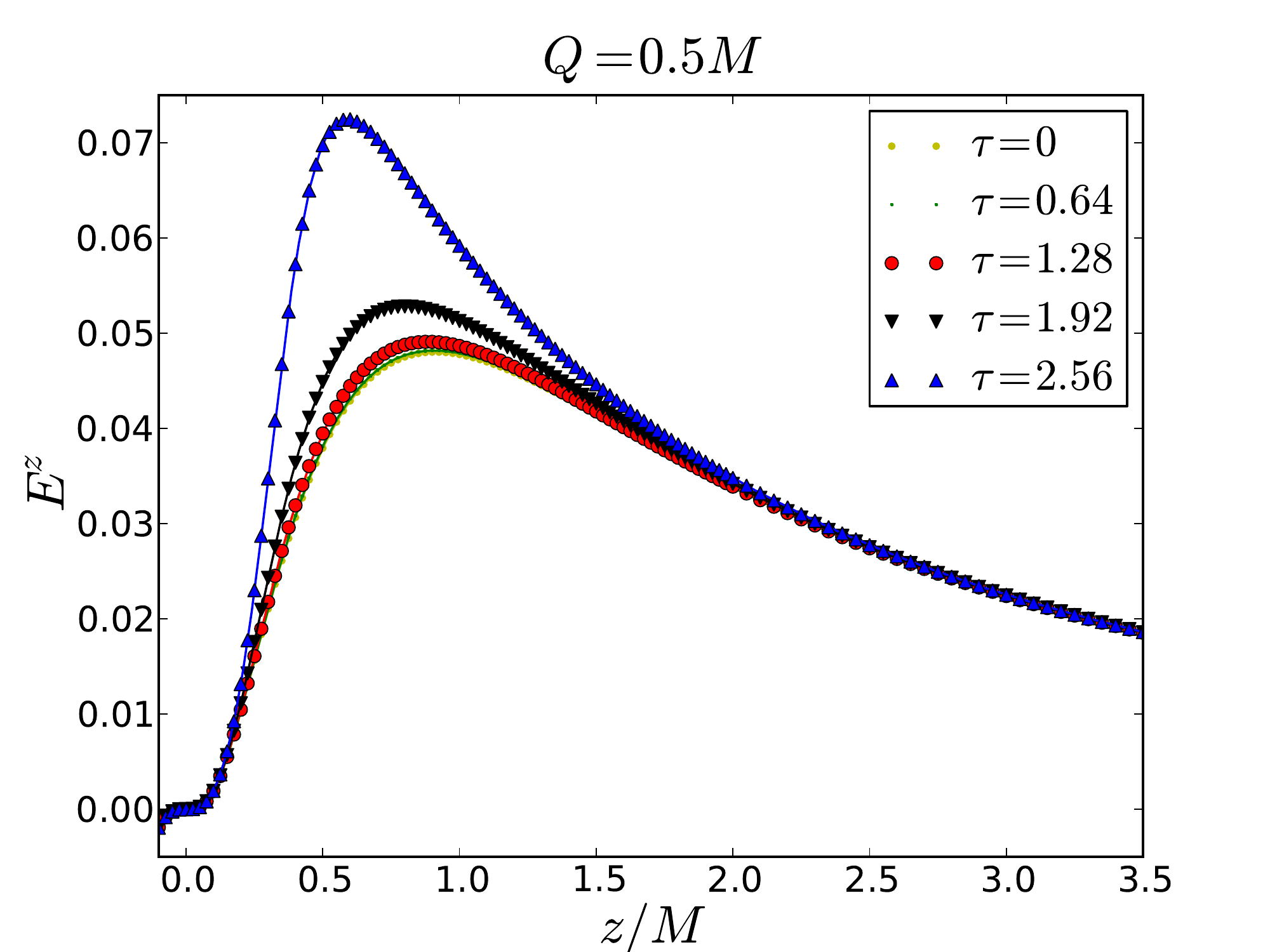}
  \caption[Numerical profiles for $\gamma_{zz}$ and $E^z$ obtained in geodesic
  slicing]{The numerical profiles for $\gamma_{zz}$ and $E^z$ (symbols) obtained
    in geodesic slicing at various times $\tau$ are compared with the
    semi-analytic results (lines).}
\label{fig:geo_slice} 
\end{figure}

The geodesic slicing condition is enforced by setting the gauge functions to
$\alpha=1$, $\beta^i=0$ throughout the evolution. The space part of the
Reissner-Nordstr{\"o}m solution in isotropic coordinates is given by
equation~(\ref{eq:inigamma}) with a conformal
factor~\cite{Graves:1960zz,Reimann:2003zd}
\begin{equation}
  \psi^2 = \left( 1+\frac{M}{2r} \right)^2 - \frac{Q^{2}}{4r^2}\ .
  \label{eq:conformalfactor}
\end{equation}
The time evolution of this solution is not known in closed analytic form, but
the resulting metric components can be constructed straightforwardly via a
simple integration procedure. 
As expected, we find a time evolution in this gauge to become numerically
unstable at times $\tau$ of a few $M$.  Before the breaking down of the
evolution, however, we can safely compare the numerical and ``analytical''
solutions. This comparison is shown in figure~\ref{fig:geo_slice} for the
$\gamma_{zz}$ component of the spatial metric and the $E^z$ component of the
electric field and demonstrates excellent agreement between the semi-analytic
and numerical results.

\begin{figure}[tbhp]
\centering
\includegraphics[width=0.45\textwidth]{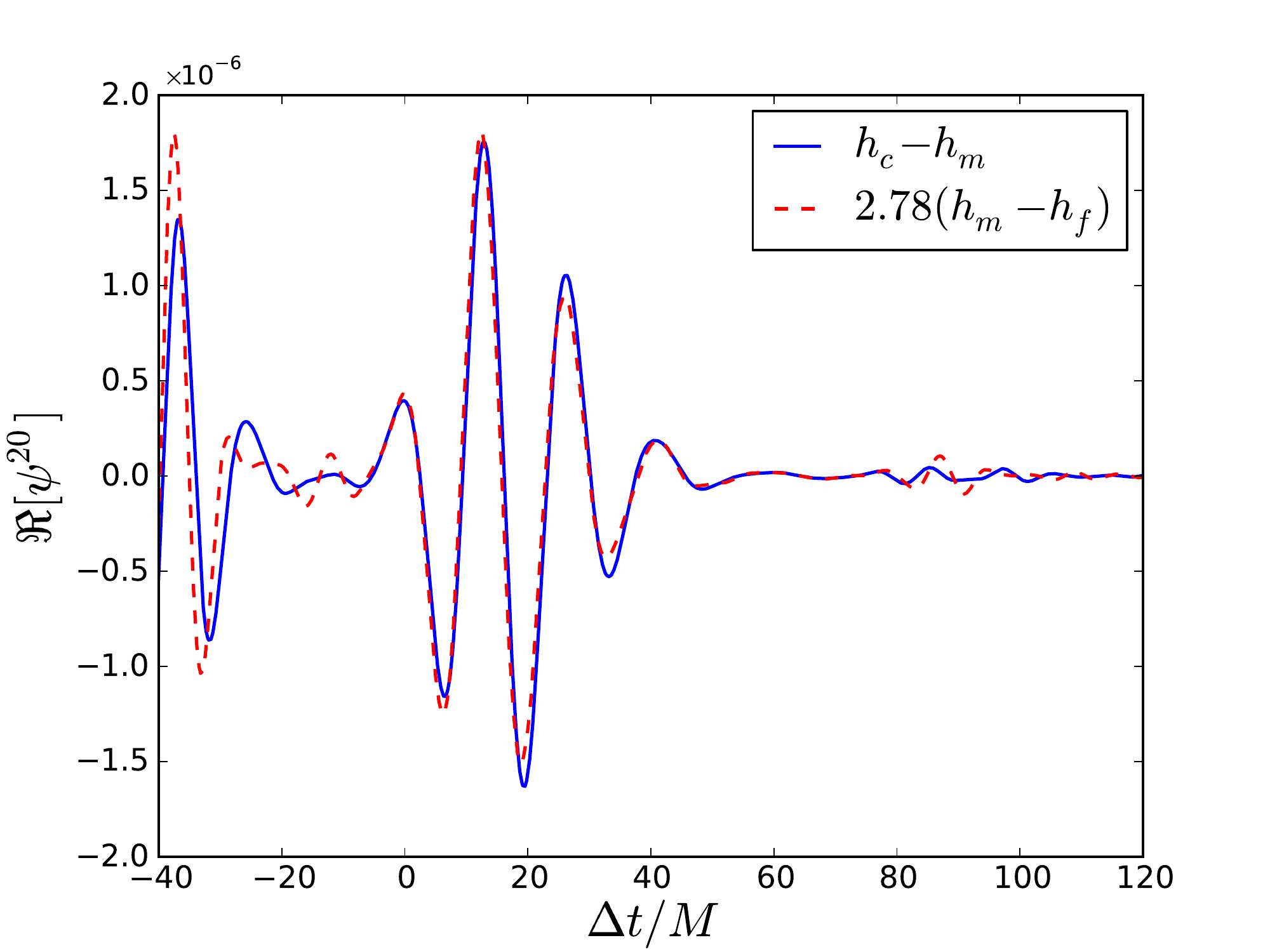}
\includegraphics[width=0.45\textwidth]{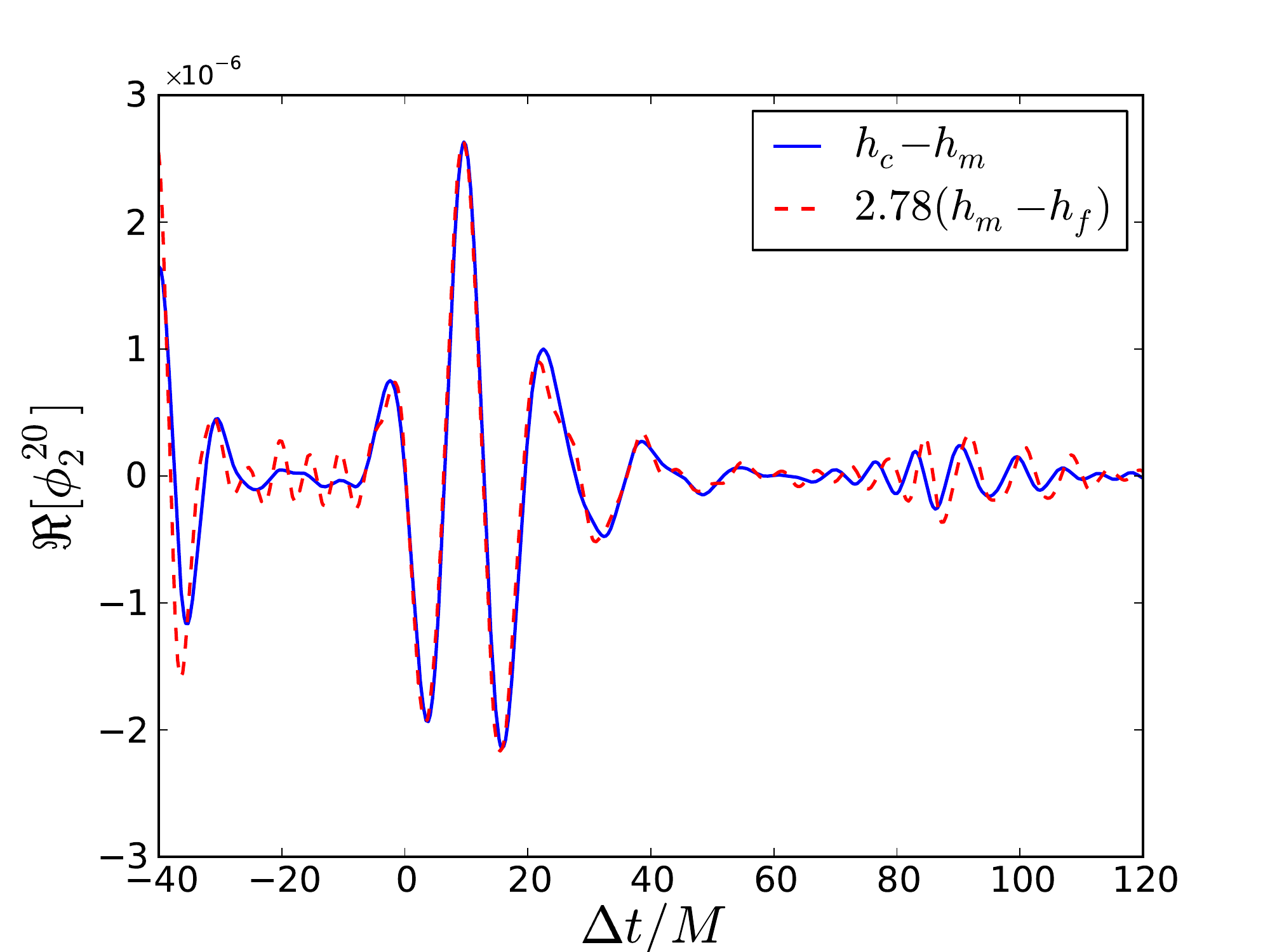}
\caption[Convergence analysis for simulation d08q05]{Convergence analysis for
  simulation d08q05 of table~\ref{tab:runs} with resolutions $h_c= M/64$,
  $h_m=M/80$ and $h_f=M/96$.  The panels show differences of the $(2,0)$
  multipoles of the real parts of $\Psi_4$ (left) and $\Phi_2$ (right) extracted
  at $R_{\rm ex}=100~M$; in each case, the high-resolution differences have been
  rescaled by a factor 2.78 as expected for fourth-order convergence.}
\label{fig:convergence}
\end{figure}

For the second test, we have evolved model d08q05 using three different
resolutions as listed in table~\ref{tab:runs} and extracted the gravitational
and electromagnetic quadrupole $(l=2,m=0)$ at $R_{\rm ex}=100M$. For
fourth-order convergence, we expect the differences between the higher
resolution simulations to be a factor $2.78$ smaller than their coarser
resolution counterparts. The numerically obtained differences are displayed with
the corresponding rescaling in figure~\ref{fig:convergence}. Throughout the
physically relevant part of the waveform, we observe the expected fourth-order
convergence. Only the spurious initial radiation (cf.~the discussion at the end
of section~\ref{wave_extraction}) at early times $\Delta t \lesssim -20$ in the
figure exhibits convergence closer to second order, presumably a consequence of
high-frequency noise contained in this spurious part of the signal. From
Richardson extrapolation of our results we estimate the truncation error of the
radiated waves to be about $1\%$.  The error due to extraction at finite
radius, on the other hand, is estimated to be 2 \% at $R_{\rm ex}=100M$.

\subsection{Collisions of two black holes: the ``static'' components and infall time}

We start the discussion of our results with the behaviour of the gravitational
and electromagnetic multipoles when the system is in a nearly static
configuration, i.e.~shortly after the start of the simulation and at late stages
after the ringdown of the post-merger hole. At these times, we expect our
analytic predictions (\ref{eq:mono}), (\ref{eq:dipole}) for the monopole and
dipole of the electromagnetic field to provide a rather accurate description.
Furthermore, the total spacetime charge $Q$ is conserved throughout the
evolution, so that the monopole component of $\Phi_1$ should be described by
\eqref{eq:mono} {\it at all times}. The quadrupole, on the other hand, is
expected to deviate significantly from the static prediction (\ref{eq:dipole})
when the black holes start moving fast.

\begin{figure}[tbhp]
\centering
\includegraphics[width=0.45\textwidth]{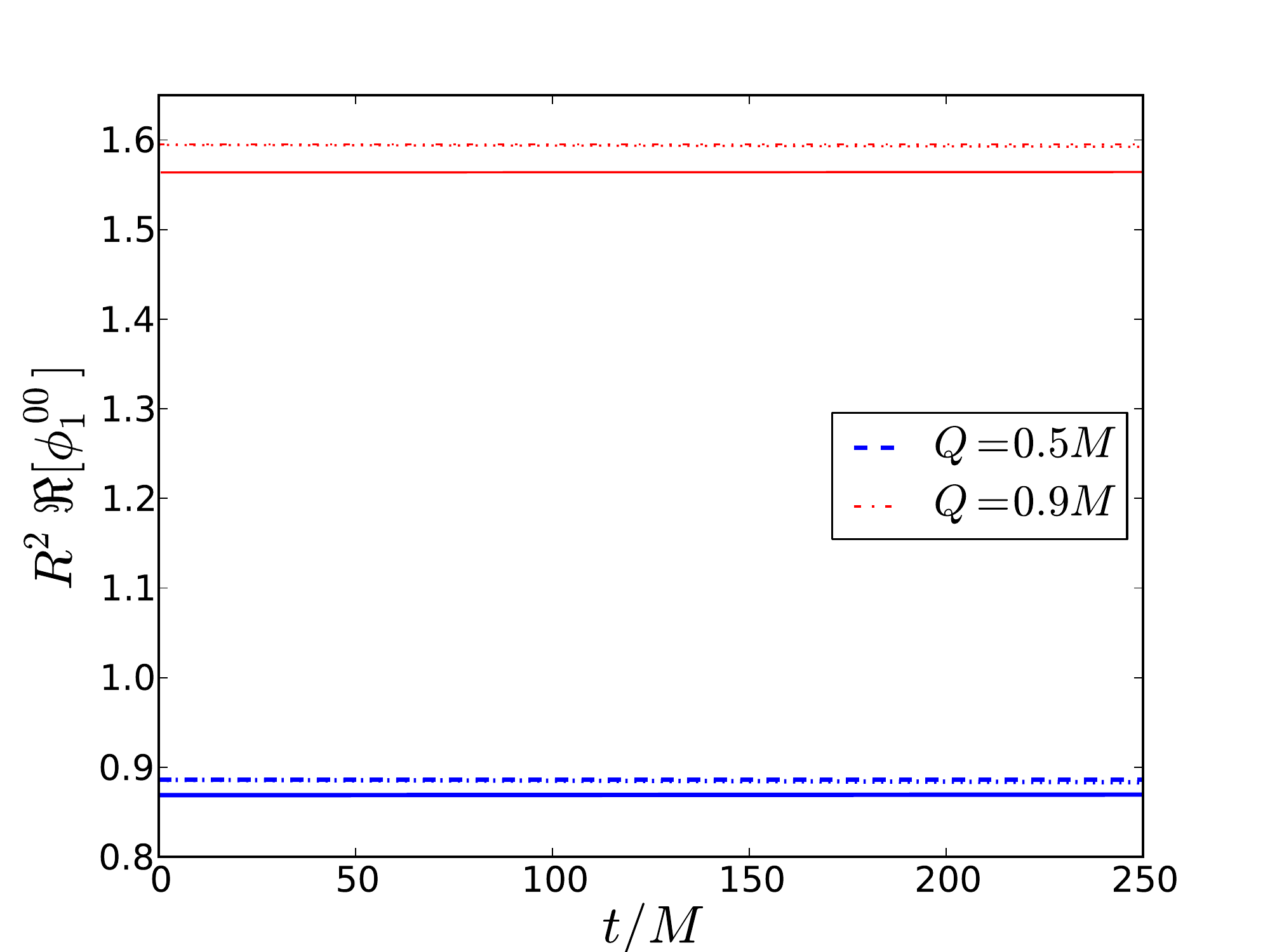}
\includegraphics[width=0.45\textwidth]{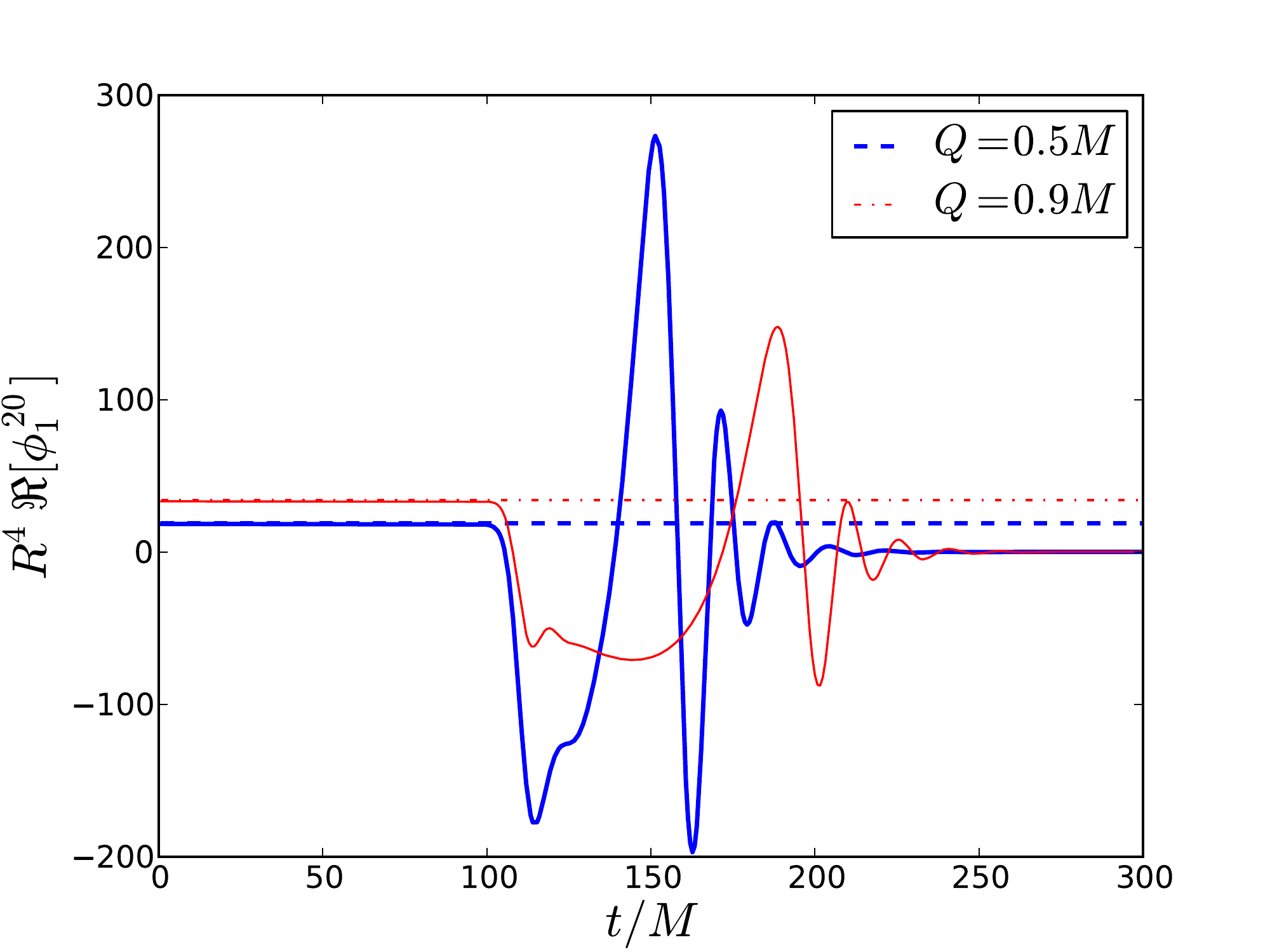}
\caption[Monopole $\phi_1^{00}$ and quadrupole $\phi_1^{20}$ of the radial part
of the electromagnetic field $\Phi_1$]{Monopole $\phi_1^{00}$ (left) and
  quadrupole $\phi_1^{20}$ (right) of the radial part of the electromagnetic field
  $\Phi_1$ extracted at $R_{\rm ex}=100M$ for simulation d08q05 of
  table~\ref{tab:runs}. The dashed curves show the predictions of
  equations~(\ref{eq:mono}), (\ref{eq:dipole}) at $R=\infty$ in the static limit. For
  the monopole case, we also added the curves obtained by extrapolating the
  results to infinite extraction radius; these curves---dotted
  lines---essentially overlap with the predictions from equation~(\ref{eq:mono}).
  \label{fig:multipoles}}
\end{figure}
As demonstrated in figure~\ref{fig:multipoles}, we find our results to be
consistent with this picture.  Here we plot the monopole and quadrupole of
$\Phi_1$. The monopole part (left panel) captures the Coulomb field and can thus
be compared with the total charge of the system. It is constant throughout the
evolution to within numerical error and shows agreement with the analytic
prediction of equation~(\ref{eq:mono}) within numerical uncertainties; we
measure a slightly smaller value for the monopole field than expected from the
total charge of the system, but the measured value should increase with
extraction radii and agree with the total charge expectation at infinity. This
is consistent with the extrapolation of the measured value to infinity as shown
in the figure.  The quadrupole part (right panel) starts at a non-zero value in
excellent agreement with equation~(\ref{eq:dipole}), deviates substantially
during the highly dynamic plunge and merger stage and eventually rings down
towards the static limit $\phi_1^{20}=0$ as expected for a spherically symmetric
charge distribution.

\begin{figure}[tbhp]
\centering
\includegraphics[width=0.7\textwidth]{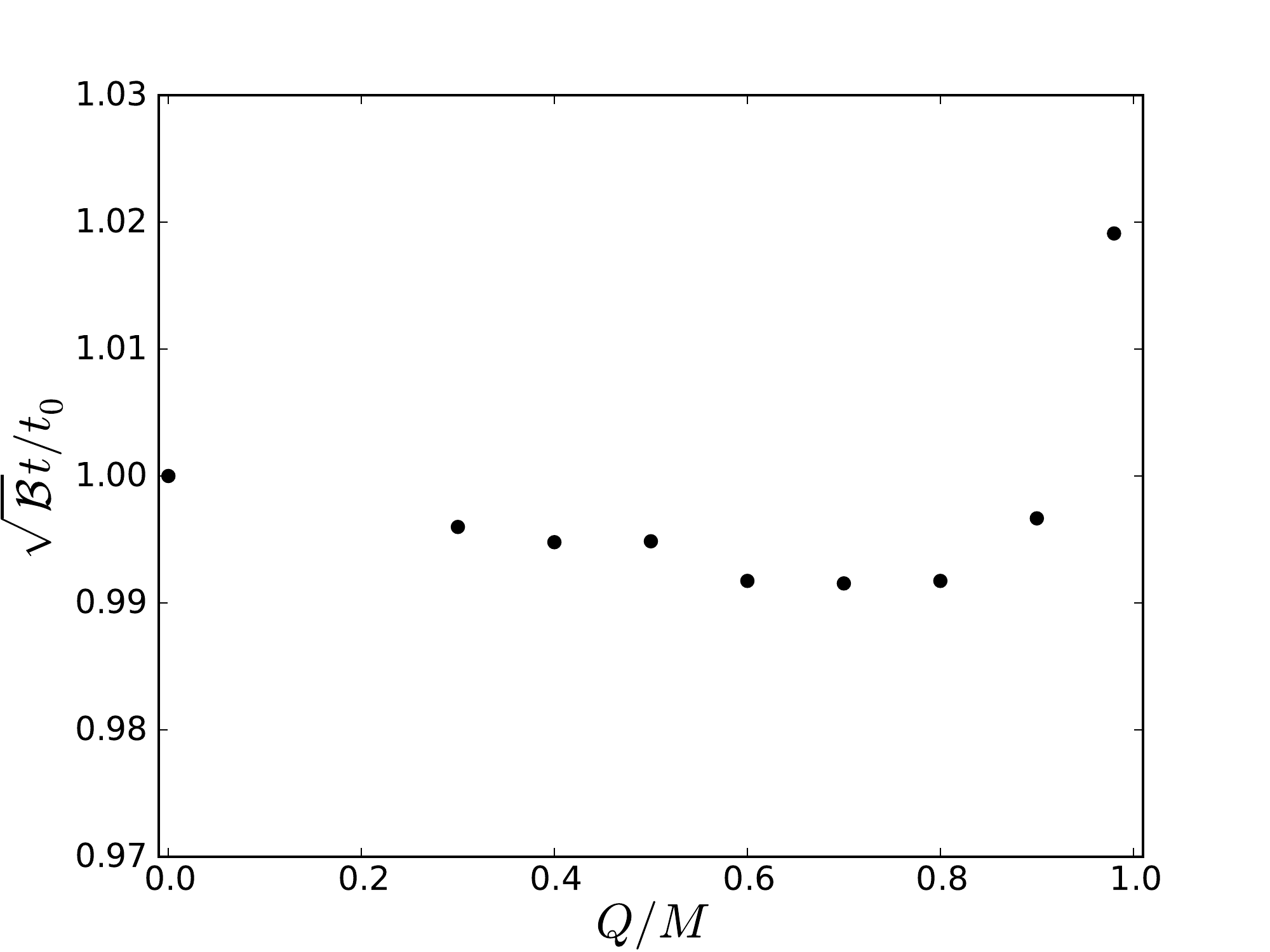}
\caption[Time for apparent horizon formation]{Time for apparent horizon formation, re-scaled by the factor
  $\sqrt{{\cal B}}$ and the apparent horizon formation time $t_0$ for an
  electrically neutral binary.
  We note that the change in the quantity we plot is only, at most, of $2\%$. 
  The coordinate time itself, however, varies by a factor 5 as one goes from $Q=0$ to $Q=0.98M$.
  \label{fig:time}
}
\end{figure}
The analytic approximation of section~\ref{classical_expectations} also predicts
a value for the time of collision (\ref{eq:timecollision}) for a given set of
initial parameters. In particular, we see from this prediction that for fixed
initial separation $d$ and mass $M$ the collision time scales with the charge as
$t_{\rm collision} \sim 1/\sqrt{\mathcal{B}}$.  In comparing these predictions
with our numerical results we face the difficulty of not having an unambiguous
definition of the separation of the black holes in the fully general
relativistic case. From the entries in table~\ref{tab:runs} we see that the
proper distance $L$ varies only mildly for fixed coordinate distance $d$ up to
$Q/M \approx 0.8$.  For nearly extremal values of $Q$, however, $L$ starts
increasing significantly as expected from our discussion at the start of this
section. We therefore expect the collision time of the numerical simulations
rescaled by $\sqrt{\mathcal{B}}/t_0$, where $t_0$ is the corresponding time for
the uncharged case, to be close to unity over a wide range of $Q/M$ and show
some deviation close to $Q/M=1$.  This expectation is borne out in
figure~\ref{fig:time} where we show this rescaled collision time, determined
numerically as the first appearance of a common apparent horizon, as a function
of $Q/M$.

\subsection{Waveforms: infall, merger and ringdown}

\begin{figure}[tbhp]
\centering
\includegraphics[width=0.45\textwidth]{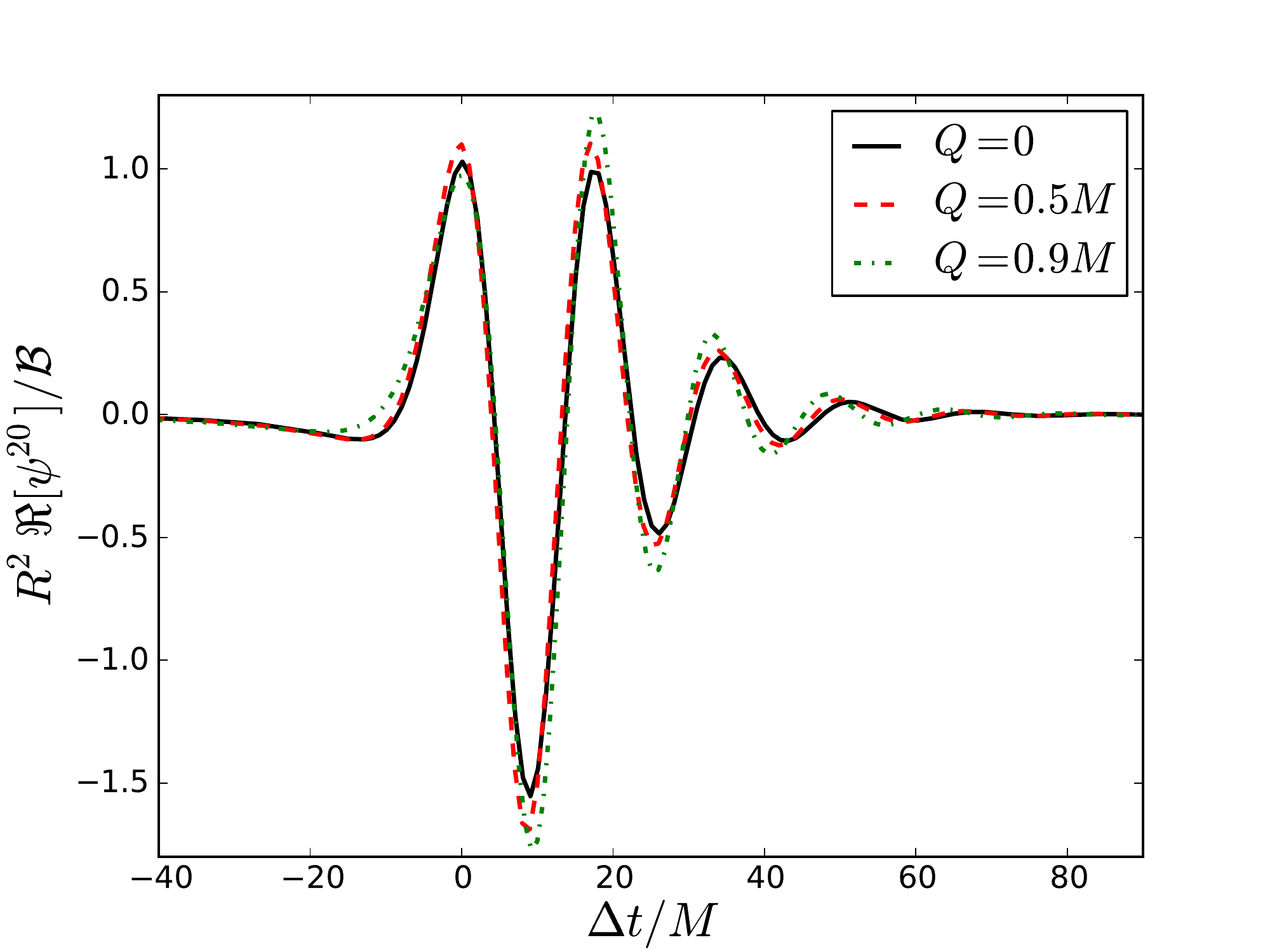}
\includegraphics[width=0.45\textwidth]{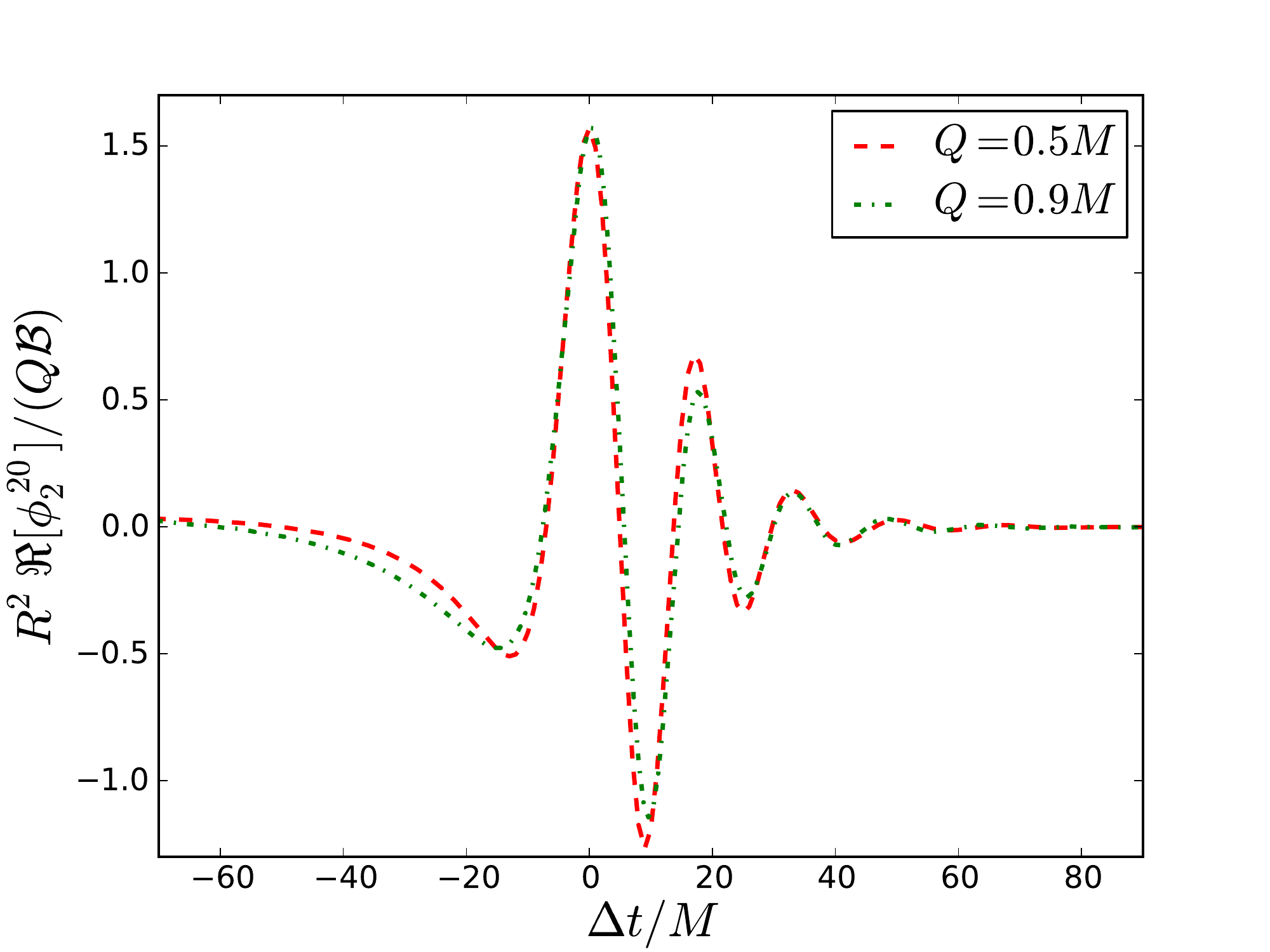}
\caption[Real part of the $(2,0)$ mode of $\Psi_4$ and $\Phi_2$]{Real part of
  the $(2,0)$ mode of $\Psi_4$ (left) and $\Phi_2$ (right panel) extracted at
  $R_{\mathrm{ex}} = 100M$.
  \label{fig:waveforms} }
\end{figure}

The dynamical behaviour of all our simulations is qualitatively well represented
by the waveforms shown in figure~\ref{fig:waveforms} for simulations d16q00,
d16q05 and d16q09.  The panels show the real part of the gravitational (left)
and electromagnetic (right) quadrupole extracted at $R_{\rm ex}=100~M$ as a
function of time with $\Delta t=0$ defined as the time of the global maximum of
the waveform.  From the classical analysis~\eqref{eq:power}, we expect the
waveforms $\Psi_4,\,\Phi_2$ to scale roughly with ${\cal B}$ and the mass or
charge of the black holes (the scaling with ${\cal B}$ is non-trivial, but both
an analytic estimate and the numerical results indicate the scaling is
approximately linear, which we shall therefore use for re-scaling the plots in
the figure).
%

The early stage of the signals are marked by the spurious radiation due to the
construction of initial data which we ignore in our analysis. Following a
relatively weak phase of wave emission during the infall of the holes, the
radiation increases strongly during the black-hole merger around $\Delta t=0$ in
the figure and decays exponentially as the final hole rings down into a
stationary state. This overall structure of the signals is rather similar for
the electromagnetic and the gravitational part and follows the main pattern
observed for gravitational-wave emission in head-on collisions of uncharged
black holes~\cite{Witek:2010xi,Witek:2010az}.

\begin{figure}[tbhp]
\centering
\includegraphics[width=0.7\textwidth]{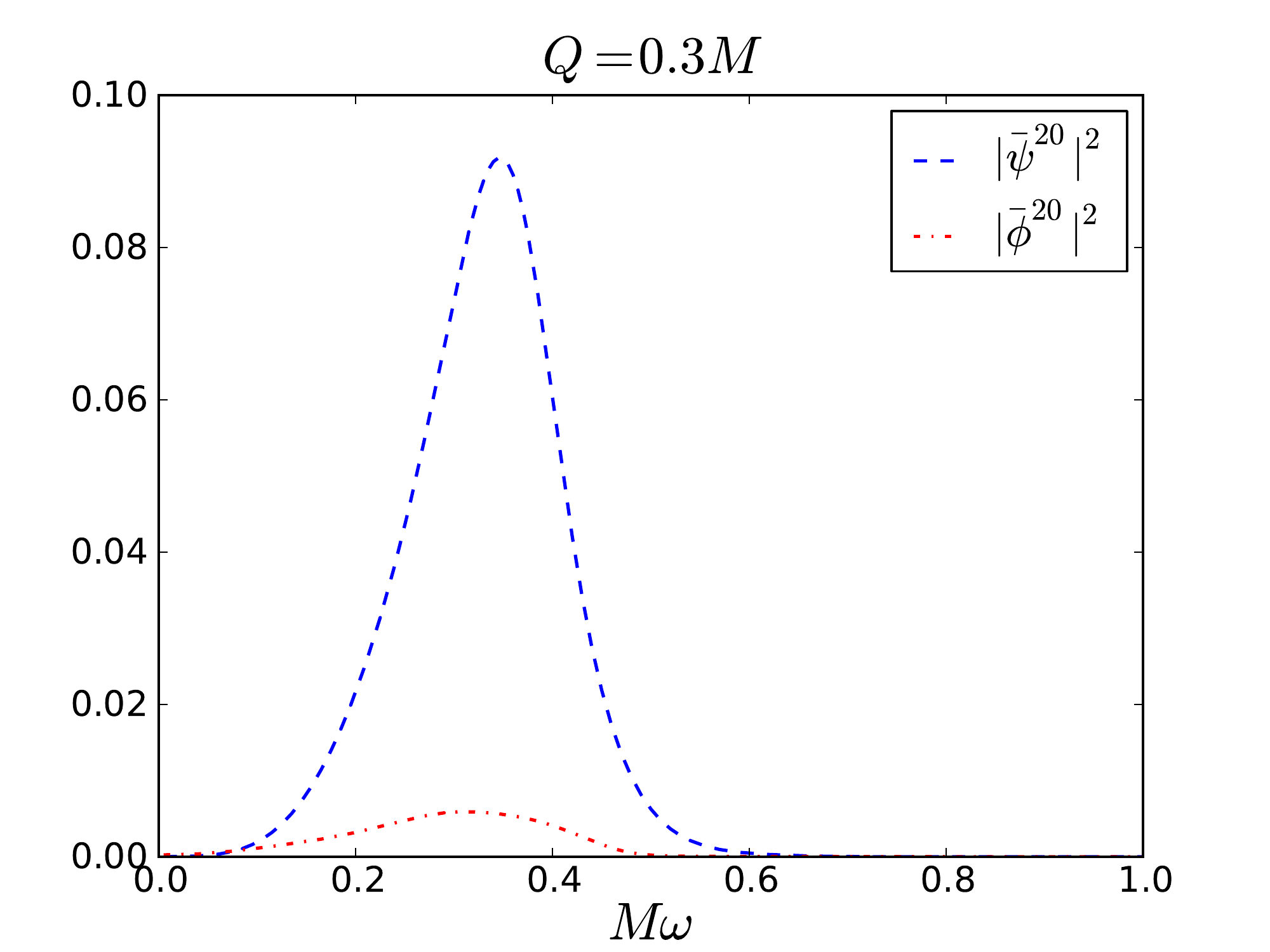}
\caption[Power spectrum for the gravitational and electromagnetic
quadrupole]{Power spectrum for the gravitational (long dashed) and
  electromagnetic (short dashed) quadrupole extracted from simulation~d08q03.
  Note that the spectrum peaks near the fundamental ringdown frequency of the
  gravitational mode; cf.~table~\ref{tab:ringdown}.}
\label{fig:fourier}
\end{figure}

\begin{table}[tbh]
  \centering
  \caption[Comparison of the ringdown frequencies]{Comparison of the ringdown
    frequencies obtained from (i) perturbative calculations~\cite{Berti:2009kk}
    and (ii) fitting a two-mode profile to the numerically extracted waveforms.
    For $Q/M=0$ the electromagnetic modes are not excited. For values of $Q/M
    \ge 0.9$ the electromagnetic mode becomes so weak that we can no longer
    unambiguously identify it in the numerical data.}
  \begin{tabular*}{0.45\textwidth}{@{\extracolsep{\fill}}ccc}
    \hline
    \hline
    $Q/M$ & $\omega_{1,2}^{\rm QNM}$ & $\omega_{1,2}^{\rm ext}$  \\
    \hline
    0       & $0.374 -0.0890i$    & $0.374 - 0.088 i $ \\
            & $0.458 -0.0950i$    &                    \\
    0.3     & $0.376 -0.0892i$    & $0.375-0.092i$     \\
            & $0.470 -0.0958i$    & $0.481-0.100i$     \\
    0.5     & $0.382 -0.0896i$    & $0.381-0.091i$     \\
            & $0.494 -0.0972i$    & $0.511-0.096i$     \\
    0.9     & $0.382 -0.0896i$    & $0.381-0.091i$     \\
            & $0.494 -0.0972i$    & ?  \\
    \hline
    \hline
  \end{tabular*}
  \label{tab:ringdown}
\end{table}
The final, exponentially damped ringdown phase is well described by perturbation
techniques~\cite{Berti:2009kk}. In particular, charged black holes are expected
to oscillate with two different types of modes, one of gravitational and one of
electromagnetic origin.  For the case of vanishing charge, the electromagnetic
modes are not present, but they generally couple for charged black holes, and we
expect both modes to be present in the spectra of our gravitational and
electromagnetic waveforms.  For verification we have fitted the late-stages of
the waveforms to a two-mode, exponentially damped sinusoid waveform
\begin{equation}
  f(t) = A_1 e^{-i\omega_1 t} + A_2 e^{-i\omega_2 t},
\end{equation}
where $A_i$ are real-valued amplitudes and $\omega_i$ complex frequencies. The
results are summarised in table~\ref{tab:ringdown} for selected values of the
charge-to-mass ratio of the post-merger black hole. Real and imaginary part of
the fitted frequencies agree within a few percent or better with the
perturbative predictions. For the large value $Q/M$, however, the wave signal is
very weak and in such good agreement with a single ringdown mode (the
gravitational one) that we cannot clearly identify a second, electromagnetic
component.  This feature is explained once we understand how the total radiated
energy is distributed between the gravitational and the electromagnetic
channels. For this purpose, we plot in figure~\ref{fig:fourier} the Fourier
spectrum of the relevant wavefunctions or, more precisely, their dominant
quadrupole contributions obtained for simulation~d08q03
$|\bar{\phi}^{20}|^2,|\bar{\psi}^{20}|^2$, where for any function $f$
\begin{equation}
  \bar{f}(\omega)=\int_{-\infty}^\infty e^{i\omega t}f(t)dt\,.
\end{equation}
It is clear from the figure that most of the energy is carried in the
fundamental gravitational-wave like mode with a peak at approximately
$\omega\sim 0.37$, close to the oscillation frequency of the fundamental
gravitational ringdown mode; see table~\ref{tab:ringdown}.

\subsection{Radiated energy and fluxes}
\label{sec:fluxes}

\begin{figure}[tbhp]
\centering
\includegraphics[width=0.75\textwidth]{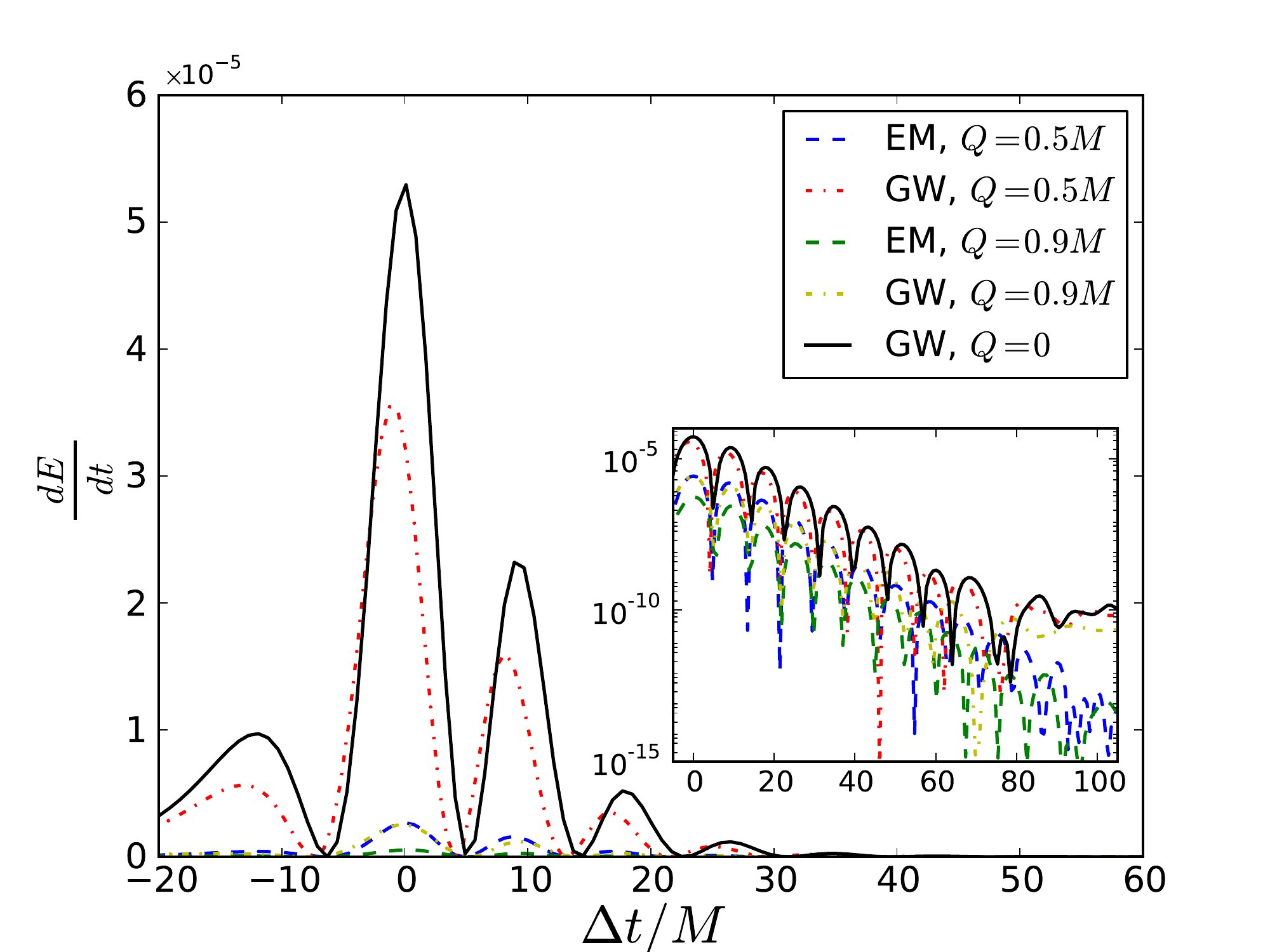}
\caption[Radiated fluxes]{Radiated fluxes for simulations d08q05, d08q09 and d08q00
  of table~\ref{tab:runs}. We have aligned the curves in time such
  that their global maximum coincides with $t=0$. The inset shows the exact same plot with the $y$-axis in logarithmic units.
  \label{fig:Flux} }
\end{figure}
The electromagnetic and gravitational wave fluxes are given by
equations~(\ref{eq:GW-flux2}) and (\ref{eq:EM-flux2}). We have already noticed
from the waveforms in figure~\ref{fig:waveforms} that the electromagnetic signal
follows a pattern quite similar to the gravitational one. The same holds for the
energy flux which is shown in figure~\ref{fig:Flux} for a subset of our
simulations with $Q/M=0$, $0.5$ and $0.9$.  From the figure, as well as the
numbers in table~\ref{tab:runs}, we observe that the energy carried by
gravitational radiation decreases with increasing $Q/M$, as the acceleration
becomes smaller and quadrupole emission is suppressed, in agreement with
prediction~\eqref{GWquadprediction}.

\begin{figure}[tbhp]
\centering
\includegraphics[width=0.75\textwidth]{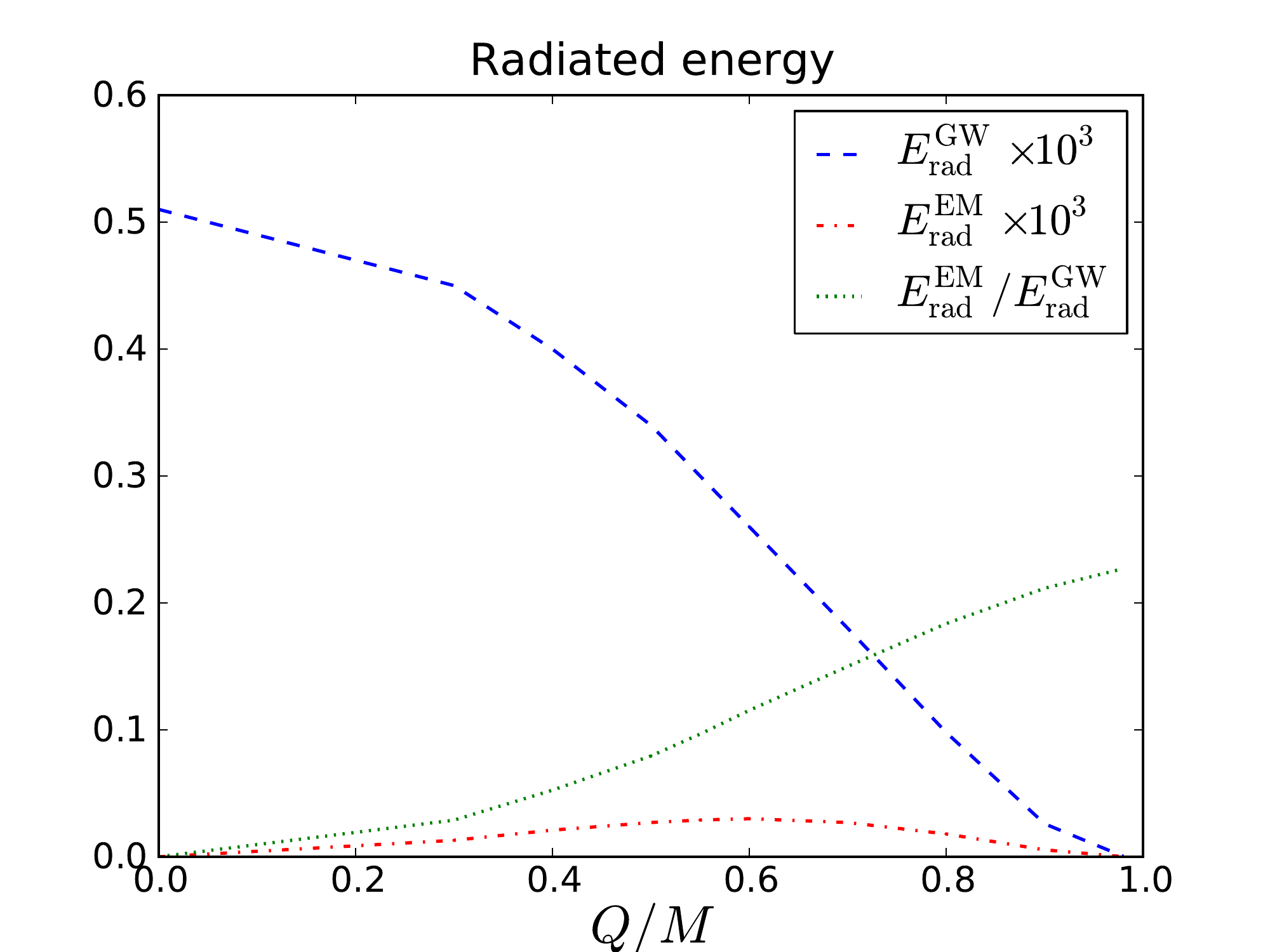}
\caption[Energy radiated in the gravitational and electromagnetic
  quadrupole]{Energy radiated in the gravitational and electromagnetic
  quadrupole as well as the ratio of the two as a function of $Q/M$.
  \label{fig:Energy_QM}}
\end{figure}

This is further illustrated in figure~\ref{fig:Energy_QM}, which illustrates the
radiated energy carried in the gravitational quadrupole and the electromagnetic
quadrupole as well as their ratio as functions of the charge-to-mass ratio
$Q/M$.
For the case of vanishing charge, the total radiated energy is already known
from the literature; e.g.~\cite{Witek:2010xi}.  The value increases mildly with
the initial separation as a consequence of the slightly larger collision
velocity but is generally found to be close to $E_{\rm rad}^{\rm
  GW}/M=0.055\%$. Our values of $0.051\%$ for $d/M\simeq 8$ and $0.055\%$ for
$d/M \simeq 16$ are in good agreement with the literature. As we increase $Q/M$,
however, $E_{\rm rad}^{\rm GW}$ decreases significantly and for $Q/M=0.9$
($0.98$) has dropped by a factor of about $20$ ($10^3$) relative to the
uncharged case. For practical reasons, we have explored the largest ratio
$Q/M=0.98$ for the smaller initial separation $d/M\simeq 8$ only; the near
cancellation of the gravitational and electromagnetic interaction and the
resulting slow-down of the collision lead to a very long infall stage with
essentially zero dynamics.

In contrast to the monotonically decreasing gravitational-wave energy, the
electromagnetic signal reaches a local maximum around $Q/M=0.6$, an expected
observation as the electromagnetic radiation necessarily vanishes for $Q/M=0$
(no charge) and $Q/M=1$ (no acceleration) but takes on non-zero values in the
regime in between.  Closer analysis of our classical, flat-space calculation
\eqref{EMquadprediction} predicts a maximum electromagnetic radiation output at
\begin{equation}
  Q_{\rm max}=\sqrt{\frac{\sqrt{329}-13}{14}}\,M\approx 0.605 M\,,
\end{equation}
in excellent agreement with the results of our simulations.

We finally consider the ratio of electromagnetic to gravitational wave energy
(dotted curve in figure~\ref{fig:Energy_QM}).  As predicted by our analytic
calculation (\ref{prediction_ratio}), this ratio increases monotonically with
$Q/M$ for fixed separation $d$.  A fit of our numerical results yields $E_{\rm
  rad}^{\rm EM}/E_{\rm rad}^{\rm GW}=0.27~Q^2/M^2$ and for our largest value
$Q/M=0.98$, we obtain a ratio of $0.227$ to be compared with $\sim 0.24$ as
predicted by equation~(\ref{prediction_ratio}). Bearing in mind the simplicity of our
analytic model in section~\ref{classical_expectations}, the quantitative
agreement is remarkable.

\section{Conclusions}
\label{sec:conclusion}

In this chapter, we performed a numerical study of collisions of charged black
holes with equal mass and charge in the framework of the fully non-linear
Einstein-Maxwell equations. Our first observation is that the numerical
relativity techniques (formulation of the evolution equations, gauge conditions
and initial data construction) developed for electrically neutral black hole
binaries can be straightforwardly extended to successfully model charged
binaries even for nearly extremal charge-to-mass ratios $Q/M \lesssim 1$.  In
particular, we notice the contrast with the case of rotating black holes with
nearly extremal spin which represents a more delicate task for state-of-the-art
numerical relativity; cf.~references~\cite{Lovelace:2011nu,Lousto:2012es} for the
latest developments on this front.  This absence of difficulties for charged
holes is not entirely unexpected.  Considering the construction of initial data,
for instance, an important difference arises in the customary choice of
conformally flat Bowen-York initial data~\cite{Bowen:1980yu} which greatly
simplifies the initial data problem. While the Kerr solution for a single
rotating black hole does not admit conformally flat slices~\cite{Garat:2000pn}
and therefore inevitably results in spurious radiation, especially for large
spin parameters, this difficulty does not arise for charged, but non-rotating
black holes; cf.~equation~(\ref{eq:conformalfactor}) and~\cite{Graves:1960zz}.

The excellent agreement between the classical calculation for the energy
emission and the numerical results reported here, allow for an investigation of
cosmic censorship close to extremality. If we take two black holes with $M_1=M_2=M/2$,
$Q_1=Q_2=(M-\delta)/2$
and we let them fall from infinity, to first order in $\delta$ we get
\begin{equation}
\begin{aligned}
Q_{\rm tot}& = M - \delta \\
M_{\rm tot}& = M - E_{\rm rad}
\end{aligned} \,.
\end{equation}
Now, the classical result~\eqref{GWquadprediction} implies that the dominant
term for the radiated energy is $E_{\rm rad} \sim \mathcal{B}^{5/2} M \sim
(\delta/M)^{5/2} M$.
Thus we get
\begin{equation}
\frac{Q_{\rm tot}}{M_{\rm tot}} \simeq 1 
                                  - \frac{\delta}{M} 
                                  + k \left(\frac{\delta}{M}\right)^{5/2}\,,
\end{equation}
where $k$ is a constant. We conclude that cosmic censorship is preserved for
charged collisions of nearly extremal holes ($\delta \ll M$), on account of the
much longer collision time, which yields much lower velocities and therefore
much lower energy output. The differences between the cases of spinning mergers
and charged collisions are interesting. In the former case, naked singularities
are avoided by radiation carrying away more angular momentum (via orbital
hangup~\cite{Campanelli:2006uy}).  In the latter case, our results suggest that
naked singularities are avoided by the smaller radiation emission, due to the
smaller accelerations involved in the infall.

We have here evolved a sequence of binaries, with equal charge-to-mass ratio starting
from rest, with $Q/M$ varying from zero to values close to extremality. Starting
with the electrically neutral case, where our gravitational wave emission
$E_{\rm rad}^{\rm GW}/M=0.055\%$ agrees well with the literature, we observe a
monotonic decrease of the emitted gravitational wave energy as we increase
$Q/M$. For our largest value $Q/M=0.98$, $E_{\rm rad}^{\rm GW}$ is reduced by
about three orders of magnitude, as the near cancellation of the gravitational
and electromagnetic forces substantially slows down the collision. In contrast,
the radiated electromagnetic energy reaches a maximum near $Q/M=0.6$ but always
remains significantly below its gravitational counterpart.  Indeed, the ratio
$E_{\rm rad}^{\rm EM}/E_{\rm rad}^{\rm GW}$ increases monotonically with $Q/M$
and approaches about $25\%$ in the limit $Q/M \rightarrow 1$. We find all these
results to be in remarkably good qualitative {\em and} quantitative agreement
with analytic approximations obtained in the framework of the dynamics of two
point charges in a Minkowski background. This approximation also predicts that
the collision time relative to that of the uncharged case scales $\sim
\sqrt{1-Q^2/M^2}$ which is confirmed within a few percent by our numerical
simulations.



\chapter{Final remarks}
\label{ch:final}

\epigraph{This probably just goes to show something, but I sure don't know
  what.}{Calvin\\ \textit{Calvin \& Hobbes}}

Numerical relativity is a fantastic tool to study and explore spacetimes whose exact form is not known.

After decades of efforts, the first stable, long-term evolutions of the orbit and merger of two black holes were finally accomplished in 2005, and since then considerable progress has been made.
This field has now reached a state of maturity, and several codes and tools exist that allow one to perform evolutions of black holes---with quite generic initial configurations---in standard four dimensional vacuum gravity.

In addition to the original (main) motivation coming from the two-body problem,
it was quickly realised that numerical relativity could be helpful for a much broader range of scenarios, with some motivation coming from fields other than gravity itself.

In this work we have thus worked to extend numerical relativity tools to new frontiers, opening a range of uncharted territory in black hole physics to be explored with contemporary numerical relativity.
In particular, we have presented the following: 
\begin{enumerate}[(i)]

\item a dimensional reduction procedure that allows the use of existing $3+1$
  numerical codes to evolve 
  higher-dimensional spacetimes with enough symmetry, including head-on
  collisions in $D \ge 5$ and black hole collisions with impact
  parameter and spin in $D \ge 6$;

\item a generalisation of the \textsc{TwoPunctures} spectral solver, allowing
  for the computation of initial data for a boosted head-on collision of black
  hole binaries in higher-dimensional spacetimes;

\item a wave extraction procedure that allows the extraction of gravitational
  radiation observables from numerical evolutions of head-on collisions of
  black holes in $D$~dimensions;

\item with the above tools, numerical simulations of black hole collisions from
  rest in five-dimensional spacetimes were successfully evolved, the
  corresponding wave forms were obtained and total energy released in the form of
  gravitational waves was computed;

\item evolutions of black holes in non-asymptotically flat spacetimes, including
  asymptotically de Sitter spacetimes, ``boxed'' spacetimes with mirror-like
  boundary conditions, and five-dimensional cylindrical spacetimes;

\item numerical evolutions of collisions of charged black holes with equal mass
  and
  charge
  , and a calculation of the energy released via emission of gravitational and
  electromagnetic radiation.

\end{enumerate}

Several open questions and research avenues remain to be explored, and we thus close with a list of natural sequels for this program:
\begin{itemize}

\item A systematic investigation of black hole collisions and dynamics in
  generic dimension. Even though the formalism here presented is valid in
  arbitrary dimension, the long-term numerical stability of the implementation
  is a different matter altogether. Currently, only the five-dimensional
  case seems to be relatively robust, with numerical instabilities occurring in
  all $D>5$ cases tried so far. It is possible that such instabilities may be
  cured with a suitable choice of gauge conditions. These issues remain under
  investigation.

\item Related to the previous point, it could be of interest to systematically
  investigate the merits and disadvantages, from the point of view of the
  numerical implementation, of dimensional reduction procedures (such as the one
  here presented) versus evolution schemes that make use of the Cartoon
  method.

\item The numbers here reported for the total energy loss for the
  five-dimensional black hole head-on collisions refer to collisions from
  rest. For the applications described in the Introduction, however, high
  velocity collisions are the most relevant ones. Such cases do not seem to be
  as robust as the analogous four dimensional systems, with numerical
  instabilities appearing when large boost parameters are
  considered. Investigation on this front is still under way.

\item For the Einstein-Maxwell study, a natural step is considering more generic
  types of initial data, in order to tackle some of the issues discussed in the
  Introduction. A non-zero boost, for instance, will allow us to study both
  binary black hole systems that will coalesce into a Kerr-Newman black hole and
  the impact of electric charge on the dynamics and wave emission
  (electromagnetic and gravitational) in high energy collisions. A further
  interesting extension is the case of oppositely charged black holes.

\end{itemize}



\appendix

\chapter{List of publications}

This thesis summarises work done in the following publications:
\begin{enumerate}

\item
{\bf ``Numerical relativity for D dimensional axially symmetric space-times: formalism and code tests''}
  \\{}M.~Zilh\~ao, H.~Witek, U.~Sperhake, V.~Cardoso, L.~Gualtieri, C.~Herdeiro and A.~Nerozzi.
  \\{}\href{http://arxiv.org/abs/1001.2302}{\tt arXiv:1001.2302 [gr-qc]}
  \\{}\href{http://dx.doi.org/10.1103/PhysRevD.81.084052}{Phys.\ Rev.\ D {\bf 81}, 084052 (2010)} 

\item
{\bf ``Black holes in a box: towards the numerical evolution of black holes in AdS''}
  \\{}H.~Witek, V.~Cardoso, C.~Herdeiro, A.~Nerozzi, U.~Sperhake and M.~Zilh\~ao.
  \\{}\href{http://arxiv.org/abs/1004.4633}{\tt arXiv:1004.4633 [hep-th]}
\\{}\href{http://dx.doi.org/10.1103/PhysRevD.82.104037}{Phys.\ Rev.\ D {\bf 82}, 104037 (2010)} 

\item
{\bf ``Numerical relativity for D dimensional space-times: head-on collisions of black holes and gravitational wave extraction''}
  \\{}H.~Witek, M.~Zilh\~ao, L.~Gualtieri, V.~Cardoso, C.~Herdeiro, A.~Nerozzi and U.~Sperhake.
  \\\href{http://arxiv.org/abs/1006.3081}{\tt arXiv:1006.3081 [gr-qc]}
\\\href{http://dx.doi.org/10.1103/PhysRevD.82.104014}{Phys.\ Rev.\ D {\bf 82}, 104014 (2010)} 

\item
{\bf ``Head-on collisions of unequal mass black holes in D=5 dimensions''}
  \\{}H.~Witek, V.~Cardoso, L.~Gualtieri, C.~Herdeiro, U.~Sperhake and M.~Zilh\~ao.
  \\\href{http://arxiv.org/abs/1011.0742}{\tt arXiv:1011.0742 [gr-qc]}
\\\href{http://dx.doi.org/10.1103/PhysRevD.83.044017}{Phys.\ Rev.\ D {\bf 83}, 044017 (2011)} 

\item
\label{paper:init-data}
{\bf ``Higher-dimensional puncture initial data''}
  \\{}M.~Zilh\~ao, M.~Ansorg, V.~Cardoso, L.~Gualtieri, C.~Herdeiro, U.~Sperhake and H.~Witek.
  \\\href{http://arxiv.org/abs/1109.2149}{\tt arXiv:1109.2149 [gr-qc]}
\\\href{http://dx.doi.org/10.1103/PhysRevD.84.084039}{Phys.\ Rev.\ D {\bf 84}, 084039 (2011)} 

\item
\label{paper:cylinders}
{\bf ``Simulations of black holes in compactified spacetimes''}
  \\{}M.~Zilh\~ao, V.~Cardoso, L.~Gualtieri, C.~Herdeiro, A.~Nerozzi, U.~Sperhake and H.~Witek.
\\\href{http://dx.doi.org/10.1088/1742-6596/314/1/012103}{J.\ Phys.\ Conf.\ Ser.\  {\bf 314}, 012103 (2011)}. 

\item
\label{paper:dS}
{\bf ``Dynamics of black holes in de Sitter spacetimes''}
  \\{}M.~Zilh\~ao, V.~Cardoso, L.~Gualtieri, C.~Herdeiro, U.~Sperhake and H.~Witek.
  \\\href{http://arxiv.org/abs/1204.2019}{\tt arXiv:1204.2019 [gr-qc]}
\\\href{http://dx.doi.org/10.1103/PhysRevD.85.104039}{Phys.\ Rev.\ D {\bf 85}, 104039 (2012)} 

\item
\label{paper:charged}
{\bf ``Collisions of charged black holes''}
  \\{}M.~Zilhao, V.~Cardoso, C.~Herdeiro, L.~Lehner and U.~Sperhake.
  \\\href{http://arxiv.org/abs/1205.1063}{\tt arXiv:1205.1063 [gr-qc]}
\\\href{http://dx.doi.org/10.1103/PhysRevD.85.124062}{Phys.\ Rev.\ D {\bf 85}, 124062 (2012)} 

\end{enumerate}
The numerical work presented in papers \ref{paper:init-data}, \ref{paper:cylinders}, \ref{paper:dS} and \ref{paper:charged} was performed by the author of this thesis.

Further publications by the author:
\begin{enumerate}

\item
{\bf ``A Double Myers-Perry Black Hole in Five Dimensions''}
  \\{}C.~A.~R.~Herdeiro, C.~Rebelo, M.~Zilh\~ao and M.~S.~Costa.
  \\\href{http://arxiv.org/abs/arXiv:0805.1206}{\tt arXiv:0805.1206 [hep-th]}
\\\href{http://iopscience.iop.org/1126-6708/2008/07/009/}{JHEP {\bf 0807}, 009 (2008)} 

\item
{\bf ``Mass inflation in a D dimensional Reissner-Nordstrom black hole: a hierarchy of particle accelerators ?''}
  \\{}P.~P.~Avelino, A.~J.~S.~Hamilton, C.~A.~R.~Herdeiro and M.~Zilh\~ao.
  \\\href{http://arxiv.org/abs/arXiv:1105.4434}{\tt arXiv:1105.4434 [gr-qc]}
\\\href{http://prd.aps.org/abstract/PRD/v84/i2/e024019}{Phys.\ Rev.\ D {\bf 84}, 024019 (2011)} 

\item
{\bf ``Mathisson's helical motions for a spinning particle: Are they unphysical?''}
  \\{}L.~F.~O.~Costa, C.~A.~R.~Herdeiro, J.~Natario and M.~Zilh\~ao.
  \\\href{http://arxiv.org/abs/arXiv:1109.1019}{\tt arXiv:1109.1019 [gr-qc]}
\\\href{http://prd.aps.org/abstract/PRD/v85/i2/e024001}{Phys.\ Rev.\ D {\bf 85}, 024001 (2012)} 

\item
{\bf ``NR/HEP: roadmap for the future''}
  \\{}V.~Cardoso, L.~Gualtieri, C.~Herdeiro, U.~Sperhake, {\it et al.}.
  \\\href{http://arxiv.org/abs/arXiv:1201.5118}{\tt arXiv:1201.5118 [hep-th]}

\end{enumerate}



\cleardoublepage
\phantomsection
\addcontentsline{toc}{chapter}{Bibliography}
\bibliographystyle{myutphys}
\bibliography{biblio}

\end{document}